\documentclass[12pt]{article}
\begin{document}

%ZZZZZZZZZ  Lemma \rref{AA27} exists term $\sigma $, $\sigma(S,i,t)=a_{i}$ %ZZZZZZZ

%ZZZZZZZLemma \rref{MP0} $T=\sum_{i=1}^{k}r^{i}$ kifejezheto egy "term"-mel
%ZZZZZ

%ZZZZZZZZ Lemma \rref{J5.1} ez az a lemma mely szerint minden function symbol
%kifejezheto egy "term"-mel egy nagyobb $\boldm_{d}$-ben ZZZZZZZZZZZ

%ZZZZZZZ Lemma \rref{K70}  componentwise termcomputation polynomially exist  %%% ZZZZZZZZZZ

%ZZZZZZZ  Lemma \rref{F1}  $e_{d,t}$ is a term ZZZZZ

%ZZZZZZ  Lemma \rref{B0} shitft can be defined by a term:  $   b=\shift_{d,t, i}(a) $ %% if $\delta =\bfegy$,  and $   b=\shift_{d,t, -i}(a) $ if $\delta =-\bfegy$. 

%%% ZZZZZZZZZZZ Lemma \rref{YB1} stretch is existential

%%ZZZZ Lemma \rref{K80} "for all $i\in 2^{v-u}$ $Q_{u,t}$" is polynomially
%%% existential if $Q_{u,t}$ is polynomially existential

%% ZZZZZZZ Lemma \rref{M6} $x\in \zo(d,t)$ is uniformly existential

%% ZZZZZZZZ Lemma  \rref{AA24} parallel operations apart from 
%%$\times, \div \bfp$ are uniformly existential

%%%%%%ZZZZZZZZ Lemma \rref{F1} $e_{d,t}$ is uniformly existential

%%%%%%ZZZZZZZZ Lemma \rref{AA73} $b[i.d]=a[\lambda(2^{u-d},i),d]$
%%%%%%%%% is polynomially existential

%%ZZZZZZ Lemma \rref{Y3}  the composition of uniformly existential families
%%% uniformly existential. 

%%ZZZZZZ Lemma \rref{AA30}  for all $ k\in 2^{d-t}$,
%%%$\calb(a_{0}[k,t],...,a_{m-1}[k,t])=c[k,t]$ is existential

%%%the composition of uniformly existential families is
%%% uniformly existential.

\tolerance=10000      %%%%%%%%%%

\def\kell{{}}
\newtheorem{theorem}{Theorem}
\newtheorem{lemma}{Lemma}
\newtheorem{claim}{Claim}
\newtheorem{corollary}[theorem]{Corollary}
\newtheorem{conjecture}{Conjecture}
\newtheorem{proposition}{Proposition}
\newtheorem{cond}{}
\renewcommand{\thecond}{\rm(\arabic{cond})}
\newenvironment{proof}{\par\noindent
 {\it Proof.\/}\enspace}{\hfill  $ blacksquare $   \medskip\par}
\newenvironment{definition}{\medskip\par\noindent
 {\bf Definition.\/}\enspace}{\hfill \par}
\newenvironment{remark}{\medskip\par\noindent
 {\bf Remark.\/}\enspace}{\hfill \par}
\def\kih{\if \count=1 \pageno=-10}

%%%\Bigl,\biggl,\Biggl
%\pagestyle{myheadings}
%\markboth{\today \hfil ${\rm \backslash
%cikk\backslash fsize  \backslash fsi} $}
%{\today \hfil  ${\rm
%\backslash cikk\backslash
%fsize \backslash fsi} $}

\def\llabel{\label}
\def\rref{\ref}
\def\nev{{}}
\def\xev#1{{}}
\def\vege{{$\sqcap\hskip-8pt\sqcup$}}%%docstyle12pt
\def\wege{\vege\end{cond}\vege\end{definition}{\vskip-15pt}}
\def\nemkell{{$\heartsuit\ $}}
\def\iev#1{#1}
\def\enp#1{{\sl Q.E.D.}{#1}}
\def\mev#1{{\bf #1}}
\def\semmi#1{}
\def\regi#1{}
\def\lapozz{{\vfill\eject}}
\def\err{{\bf R}}
\def\calq{{\cal Q}}
\def\cala{{\cal A}}
\def\calb{{\cal B}}
\def\cald{{\cal D}}
\def\calh{{\cal H}}
\def\cali{{\cal I}}
\def\calu{{\cal U}}
\def\calm{{\cal M}}
\def\caln{{\cal N}}
\def\calv{{\cal V}}
\def\calw{{\cal W}}
\def\cale{{\cal E}}
\def\calg{{\cal G}}
\def\calc{{\cal C}}
\def\calj{{\cal J}}
\def\calk{{\cal K}}
\def\calp{{\cal P}}
\def\calo{{\cal O}}
\def\calq{{\cal Q}}
\def\calr{{\cal R}}
\def\cals{{\cal S}}
\def\calf{{\cal F}}
\def\call{{\cal L}}
\def\calx{{\cal X}}
\def\calt{{\cal T}}
\def\calz{{\cal Z}}
\def\caly{{\cal Y}}
\def\bolda{{\bf A}}
\def\boldb{{\bf B}}
\def\boldc{{\bf C}}
\def\boldd{{\bf D}}
\def\bolde{{\bf E}}
\def\boldf{{\bf F}}
\def\boldg{{\bf G}}
\def\boldh{{\bf H}}
\def\boldp{{\bf P}}
\def\boldq{{\bf Q}}
\def\boldi{{\bf I}}
\def\boldj{{\bf J}}
\def\boldk{{\bf K}}
\def\boldl{{\bf L}}
\def\boldm{{\bf M}}
\def\boldn{{\bf N}}
\def\boldr{{\bf R}}
\def\bolds{{\bf S}}
\def\boldz{{\bf Z}}
\def\boldt{{\bf T}}
\def\boldu{{\bf U}}
\def\boldv{{\bf V}}
\def\boldw{{\bf W}}
\def\boldx{{\bf X}}
\def\boldy{{\bf Y}}
\def\bfa{{\bf a}}
\def\bfb{{\bf b}}
\def\bfc{{\bf c}}
\def\bfd{{\bf d}}
\def\bfe{{\bf e}}
\def\bff{{\bf f}}
\def\xbff{{\bf f}}
\def\bfg{{\bf g}}
\def\bfh{{\bf h}}
\def\bfk{{\bf k}}
\def\bfj{{\bf j}}
\def\bfl{{\bf l}}
\def\bfs{{\bf s}}
\def\bfp{{\bf p}}
\def\bfq{{\bf q}}
\def\bfr{{\bf r}}
\def\bft{{\bf t}}
\def\bfu{{\bf u}}
\def\bfv{{\bf v}}
\def\rcp#1{{\frac{1}{#1}}}
\def\godel#1{{{}^{\lceil}{#1}{}^{\rceil}}}
\def\phi{\varphi}
\def\theta{\vartheta}
\def\epsilon{\varepsilon}
\def\barxi{\bar{\xi}}
\def\barf{\bar{f}}
\def\bard{\bar{d}}
\def\barq{\bar{q}}
\def\baru{\bar{u}}
\def\bars{\bar{s}}
\def\bcalf{\bar{\calf}}
\def\bcalg{\bar{\calg}}
\def\bupsilon{\bar{\Upsilon}}

\def\Sup{{\tt Sup}}
\def\cl{{\tt cl}}
\def\length{{\tt length}}
\def\flength{{\tt flength}}
\def\state{{\tt state}}
\def\range{{\tt range}}
\def\domain{{\tt domain}}
\def\func{{\tt func}}
\def\dir{{\tt dir}}
\def\int{{\tt int}}
\def\Int{{\tt Int}}
\def\padlo#1{{\lfloor #1\rfloor}}
\def\seq{{\tt seq}}
\def\sp{{\tt sp}}
\def\dist{{\tt dist}}
\def\store{{\tt store}}
\def\standard{{\tt standard}}
\def\prob{{\tt prob}}
\def\mod{{\rm mod}}
\def\base{{\tt base}}
\def\rank{{\tt rank}}
\def\pow{{\tt pow}}
\def\universe{{\tt universe}}
\def\univ{{\universe}}
\def\sseq#1#2{{\seq(#1,\lbrace 0,1\rbrace^{#2})}}
\def\ppp#1#2{{#1}_{0},\ldots,{#1}_{#2-1}}
\def\pprob{{\prob_{p,A}}}
\def\pprrob{{\prob_{p,\lbrace 1,\ldots ,n\rbrace }}}
\def\ppprob{{\prob_{p,A}\times \prob_{p,A}}}
\def\bcks{{\backslash}}
\def\blambda{{\bar{\lambda}}}
\def\bdelta{{\bar{\delta}}}
\def\bgamma{{\bar{\gamma}}}
\def\bnu{{\bar{\nu}}}
\def\brho{{\bar{\rho}}}
\def\bphi{{\bar{\phi}}}
\def\btau{{\bar{\tau}}}
\def\biota{{\bar{\iota}}}
\def\bsigma{{\bar{\sigma}}}
\def\balpha{{\bar \alpha}}
\def\tsigma{{\tilde{\sigma}}}
\def\trho{{\tilde{\rho}}}
\def\lattice{{\tt lattice}}
\def\ttg{{\tt g}}
\def\ttv{{\tt v}}
\def\ttc{{\tt c}}
\def\abstl{{\tt abstl}}
\def\allattice{{\tt allattice}}
\def\basis{{\tt basis}}
\def\vol{{\tt vol}}
\def\repr{{\tt repr}}
\def\inp{{\tt input}}
\def\outp{{\tt output}}
\def\set{{\tt set}}
\def\ind{{\tt ind}}
\def\prm{{\tt prm}}
\def\Skolem{{\tt Skolem}}
\def\bfn{{\bf n}}
\def\bfo{{\bf o}}
\def\bfi{{\bf i}}
\def\bolm{{\boldl_{\omega}(M)}}
\def\bbolm{{\bar\boldl_{\omega}(M)}}
\def\var{{\tt var}}
\def\fwrel{{\tt fwrel}}
\def\fwfunc{{\tt fwfunc}}
\def\rel{{\tt rel}}
\def\frel{{\tt frel}}
\def\height{{\tt height}}
\def\fin{{\tt fin}}
\def\sink{{\tt sink}}
\def\graph{{\tt graph}}
\def\depth{{\tt depth}}
\def\rfsymb{{\tt rfsymb}}
\def\theory{{\tt theory}}
\def\skolem{{\tt skolem}}
\def\subst{{\tt subst}}
\def\constituent{{\tt constituent}}
\def\diag{{\tt diag}}
\def\Diag{{\tt Diag}}
\def\Def{{\tt Def}}
\def\bv{{\tt bv}}
\def\tree{{\tt tree}}
\def\free{{\tt free}}
\def\bound{{\tt bound}}
\def\var{{\tt var}}
\def\nonvar{{\tt nonvar}}
\def\fv{{\tt fv}}
\def\fv{{\tt fv}}
\def\elka{{\boldl\boldk}}
\def\order{{\tt order}}
\def\symb{{\tt symb}}
\def\size{{\tt size}}
\def\full{{\tt full}}
\def\cont{{\tt cont}}
\def\weight{{\tt weight}}
\def\unif{{\tt unif}}
\def\spl{{\tt spl}}
\def\dspl{{\tt dspl}}
\def\tsum{{\bolds}}
\def\rsum{{\bolds^{-1}}}
\def\gate{{\tt gate}}
\def\inn{{\tt inn}}
\def\node{{\tt node}}
\def\branch{{\tt branch}}
\def\snoop{{\tt snoop}}
\def\avoid{{\tt avoid}}
\def\powerset{{\tt powerset}}
\def\invisible{{\tt invisible}}
\def\bcmt{{\boldc_{m}^{(\times)}}}
\def\snode{{\tt snode}}
\def\brhom{{C_{\brho,m}}}
\def\rand{{\tt rand}}
\def\det{{\tt detin}}
\def\bdet{{\tt b.detin}}
\def\bdetin{{\tt b.detin}}
\def\leak{{\tt leak}}
\def\lathatatlan{{\overline{\invisible}}}
\def\parity{{\tt parity}}
\def\extension{{\tt extension}}
\def\legyen{{:=\ }}
\def\image{{\tt image}}
\def\comp{{\tt comp}}
\def\rv{{\tt rv}}
\def\inset{{\tt inset}}
\def\outset{{\tt outset}}
\def\boutset{{\tt b.outset}}
\def\exit{{\tt exit}}
\def\entrance{{\tt entrance}}
\def\bigcomp{{\tt bigcomp}}
\def\out{{\tt out}}
\def\inn{{\tt in}}
\def\io{{\tt io}}
\def\cq{{C^{(Q)}}}
\def\block{{\tt block}}
\def\leaky{{\tt leaky}}
\def\lin{{\tt lin}}
\def\blin{{\overline{\tt lin}}}
\def\poly{{\tt poly}}
\def\subspace{{\tt subspace}}
\def\shadow{{\tt shadow}}
\def\filter{{\tt filter}}
\def\bfilter{{\overline{\tt filter}}}
\def\wspl{{\tt wspl}}
\def\pre{{\tt pre}}
\def\post{{\tt post}}
\def\perc{{\tt perc}}
\def\history{{\tt history}}
\def\cell{{\tt cell}}
\def\mem{{\tt space}}
\def\conceded{{\tt conceded}}
\def\visible{{\tt visible}}
\def\ido{{\tt time}}
\def\fseq{{\lbrace 0,1\rbrace }^{<\infty}}
\def\infseq{{\lbrace 0,1\rbrace }^{\infty}}
\def\bias{{\tt bias}}
\def\even{{\tt even}}
\def\odd{{\tt odd}}
\def\calmq{{\calm_{q}}}
\def\distance{{\tt distance}}
\def\kesz{{\tt end}}
\def\ext{{\tt ext}}
\def\kond{{\tt cond}}
\def\DET{{\tt DET}}
\def\Form{{\tt Form}}
\def\bidiv {{\tt bidiv}}
\def\bfnull {{\bf 0}}
\def\bfegy {{\bf 1}}
\def\coeff {{\tt coeff}}
\def\res {{\tt res}}
\def\sfp {{\tt sfp}}
\def\SFP {{\tt SFP}}
\def\num {{\tt num}}
\def\bks{{\boldk=\langle \boldk_{d} \mid
d\in \omega\rangle }}
\def\iter {{\tt iter}}
\def\alt {{\tt alt}}
\def\integer{{\tt integer}}
\def\relation{{\tt relation}}
\def\Seq{{\tt Seq}}
\def\mult{{\tt mult}}
\def\ltort{{\lbrack\!\lbrack}}
\def\rtort{{\rbrack\!\rbrack}}
\def\lfrac{{\langle\!\langle}}
\def\rfrac{{\rangle\!\rangle}}
\def\matrix{{\tt matrix}}
\def\column{{\tt column}}
\def\aalpha{{\alpha}}
\def\shift{{\tt shift}}
\def\tg{{\tilde g}}
\def\pos{{\tt pos}}
\def\zo{{\tt zo}}
\def\minus{{\tt minus}}
\def\perf{{\tt perf}}
\def\prop{{\tt prop}}
\def\struc{{\tt struc}}
\def\rlt{{\tt rlt}}
\def\tcalq{{\tilde \calq}}
\def\interp{{\tt interp}}
\def\language{{\tt language}}
\def\width{{\tt width}}
\def\length{{\tt length}}
\def\transition{{\tt transition}}
\def\ttime{{\tt ttime}}
\def\oracle{{\tt oracle}}
\def\conn{{\tt conn}}
\def\Cont{{\tt Cont}}
\def\prmt{{\tt prmt}}
\def\nat{{\tt nat}}
\def\formula{{\tt formula}}
\def\place{{\tt place}}
\def\quadr{{\tt quadr}}
\def\hcalv{{\tilde \calv}}
\def\idez{{``}}
\def\fnc{{\tt fnc}}
\def\cycle{{\tt cycle}}
\def\zext{{\tt zext}}
\def\term{{\tt term}}
\def\str{{\tt str}}
\def\compr{{\tt compr}}
\def\ndig{{\tt ndig}}
\def\mapm{{\tt mapm}}
\def\mstr{{\tt mstr}}
\def\quant{{\tt quant}}
\def\caldnull{{\cald_{0}}}
\def\boolean{{\tt boolean}}
\def\circuit{{\tt circuit}}
\def\pair{{\tt pair}}
\def\tuple{{\tt tuple}}
\def\mseq{{\tt mseq}}
\def\fsymb{{\tt fsymb}}
\def\Circ{{\tt Circ}}
\def\csize{{\tt csize}}
\def\ladder{{\tt ladder}}
\def\sgnt{{\tt sgnt}}
\def\vx{{\vec x}}
\def\va{{\vec a}}
\def\vy{{\vec y}}
\def\vgamma{{\vec\gamma}}
\def\vchi{{\vec\chi}}
\def\circode{{\tt circode}}
\def\quantp{{\tt quantp}}
\def\gseq{{\tt gseq}}
\def\SForm{{\tt SForm}}
\def\brho{{\bar\rho}}
\def\numst{{\tt numst}}
\def\Turing{{\tt Turing}}
\def\Start{{\tt Start}}
\def\next{{\tt next}}
\def\vY{{\vec Y}}
\def\enc{{\tt enc}} 
\def\incr{{\tt incr}}
\def\decr{{\tt decr}}
\def\bit{{\tt bit}}
\def\Incr{{\tt Incr}}
\def\Decr{{\tt Decr}} 
\def\chr{{\tt chr}}
\def\short{{\tt short}}
\def\rem{{\tt rem}}
\def\id{{\tt id}}
\def\val{{\tt val}}
\def\cert{{\tt cert}}
\def\balpha{{\bar \alpha}}
\def\lz{{(\! (}}
\def\rz{{)\! )}}
\def\tape{{\tt tape}}
\def\tpno{{\tt tpno}}
\def\tplength{{\tt tplength}}
\def\aut{{\tt aut}}
\def\head{{\tt head}}
\def\bin{{\tt bin}}
\def\Cell{{\tt Cell}}
\def\trfunc{{\tt trfunc}}
\def\bcalc{{\bar\calc}}
\def\barell{{\bar \ell}}
\def\bcalt{{\bar{\cal T}}}
\def\Width{{\tt Width}}
\def\CELL{{\tt CELL}}
\def\CONT{{\tt CONT}}
\def\xcalm{{\calm}}
\def\nbl{{\nabla}}
\def\atn#1{{}}
%\def\atn#1{{\hrule#1}}
%%%%%%kkkkkkkkkkkkkkkkkkkkkkkkkkkkkkkkkkkkk%kezdodik

\title{Lower Bounds for RAMs and Quantifier Elimination} \author{Mikl\'os
Ajtai\\  IBM Research, Almaden Research Center}
\maketitle

{\bf Abstract.}
For each natural number $d$ we consider a finite structure $\boldm_{d}$
whose universe is the set of all $0,1$-sequence of length $n=2^{d}$,
each
representing a natural number in the set $\lbrace 0,1,...,2^{n}-1\rbrace
$ in binary form. The operations included in the structure are the
four constants $0,1,2^{n}-1,n$, multiplication and  addition modulo
$2^{n}$,
the unary function $ \min\lbrace 2^{x}, 2^{n}-1\rbrace$, the binary
functions  $\lfloor x/y\rfloor $ (with $\lfloor x/0 \rfloor =0$),
$\max(x,y)$, $\min(x,y)$, and  the  boolean vector  operations
$\wedge,\vee,\neg$ defined on $0,1$ sequences
of length $n$, by performing the operations  on all components
simultaneously. These are
essentially the arithmetic  operations that  can be performed on a
 RAM, with wordlength $n$, by a single instruction. We show that there exists 
an $\epsilon>0$ and a term
(that is, an algebraic expression)  $F(x,y)$ built
up
from the mentioned operations, with the only free variables $x,y$, such
that  if $G_{d}(y)$, $d=0,1,2,...$, is a sequence of terms, and for all $d=0,1,2,...$,
$\boldm_{d}\models \forall x, [G_{d}(x)=0\leftrightarrow \exists y, F(x,y)=0]   $, then
for infinitely many integers $d$, the depth of the term $G_{d}$, that is, the 
maximal number of nestings of the operations in it, is at least $\epsilon (\log d)^{\frac{1}{2}}=
\epsilon (\log \log n)^{\frac{1}{2}}$.
 
The following is a consequence. We are considering RAMs  $N_{n}$,  with wordlength 
$n=2^{d}$, whose arithmetic instructions are the arithmetic operations listed above, and also 
have 
the usual other RAM instructions. The size of the memory is restricted only by the address 
space, that is, it is  $2^{n}$ words.   The RAMs has a 
finite 
instruction set, each instruction is encoded by a fixed natural number independently of $n$. Therefore a 
program $P$ can run on each machine 
$N_{n}$, if $n=2^{d}$ is sufficiently large. We show that there exists an $\epsilon>0$ and a 
program $P$, such that it satisfies the following two conditions.\\
\indent (i) For all sufficiently large $n=2^{d}$, if $P$ running on $N_{n}$ gets an input consisting of 
two words $a$ and $b$, then, in constant time, it gives a 
$0,1$ output $P_{n}(a,b)$.  \\
\indent (ii) Suppose that $Q$ is a program such that for each sufficiently large $n=2^{d}$, if $Q$,
running on $N_{n}$, 
gets  a word
$a$ of length $n$	as an input, then it decides whether there exists a word $b$ of length $n$ 
such that  
$P_{n}(a,b)=0$. Then, for infinitely many positive integers $d$, there exists a word $a$ of length 
$n=2^{d}$,  
such that the running time of 
$Q$ on $N_{n}$ at input $a$ is at least $\epsilon (\log d)^{\frac{1}{2}} (\log \log d)^{-1}\ge  
 (\log d)^{\frac{1}{2}-\epsilon}= (\log \log n)^{\frac{1}{2}-\epsilon}$.

\section{\llabel{introduction} Introduction}

\subsection{Motivation, historical background}

One of the central questions of complexity theory is the
comparison of the computational resources needed for deterministic and
nondeterministic computation. Namely, assume that we want to find a
$0,1$-sequence satisfying a test $T$. Is it true, under some natural
assumptions                               
on the test and on the algorithm searching for $x$, that to find $x$
requires essentially more computation, than checking that a given $x$
really satisfies  $T$?  In the case when both the test and the searching
algorithm  must be performed in polynomial time (in the length of $x$)
by a turing machine, this leads to the $P=NP?$ question.

 In an earlier paper \cite{Ajt0}  the author has shown 
that if both the test and the computation consist of an evaluation of an algebraic
expression made from the operations described in the abstract, and the length of the 
algebraic expressions are constant then deterministic an nondeterministic computations can be
separated. An equivalent formulation in term of RAMs is that there exists a constant time test
$P$  in the sense described in the abstract, such that there exists no constant time program 
$Q$, which decides for  all $n$  and for all words  $a$ of lengths $n$, while running on $N_{n}$,   
whether there exists a  word $b$ of length $n$ with $P_{n}(a,b)=0$. The main motivation of the 
present paper is to improve the time lower bound on $Q$. The methods in \cite{Ajt0} show only
that a $Q$ with the given properties cannot work in constant time but do not give any 
specific unbounded function $f(n)$ as a lower bound.

 First we compare our results to other  theorems, where nonlinear lower bounds were
given,  or deterministic and nondeterministic computation were separated in  general 
computational 
models.  
Some of these proofs were based on diagonalization arguments. In fact the high level structure of 
the present proof   and the proof in \cite{Ajt0} is very similar to the structures of the proofs
given in \cite{HPV}, \cite{PPST}, or \cite{Fortnow}. 
The technical details however are
completely different. 

 For multi-tape turing machines linear time
nondeterministic and deterministic computations were separated in \cite{PPST} by
Paul, Pippenger, Szemer\'edi, and Trotter in 1984. Their theorem and the present result are not 
comparable in the sense, that none of them follows from the other, since in the turing machine model 
longer bitwise computations can be done than in our RAM model with the given time limit, but the 
RAM 
model 
allows arithmetic operations
e.g., multiplication, and  division of $n$ bit numbers, and it is not known whether these 
operations 
can be computed
on a multitape  turing machine in linear time. For uniform computational models where the 
working
memory is smaller than the input, Fortnow gave nonlinear lower bounds in \cite{Fortnow}. 
In a similar sense as in the case of \cite{PPST} our results and the results of \cite{Fortnow}
are not comparable. 
The highlevel structures of the proofs in both \cite{PPST} and \cite{Fortnow} however are very 
close to the highlevel
structure of the present proof.  The argument which forms the highlevel
structure of all of these proofs was
used by Hopcroft, Paul, and Valiant (see \cite{HPV}) in 1977. 
 In this paper  we will use the outline of the proofs in \cite{PPST}  as 
a model while giving the sketch of the present
proof.
 There are also nonlinear lower bounds for nonuniform models of computations see 
\cite{Ajt}, \cite{BJS1},
\cite{BJS2}, but the results are also incomparable to the present ones and even the high level structures
of the proofs are  completely different.

We can say  that the difference between these already existing lower bounds and the 
ones in the present paper and in \cite{Ajt0} is that they are based on different properties of the 
computational models.
 Both in the case of the turing machine model, and in the models with small working memory, a 
lower bound proof is possible because of the organization of the memory, which in the second 
case 
includes the input. In both cases there is some  restriction on the structure/use of the memory 
that 
is the crucial property used in the proof. 
In contrast, our present proofs, or the proofs  in \cite{Ajt0},  are  not based on  properties of 
the memory structure or the 
memory access, but on properties of  the set of
arithmetic instructions. Therefore 
our results say something about the set of  arithmetic 
operations multiplication, addition etc., which is used in the usual random access machines.

As an additional motivation we can say that
solving several search problems, each within the framework
 of our theorem, frequently occurs as part of computational problems 
to be solved on a RAM. Of course our lower bound does not imply a lower bound 
for the solution for all of the search problems together, still it may show that we cannot hope for a 
fast solution by solving each of these search problems separately.

\subsection{The formulation of the results}

First we formulate our result about RAMs.  
 For each positive integer $n$ we define 
a von Neumann type machine $N_{n}$ with word length $n$. (See also \cite{Ajt1}.) These machines have
a common finite instruction set. Each instruction has a name, which is a natural number.
We consider only the machines $N_{n}$ for,  say, $n>10$, where such a name fits into a memory 
cell. The set of these names will be denoted by $\cali$. A program $P$ is a sequence from
the elements of $\cali$. When we say that the machine $N_{n}$ executes the program $P
$ of length $k$, we mean that the machine starts to work from the state where the first
$k$ memory cells contains the elements of $P$ in their natural order and the contents of the
other memory cells are zeros. The total number of memory cells is restricted only by
the address space, say, it is $2^{n}$.  The instruction set contains (i)
 arithmetic instructions: addition and multiplication modulo $2^{n}$, the constants 
$0,1,n,2^{n-1}$, 
the unary function $ \min\lbrace 2^{x}, 2^{n}-1\rbrace$, the binary functions $\lfloor x/y\rfloor 
$ with $\lfloor x/0 \rfloor=0$,
$\max(x,y)$, $\min(x,y)$, and  the  boolean vector  operations
$\wedge,\vee,\neg$ defined on $0,1$ sequences
of length $n$. (ii) read, write instructions, (iii) control transfer instructions, (iv)
input/output instructions, (v) halt instruction.    

Assume that $c,k$ are positive integers. 
 A  program $P$ will  be called a $c$-size $k$-ary test,  if $\length(P)\le c$, $k\le c$ and for all 
positive integers $n>10$, and for 
all integers
$x_{1},...,x_{k}\in [0,2^{n}-1]$, the program $P$ on machine $N_{n}$, at input 
$x_{1},...,x_{k}$  uses only the first $c$ memory cells, and 
produces an output $P_{n}(x_{1},...,x_{k})\in \lbrace 0,1\rbrace $. The time requirement of 
$P$ on $N_{n}$ is the smallest integer $t$ such that for all integers $x_{1},...,x_{k}\in [0,2^{n}-1]$, the program
$P$ at input $x_{1},...,x_{k}$ provides an output in time at most $t$.  

\begin{theorem} \llabel{TT1} There exist an $\epsilon>0$, a positive integer $c$ and a $c$-size 
binary test $P$, 
with time requirement at most $c$ on each machine $N_{n}$,
such that for all $c'>0$,  and  for all $c'$-size unary tests $Q$ the following holds. 
Suppose that for all sufficiently large positive integers $n$,
 and for all $a\in [0,2^{n}-1]$, the following two 
statements are equivalent:
 \\ \indent  (i) \ \ \ $\exists x\in [0,2^{n}-1]$,  $P_{n}(x,a)=0$,\\
\indent    (ii)  \ \  $Q_{n}(a)=0$.\\ 
Then for infinitely many positive integers	 $n$,  the time requirement of  $Q$ on $N_{n}$,
 is at 
least \\
\centerline{ $ \epsilon (\log \log n)^{\frac{1}{2}} (\log \log \log n)^{-1}$}
\end{theorem}

 In other words, there exists a constant time test $P(x,a)$, depending on a 
parameter $a$, such that the question whether it  has a solution in $x$ or not, cannot be decided for all $n$
by a constant size program $Q$ 
which gets $a$ as an input, even if the time used by $Q$ on $N_{n}$ can be 
as large as $\epsilon (\log d)^{\frac{1}{2}} (\log \log d)^{-1}$, where $n=2^{d}$. 
The theorem remains true even in the following stronger nonuniform  version. Suppose that the 
sequence $Q_{n}$, $n=1,2,...$ is a sequence of programs, and $f,g$ are functions defined on the 
the set of natural numbers with real values. We say that the sequence $Q_{n}$ is a  family of 
unary tests with size bound $f$ and  time limit $g$, if for each sufficiently large $n$, $Q_{n}$
is a program, that is, a sequence from the elements of $\cali$, of length at most $f(n)$, and for 
each $a\in [0,2^{n}-1]$,
$Q_{n}$, while  running on the machine $N_{n}$ at input $a$, gives a $0,1$ 
output  $Q_{n}(a)$	in 
time at most  $g(n)$.

\begin{theorem} \llabel{TT2} There exist an $\epsilon>0$, a positive integer $c$ and a $c$-time 
binary test $P$, 
with time requirement  at most $c$,
such that for all families of unary tets $Q_{n}$, $n=1,2,...$, with both size  bound and time limit 
$\epsilon (\log \log n)^{\frac{1}{2}} (\log \log \log n)^{-1}$ the following holds. For
infinitely many  positive integers $n$, there exists an $a\in [0,2^{n}-1]$, such that  the following 
two statements 
are  not equivalent:
 \\ (i) \ \ \ $\exists x\in [0,2^{n}-1]$,  $P_{n}(x,a)=0$,\\
(ii) \ \  $Q_{n}(a)=0$.
\end{theorem}

The proof of these theorems  will be based on a theorem about the structures $\boldm_{d}$ described in the abstract. Our next goal is to  formulate that result.

\begin{definition} 1. The set of all natural numbers
will be denoted by  $   \omega $,      that is,  $   \omega=\lbrace 0,1,2,\ldots \rbrace
 $.
Each natural number  $   n $    is considered as the set of all
natural
numbers
less than  $   n $,      that is,  $   n=\lbrace 0,1, \ldots ,n-1\rbrace  $
and  $   0=\emptyset $,       $   1=\lbrace \emptyset\rbrace  $,       $   2=\lbrace
\emptyset, \lbrace \emptyset\rbrace  \rbrace  $,      etc.

2. Assume that  $   a,b\in\omega $,       $   b\ge 2 $.     The natural
number  $   a $    can be written in a unique way in the form of
\xev{184a}
 $   \sum_{i=0}^{\infty}\alpha_{i}b^{i} $,      where  $   \alpha_{i}\in
\lbrace
0,1, \ldots ,b-1\rbrace  $    for  $   i\in \omega $.     The integer
 $   \alpha_{i} $    will
be denoted by  $   \coeff_{i}(a,b) $.     In other words
 $   \coeff_{i}(a,b) $    is
the  $   i $th {\idez}digit"   \xev{204d}
of  $   a $    in the numeral system with base  $   b $.        We extend the definition
of  $   \coeff_{i}(a,b) $    for negative integers  $   i $    as well, by
 $   \coeff_{i}(a,b)=0 $    for all  $   i=-1,-2, \ldots  $.

3.  $   \calm $    will denote a first-order
language
with
\xev{056a}
equality, which does not contain any other relation symbols,
and contains the following function and constant symbols. (We
consider
constant symbols as  $   0 $-ary function symbols as well.)

Constant symbols:  $   {\bf 0},{\bf 1}, {\bf -1}, \bfn $.

Unary function symbol:  $   \caln $,       $   \bfp $,      ( $   \caln $    stands for
``{\it n}egation",
 $   \bfp $    stands for {\idez}{\it p}ower").

Binary function symbols:  $   + $,       $   \times $,
       $   \div $,        $   \max $,       $
\min $,       $   \cap $.

4. Since $\calm $ is a language with equality, in the interpretations defined below,
we do not define the interpretation of the relation ``$=$", it is already given as ``equality". Assume that  $   d\in \omega=\lbrace 0,1,2,\ldots \rbrace  $
and
 $   n=2^{d} $.
  $   \boldm_{d} $
will denote the following interpretation of the language
 $   \calm $   :
 $   \universe(\boldm_{d})=\lbrace 0,1,\ldots
,2^{n}-1\rbrace=2^{n}
 $    and for
all
 $   x,y,z\in \universe(\boldm_{d}) $,
  \\
 ( $   \boldm_{d}\models +(x,y)=z $   ) iff
 $   x+y \equiv z $     $   (\mod \ 2^{n}) $,
\\
 $   (\boldm_{d}\models \times(x,y)=z $   ) iff  $   xy
\equiv z $     $   (\mod \ 2^{n}) $,
\\
\semmi{( $   \boldm_{d}\models \bfp(x,y)=z $   ) iff {\idez}either
( $   x^{y}\ge2^{n}\wedge z=2^{n}-1 $   ) or ( $   x^{y}< 2^{n} \wedge
z=x^{y} $   )"}
( $   \boldm_{d}\models \bfp(x)=z $   ) iff  $   z=\min\lbrace
2^{x},2^{n}-1\rbrace
 $,
\\
 $   (\boldm_{d}\models z=\div(x,y)) $    iff  $  (y\not=0 \wedge z=\lfloor {x/y}
\rfloor) \vee(y=0 \wedge z= 0)  $    \\
 $   (\boldm_{d}\models z={\bf 0}) $    iff  $   z=0 $,     \\
 $   (\boldm_{d}\models z={\bf 1}) $    iff  $   z=1 $,     \\
 $   (\boldm_{d}\models z={\bf n}) $    iff  $   z=n $,     \\
 $   (\boldm_{d}\models z={\bf -1})  $    iff  $   z=2^{n}-1 $,     \\
 $   (\boldm_{d}\models z=\max(x,y)) $    iff  $   z=\max\lbrace
x,y\rbrace
 $,     \\
 $   (\boldm_{d}\models z=\min(x,y)) $    iff  $   z=\min\lbrace
x,y\rbrace
 $,      \\
 $   (\boldm_{d}\models z=x\cap y) $    iff \\
 $   \coeff_{i}(z,2)=\min(\coeff_{i}(x,2),\coeff_{i}(y,2)) $
for
 $   i=0,1,\ldots ,n-1 $,      \\
 $   (\boldm_{d}\models z=\caln(x)) $    iff
 $   \coeff_{i}(z,2)=1-\coeff_{i}(x,2) $    for
 $   i=0,1,\ldots ,n-1 $.

We will call the interpretations  $   \boldm_{d}$, $ d\in \omega $
of
 $   \calm $
the standard 
interpretations of  $   \calm $.

5. Motivated by the definition of the standard
interpretations
we
will use the following notation as well when we use the
functions
symbols
of  $   \calm $   :  $   +(x,y)=x+y $,       $   \times(x,y)=x\times y =xy$,
 $   \bfp(x)=2^{x} $.
      Generally we will use this notation only
 if it is clear from the context  the we mean the
function symbol interpreted in a structure  $   \boldm_{d} $,
otherwise
 $   x+y,xy,2^{x} $    retain their usual meaning as operations
among
real
numbers.  Although the relation $\le$ is not included in the language $\calm
$ sometimes we will write $\boldm_{d} \models a\le b$ as an abbreviation
 for $\boldm_{d} \models a=\min(a,b)$.

  6. When   we use  the
function
symbols of  $   \calm  $    we will write  $   x-y $    for  $   x+(-\bfegy)y $
and
 $-y $
for  $   (-\bfegy)y $.   

7. Assume that  $   F(x,y) $    is a term of  $   \calm $   
and  $   G= \langle G_{d}(y) \mid d\in \omega \rangle $,       $   d\in
\omega $    is a
sequence of terms of $\calm$. We will say that the sequence  $   G $    decides whether
\xev{179a}
there exists a solution for  $   F $,    if for all sufficiently
large  $   d\in \omega  $,      we have $$     \boldm_{d} \models \forall
y,
[G_{d}(y)=\bfnull \leftrightarrow \exists x, F(x,y)]$$

8.  The length of a term $\tau$ of a first-order
language $\call$  is the
total number of symbols (counted with multiplicity) in it. This number  will be denoted
by $\length(\tau)$. The depth of the term $\tau$, that will be denoted by 
$\depth(\tau)$, is the maximal number of  nestings of the function symbols in it. 
 \vege\end{definition}

\begin{theorem} \llabel{TT3}   There exists an  $   \epsilon>0 $
and a term
  $   F(x,y) $    of  $   \calm $   such that the following holds.
Assume that  $   G=\langle G_{d}(y) \mid d\in \omega\rangle  $    is
a
sequence
of terms of  $   \calm $    such that  $   G $    decides whether there
exists a solution for  $   F $.  Then for infinitely many  $   d\in
\omega $,      the depth of  $   G_{d} $    is at least 
 $ \epsilon (\log d)^{\frac{1}{2}} = \epsilon (\log\log n)^{\frac{1}{2}}$,
where $n=2^{d}$. \end{theorem}

The depth of a propositional formula is the maximal number of nestings of function 
symbols and 
boolean operations together.
It is easy to show  that there exists a $c\in \omega$ such that for each propositional 
formula 
$P(x_{1},...,x_{k})$ of $\calm$, there exists a term 
$F(x_{1},...,x_{k})$ 
of $\calm$, such that
$\depth (F) \le c \hskip 1pt \depth (P)$,  and 
for all $d\in \omega$, $\boldm_{d} \models \forall x_{1},...,x_{k}, P(x_{1},...,x_{k}) 
\leftrightarrow F(x_{1},...,x_{k})=0$. This implies 
 the following equivalent form of
Theorem \rref{TT3}. The theorem says that  over the structures $\boldm_{d}$ 
quantifier 
elimination 
is not
possible in a strong quantitative sense.

\begin{theorem} \llabel{TT4}
There exists an $\epsilon>0$  and an existential formula $\psi(y)$ of $\calm$ 
containing a single 
existential quantifier,  such that, if 
$P_{d}(y)$, $d\in \omega$ is a sequence of propositional formulas and for all $d\in 
\omega$,
$\boldm_{d} \models \forall y , \psi(y) \leftrightarrow P_{d}(y) $, then for infinitely 
many $d\in 
\omega$, $\depth(P_{d} 
)\ge  \epsilon (\log d)^{\frac{1}{2}}$. \end{theorem}

Weaker versions of Theorems \rref{TT3} and Theorem \rref{TT4} were proved in 
\cite{Ajt0}. 
E.g.,
the weaker version of Theorem \rref{TT3}
is equivalent 
to the statement  that
if  $G=\langle G_{d} \mid d\in \omega\rangle $ is a sequence of terms of
$\calm$ such that $G$	decides whether there exists a solution for $F$ then there
exists a sequence $d_{0},d_{1},...$  of natural numbers such that $\lim_{i\rightarrow 
\infty} \depth(G_{d_{i}})=\infty$.
The motivation for the formulation of Theorem \rref{TT3}, apart from the fact the it is 
used
in the proofs about RAMs, is that it is a natural continuation of a long chain of results 
in 
mathematics which say that certain search problems, e.g., equations, cannot be solved 
by the same 
operations as were used in their formulation. For example Galois' theorem about the 
unsolvability 
of equations of degree five by algebraic operations belong to this category. (Sevaral 
other 
examples of this nature is described \cite{Ajt0}). In the present case we give such a 
 Theorem in a quantitative form by giving a lower bound on the depth on the algebraic 
expression
which could compute a solution.

There are many important first-order structures where quantifier elimination is
possible (e.g., the field of real numbers, field of complex numbers) and also where it is 
not 
possible (e.g., Peano 
Arithmetic).  Theorem \rref{TT4}  gives us an example where quantifier elimination is not 
possible, moreover the statement is true in a quantitative form. The particular 
choice of the structures involved in the theorem is motivated by the connection with random 
access machines.   

The first-order properties of structures similar to the structures $\boldm_{d}$  were studied for a
long time in the theory of Fragments of Peano Arithmetic. In that case however the set of 
operations defined by function symbols is usually more restricted (although sometimes 
exponentiation in some restricted form is allowed).  In that theory the basic structure is usually
not a finite set as in the case of $\boldm_{d}$, but rather an infinite initial segment of a
nonstandard model of Peano Arithmetic, which is closed under addition, multiplication and 
sometimes  under the operation $x ^{\lfloor \log y \rfloor}$.  The advantage of this is that 
instead of
speaking about  an infinite sequence of structures the results can be formulated in a single 
structure.
A similar solution may be possible in our case too, but then the  connections with RAMs
would be much more complicated than with the present formulation of the result. Namely, it
would be difficult to maintain a fixed upper 
bound on the sizes of memory cells since each multiplication would double the  number of bits 
in a word.

The following theorem  shows that the lower bounds that we proved in the four theorems 
described above are
probably very far from the truth.

\begin{definition} Assume that $F(x,y)$ is a term of $\calm$.  We describe a problem
in $NP$,  which will be called ``the solution of the equation $F(x,a)=0$ in $x$".

If the size of he problem is $n$, where we assume that $n=2^{d}$, then the input of the 
problem is an integer $  a \in 2^{n} $. 
 An integer $b\in \lbrace 0,1,\ldots ,2^{n}-1 \rbrace$ is  a solution of the problem 
if $\boldm_{d} \models F(b,a)=\bfnull$, where $n=2^{d}$.  
\vege\end{definition}

\begin{theorem} \llabel{TT5} There exists a term
  $   F(x,y) $    of $\calm$, such that the solution of the equation $F(x,a)=0$ in $x$
is  an $NP$-complete problem. \end{theorem}

The following two theorems are important steps in the proof of Theorem \rref{TT3} and
Theorem \rref{TT4}. 

 If $t\in \omega$ then the $i$th $2^{2^{t}}$-ary digit of a natural number $a$
will be denoted by $a[i,t]$, that is, $a=\sum_{i=0}^{\infty} a[i,t]2^{i2^{t}}$.
Assume that $d,t\in \omega$,
$d\ge t$. We may consider the elements of $\boldm_{d}$ as $2^{d-t}$ dimensional vectors
whose components are in $\boldm_{t}$, namely the integer $a\in \boldm_{d}$ will represent the 
vector $\langle a[0,t],a[1,t],...,a[2^{d-t}-1,t]\rangle$, that is the $2^{2^{t}}$-ary digits of
$a$ are the components of the vector represented by $a$.

Let $\bff$ be a $k$-ary function symbol of $\calm$ for some
$k\in \lbrace 0,1,2\rbrace $. For all $d,t\in \omega$ with $d\ge t$, we define
a $k$-ary  function $\bff_{d,t}$ on the universe of $\boldm_{d}$ in the following 
way.
Assume that $d,t\in \omega $ is fixed with $d\ge t$ and $a_{0},...,a_{k-1} \in \boldm_{t}$. Then
 $\bff_{d,t}(a_{0},...,a_{k-1}) $ is the unique element $b\in 
\boldm_{d}$ with the property that for all $i\in 2^{d-t}$, we have $\boldm_{t} 
\models \bff(a_{0}[i,t],...,a_{k-1}[i,t])=b[i,t]$. In other words we consider each element 
of 
$a\in \boldm_{d}$ as a vector $\langle a[0,t],...,a[2^{d-t}-1,t]\rangle$ and perform the 
operation $\bff$  component-wise in $\boldm_{t}$. The following 
theorem states that,  if $\bff$ is a function symbol and $\bff\notin \lbrace \times,
\div,\bfp\rbrace $, then the function $\bff_{d,t}$
can be defined by  an existential formula in $\boldm_{d}$, that is, vector operations
apart from the exceptions of multiplication, division, and  exponentiation are existentially 
definable.

\begin{theorem} \llabel{TT6}  Assume that $\bff$ is a $k$-ary function
symbol of $\calm$	for some $k\in \lbrace 0,1,2\rbrace $ and $\bff\notin \lbrace\times,\div,\bfp \rbrace  $. Then there exists an 
existential first-order formula $\psi
(x_{0},...,x_{k-1},y,z)$ of $\calm$ such that for all
$d,t\in \omega$ with $d\ge t$, and for all $a_{0},...,a_{k-1},b\in \boldm_{d}$, 
the following two 
conditions are equivalent: \\ \smallskip
(i) \ \ $ \bff_{d,t}(a_{0},...,a_{k-1})=b $,\\  \smallskip
(ii) \ $\boldm_{d} \models \psi(a_{0},...,a_{k-1},b,t)$.
\end{theorem}

 For the exceptional function symbols $\times,\div,\bfp$ we do not know
whether the statement in Theorem \rref{TT6} holds. We may define the vector 
operations for these function symbols too in a somewhat larger structure 
$\boldm_{v}$ by an existential formula $\psi$,  if  $v\ge t+c(d-u)$, where $c$ is a 
sufficiently 
large constant.

\begin{theorem} \llabel{TT7}  Assume that $\bff$ is a $k$-ary function
symbol of $\calm$ for some $k\in \lbrace 0,1,2\rbrace $. Then there exists a $c\in 
\omega$ and an existential first-order formula $\psi
(x_{0},...,x_{k-1},y,z,w)$ of $\calm$, such that for all
$d,t\in \omega$ with $d\ge t$, and for all $a_{0},...,a_{k-1},b\in \boldm_{d}$,  
if $v\in \omega$, $v\ge t+c(d-t)$, then
the following two 
conditions are equivalent: \\ \smallskip
(i) \ \ $ \bff_{d,t}(a_{0},...,a_{k-1})=b $,\\ \smallskip
(ii) \ $\boldm_{v} \models \psi(a_{0},...,a_{k-1},b,d,t)$.
\end{theorem}

This theorem motivates the following definition. Assume that for all $d,t\in \omega$
with $d\ge t$, $F_{d,t}(x_{0},...,x_{i-1})$ is an $i$-ary function defined on 
$\boldm_{d}$ and with values in $\boldm_{d}$. We will say that the family
of functions $F=\langle F_{d,t} \mid d,t\in \omega, d\ge t\rangle$ is polynomially 
existential in $\boldm$ if there exist a $c\in \omega$ and an existential first-order formula
$\psi(x_{0},...,x_{i-1},y,z,w)$ of $\calm$ such that for all $d,t\in \omega$ with 
$d\ge t$ and for all $a_{0},...,a_{i-1},b\in \boldm_{d}$ if $v\in \omega$, 
$v\ge t+c(d-t)$ then
 the following two
conditions are equivalent:
\\ \smallskip
(i) \ \ $ F_{d,t}(a_{0},...,a_{k-1})=b $,\\ \smallskip
(ii) \ $\boldm_{v} \models \psi(a_{0},...,a_{k-1},b,d,t)$.

Therefore Theorem \rref{TT7} says that for each function symbol $\bff$ of
$\calm$ the family of functions $\bff_{d,t}$, $d,t\in \omega$, $d\ge t$
is polynomially existential. We use the word polynomially because of the following reason. In Theorem \rref{TT7}  we 
consider the elements of $\boldm_{d}$ as vectors with $2^{d-u}$ components, 
where
each component is an element of $\boldm_{t}$. In the formula  $\psi$ 
we existentially quantify elements of $\boldm_{v}$ which can be considered as
vectors with $2^{c(d-t)}$ components which are in $\boldm_{t}$. Therefore the 
number of components of the existentially quantified vectors is a polynomial 
of the number of component in the arguments of the function. This is true not only
for Theorem \rref{TT7} but in  general for polynomially existential families. In addition to this, as the following theorem will show the notion of polynomially existential families of functions
is closely related to the notion of  polynomial time computation.

For the following definition recall that $N_{m}$ is a random access machine
with word length $m$ and with $2^{m}$ memory cells. In case $m=2^{d}$, for some 	
$d\in \omega$, the machine can compute each $\calm$ operation in $\boldm_{d}$ 
by a single instruction.

\begin{definition} Suppose that $F=\langle F_{d,t} \mid d,t\in \omega, d\ge t\rangle$ is a family of
  $k$-ary functions, where each function $F_{d,t}$, $d,t\in \omega$, $d\ge t$ is a
$k$-ary function defined on 
$\boldm_{d}$ with values in $\lbrace 0,1\rbrace $. We will say that the family $F$
is polynomial time computable with respect to $\boldm$ if there exist a 
$\gamma_{1}\in
 \omega $ and a program $P$ for the family of RAMs $N_{n}$ such the
the following  holds,

\begin{cond} \llabel{Y25.2}
for all sufficiently large $d\in \omega$, for all $t\in \omega $ with $d\ge t$, and for all
$a_{0},...,a_{k-1}\in \boldm_{d}$,   
 the machine $N_{m}$,   where $m=2^{d}$, 
with  program $P$ and input $k,d,t, a_{0},...,a_{k-1}$, using only the first $2^{\gamma_{1}(d-
t)}$ memory cells in time $2^{\gamma_{1}(d-t)}$ computes  $F_{d,t}(a_{0},...,a_{k-1})$. 
\end{cond}

We assume that at time $0$ the program and the input
is in the  first $\length(P)+3+k$ memory cells, the program is in the first
$\length(P)$ cells and the input $k,d,t,a_{0},...,a_{k-1}$ is in the 
next $3+k$ cells in the given order. 
 \vege\end{definition}

\begin{remark} 1. Since $d$ is sufficiently large we may assume that $2^{2^{d}}$ the 
total number
of memory cells of $N_{m}$ is larger than $2^{\gamma_{1}(d-t)} $ the number
of memory cells required for the computation, and $2^{2^{d}}$ is also larger
than $\length(P)+k+3$ the number of 
memory cells required for the input.

2. We stated the definition for  functions with $0,1$-values. In fact, we may  
allow any value in $\boldm_{d}$ or even a sequence of length $2^{d-t}$ in 
$\boldm_{d}$, and everything that we prove about this notion  remains
true. We will use however this notion in a nondeterministic setting where the 
$0,1$-valued functions are sufficient for our purposes.
\vege\end{remark}

\begin{theorem}  \llabel{TT8}  Suppose that $F=\langle F_{d,t} \mid d,t\in \omega, d\ge t\rangle$ is a family of
  $k$-ary functions, where each function $F_{d,t}$, $d,t\in \omega$, $d\ge t$ is a
$k$-ary function defined on 
$\boldm_{d}$ with values in $\lbrace 0,1\rbrace $.
Assume further that  the family $F$ is 
polynomial 
time computable with respect to $\boldm$. 
Then the family $F$  is
polynomially existential in $\boldm$.
\end{theorem}

\section{\llabel{sketch} Sketch of the proof of Theorem \rref{TT3}.}

\subsection{\llabel{overview} Overview} As we have mentioned already, a weaker version 
of Theorem \rref{TT3} was proved in \cite{Ajt0}, namely it has been shown  that
if  $G=\langle G_{d} \mid d\in \omega\rangle $ is a sequence of terms of
$\calm$ such that $G$	decides whether there exists a solution for $F$ then there
exists a sequence $d_{0},d_{1},...$  of natural numbers such that $\lim_{i\rightarrow 
\infty} \depth(G_{d_{i}})=\infty$.
 We will refer to this theorem as Theorem 
$\bolda$.
The proof of Theorem $\bolda$  did not 
provide any  unbounded function
$f(x)$ such that we could conclude that for infinitely many $d$, the depth of $G_{d}$ 
is at least $f(d)$. It seems that the lack of such a function $f$ is a consequence of the 
nature of the indirect proof given in \cite{Ajt0}. The paper 
\cite{Ajt0} also described a 
generalized version of Theorem $\bolda$, which essentially abstracted those properties 
of the
structures $\boldm_{d}$ which were needed in the proof. For the present  proofs 
 these properties are not sufficient.
(It is possible that the improved lower bounds hold for the generalized version
of Theorem $\bolda$ and can be proved by different methods.)
 We have to go back to the original definition
of the structures $\boldm_{d}$ in terms of its arithmetic operations, and formulate 
new additional properties which will be used in the proof of Theorem 
\ref{TT3}.

 We start sketching the proof of Theorem \ref{TT3}
by comparing it to the proof of Paul, Pippenger, Szemer\'edi, and Trotter about the 
separation of deterministic and non-deterministic linear time computation on
multitape turing machines  (see \cite{PPST}). We will refer to their theorem as
the PPST Theorem. We will point out which are those steps  in the proof of the PPST 
Theorem 
which has an 
analogue in the present paper. 
 
The outline of the proof of the PPST Theorem is, roughly speaking, the following.
The proof has three parts that we will call Collapsing, Simulation, and Diagonalization.
The roles of these parts can be summarized this way.

{\it Collapsing}. This is an indirect argument.
 Assuming that  the PPST theorem is not true
it is shown that the alternating hierarchy of linear time computation on multitape turing 
machines 
is 
collapsing, that is, for each $k$ there exists a $c$ such that each computation with $k$ 
alternation and time $n$ can be also performed by a machine with no alternations and 
in time $cn$.

{\it Simulation}. It is shown, without the indirect assumption, that
any computation  performed by a multitape turing 
machine  (without alternations) in time $n$,  can be also performed on an alternating 
machine with 
four alternations 
in time $\epsilon_{n}n$, where $\lim_{n\rightarrow \infty} 
\epsilon_{n}\rightarrow 0$.

{\it Diagonalization}. Assume that the PPST theorem is not true. The Collapsing and 
Simulation results described above lead to a contradiction through a diagonalization
argument.

First we describe what is the concept of ``computation" in our case. We do not
define a machine which performs the computation we only describe functions 
that we want to compute. We may think that the process of
evaluating a term or a first-order formula is the computation. (The RAM model, 
described earlier,
is not equivalent to this model of computation if the depth of the formulas can be 
larger than 
constant.)
 The analogue of non-alternating turing machine is the 
following. A term $\tau  \in \calm$ is given and an $n\in \omega$, $n=2^{d}$ is fixed. 
We want to 
compute the function which assigns to each $a\in \boldm_{d}$ the truth value of
 $
\boldm_{d}\models \tau(a)=\bfnull$.  

 The analogue of a turing machine with $k$ alternation is the following. A 
$\Sigma_{k}$ or $\Pi_{k}$ first-order formula $\phi$ of $\calm $ is given  and an 
$n\in 
\omega$,
$n=2^{d}$ is 
fixed. We want to compute the function which assigns to each $a\in \boldm_{d}$, the 
truth value of
 $
\boldm_{d}\models \phi(a)$.  

The term $\tau$ and the formula $\phi$ in the ``computations"
described above will be taken from sets depending on $n$. Namely, let $\calt_{n}$
be the set of all terms $\tau $ of $\calm$ which can be computed by an algebraic 
circuit (whose gates perform $\calm$-operations in $\boldm_{d}$, $n=2^{d}$) of size 
at most $2^{d+\log d+ 3}$. 
$\calh'_{n}$ will be  a  set containing only $\Sigma_{m} $ formulas, where
$m=c (d+\log d)^{1\over 2}$ for a constant $c$. (We will say more about  it later.) 
$\calh_{n}$ will be a similar but somewhat larger set of  first-order formulas of $\calm$ 
with the property that if  we perform a constant number of boolean operations or 
variable changes on the elements of 
  $\calh'_{n}$  then we get an element of  $\calh_{n}$.  With these definitions we can 
give a 
short 
description of the three parts of the present proof, which are analogues of the three 
parts in the 
proof of the PPST theorem.

{\it Collapsing.} Assuming that the theorem is not true we show that for each fixed $n=2^{d}$ 
there exists a term $\tau(x,y) 
\in  \calt_{n}$ and there  exists a 
function 
$\bfg $ (an 
analogue of
the G\"odel numbering) which assigns to each element of $\phi\in \calh_{n}$ an integer
$\bfg(\phi)\in \boldm_{d}$ such that if $q= \lfloor d+\log_{2} d \rfloor $ then for all
$a\in \boldm_{d}$,
$\boldm_{d} \models \phi(a) \leftrightarrow \boldm_{q} \models \tau(a,\bfg(\phi))=\bfnull$.

{\it Simulation.} We show that for each $\tau\in \calt_{n}$, there exists a $\lambda_{\tau}
\in \calh_{n}'$ such that for all $a,b\in \boldm_{d}$,  $\boldm_{d}\models
\lambda_{\tau}(a,b)$ is equivalent to
$\boldm_{q}\models \tau(a,b)=\bfnull $, where $q= \lfloor d+ \log d\rfloor$.

{\it Diagonalization}. Using the Collapsing and Simulation statements we show that
there exists a formula $\mu(x,y)$ of $\calh_{n}$ such that	for all $\phi \in \calh_{n}$,
and for all $a\in \boldm_{d}$, $\boldm_{d} \models \phi(a)$ iff $\boldm_{d}\models
\mu(a,\bfg(\phi))$, that is, the truth, at least for the formulas in $\calh_{n}$, are definable
in $\boldm_{d}$. This leads to a contradiction.

We give now a more detailed description of the various parts of the proof. We  start with the 
diagonalization since it has the simplest proof.

\subsection{\llabel{diagonalizatoin} Diagonalization.} This is similar to the argument in 
G\"odel's incompleteness theorem or, more closely, to Tarski's
proof about the non-definability of truth functions.

Starting with an arbitrary formula $\phi(x)\in \calh_{n}$ and the statement formulated 
in  
``Collapsing" we get a $\tau\in \calt_{n}$ with
$\boldm_{d} \models \phi(a) \leftrightarrow \boldm_{q} \models \tau(a,\bfg(\phi))$ for all $ a\in 
\boldm_{d}$. It is important that $\tau$ does not
depend on $\phi$. 
Next by the ``Simulation" statement we get, that there exists a  first-order formula
 $\lambda_{\tau}(x,y)\in 
\calh_{n}'$ for this particular $\tau$. Clearly if $\mu(x,y)
\equiv \lambda_{\tau}(x,y)$ then for all $\phi\in \calh_{n}$
and for all $a\in \boldm_{d}$, $\boldm_{d} \models \phi (a)$ 
iff $\boldm_{d} \models \mu(a,\bfg(\phi))$. 

Now we apply this for $\phi(x)\equiv \neg \mu(x,x)$. Since $\mu$ is in $\calh'_{n}$,
our assumptions about $\calh_{n}$ imply that $\phi\in \calh_{n}$. With the choice
$a\legyen \bfg(\phi)$ we get $\boldm_{d}\models \phi(\bfg(\phi)) \leftrightarrow
\mu(\bfg(\phi),\bfg(\phi)) $, that is, we have $\boldm_{d}\models 
\neg \mu(\bfg(\phi),\bfg(\phi)) \leftrightarrow
\mu(\bfg(\phi),\bfg(\phi))  $ a contradiction.

\subsection{\llabel{collapsing} Collapsing.} 
First we give the definitions of the sets $\calh_{n}'$ and $\calh_{n}$. Assume that $\phi$
is prenex first-order formula of
 $\calm$. We form blocks from the quantifiers of $\phi$, such that 
(a) each block is an interval of consecutive quantifiers of identical types, that is, existential or
universal and  (b) two consecutive quantifiers of identical type is always in the same 
block. Suppose that $\phi$ has $k$ blocks and the number of quantifiers in the blocks 
are
$\iota_{1},...,\iota_{k}$. We will say that the sequence $\langle \iota_{1},...,\iota_{k} \rangle  $
is the quantifier pattern of $\phi$. (We do not identify which are the universal and existential  
quantifiers.)

We describe now the sets $\calh_{n}$, $\calh'_{n}$.
 Assume that $M,j_{1},...,j_{m}$ are positive integers. The set of all prenex 
first-order formulas
 $\phi $	 of $\calm$  satisfying the following two conditions  will be denoted by 
$\boldl(M,j_{1},...,j_{m})$: 

{\sl (i)  if the quantifier pattern of $\phi$ is $\langle \iota_{1},...,
\iota_{k}\rangle$ then $k\le m$ and $\iota_{i}\le j_{i}$ for all i=0,...,k-1.

 (ii)  if  $\phi\equiv Q_{r}x_{r},...,Q_{1}x_{1},P(x_{r},
...,x_{1})$, where $Q_{r},...,Q_{1}$ are quantifiers and $P$ is a propositional formula
of $\calm$ then $\length(P(x_{1},...,x_{r}))\le M$, where $\length (P)$ is the number of 
symbols in $P$.}

 Let $\delta= \epsilon (\log d)^{\frac{1}{2}}$,  $m=\lfloor c \delta
\rfloor$. The exact definitions of $\calh_{n}'$,
and $\calh_{n}$ are too technical to describe them in this sketch, but we may think
that they are  essentially of the following form $\calh_{n}=\boldl(c^{\delta}, c,c^{2},...,c^{m})$, 
$\calh'_{n}=\boldl(c_{1}^{\delta}, c_{1},c_{1}^{2},...,c_{1}^{m}  )$, where $c>2$ and 
$c_{1}>2$
are constants, and $ c$ is sufficiently large with respect to $c_{1}$.
The essential feature of these formulas are that there are upper bounds on the number of 
quantifier 
blocks,
 the lengths of the formulas, and the sizes of the quantifier blocks starting from $c$ or
$c_{1}$
can grow only exponentially.

Naturally the starting point of the collapsing argument is
that, by the indirect assumption, if a first-order formula $\phi$ contains a subformula $\exists x, F(x,y)=\bfnull $ then
it can be replaced by the formula $G_{d}(y)=\bfnull$, and by this replacement we have decreased
 the number of quantifiers in $\phi$. Unfortunately it may happen that  such a subformula
does not exist. Indeed, if the prenex form of $\phi$ is $Q_{1}y_{1},...,Q_{k} y_{k}, \exists x, 
F(x,y_{1},...,y_{k})=\bfnull $,  where $Q_{0},...,Q_{k-1}$ are quantifiers, 
then for $k>1$ the indirect assumption is
not applicable since  $F$ depends on too many parameters. In 
this case  however we may consider the formula not in $\boldm_{d}$ but in $\boldm_{d+r}$
for $r=\lceil \log_{2} k\rceil$, where the sequence $y_{1},...,y_{k}$ from the 
elements of $\boldm_{d}$ can be encoded by a single element of $\boldm_{d+r}$.
This is done in the proof of Theorem~$\bolda$, and can be done in the present case as well. 

There is however another difficulty. The term $F(x,y)$ in the indirect assumption
is of constant size and $n=2^{d}$ can be arbitrarily large. Therefore the 
indirect assumption is not applicable if the size of $F(x,y)$ is not constant. Actually
the definition of the set $\calh'_{n}$ allows formulas whose sizes grow with $n$.
This cannot be avoided since the terms $G_{d}$ may have sizes growing with $n=2^{d}$ so 
after a single application of the indirect assumption, when we replace
$\exists x, F(x,y) $ by $G_{d}(y)$, we may get a formula containing a term of
size $\epsilon (\log d)^{1/2}$. 
 This problem did not arise in the proof of Theorem~$\bolda$ since there the terms
$G_{d}$ were  of constant sizes. (Another similar problem however arose since
after we reduced $\boldm_{d} \models \phi(a)$ to $\boldm_{d+c}\models \tau(a)=\bfnull$, the 
size 
of term $\tau $, although did not depend on $d$, but it still
did depend on $\phi$. The solution of that problem given in \cite{Ajt0} is not applicable to the 
present case.)
 
The solution of the problem, caused by the non-constant size of a term $F$,  is the 
main 
part of the proofs in this paper. For the solution we will use a lemma which says that the 
evaluation in $\boldm_{t}$
of  an algebraic circuit $C$ with $\calm$-operations  ($\calm$-circuits) can be evaluated by an existential  (or a universal) formula 
in
$\boldm_{v}$, provided $v\ge t+ c\log |C|$, where $c\in \omega$ is a sufficiently large
constant. In other words the input-output relation 
of such a circuit can be defined by an existential formula. 
To give a rigorous formulation of this lemma we will encode each $\calm$-circuit
by two integers.  (It is possible to encode the circuits by one integer, we use two 
integers only because it is more convenient.) We do this in the following way. We 
consider an 
$\calm$-circuit
as a directed graph  whose nodes are the gates (and the input nodes) and are labeled with the name of the corresponding operations. For the sake of simplicity we assume now
that the arity of each $\calm$ operation is $2$. 
At each  node $x$ which is not
an input node  there are exactly two incoming edges with tails, say $y,z$, one labeled with $0$
the other labeled with $1$. At $x$ we perform an $\calm$ operation assigned to $x$
on the elements which are the outputs of the gates at nodes $y$ and $z$.

	Suppose that the  $\calm$-circuit $C$ has  $m$ nodes, and the 
set of nodes is the set $\lbrace 0,1,...,m-1\rbrace $. Then $C$
can be described by three sequences. Sequence $j$ for $j=0,1$ is defined in the following way. The $i$th
element of  sequence $j$ is the tail of of the edge labeled by $j$ whose head is 
the node $i$. (If there are no incoming edges at node $i$, that is, $i$ is an input node
then the $i$th element is $0$). The definition of sequence $2$: the $i$th elements of sequence $2$ is a label which shows which $\calm$ operation must be executed at node $i$, or whether
node $i$ is an input node.  To encode the three  sequences by two  integers,
first we choose  the smallest  integer $d$ such that $m<2^{2^{d}}$ and $2^{2^{d}}$ is  also larger
  than the  number of $\calm $ operations. Then we encode
the three sequences of length $m$ by a single integer $a$ with $3 m$ digits in the $2^{2^{d}}$-ary numeral system, such that the digits of $a$ form the three
sequences. This way the $\calm$-circuit $C$ is characterized by $2$ integers the integer
$d$ that we  denote by $\Circ_{0}(C)$ and the integer $a$ that we denote by 
$\Circ_{1}(C)$.  It is important that an $\calm$-circuit $C$ can be evaluated in 
any structure $\boldm_{t}$ with $t\in \omega$, but the encoding $\Circ_{0}(C)$,
$\Circ_{1}(C)$ does not depend on the choice of the structure $\boldm_{t}$. 

We also need a way the encode the input of the circuit. Assume that we want to 
evaluate an $\calm$-circuit $C$ in the structure  $\boldm_{t}$ for some $t\in 
\omega$, and
the number of input nodes of $C$ is $k$ and the input is the sequence
$a_{0},...,a_{k-1}\in \boldm_{t}$. Then we encode this input with the single
integer $\enc_{k,t}(a_{0},...,a_{k-1})=\sum_{i=0}^{k-1}a_{i}2^{i2^{t}} $.
Here the arithmetic operations are performed as among the integers so the sum is not necessarily
in $\boldm_{t}$. 

Now we can formulate the lemma which says that the input output relation of a
$\calm$-circuit can be defined by 
an existential (or universal) formula of $\calm$ in a not too large structure 
$\boldm_{v}$. \smallskip

 {\bf Circuit Simulation Lemma.} (See Lemma \rref{E33})   { \it
There exists an existential formula $\phi(x_{0},...,x_{4})$ of $\calm$  with the 
following 
property. For  all sufficiently large $c\in \omega$,  for all
$\calm$-circuits $C$ with $k$ inputs, and  for all $t,v\in \omega$,
if 
$v\ge t+c \log|C|$ then for all $a_{0},\ldots ,a_{k},b\in \boldm_{t}$,  we 
have
that $\Circ_{0}(C)\in \boldm_{v} $,  $\Circ_{1}(C)\in \boldm_{v} $,
$\enc_{k,t}(a_{0},\ldots ,a_{k-1})=
\sum_{i=0}^{k-1}a_{i}2^{i2^{t}} \in \boldm_{v}$, and
$$\boldm_{t} \models C(a_{0},\ldots ,a_{k-1})=b \ \ \leftrightarrow \ \
\boldm_{v} \models \phi(\enc_{k,t}(a_{0},\ldots ,a_{k-1}),b ,t,\Circ_{0}(C),
\Circ_{1}(C)  ) $$} \smallskip

The lemma only states the existence of an {\it existential} formula with the required
property, but if we can define a function with an existential formula then we can also
define it by a {\it universal} formula  by simply saying, that the function does not take any 
other values.

The smallest possible $v$ guaranteed by the lemma namely $v=t+c \log|C|$ has the 
following significance. Let $m=|C|$.  Then the input of the circuit is a sequence of 
length
at most $m$ from the elements of $\boldm_{t}$. The elements of $\boldm_{v}$
can be considered as sequences of length $2^{v-t}=m^{c}$ from the elements of
$\boldm_{t}$. That is, the lemma says that if we are allowed to quantify existentially
sequences from $\boldm_{t}$ whose length is a polynomial in $m$ then we can 
define the input-output relation of the circuit $C$.  
We will prove this by showing the input-output relation of  any computation done in time 
polynomial in $m$, can be defined in $\boldm_{v}$ by an existential formula, provided
that the computation is done on a RAM with word length $2^{t}$. (That is, each word
is an element of $\boldm_{t}$.)  If $t$ is so small the with words
of length $t$ we cannot address $m $ memory cells then the computation with time
polynomial in $m$ is done on a turing machine. 

Now we may return to the sketch of the ``Collapsing" part of the proof, which was interrupted because  we needed a tool (the Circuit Simulation Lemma) to handle the 
problem with the size of the terms in  the formula $\phi$.
Assume now that the first-order formula $\phi$ contains 
a propositional formula $H(x_{1},...,x_{k-1})=0$ whose 
size depends on $d$. The Circuit Simulation Lemma with $F\legyen H$ makes it possible 
to replace the formula 
$H(x_{1},...,x_{k-1})=0$ in $\phi$
 by an existential or universal formula $\psi$ of constant size. Since the new 
quantifier can be included in the previous quantifier block, the number of blocks is not growing.

 After these changes in $\phi$ we will get a formula  $\phi'$ which is equivalent to $\phi $ 
(at least in a larger structure $\boldm_{v}$). The formula 
 $\phi'$  contains a subformula  of the form $\exists x, F(x,y)$. By the 
indirect assumption this can be replaced by a formula $G(y)=0$ and this way we decreased the 
number of quantifier changes in $\phi$. (The formula $\phi'$ is not in prenex form, 
because of the encoding problems, but after the
replacement we take it to prenex form again.)
 
 After repeated use of the Circuit 
Simulation Lemma we have a sequence of formulas $\phi=\phi_{0}, \phi _{1},...,\phi_{k}$ 
and a 
sequence of integers $d=v_{0},v_{1},...,v_{k}$ such that  the number of quantifier blocks in 
the formulas $\phi_{i}$ is strictly decreasing with $i$, and for all $i=0,1,...,k$ we have the 
following:
for all $a\in \boldm_{d} $, $\boldm_{d}\models \phi(a)$ iff $\boldm_{v_{i}} \models \phi_{i}
(a_{i},\bfg_{i}(\phi))$.
The integer $\bfg_{i}(\phi)$ encodes the parameters of the formulas $\phi_{j}$,
$j\le i$, which arose at the
applications of the circuit simulation lemma till that stage of the proof.
 Meanwhile we are maintaining reasonable bounds on $v_{i}$ and $\length (\phi_{i})$.
Let $k$ be the smallest integer such that $\phi_{k}$ has a single block of quantifiers, that is,
$\phi_{k}$ is either universal or existential. We may assume that $\phi_{k}$ is existential 
otherwise we may work with its negation. We  may   also assume, based on the techniques 
mentioned earlier, that $\phi_{k}$ has a single
existential quantifier.  The formula $\phi$ was chosen from the set $\calh_{n}$, where 
$n=2^{d}$.
Using the upper bounds in the definition of $\calh_{n}$, on the number of quantifier blocks, their 
sizes and the length of $\phi$, and using the upper bounds in the circuit simulation lemma, we 
get that that $ v_{k}\le d+\log d$. It is easy to see that
 we may assume that $v_{k}=\lfloor d+\log d\rfloor$.
 Since $\phi_{k}$ is of the form $\phi_{k}(y) \equiv
\exists x, F_{0}(x,y)=0$ we may apply again the indirect assumption and get  a term $\tau$ of
$\calm $ such that $\depth(\tau)\le \epsilon (\log d)^{\frac{1}{2}}$, and 
for all $a\in \boldm_{d}$ the following three statements are equivalent: \\
\indent (i) \ \ $\boldm_{d} \models \phi(a)$ \\
\indent (ii) \ $\boldm_{v_{k}} \models \phi_{k}(a))$ \\
\indent (iii) $\boldm_{v_{k}} \models \tau(a,\bfg(\phi))=\bfnull $, where 
$\bfg(\phi)=\bfg_{k}(\phi)$\\
which completes the sketch of the collapsing argument.

\subsubsection{\llabel{skcircuit} Sketch of the Proof of the Circuit Simulation Lemma.}

The Circuit Simulation Lemma is an easy consequence of 
 Theorem \rref{TT8}. Theorem \rref{TT8} essentially says  that the result of  a computation
done on the machine $N_{2^{d}}$ with word length $2^{d}$ and in space and time
polynomial in $2^{d-t}$, where $d\ge t$, can be expressed by an 
existential formula of $\calm$ in the structure $\boldm_{v}$, if   $v\ge t+c(d-t)$ and $c$ is a 
sufficiently large constant.
We will apply this for the proof of the circuit simulation lemma with the parameter $t$
given in the lemma and with  $d\legyen t+c'\Circ_{0}(C)$, where $c'>0$ is a sufficiently large constant. It is an easy consequence of the definition of $\Circ_{0}(C)$
that $2^{\Circ_{0}(C)}\ge \frac{1}{2} \log|C| $.  (See Lemma \rref{EW51}.)
Therefore the evaluation of the circuit $C$ in the structure $\boldm_{t}$ trivially can be done on the RAM $N_{2^{d}}$, since an element of $\boldm_{t}$ can be stored
in a single memory cell, 
and also an element of $C$ can be stored in a single memory cell, 
the operations of $\calm_{t}$ can be performed in constant time, and the time and memory  that can be used is at least a sufficiently large polynomial of $|C|$.
So circuit evaluation (or checking that a guessed output is correct)  is polynomial time computable with respect to $\boldm $, and Theorem \rref{TT8} implies the existence of the formula with the properties required by the Circuit Simulation Lemma.

\subsubsection{\llabel{theorems678} Motivation for Theorem  \rref{TT6} and 
\rref{TT7}.}

Our final goal is to prove Theorem 
\rref{TT8}.  Theorem \rref{TT6} and \rref{TT7} can be considered as steps
in this proof.  Therefore as a motivation we look again at the statement of Theorem
\rref{TT8} which says  that the result of  a computation
done on the machine $N_{2^{d}}$ with word length $2^{d}$ and in space and time
polynomial in $2^{d-t}$, where $d\ge t$, can be expressed by an 
existential formula $\phi$ of $\calm$ in the structure $\boldm_{v}$, if   $v\ge t+c(d-t)$ and $c$ is a 
sufficiently large constant.
If we just think about what can we quantify in such a formula $\phi$ Theorem \rref{TT8} is not surprising. Indeed 
 with such a formula $\phi$ we can existentially quantify a
sequence from the elements of $\boldm_{d}$ which is of length $2^{c(d-t)}$, so which can be the whole history of the mentioned  a polynomial time computation. 
The problem arises when we want to verify that a given sequence of elements
of $\boldm_{d}$
is really a history of a computation. Such a verification have to check that we get certain elements of the sequence by arithmetic operations from other elements. This is the 
main motivation for Theorem \rref{TT6} and Theorem \rref{TT7}, since they
say that with a certain type of fromula we can simultaneously perform a large number
of arithmetic operations. There will be another problem too, namely performing 
parallel arithmetic operations as in theorems \rref{TT6} and \rref{TT7}
   is not enough, we also
have to be able  to rearrange the sequence somehow such that the operands
of the arithmetic operations are at the right places.  This problem arises 
at several different parts of the proofs so we will discuss it in more detail there.

We were speaking about sequences formed from the elements
of a structure $\boldm_{d}$. If we want to speak about such a sequence in a larger structure $\boldm_{v}$, then we have to represent it there as a single element of 
$\boldm_{v}$. We will represent a sequence $a_{0},...,a_{k-1}\in \boldm_{d}$
 by the integer $a=\sum_{i=0}^{k-1} a_{i}2^{i 2^{d}}$.
Therefore   the elements of the sequence $a_{0},...,a_{k-1}$ are the
``digits" of the integer $a$ in the  $2^{2^{d}}$-ary numeral system. The $i$th
digit of the integer $a$ in the $2^{2^{d}}$-ary system will be denoted by $a[i,d]$,
so in our example $a_{i}=a[i,d]$ for $i\in k$ and $a[i,d]=0$
for all $i\ge k$. In particular the $i$th binary bit of the natural number
 $a$ will be denoted by $a[i,0]$. 

Our first results clarify what can we define by propositional or first-order existential
formulas with such sequences. These results are all preparations for the proofs
of Theorem \rref{TT6} and Theorem \rref{TT7}

\atn{May30,1:22AM}

\subsubsection{Basic results about propositional and existential definitions
in $\boldm_{d}$.}

Let $R=\langle R_{d} \mid d\in \omega \rangle$ be a family of $k$-ary relations
where for each $d\in \omega$,  $ R_{d}$ is a $k$-ary relation on $\boldm_{d}$.
We will say that the family $R$ is uniformly propositional/existential in $\boldm$,
if the there exists a propositional/existential formula $\phi$ of $\calm$ such that for 
all $d\in \omega$
 and for all $a_{0},...,a_{k-1}\in \boldm_{d}$, $R_{d}(a_{0},...,a_{k-1})$
is equivalent to $\boldm_{d}\models \phi(a_{0},...,a_{k-1})$.  A family of 
$k$-ary	functions
$f=\langle f_{d} \mid d\in \omega\rangle$ is uniformly propositional/existential if the 
family of relations $R_{d}(x_{0},...,x_{k-1},y) \leftrightarrow R_{d}
(x_{0},...,x_{k-1})=y$ is uniformly propositional/existential. Of course
if $f(x)=y$ can be defined by an existential formula then it also can be defined
by the universal formula $\forall z, z=y\vee \neg(f(x)=y)$, so a uniformly existential
function is also uniformly universal, so we could call it a uniformly $\Delta_{1}$-function.

Note that the notion of uniformly existential is a stronger one than the notion polynomially 
existential  relations defined earlier, since here we have to define the relation
 $R_{d}$ in the structure $\boldm_{d}$ and not in a larger extension $\boldm_{v}$.
It is easy to see that every uniformly existential family is also polynomially existential.

 Assume that $f=\langle f_{d} \mid d\in \omega \rangle$ is a family of $k$-ary
function. We will say that the term $\tau$ uniformly defines the family $f$, if for
all 
$d\in \omega$, and for all $a_{0},...,a_{k-1},b\in \boldm_{d}$, we have $f_{d}
(a_{0},...,a_{k-1})=b$ iff $\boldm_{d} \models \tau(a_{0},...,a_{k-1})=b$.

This part of the proof builds up tools which make it possible to prove about 
more and  more specific families of functions and  relations that they are uniformly 
propositional or existential or uniformly can be defined by a term. Frequently 
the proof is only the simple application of one or two arithmetic operations 
of $\calm$. For example if we consider and element $a$ of $\boldm_{d}$ 
as  a $0,1$ sequences of length $2^{d}$, formed from its binary bits, that is,
$a$
is represented by the sequence $\langle a[0,0],a[1,0],...,a[2^{d}-1,0]\rangle$,
then we can perform
boolean operations component-wise on these sequences.
Moreover we may also shift such a seqeunce by a given amount in either direction
using a term. That is, there exists a term $\sigma $ such that for all 
$a\in \boldm_{d}$, $i\in 2^{d}$ and we get $\sigma(a,i)$ from $a$ by shifting $a$ with
$i$ places toward the more significant digits (and putting zeros into the empty places).
The term $\sigma(a,i)= a \bfp(i)$ (in this case $\bfp(i)=2^{i}$) is good for this purpose. If we want to shift $a$ in the other direction then we can use the term
$\div(a, \bfp(i))$.

For each $d,t\in \omega$ we will denote by $e_{d,t}$ the unique element of 
$\boldm_{d}$ with $e_{d,t}[i,t]=1$ for all $i\in 2^{d-t}$. There exists a term 
$\tau$ of $\boldm_{d}$ such that for all $d,t\in \omega$, with $d\ge t$,
 $\boldm_{d}\models\tau(t)=e_{d,t}$. For the proof of this fact we have to use only
the closed form  of the  sum of a finite  geometric
series. (See Lemma \rref{F1}.)

Using these simple facts,  about the binary component-wise boolean operations,
about the various types of shifts and about the element $e_{d,t}$ we already can 
prove about more interesting functions that they are uniformly propositional or 
existential.   A trivial but very important observation is the following. Let 
$\calb(x_{0},...,x_{k-1})$ be a boolean expression with
$k$-variables and let $a_{0},...,a_{k-1}\in \boldm_{d}$.

\begin{cond} \llabel{Y30}
 Then the relation $R_{d}$ defined by $\forall i \in 2^{d}, \calb(a_{0}[i,0],...,a_{k-1}
[i,0])$ is propositional.  \end{cond}

Indeed, if the term $\tau$ is built up from the $\calm$  operations, $\cap$ and $\caln$
the same way as $\calb$ from the boolean operations $\wedge$ and $\neg$, then
$R_{d}(a_{0},...,a_{k-1})$ holds iff $\boldm_{d} \models \tau(	a_{0},...,a_{k-1})=
-\bfegy$, since all of the binary bits of $-\bfegy$ is $1$.
This way we expressed a universal statement  about the components of  elements
in $\boldm_{d}$ 
by a propositional formula. This elimination of universal quantifier
will be very important in the proofs.  

This argument about boolean expression can be mixed with the operation shift.
As a result

\begin{cond} \llabel{Y29}
 we can express by a propositional formula a relation defined by
 $$\forall i \in 2^{d}, \calb(a_{0}[i+j_{0},0],...,a_{k-1}[i+j_{0},0],..., a_{0}
[i+j_{r-1},0],...,a_{k-1}[i+j_{r-1},0] )$$
where $\calb $ is a boolean expression with $r k$ variables, and $j_{0},...,j_{r-1}$
are integers. \end{cond}

If $u\le d$ then with this type of propositional formula we can say that
the sequences $a_{i}[0,0],...,a_{i}[2^{d}-1,0]$, $i=0,...,l-1$ describe the history of a turing machine
with a tape with $2^{u}$ cells, each containing a $0,1$ bit which works from time
$0$ till time $2^{d-u}-1$ and whose finite automaton $\cala$, directing the 
movement of the head etc., has $2^{l-2}$. states. The  contents of the cells of the tape
at time $t$,  will be given by the sequence $a_{0}[t 2^{u},0],...,a_{0}[t 2^{u}+2^{u}-1,0]$. If at time $t$ the head is a cell $j$ for some $j\in 2^{u} $, then 
$a_{1}[t 2^{u}+j, 0]=1$, and $a_{1}[t 2^{u}+i, 0]=0$
for all $i\in 2^{u}$, $i\not= j$. Finally the state of the finite automaton $\cala$ at time $t$
will be determined by the $l-2$ bits $a_{2}[t 2^{u}+j,0],...,a_{l-1}[t 2^{u}+j,0]$, where the head is at cell $j$ at time $t$. (For all $i\in 2^{u}$, $i\not=j$ we have 
$a_{2}[t 2^{u}+j,0]=...=a_{l-1}[t 2^{u}+j,0]=0$.) It is easy to see that there exists a Boolean expression $\calb$ as in statement \rref{Y29}, with $r=6$, 
$j_{0}=-1$, $j_{1}=0$, $j_{2}=1$, $j_{3}=2^{u}-1$, $j_{4}=2^{u}$, 
$j_{5}=2^{u+1}$, $k=l+1$,  $a_{0}\legyen a_{0},...,a_{l-1}\legyen a_{l-1}$,
$a_{k-1}\legyen e_{d,u}$ such that 
 $$\forall i \in 2^{d}, \calb(a_{0}[i+j_{0},0],...,a_{k-1}[i+j_{0},0],..., a_{0}[i+j_{r-1},0],...,a_{k-1}[i+j_{r-1},0] )$$
holds iff the sequence  $a_{0}[2^{u}t+j,0]$, $j\in 2^{u}$, $t\in 2^{d-u}$ is the 
history of a  turing machine with the finite automaton $\cala$, in the sense that at 
time $t$ the content of $\cell$ number $j$ is $a_{0}[2^{u}t+j,0]$.
The reason is that if at time $t$ the head is at cell $j$, then the changes from
time $t$ to $t+1$ may involve only the contents of cell 
$j-1,j$ or $j+1$ and the state of the head.
Therefore the rule defined by the finite automaton $\cala$ involves only the bits of
$a_{i}[t 2^{u}+j+\delta]$ and 	$a_{i}[(t+1) 2^{u}+j+\delta]$ for $i\in k$ and 
$\delta\in \lbrace -1,0,1\rbrace $ (not all of them are needed) and this can be 
expressed by a boolean expression $\calb$. The role of the integer $e_{d,u}$ is that
it signals if the head is at the end of the tape, where the rules of the head movement may be different than at other locations.

This argument is sufficient  the prove the NP-completeness result but we also 
use it for  other purposes. In the case of $NP$-completeness it must be aplied with 
$u=\lfloor d/c\rfloor$ where $c>1$ is a constant. The fact that ``to be a history 
of turing machine with a fixed finite automaton $\cala$" is uniformly propositional in 
$\boldm$ implies that the input-output relation for the same turing machine is 
uniformly existential, if stated in the same structure $\boldm_{v}$,	and naturally this 
remains true for the history of a nondeterministic turing machine. (If we are speaking 
about turing machines with tape length $\ell$
that are working till time $T$ then the existential formula that defines the input/output 
relation is formulated in $\boldm_{v}$ where $v\ge \log_{2} \ell +\log_{2} T$.)

For the proof  Theorem \rref{TT6} we do not use turing machines, but the techniques are 
similar to the one that were used for the proof related to them. One important difference will, be
that in the case of turing machines we needed only the binary bits, that is, $a[i,0] $
of various elements $a\in \boldm_{d}$, while in the proof of Theorem \rref{TT6}
we will need that $2^{2^{u}}$-ary digits, that is, $a[i,u]$ for some $u\le d$.  
We do not give here an outline of the proof of Theorem \rref{TT6} because it consists
of several independent lemmas related to the various operations of $\calm$. The only 
common idea in these result is the one that we have illustrated in the case of turing machines.

\subsubsection{Sketch of the proof of Theorem \rref{TT7}}

 The 
statement  of Theorem  \rref{TT7} follows from 
Theorem \rref{TT6} if $\bff\notin \lbrace \times, \div, \bfp\rbrace $.
The most important case is $\bff=\times$, once we have the theorem for $\bff=\times$ 
the $\bff=\div$ 
case is relatively easy. For $\bff=\bfp$ we have a somewhat longer proof 
 but it is conceptually simpler.  

We sketch here the basic idea of the proof of Theorem \rref{TT7} for $\bff=\times$.  

We have to show that there exists an existential formula $\psi(x_{0},x_{1},y,z,w)$ 
of $\calm$  and a $c\in 
\omega$  such that for all
$d,u\in \omega$ with $d\ge u$, and for all $a_{0},a_{1},b\in \boldm_{d}$,  
if $v\in \omega$, $v\ge u+c(d-u)$, then
the following two 
conditions are equivalent: \\ \smallskip
(i) \ \ $ \times_{d,u}(a_{0},a_{1})=b $,\\ \smallskip
(ii) \ $\boldm_{v} \models \psi(a_{0},...,a_{k-1},b,d,u)$.

 To get $\times_{d,u}(a,b) $ in 
$\boldm_{d}$ we want  to use $\times(a,b)$ that is $a b$ in $\boldm _{d}$. 
We want to get an element $h$ of 	$\boldm_{d}$, such that  $ h[i,u]=a[i,u] b[i,u]$
for all $i\in  2^{d-u}$. The choice  $h= a b$ is obviously not good since $ (a b)[i,u]$
is a linear combination of various products $a[k,u] b[l,u] $.  To separate the  products  $a[k,u]b[k,u]$ that we need from the products $a[i,u]b[j,u]$, $i\not=j$
that we do not need
we
 replace $a$ and $b$ by two other integers $F_{0}(a)$ and $F_{1}(b)$ so that they have the 
same $2^{2^{u}}$-ary digits as $a$ and $b$ only these digits are  stretched out on  longer intervals. We hope that
 this way all of the products $a[i,u]b[j,u]$ will contribute to
different digits of $F_{0}(a)F_{1}(b)$. 
 Let $s=2^{d-u}$.  We may try first  
 $F_{0}(a)=\sum_{i=1}^{s-1}a[i,u]2^{is2^{u}} $, and  $F_{1}(b)=b$,	so the distance
of the  $2^{2^{u}}$-ary  $a[i,u]$ digits in $F_{0}(a)$ is $s$. The integer $b$ is smaller than 
$2^{s2^{u}}$ therefore each product $a[k,u]b[l,u]$ will contribute to at most one digit of
$F_{0}(a)F_{1}(b)$, if we disregard the carryover. The carryover however is a problem since a
product $a[k,u]b[l,u]$ is a two digit $2^{2^{u}}$-ary number so it contributes both to  the $sk+l$th digit and
the $sk+l+1$th digit of $F_{0}(a)F_{1}(b)$.

To avoid the complications caused by the carryover problem,  we stretch out the  
sequence of digits of 
$a$ by a factor of $2s$ and  the  
sequence of digits of $b$ by factor of $2$.
That is, we have  $F_{0}(a)=\sum_{i=1}^{s-1}a[i,u]2^{2s i2^{u}}\in \boldm_{q+
(q-u)+1}$, $F_{1}(b)=\sum_{i=1}^{s-1}b[i,u]2^{2i2^{u}}\in \boldm_{q+1} $.  Now the carryover is not a problem since we care about only
the values of the digits of $F_{1}(a)F_{1}(b)$ at even numbered places, while the carryovers 
influence only digits
at  odd  numbered places.   Note here that the functions $F_{0}, F_{1}$ were defined by 
moving the $2^{2^{u}}$-ary digits into new places. We will have to show that
the functions $F_{0}$ and $F_{1}$ can be defined by an existential formula in 
$\boldm_{v}$, where $v\ge u+c(d-u)$. For the moment we accept that this 
can be done somehow and continue the computation of $\times_{d,u} $, but later we will return to this 
question. 

We have that if $w=F_{0}(a)F_{1}(b)$ then $w< 2^{(2s^{2}+2s)2^{u}}\in \boldm_{v'}$,  where $v'=u+4(d-u)$
and for all $k\in s$, $(a[k,u]b[k,u])_{\boldm_{u}}=w[2ks+2k,u]$, where
$(x y)_{\boldm_{u}}$ means that we have to take the product in $\boldm_{u}$,
that is, modulo $2^{2^{u}}$.

We define a function $F_{2}$  by 
$F_{2}(p)=\sum_{k=0}^{s-1}  p[2s k+2k,u]2^{k2^{u}} $ for all
 $p\in \boldm_{v'}$. Clearly   we have  $(F_{2}(w))[k,u]=a[k,u] b[k,u]$.
So we have shown that $F_{2}(F_{0}(a) F_{1}(b))=\times_{d,u} (a,b)$.
The function $F_{2}(p)$ is also defined by moving the $2^{2^{u}}$-ary digits 
of the integer $p$ to other places, and turning some of the digits into zeros. 
As in the case of the functions $F_{0},F_{1}$ we have to show that $F_{2}$
can be defined by an existential formula in $\boldm_{v}$. Finally it is easy
to prove that if all of the three functions $F_{0},F_{1},F_{2}$ are defined 
by an existential formula in $\boldm_{v}$, then their composition $
F_{2}(F_{0}(a)F_{1}(b))$ can be also defined by an existential formula 
in $\boldm_{v}$.

Now we show that the functions $F_{i}$, $i=0,1,2$ described above can be defined
by an existential formula in $\boldm_{v}$ provided that  $v\ge u+c (d-u)$
and $c\in \omega$ is a sufficiently large constant.
Recall that for each $0,1,2$ the value $F_{i}(p)$  was defined in the following 
way. We took the $2^{2^{u}}$-ary digits of $p$ and replaced some of them by
$0$ and took the others to new places, to get the $2^{2^{u}}$-ary form of
$F_{i}(p)$. All of this, which digits must be replaced by zero, and  where we
put 
the remaining digits,   was explicitly described in the sense that we could compute 
it in time polynomial in $s=2^{d-u}$ by a turing machine which needed only the input
$s$. Motivated by this we prove here a general statement which says that functions with this property are polynomially existential which will imply in our case that
$F_{i}$, $i=0,1,2$ are existentially definable in $\boldm_{v}$ is $c$ is a sufficiently large constant. Later we will need this result for the proof of Theorem \rref{TT8} as
well, where its very general nature will be fully used.

The result in a simple form which is sufficient for proving the required properties of 
$F_{i}$, $i=0,1,2$ is the following.

{\bf Digit Relocation Lemma} (See  also Lemma \rref{AA73})
 {\it Assume that $\lambda(x,y)$ is a function
defined for all $x\in \omega$, $y\in x$ such that the 
value of $\lambda$ is in the set $\lbrace 0,1,...,x\rbrace $, and
given the input $x,y$,
$\lambda(x,y)$ can be computed by a turing machine in time and space polynomial
in $x$. Then the family  $R= \langle R_{d,u} \mid d,u\in \omega, d\ge u\rangle$ of  binary relations is polynomially existential, where for all $d,u\in \omega$ with $d\ge u$
and for all $a,b\in \boldm_{d}$, $R_{d}(a,b)$ holds if for all
$i\in 2^{d-u}$, $b[i,u]=a[\lambda(2^{d-u},i),u]$. }

According to this lemma the integer $b$ is defined in a way that its 
$2^{2^{u}}$-ary digits are selected from the $2^{2^{u}}$-ary digits of the integer
$a$. The selection is made by a turing machine but without the knowledge of the 
integer $a$.  According to the assumptions of the lemma  the value of $\lambda(x,y)$ can be $x$. In this case if $\lambda(2^{d-u},i)=2^{d-u} $ then we get that
$b[i,u]=a[2^{d-u},u]=0$, since $a<2^{2^{d}}$. Therefore $0$ is always among the 
digits of $a$ which can be used as digits of $b$ (this is important in the case of the
functions $F_{j}$, $j=0,1,2$).  (In Lemma \rref{AA73} we formulate a somewhat more general form of this result namely we allow $\lambda $ depend on a parameter
which is an element of $\boldm_{d}$ representing a $0,1$ sequence of length $2^{d-u}$.) 

In the proof  of the Digit Relocation Lemma we will construct the integer $b$ from the integer $a$ by constructing
a sequence $\alpha_{0}=a, \alpha_{1},...,\alpha_{\nu}=b$, where $\alpha_{i}\in
\boldm_{d}$ and we get each $\alpha_{i+1}$ from $\alpha_{i}$ by one of the 
following  operations $\eta_{i,\iota}$ defined below.   In describing these operations we will consider an 
element $w$ of $\boldm_{d}$ as the sequence
$\langle  w[0,u],w[1,u],...,w[s-1,u]\rangle$
 of length $s=2^{d-u}$ from the elements
of $\boldm_{u}$. Therefore we define operations acting on such sequences
and they induce a corresponding operations on $\boldm_{d}$ as well. For each
$i\in s$, $\iota\in 4$ we define an operation $\eta_{i,\iota}$, which applied
to the sequence $x=\langle x_{0},...,x_{s-1}\rangle \in (\boldm_{u})^{s}$
gives the following:

$\eta_{i,0}(x)=\langle y_{0},...,y_{s-1}\rangle$,
where for all $j \in \lbrace 0,1,...,s-2\rbrace \bcks \lbrace i\rbrace 
$, $y_{j}=x_{j}$, and $y_{i}=0$. That is, we get $\eta_{i,0}
(x)$ from $x$ by  replacing $x_{i}$ with $0$.

 $\eta_{i,1}(x)=\langle y_{0},...,y_{s-1}\rangle$,
where for all $j \in \lbrace 0,1,...,s-2\rbrace \bcks \lbrace i,i+1\rbrace 
$, $y_{j}=x_{j}$, and $y_{i}=x_{i+1}$, $y_{i+1}=x_{i}$. That is, we get $\eta_{i,1}
(x)$ from $x$ by swapping $x_{i}$ and $x_{i+1}$.

$\eta_{i,2}(x)=\langle y_{0},...,y_{s-1}\rangle$,
where for all $j \in \lbrace 0,1,...,s-2\rbrace \bcks \lbrace i\rbrace 
$, $y_{j}=x_{j}$, and $y_{i}=x_{i+1}$. That is, we get $\eta_{i,0}
(x)$ from $x$ by replacing $x_{i}$ with $x_{i+1}$.

$\eta_{i,3}(x)=x$ that is the sequence remain unchanged. (In the detailed proof
this operation will be missing because we will reach the same effect in a different way.)

Therefore our assumption is  that a turing machine computes in 
polynomial time a sequence  of pairs $\langle i_{m},\iota_{m}\rangle$, for 
$m=0,1,...,\nu-2$ and $\alpha_{m+1}=\eta_{i_{m},\iota_{m}} \alpha_{m}$
for $m=0,1,...,\nu-2$. In other words 
$$b=\eta_{i_{\nu-2},\iota_{\nu-2}}(\ldots \eta_{i_{0},\iota_{0}}(a) \ldots )$$

To show that this whole construction is uniformly existential
we  need a turing machine, in some generalized sense, which can perform the operations  $\eta_{i,\iota}$ if the 
sequence $x=\langle x_{0},...,x_{s-1}\rangle$ is the sequence of contents of the cells.
More precisely we will consider a turing machine $\calt$ with a fixed tape length 
$s=2^{d-u}$
such that each cell contains a pair $\langle \delta, w\rangle$, where $\delta$ is  
$0,1$-sequence $\delta=\langle \delta_{0},...,\delta_{k-1}\rangle$ of length $k$, where 
$k$ is a constant and $w\in \boldm_{u}$. The finite automaton $\cala$ directing 
the head movement and the changes in the contents of the cell from time
$t$ to time $t+1$ works in the following way.
If the head is at time $t$ at cell $j$ whose content is the pair $\langle \delta,w\rangle$
$\delta\in \lbrace 0,1\rbrace_{k} $, $w\in \boldm_{u}$ then the input of the finite
automaton  $\cala$ is $\delta$, that is, the finite automaton simply does {\it not}
see the element $w$ of $\boldm_{u}$. 
Suppose that at time $t$ the content of cell $j$ is $\langle \delta^{(t)}, w_{t,j}\rangle$. We will denote the sequence $\langle w_{t,0},...,w_{t,s-1}\rangle$
by $W_{t}$. (In the detailed proof we do not   allow the 
cells to contain elements in $w\in \boldm_{u}$. We let only the turing machine compute the sequence of operations and then execute the operations on sequences
of elements from $\boldm_{u}$, and show that both steps are existentially definable. The two versions has the same basic idea, and the one that we 
sketch here is perhaps more intuitive, but we need the version of  turing machines
that we describe in the detailed proof for other purposes as well.)

 Suppose that the head is at cell
$j$ at time $t$. Then depending on the input  of $\cala$ described above, it gives an 
output which consists of three different things:

(i)
$\cala$ directs the head either to change the content of cell $j$ or leave it unchanged,

(ii) $\cala$ directs the head to stay at cell $j$ or to move either 
to cell $j+1$ or to cell $j-1$ (if the destination cell does not exist then the
head does not move)

(iii)
$\cala$ also give as an output an integer $\iota\in \lbrace 0,1,2,3\rbrace $. If the
head is at time $t$ at cell $s-1$ then $W_{t+1}=W_{t}$. If the head is is at time
$t$ at cell $i$, where $i\not= s-1$ then $W_{t+1}=\eta_{i,\iota}(W_{t})$.

This completes the definition of the turing machine $\calt$ that we will call a generalized turing machine so the word turing machine in itself will mean a turing machine
in its original sense. If the finite automaton 
$\cala$ and its initial state at time $0$ and the contents of the cells at time $0$
are given, the  rules described above uniquely determine the history of the generalized turing machine $\calt$. We will consider a generalized turing machine $\calt $ of this type  where  $W_{0}=\langle a[0,u],...,a[s-1,u]\rangle$ and the generalized turing machine
determines the type (iii) output of the automaton $\cala$ in a way that
$W_{s^{\gamma}-1}= \langle b[0,u],...,b[s-1,u]\rangle$. Since the function
$\lambda$ is computable in time polynomial in $s$, such a generalized turing machine exists
if $\gamma\in \omega$ is a sufficiently large constant.

We claim that the same way as  we have seen earlier with a  turing machine, 
where each cell contained only  a single $0,1$ bit, the history of the generalized turing machine  also can be defined by an existential
formula in $\boldm_{v}$, where $v=u+(\gamma+2)(d-u)$.
This is true in the following sense.   

When we proved the existential definability of the history of a turing machine,
then we encoded the the $0,1$ bits occurring in the cells of the machine at various
times as the binary bits of an integers $a_{i}\in \boldm_{d} $, $i \in l$ for a suitably chosen $d\in 
\omega$. Even when the integers $a_{i}$ encoded the position of the head and
the state of the head at each time, the integer $l$ remained a constant.
Now however the situation is changed since the contents of the cells are elements
of $\boldm_{u}$ so we cannot encode them with a constant number of $0,1$,
bit. To keep  the advantages   of the $0,1$-bits that can be the arguments of 
boolean expressions, and at the same time allow the encoding of sequences from the
elements  of $\boldm_{u}$, we will do the following.
The history of the generalized turing machine with tape length  $\ell=2^{d-u}$, which
works  till time $T=2^{\gamma(d-u)}$ will  be encoded
by the integers in $\boldm_{v}$, where $v=u+(\gamma+1)(d-u)$.
For example the  $\delta_{i}$ in cell $j$ at $t$ time
will be encoded by the $a_{0,i}[t \ell +j]$, where $a_{0,i}\in \boldm_{v}$. This
way even for encoding $0,1$ bits we use the $2^{2^{u}}$-ary form of integers.
Therefor to define the history in an existential way
we have to define the set of integers  $a$ in $\boldm_{v}$ whose $2^{2^{u}}$-ary bits are
all zeros and ones by an existential formula. This is not a problem since $a$ has this 
property  iff  $a\le_{v,u} e_{v,u}$, and Theorem \rref{TT6} implies that the 
family of relations $\le_{v,u}$  is uniformly
existential.

We also need to encode the position of the head and the state of the finite automaton
this will be done, as before, by the integers $a_{1},a_{2},...,a_{l-1}$, but now using their
$2^{2^{u}}$-ary digits which are only ones and zeros. Consequently if at time 
$t$ the head is at cell $j$, then $a_{1}[t \ell+j,u]=1$, $a_{1}[t\ell+j',u]=0$ for all $j'\in \ell \bcks \lbrace j\rbrace $. If $a_{1}[t \ell+j,u]=1$
$a_{2}[t2^{d-u}+j,u],...,a_{l-1}[t2^{d-u}+j,u]$ determines the
state of the finite automaton at time $t$ while  $a_{2}[t2^{d-u}+i,0]=...=a_{l-1}[t2^{d-u}+i,0]=0$ for all $i\not=j$.

In the generalized  turing machine the content of a cell is a pair
$\langle \delta ,w\rangle$ where the content of each cell is a
$0,1$-sequence $\delta=\langle \delta_{0},...,\delta_{k-1}\rangle \in \lbrace 0,1
\rbrace ^{k}$ and $w\in \boldm_{u}$.
The history of the contents $w$ of the cells
will be represented by an integer $\beta\in \boldm_{v}$ such that if at time
$t$ the content of cell $j$ is $w_{t,j}$ then $\beta[t 2^{d-u}+j,u]=w_{t,j}$.

We define
the family of $k+l+1$-ary relations relation $R_{\cala}=\langle R_{v,\cala}
\mid v\in \omega \rangle$ by: for all $v\in \omega$, and for all
$a_{0,0},...,a_{0,k-1},a_{2},...,a_{l-1},\beta,d,u\in \boldm_{v}$,
 $R_{v,\cala}
(a_{0,0},...,a_{0,k-1},a_{2},...,a_{l-1},\beta,d,u)$ iff $v\ge d\ge u$, $v=u+(\gamma+1)(u-d)$ and the sequence 
$a_{0,0},...,a_{0,k-1},a_{2},...,a_{l-1},\beta$ describes a history of the turing machine with the finite
automaton $\cala$. 

The proof of the fact that the  family of relations $R_{\cala}$ is uniformly existential
is almost the same as in the case of (non-generalized) turing machines,
since $a_{0,i}[r,u]\in \lbrace 0,1\rbrace $ for $i\in k$, $r\in \ell T$ and $a_{\nu}[r,u]
\in \lbrace 0,1\rbrace $ for $\nu=,1...,l-1$, $r\in \ell T$,
 we are able to express the
the rules defining the turing machine by boolean expressions, provided that twe 
disregard the elements $w_{t,j}\in \boldm_{u}$.
Instead of using the elements $w_{t,j}$ directly we will use the elements 
$D_{t,j,t,',j'}$ defined by $D_{t,j,t',j'}=1$ if $w_{t,j}=w_{t',j'}$  and
$D_{t,j,t',j'}=0$  if $w_{t,j}=w_{t',j'}$. Since these elements take only $0,1$
values we will be able to express evrything about the working of the turing machine by 
boolean expressions.  For the description of the rules
defining the turing machine we need $D_{t,j,t',j'}$ only in the special
cases $t'\in \lbrace t,t+1\rbrace $, $j'\in H_{j}=\lbrace j-2,j-1,j,j+1,j+2\rbrace $.
In terms of the integer $\beta$ this means that we need to know
the boolean values of the statements
$\beta[t 2^{d-u}+j, u]=\beta[t' 2^{d-u}+j']$ for $t'\in \lbrace t,t+1\rbrace $,
$j'\in H_{j}$. 
 We define an element 
$\beta_{\xi,\eta}=2^{\eta + \xi(d-u) 2^{u}}\in \boldm_{v}$, for $\xi\in \lbrace 0,1\rbrace $, $\eta\in H_{0}$. 
We get that the sequence of  $2^{2^{u}}$-ary digits of $\beta_{\xi,\eta}$ by shifting the digit sequence of $\beta$ toward the more significant digits (for negative values of $\eta$ it means shifting to the opposite direction.)
As we have seen for all relevant  values of $\xi,\eta$ there exists a term 
$\tau_{\xi,\eta}$ such that $\boldm_{v}\models \beta_{\xi,\eta}=\tau_{\xi,\eta}(\beta)$. Therefore in the existential formula that will define the relation $R_{d,\cala}$
in  $\boldm_{v}$ we can use the integers $\beta_{\xi,\eta}$, $\xi\in \lbrace 0,1\rbrace 
$, $\eta\in H$. Therefore we get the required $0,1$-bits $D_{t,j,t',j'}$, $t'\in \lbrace 
t,t+1\rbrace $, $j'\in H_{j}$ as the $2^{2^{u}}$-ary digits of the integers $\rho_{\xi,\eta,\xi'\eta'}$, where $\rho_{\xi,\eta,\xi',\eta'}$, $\xi,\xi'\in \lbrace 0,1\rbrace $,
$\eta,\eta'\in H_{0} $ is defined by
$$\boldm_{v} \models \rho_{\xi,\eta,\xi'\eta'}=
{\min}_{v,u}(2^{2^{v}}-1, \beta_{\xi,\eta}-\beta_{\xi',\eta'})$$
Since the operation ${\min}_{v,u}$ can be defined by an existential formula in 
$\boldm_{v}$ this is also true for the integers $ \rho_{\xi,\eta,\xi'\eta'}$.
Finally using the integers  $\rho_{\xi,\eta,\xi'\eta'}$ and also the integers  
$a_{0,0},...,a_{0,k-1},a_{2},...,a_{l-1}$ we are able to express the rules defining
the turing machine in the form of
$$\forall i\in 2^{v-u} \calb(a_{0,0},...,a_{0,k-1},a_{2},...,a_{l-1},\vec \rho_{\xi,\eta,\xi',\eta'}  )  $$
where $\vec \rho_{\xi,\eta,\xi',\eta'}$ is the sequence of all expressions $\rho_{\xi,\eta,\xi',\eta'}$, $\xi,\xi'\in \lbrace 0,1\rbrace $,
$\eta,\eta'\in H_{0} $.

This implies, as we have seen in statement \rref{Y30}, that the relation $R_{v,\cala}$ 
can be expressed by an existential formula in $\boldm_{v}$. Form this it is easy to get
the statement of the Theorem \rref{TT7}, since we have to say only that there exists
a history of the generalized turing machine with a given initial and final states.

\subsubsection{Sketch of the proof of Theorem \rref{TT8}}

The assumption of the theorem is that the functions $F_{d,t}$ are computable
on the RAM $N_{2_{d}}$ in time $2^{\gamma_{1}(d-t)}$ using only the first
$2^{\gamma_{1}(d-t)}$ memory cells.

Such a computation can be performed also by a circuit $C$ of size
$2^{\gamma_{2}}(d-t)$ whose gates perform operations in $\boldm_{d}$, where
$\gamma_{2}$ depends only on $\gamma$ and $C$ is given independently of the
program and input of the machine $N_{2^{d-u}}$. Moreover the circuit $C$ can be 
chosen in a way that it is computable by a turing machine $\calt$ with the intput
$2^{d-t}$ in time $2^{\gamma_{3}(d-t)}$ where $\gamma_{3}$ depend only on 
$\gamma_{1}$. 

If we want to define existentially the $F_{d,t}(a_{0},...,a_{k-1})$ then we may guess
what will be the outputs of the gates of $C$ and then verify by an existential formula
that these values are consistent with each other and the input. The verification has two
steps. Suppose that at each gate $G$ both the guessed output of gate $G$ and  given  
the (guessed) inputs of the gate $G$ are given.
We associate each gate with one of the natural numbers $0,1,...,|C|-1$, and 
for example, the sequence of outputs is represented by the integer  $\sum_{i=0}^{|C-1|}b_{i}2^{i 2^{t}}$, where $b_{i}\in \boldm_{t}$ is the guessed output 
at gate number $i$. The two (or less) inputs at each gate are encoded in a similar way.  
 Then we have to verify that the values 
which are given
more than once as inputs and outputs are the same. Since the structure of the circuit 
can be calculated by a turing machine this verification can be done using the 
Digit Relocation Lemma. The other step in the verification is that each gate performs 
correctly the operation assigned to it. Here we assume that at each gate together with
the inputs and output also the name of the operation is also given (where such a name a  natural number in the set 
$\lbrace 0,1,...,\bfk\rbrace $, where $\bfk$ is the number of $\calm$ operations
).  The assignment of the operations to the gates is encoded by an integer
$\sum_{i=0}^{|C-1|}q_{i}2^{i2^{t}}$ where $q_{i}\in \bfk$ is the name of the operation at gate $i$. (Here we assume that $2^{2^{t}}> \bfk$ but this is only a 
technical  problem that can be avoided easily.) 

Using Theorem \rref{TT7} we can perform parallel all of the operations at the all
of the gates, in the sense the we can define the result by an existential formula. Now
we have the results of all of the operations performed on the twp inputs at each gate.
With another existential formula we can check that the output at $G$ is identical
to the result which corresponds to the name of the operation assigned to gate $G$.
For this checking we use the same technique as was used to conclude the proof of 
Theorem \rref{TT7}, that is, first we express the equalities between that
various integers at gate $G$ in terms of $0,1$-bits and then use statement   
\rref{Y30}.

\subsection{\llabel{simulation} Simulation}  Our goal is to show that 
{\sl for each $\tau\in \calt_{n}$, there exists a $\lambda_{\tau}
\in \calh_{n}'$ such that for all $a,b\in \boldm_{d}$,  $\boldm_{d}\models
\lambda_{\tau}(a,b)$ is equivalent to
$\boldm_{q}\models \tau(a,b)=\bfnull $, where $q= \lfloor d+ \log d\rfloor$.}

Recall that $\calt_{n}$ is a set of terms of $\calm$ with some bounds on their sizes
and $\calh_{n}'$ is a set of first-order formulas with some restriction on the number of
quantifier changes and on the sizes  of the formulas. For the moment we 
 disregard the quantitative bounds on the term $\tau$ and the formula $\lambda$, we
consider only the following general question.

Assume that $\tau(x,y)$ is a term
of $\calm$, $a,b\in \boldm_{d}$
and we want to know whether $\boldm_{q}\models \tau(a,b)=\bfnull$, but we are 
allowed only to evaluate first-order formulas in $\boldm_{d}$, 
where $q= \lfloor d+ \log d\rfloor$. How can we do this?
Since the structure $\boldm_{d}$ is much smaller than $\boldm _{q}$ we cannot
simply perform the computation of $(\tau(a,b))_{\boldm_{q}}$
in $\boldm_{d}$. It is true that the starting points  for the computation of 
$(\tau(a,b))_{\boldm_{q}}$, namely the elements $a,b$ are in $\boldm_{d}$,
but during the computation we may get partial results 
which are in $\boldm_{q}$ but  not in $\boldm_{d}$.

In spite of this difficulty our plan is to follow the computation of 
$(\tau(a,b))_{\boldm_{q}}$ step by step in $\boldm_{d}$, that is, we want
to simulate the computation of $\tau(a,b)$ in $\boldm_{q}$ by doing something in
$\boldm_{d}$. During this simulation we have to represent the partial results, which 
are elements of $\boldm_{q}$, in  some way in $\boldm_{d}$. The structure
$\boldm_{d}$ does not have enough elements for this. So we will represent each
element $h$ of $\boldm_{q}$ by a binary relation $\eta^{(h)}$ on $\boldm_{d}$, and we will do it
in a way that all of the elements $w$ of $\boldm_{q}$ which occur as partial result 
during the computation of  $(\tau(a,b))_{\boldm_{q}}$ will be represented
by a binary relation $\eta^{(w)}$ which can be defined by a first-order formula $\phi_{w}$ on $\boldm_{d}$ in the sense that for all $x,y \in \boldm_{d} $,
we have $\eta^{(w)}(x,y)$ iff $\boldm_{d} \models \phi_{w}(x,y)$.
At the beginning of the computation that is if $w=a$ or $w=b$ or $w$ can be defined by  a constant symbol in $\boldm_{q}$, then the formula $\phi_{w}$ will be of 
constant size. As we will proceed with the computation of  $(\tau(a,b))_{\boldm_{q}}$, the formulas  $\phi_{w}$ corresponding to the partial
results $w$ will be larger and larger, at each step the number of quantifier alternations 
in $\phi_{w}$ will grow by an additive constant and the size of the formula $\phi_{w}$ by a multiplicative constant.

We will define the formulas $\phi_{w}$ in the following way. First we define $\phi_{w}$ for each
element  $w\in \boldm_{d}$, and when $\boldm_{q}\models w=\bfc$, where 
$\bfc$  is a constant symbol of $\calm$.  Then we give a general rule such that in the knowledge
of $\phi_{w}$ and $\phi_{w'} $ it will be possible to construct $\phi_{\bff(w,w')}$ or $\phi_{\bfg(w)}$
for all function binary function symbols  $\bff$ and unary function symbols $\bfg$
of $\calm$.

First we indicate how can we do this with unary relations.
The simplest solution is to represent each element of $\boldm_{q}$ by a unary relation using the binary form of $w$. For each $w\in \boldm_{q}$ let 
 $\xi^{(w)}$ be the unary relation on $\boldm_{d}$ defined by ``for all
$ i\in \boldm_{d}$,  $\xi^{(i)}(a)$ iff $w[i,0]=1$", that is, $\xi^{(i)}(a)$
holds iff the $i$th binary bit of $w$ is $1$. If $d$ is sufficiently large and 
 $q=\lfloor d+ \log d\rfloor$
we have that  $w \rightarrow \xi^{(w)}$  is a one-to-one map 
of $\boldm_{q}$ into the set of unary relations on $\boldm_{d}$.
The next question is that if we have the relations $\xi^{(w)}$ and $\xi^{(w')}$
how can we get from them the relations $\xi^{(w+w')}$, $\xi^{(ww')}$,
$\xi^{\div(w,w')}$ etc. We will see that these relations can be defined by a first-order formulas from the relations $\xi^{(w)}$ and $\xi^{(w')}$. 

We will use binary not unary relations to represent the elements of $\boldm_{q}$
but we do this only because it is technically more convenient  but the 
binary representation
that we will use is, in some sense, equivalent to the unary representation
described above.

We may think that the structures $\boldm_{t}$, $t=0,1,...$ are constructed in this order. The property of the operations of $\calm$ on these structure that we sketched 
above and will define below means that at the time when we have 
 constructed the structure $\boldm_{d}$ we are able to ``predict"
what will be the result of various operations in $\boldm_{q}$  for some $q>d$.
This motivates the term ``predictive" that we will use in the following definition.
(Recall that $\coeff_{i}(a,b)$ is the $i$th digit of $a$ in the $b$-ary numeral system.)

\begin{definition}  
1. The set of functions symbols of $\calm$ (including the constant
symbols)  will be denoted by $\fsymb(\calm)$

2. Let $\calj$ be a function. We will say
that $\boldm$ is $\calj$-predictive if the following conditions are satisfied.

\begin{cond} \llabel{*EH50.1} The function $\calj$ is a monotone increasing function 
defined on $\omega$  and with
values in $\omega$. 
\end{cond}

\begin{cond} \llabel{*EH50.2} 
For all  sufficiently large $d \in \omega$, $\calj(d)\in \boldm_{d}$ and $\calj(d)>d $. 
\end{cond}

\begin{cond} \llabel{*EH50.3}
There exists a function  defined on $\fsymb(\calm)$ assigning to each function
symbol
$f(x_{0},\ldots ,x_{k-1})$ of $\calm$,
a  first-order formula $\Phi_{f}(x,y,z,Y_{0},\ldots ,Y_{k-1})$ of $\calm$,
where $x,y,z$ are  free first-order variables and $Y_{0},\ldots ,Y_{k-1}$ are
free variables for binary relations, such that  the following holds.
For all  $d,r\in \omega$ with $d+ r\le \calj({d})$
\xev{092f}
there exists a map $a \rightarrow\eta_{d,r}^{(a)}$ of
$\universe(\boldm_{d+r})$ into the set of binary relations on
$\universe(\boldm_{d})$ with the following properties:

(i) For each  $a,u,v\in \boldm_{d}$,  we have $\eta_{d,r}^{(a)}(u,v)$
iff ``$u=0$ and $\coeff_{v}(a,2)=1$".

(ii) Suppose that $f(x_{0},\ldots ,x_{k-1})$ is a $k$-ary function symbol
of
$\calm$,   for some $k=0,1,2$ (including the constant symbols for $k=0$),
$\bar f = (f)_{\boldm_{d+r}}$, and 
$a_{0},\ldots ,a_{k-1} \in \boldm_{d+r} $.  Then for all $
u,v \in \boldm_{d} $, \xev{092g}
$\eta_{d,r}^{(\bar f(a_{0},\ldots ,a_{k-1}))}(u,v)$ iff  \\ \centerline{$\boldm_{d} \models
\Phi_{f}(u, v, r,\eta_{d,r}^{(a_{0})},\ldots ,\eta_{d,r}^{(a_{k-1})}) $.}
\vege\end{cond}
\end{definition}

The proof of the simulation statement is based on the following lemma. 

\begin{lemma} \llabel{*EF50}
{\sl Assume that $c>0$ is a real, and  $\calj(x)= \lfloor 
x+c\log x \rfloor $. Then $\boldm$ is $\calj$ predictive.} \end{lemma}

In \cite{Ajt0} a weaker result of similar nature is proved which implies 
that there exists a function $g(x)$ with $\lim_{x\rightarrow \infty }g(x)=\infty$,
such that if $\calj_{0}=x +g(x)$ then $\boldm$ is $\calj_{0}$-predictive. 
Some of the partial results of the proof given there were stronger 
than 
what was needed for the theorem formulated in \cite{Ajt0}. We get Lemma 
\ref{*EF50} by 
using the full strength of these partial results in particular about the first-order 
definability of the bits of the results of multiplication and
division between large numbers.
The proof is given in section  \rref{predictivity}.
This completes the sketches of the various parts of the theorem.

\eject
\section{\llabel{existential}  Existential and propositional families of relations on  $   
\boldm $}

A large part of our proofs consists of constructions of first-order formulas of
$\calm$ which define relations or functions in the structures $\boldm_{d}$, $d\in 
\omega$ with properties which are useful in proving our theorems. For example the 
``Collapsing
statement" and ``Simulation statement"  claim the existence of certain first-order 
formulas of $\calm $ that we will construct during our proofs.  	In spite of the fact
that Theorem \rref{TT3} and Theorem \rref{TT4} are ``non-existence" statements,
which claim that formulas of $\calm$ with given properties  do not exist
for their proof we use statements which claim the existence of formulas of $\calm$ with
other properties. (This is a typical situation in lower bound proofs.)

In the Collapsing and Simulation statements we are speaking about first-order formulas
which are interpreted in a structure $\boldm_{d}$ where $d\in \omega$. The sizes of 
these formulas may depend on $d$. In this section we consider a simpler question 
where there is only one formula. More precisely let $R_{d}$, $d\in \omega$ be a family 
of
 relations, where for all $d\in \omega$, $R_{d}$ is a $k$-ary relations on $\boldm_{d}$. If there exists a 
first-order formula $\phi(x_{0},...,x_{k-1})$ of $\calm$ such that for all
$d\in \omega$ and  for all $a_{0},...,a_{k-1}$, $R_{d}(a_{0},...,a_{k-1})$ iff
$ \boldm_{d} \models \phi(a_{0},...,a_{k-1}) $ then we will say that the formula
$\phi$ defines the family $R_{d}$. This means that the whole family can be defined
by a single first-order formula. We will say that such a family of relations is uniformly 
first-order definable. The special cases when $\phi$ is propositional or existential will be 
very important and then the corresponding families of relations will be called uniformly
propositional and uniformly existential.
  We will use similar definitions for families of functions as well.

  The importance of these notions is that there is a large number of explicitly 
defined relations and functions, which are either uniformly propositional or
uniformly existential, and we use them in the proofs of the Collapsing  
statement. This  section contains the  formulations and proofs of results of these 
types.
  
It is easy to see that for each propositional formula $P(x_{0},...,x_{k-1})$ there exists 
a term $t(x_{0},...,
x_{k-1})$ of $\calm$ such that for all $d\in \omega$, $\boldm_{d} \models \forall 
x_{0},...,x_{k-1},  P(x_{0},...,x_{k-1})\leftrightarrow t(x_{0},...,x_{k-1})=\bfnull$.
(See Lemma  \rref{AA44} below.) This implies that for all  uniformly propositional family of relations 
$R_{d}(x_{0},...,x_{k-1})$, $d\in \omega$, there exists a term $t$ of $\calm $ such that for all $a_{0},...,a_{k-1}\in \boldm_{d}$, $\boldm_{d}\models t(a_{0},...,a_{k-1})=\bfnull$
is equivalent to $R_{d}(a_{0},...,a_{k-1})$. In particular if $f_{d}$, $d\in \omega$
is a family of $k-1$-ary functions such that $R_{d}(a_{0},...,a_{k-1})$
holds iff $f_{d}(x_{0},...,x_{k-2})=x_{k-1}$ then for all  $a_{0},...,a_{k-1}\in 
\boldm_{d}$, $\boldm_{d}\models t(a_{0},...,a_{k-1})=\bfnull$
is equivalent to $f_{d}(a_{0},...,a_{k-2})=a_{k-1}$. In this case we say that the family
of functions $f_{d}$ is propositional. We will be also interested in families where
each functions	$f_{d}$ is computable by the same term.  This  requirement can be  
formulated as follows.

{\sl  There exists a term $s(x_{0},...,x_{k-2})$ of 
$\calm$ such that for all $d\in \omega$ and for all $a_{0},...,a_{k-2},a_{k-1}\in 
\boldm_{d}$, $\boldm_{d}\models s(a_{0},...,a_{k-1})=f_{d}
(a_{0},...,a_{k-2})=a_{k-1}$.}

If this last condition is satisfied then we will say that the family function $f_{d}$, $d\in 
\omega$ can be uniformly 
expressed by a term. This clearly implies
that the family of function $f_{d}$ is propositional.  In some cases we will need this 
stronger property. (When we say stronger we mean only that the definition is formally
stronger but we do not know  whether there exists a uniformly propositional family of 
function which cannot be expressed uniformly by a term.)

If $f(x_{0},...,x_{k-1})$
is a function which is defined by an existential formula $\phi(x_{0},...,x_{k-1},y)$ in  
$\boldm_{d}$, that is, for all $a_{0},...,a_{k-1},b$ we have $\boldm_{d} \models
f (\ppp{a}{k})=b \leftrightarrow \phi (\ppp{a}{k},b)$ then the function $f$
can be also defined by a universal formula namely
$\boldm_{d} \models
f (\ppp{a}{k})=b \leftrightarrow \forall x, x=b \vee \neg \phi (\ppp{a}{k},b)$.
Therefore each existential family of functions is also universal.

Most of the families of relations and functions where we will prove that they are 
uniformly existential or propositional are related to the notion of $``digits"$  
of integers in various numeral systems. For example it is easy to show that the 
family of relations $R_{d}(u,i,a,t)$ defined by
``$t\le d$ and $u$ is that $i$th digit of $a$ in the numeral system with base 
$2^{2^{t}}$" is 
propositional. The numeral systems with base $2^{2^{t}}$ has a particular importance 
for us since if $d\le t$ the a sequence $u_{0},...,u_{k-1}\in \boldm _{}$, $k=2^{d-
t}$, 
can be encoded as the sequence of $2^{2^{t}}$-ary digits of a single element $a\in 
\boldm_{d}$, namely $a=\sum_{i=0}^{k-1}u_{i}2^{2^{it}}$. 
As we have mentioned already in section \rref{sketch}
we will frequently need to encode sequences from the elements of a structure 
$\boldm_{t}$ by a single element of a larger structure $\boldm_{d}$.
A very important and characteristic example is the Circuit Simulation Lemma
whose intuitive statement was described in section \rref{sketch}. Here the a circuit,
by definition, will be the sequence of its nodes with various labelings which describe
the operations and the 	``wires" between the nodes. A circuit given this way
will be encoded by a single element  $a$ of a sufficiently large structure $\boldm_{d}$
and  the $2^{2^{t}}$-ary digits of the integer $a$ will define the sequences of nodes 
and labelings, for a suitably chosen positive integer $t<d$.

We will need also to perform operations on sequences of integers which are encoded
as the $2^{2^{t}}$-ary digits of an integer $a\in \boldm_{d}$. This is very important  
for the  Vector Property Lemma (see the formulation of this lemma in section \rref{sketch} and the explanation before the statement of the lemma).
In this section we prove the Vector Property Lemma for each of the operations
$\bff\notin \lbrace \times, \div, \bfp\rbrace $. We actually get in these cases a stronger version of the lemma with $c=0$. 
Th most problematic cases proved in this section will be the operations $\min$ and 
$\max$.

\subsection{\llabel{existential1}  Existential and propositional formulas in $\boldm_{d}$, basic properties}

\begin{definition}
 $   \func(A,B) $    will denote the set
of all functions defined on the set  $   A $    with values in the
set
 $   B $. \vege\end{definition}

\begin{definition} 
1.   Assume that $k\in \omega$ and  $R=\langle R^{(d)} \mid d\in \omega\rangle$
is a family of $k$-ary relations, such that for each $d\in\omega$, $R^{(d)}$ is a
$k$-ary relation on $\boldm	_{d}$. Then we will say that $R$ is a family of $k$-ary
relations on $\boldm$.

2. Suppose that $k\in \omega$ and
 $   R=\langle R^{(d)}
\mid d \in \omega\rangle  $    is a family of  $   k $-ary
relations on $\boldm$.
 We will \xev{204c}
say that  the family  $R$    is uniformly propositional on $\boldm$
  if there exists a propositional  formula
 $   P(x_{0},\ldots ,x_{k-1}) $    of  $   \calm$    such that,
for all $  d\in \omega $, and
for all  $   a_{0},\ldots ,a_{k-1}\in
\boldm_{d}  $,
 $   R^{(d)}(a_{0},\ldots ,a_{k-1})$ is equivalent to
$ \boldm_{d} \models P(a_{0},\ldots ,a_{k-1}) $.
In a similar way the family $R$ is called uniformly existential on $\boldm$
 if there exists an existential first-order formula
 $   \phi(x_{0},\ldots ,x_{k-1}) $    of  $   \calm$    such that,
for all $  d\in \omega $, and
for all  $   a_{0},\ldots ,a_{k-1}\in
\boldm_{d}  $,
 $   R^{(d)}(a_{0},\ldots ,a_{k-1})$ is equivalent to
$ \boldm_{d} \models \phi(a_{0},\ldots ,a_{k-1}) $.
\vege\end{definition}

\begin{definition} \kell
1.  Assume that  $f=\langle f^{(d)} \mid d\in
\omega\rangle$
is a family of $k$-ary functions, such that for each $d\in\omega$, $f^{(d)}$ is a
$k$-ary function defined on $\boldm_{d}$ and with values in $\boldm_{d}$.
Then we will say that $f$ is a family of $k$-ary
functions on $\boldm$.

 \kell  2.  Suppose that $k\in \omega$ and
 $   f=\langle f^{(d)}
\mid d \in \omega\rangle  $    is a family of  $   k $-ary
functions on $\boldm$.
 We will \xev{204c}
say that  the family  $f$    is uniformly propositional on $\boldm$
  if there exists a propositional  formula
 $   P(x_{0},\ldots ,x_{k-1},y) $    of  $   \calm$    such that,
for all $  d\in \omega $, and
for all  $   a_{0},\ldots ,a_{k-1},b\in
\boldm_{d}  $,
 $   f^{(d)}(a_{0},\ldots ,a_{k-1})=b$ is equivalent to
$ \boldm_{d} \models P(a_{0},\ldots ,a_{k-1},b) $.
In a similar way the family $f$ is called uniformly existential on $\boldm$
 if there exists an existential first-order formula
 $   \phi(x_{0},\ldots ,x_{r-1},y) $    of  $   \calm$    such that,
for all $  d\in \omega $, and
for all  $   a_{0},\ldots ,a_{k-1},b\in
\boldm_{d}  $,
 $   f^{(d)}(a_{0},\ldots ,a_{k-1})=d$ is equivalent to
$ \boldm_{d} \models \phi(a_{0},\ldots ,a_{k-1},b) $.
                                   \vege\end{definition}

\begin{definition}
1. If  $   a,t\in \omega $    and  $   i $    is an integer then we will use
the
notation  $   a[i,t]=\coeff_{i}(a,2^{2^{t}}) $.     (By the
definition of
the
function  $   \coeff $    if  $   i $    is negative then  $   a[i,t]=0 $.    ) E.g.,
the
 $   i $th
binary bit of the natural number  $   a $    is  $   a[i,0] $.

2. The elements of  $   \boldm_{d} $    are
natural numbers, in the set  $   \lbrace 0,1, \ldots
,2^{2^{d}}-1\rbrace  $,      but
sometimes it is useful to represent them as sequences of
various
types. For each fixed  $   d\in \omega $    and  $   p\in \lbrace
0,1, \ldots ,d\rbrace  $    each element  $   a\in \boldm_{d} $    will
be
represented
by a sequence     \xev{201b}
 $   \langle
a[0,p],a[1,p], \ldots ,a[2^{d-p}-1,p]\rangle  $,      that we will denote by
 $   \ltort a,d,p\rtort  $.     This is a sequence of length
 $   2^{d-p} $
whose
elements are the  $   2^{2^{p}} $-ary digits of the natural
number  $   a $.
 \vege\end{definition}

\begin{lemma} \llabel{B0.55}
There exist binary terms  $   h,g $    of
 $   \calm $    such that for all  $   d\in \omega $    and for all  $   a,b $    in $
\boldm_{d} $,

\begin{cond} \llabel{B0.56}
  $   a=b $    implies  $   \boldm_{d} \models
g(a,b)=\bfegy $
and
 $   a \not= b $    implies  $   \boldm_{d} \models g(a,b)=\bfnull $.  \end{cond}

\begin{cond} \llabel{B0.57}
  $   a<b $    implies  $   \boldm_{d} \models
h(a,b)=\bfegy $
and
 $   a \ge b $    implies  $   \boldm_{d} \models h(a,b)=\bfnull $.  \end{cond}
  \end{lemma}

Proof.  The definitions of the terms are $g(x,y)=\bfegy - \min(\bfegy, x-y)$, and\\
 $h (x,y)=(\bfegy -g(x,y))(\bfegy -g(\min(x,y),y))$.    \enp{(Lemma \rref{B0.55})}
\vskip 5pt

For the following definition recall that the interpretation of the function symbol 
$\bfp(x)$ in the structure $\boldm_{d}$ was the function $\min(2^{n}-1,2^{x})$,
where $n=2^{d}$.

\begin{definition} We define a term $ \bfq(x)$ of $\calm$ by
$ \bfq(x)= h(x, \bfn)\bfp(x)$ where $h(x,y)$ is the term defined in Lemma
	\rref{B0.55}.  This definition implies that for all $x,z\in \boldm_{d}$, if 
$n=2^{d}$ then  \vskip 3pt
( $   \boldm_{d}\models \bfq(x)=z $   ) iff  ``either $x<2^{d}$ and  $z=\min\lbrace
2^{x},2^{n}-1\rbrace=2^{x}
 $, or $x\ge 2^{d} \wedge z=0$.\\
Therefore the terms  $\bfq(x)$ and $\bfp(x)$ take different values only if $x\ge 
2^{d}$. In this case $\bfp(x)=2^{2^{d}}-1$ and $\bfq(x)=0$. 
\vege\end{definition}

\begin{lemma} \llabel{AA44}
For each propositional formula $P(x_{0},...,x_{k-1})$ there exists a term $t(x_{0},...,
x_{k-1})$ of $\calm$ such that for all $d\in \omega$, 
$\boldm_{d} \models \forall 
x_{0},...,x_{k-1},   (P(x_{0},...,x_{k-1})\rightarrow t(x_{0},...,x_{k-1})=\bfnull)\wedge 
(\neg P(x_{0},...,x_{k-1})\rightarrow t(x_{0},...,x_{k-1}) =\bfegy)$.
\end{lemma}

We need this lemma so frequently that we will use it without a reference. \vskip 2 pt

Proof of Lemma \rref{AA44}.  We prove the lemma by induction on the number of 
logical connectives in the formula $P$. If $P$ is an atomic formula then it is of the 
form
$t_{0}=t_{1}$, where $t_{0},t_{1}$ are terms of $\calm$. If $g$ is a term with the 
properties described in Lemma \rref{B0.55} then the term $t=\bfegy- g(t_{0},t_{1})$
meets the requirements of the lemma. 

Suppose that $P=\neg P'$ and the term $t'$ is chosen so that  the conditions of the 
Lemma are
satisfied with $P\legyen P'$, $t \legyen t'$. Then $t=\bfegy- t'$ is the required term.
Assume that $P=P_{0}\vee P_{1}  $, and the terms $t_{i}$, $i=0,1$ are chosen so 
that  the conditions of the Lemma are
satisfied with $P\legyen P_{i}$, $t \legyen t_{i}$ for $i=0,1$. Then $t=t_{0}t_{1}$
meets the requirements of the lemma. The remaining logical connectives can be 
expressed as combinations of $\neg$ and $\vee$.
 \enp{(Lemma \rref{AA44})}

\begin{lemma} \llabel{AA34} There exists a term $\kappa(x)$ of $\calm$ such that
for all $d,t\in \omega$ if $t\le d$ and  $b=2^{2^{t}}-1$ then $\boldm_{d} \models
\kappa(t)= b$.
\end{lemma}

Proof of Lemma \rref{AA34}.
The term $\bfq(\bfq(x)) - \bfegy$ satisfies the requirements of the lemma.
  (The choice $\kappa(x)=\bfp(\bfp(x))-1$ is not satisfactory  since 
in the $t=d$ case the definition of $(\bfp)_{\boldm_{d}}$ would imply
$\boldm_{d}\models \kappa(d)=r$, where   $r=\min (2^{2^{d}}-1, 
2^{2^{d}}-1-1)=
2^{2^{d}}-2$.) 
\enp{(Lemma \rref{AA34})}

\begin{lemma} \llabel{Q0}  There exists a term $\sigma $ of $\calm$
such that for all $d,b,a,j,k, \alpha_{0},...,\alpha_{k-1}\in \omega$,
 if $\alpha_{0},...,\alpha_{k-1}< 2^{b}$, $2^{k b}-1\in \boldm_{d}$,
$a=\sum_{i=0}^{k-1}\alpha_{i}2^{i b}$, and $j\in k$, then 
$\boldm_{d}\models \alpha_{j}=\sigma(a,b,j) $.
\end{lemma}

Proof of Lemma \rref{Q0}. We have that for all $l\in k$ if $S_{l}=\sum_{i=0}^{l-1} 
\alpha_{i} 2^{i b}$, then $\boldm_{d}\models S_{l}=a-\div(a,2^{l b})$.
Since $\alpha_{j}=\frac{S_{j}-S_{j-1}}{2^{(j-1)b}}$ this implies our statement.
\enp{(Lemma \rref{Q0})}
\vskip 5pt

In the following definition if $0\le t\le d$, then  for each an element  $a\in 
\boldm_{d}$ we consider
the sequence $\ltort a,d,t \rtort$, that is, the sequence
of  $2^{2^{t}}$-ary digits of the integer $a$. We define a unary operation 
$\shift_{d,t,i}(a)$ which shifts this sequence by $i$ places toward the more
significant places if $i\ge 0$. Those which would represent a number larger that 
$2^{2^{d}}$ 
will disappear and on the other end of the sequence the new elements will be zeros.
If $i<0$ then the shift is in the other direction with similar rules. First we consider the special case when $t=0$, the corresponding function will be denoted by $\shift_{d,i}$.

\begin{definition} Assume that  $   d\in \omega $    and  $   i $    is an
integer. We
define
a function  $   \shift_{d,i} $    on  $   \boldm_{d} $.     For all  $   a,b\in
\boldm_{d} $,
 $   \shift_{d,i}(a)=b $     iff for all   $   k\in 2^{d} $,       $
b[k,0]=a[k-i,0] $.     (Recall that $a[j,0]$ is defined for all integers
$j$, and  if  $   j $    is
negative
then  $   a[j,0]=0 $.)  For each  $   d,t \in \omega  $,       $   d\ge t $    and
integer  $   i  $,    we define a function  $   \shift_{d,t,i} $    on  $
\boldm_{d} $   :
for all  $   a,b\in \boldm_{d} $,
 $   \shift_{d,t,i}(a)=b $    iff for all  $   k\in 2^{d-t} $,       $    b[k,t]=a[k-i,t]$.     These definitions
imply that
for each  $   d,t\in \omega $    with  $   t\le d $,      if  $   i $    is an integer then
 $   \shift_{d,i}\in \func(\boldm_{d},\boldm_{d})  $
and  $   \shift_{d,t,i} \in
\func(\boldm_{d},\boldm_{d})
  $.
\vege\end{definition}

The next two lemma shows that the function shift can be defined uniformly with a 
term.
In the formulation of  Lemma \rref{Q1} it is important that we have defined the interpretation 
of the operation 
$\div$ such that for all $d\in \omega$, and $a\in \boldm_{d}$, we have $\boldm_{d} 
\models\div(a,\bfnull)=\bfnull$.

\begin{lemma} \llabel{Q1} Assume that  $   d,t\in \omega $,
 $   d\ge
t $,
 $   a,i\in
\boldm_{d} $. Then
 $   \shift_{d,t,i}(a) $    is the unique element  $   b $    of
 $   \boldm_{d} $
such that
 $   \boldm_{d} \models b=a \bfq(i)$, and  the integer  
 $   \shift_{d,t,-i}(a) $    is the unique element  $   b $    of
 $   \boldm_{d} $
such that
 $   \boldm_{d} \models   b=\div(a, \bfq(i))$.
\end{lemma}

Proof of Lemma \rref{Q1}. The statement of the lemma is an
immediate
consequence of the definitions of the function
 $   \shift_{d,t,i} $
and the
structure  $   \boldm_{d} $.      \enp{(Lemma \rref{Q1})}

\begin{lemma} \llabel{B0} \nev
There exists a  term $\tau$ of $\calm$ such that for all
 $  d,t\in \omega $, with $d\ge t$,   and for all    $   a,i\in \boldm_{d} $,     $\delta
\in
\lbrace
-\bfegy, \bfegy\rbrace
 $,    the following holds. Assume that and  an integer
$b$ is defined by   $   b=\shift_{d,t, i}(a) $ if $\delta =\bfegy$,  and
$   b=\shift_{d,t, -i}(a) $ if $\delta =-\bfegy$.  Then
 $   \boldm_{d}
\models  b=  \tau(a,i,t,\delta)  $. 
\end{lemma}

\begin{remark} In this lemma the possible values of $\delta $ are terms of $\calm$.  
Since the integer $-1$ is not in $\boldm_{d}$, we cannot use it  as 
an argument for the term $\sigma$.  The term $-\bfegy $ takes  the value 
$2^{2^{d}}-1$  in 
$\boldm_{d}$, which plays the role of $-1$.  
\vege\end{remark} \vskip 5 pt

Proof of Lemma \rref{B0}. Lemma \rref{Q1} implies that there exists a term
$\sigma_{+}$ which meets the requirements with the lemma with $\sigma \legyen 
\sigma_{+}$  provided that $\delta= \bfegy$. In a similar way there exists another 
term $\sigma_{-}$ that meets the requirements of the lemma if $\delta=-\bfegy$.
Consequently the term $\sigma= g(\delta,\bfegy) \sigma_{+}+ g(\delta,-\bfegy) 
\sigma_{-}$ meets the requirements of the lemma in all cases, where $g$ is the term
defined in Lemma \rref{B0.55}.
\enp{(Lemma
\rref{B0})}

\begin{lemma} \llabel{B0.5}   There exists a term
 $   \sigma(x,y,z,w) $    of
 $   \calm $    such that if
  $   d,i,j\in \omega $,       $   t\in d $,       $   0\le i\le
j
<2^{d-t} $,       $   a\in \boldm_{d} $,
 $   a=\sum_{k=0}^{2^{d-t}-1}a_{k}(2^{2^{t}})^{k} $
and  $   b=\sum_{k=i}^{j}a_{k}(2^{2^{t}})^{k} $    then
 $   \boldm_{d}\models b=\sigma(a,i,j,t) $.    \end{lemma}

Proof of Lemma \rref{B0.5} The statement of the lemma is a
consequence of Lemma \rref{B0}. We shift the digits of  $   a $
first
toward
the more significant digits,  in a way that some of
the
digits
which are not needed in  $   b $    disappear. Then we repeat this
in
the other
direction. More precisely. Let  $   q=2^{d-t}-j-1 $,       and let
 $   b_{0}=\shift_{d,-q,t}(\shift_{d,q,t}(a)) $.
Then  $   b=\shift_{d,-i,t}(\shift_{d,i,t}(b_{0})) $.  
By Lemma \rref{B0} the function $\shift$ can be expressed uniformly by a term of 
$\calm$.   \enp{(Lemma\rref{B0.5})}

\begin{lemma} \llabel{AA27} There exists a term $\sigma$ of $\calm$ such that 
the following holds.
Assume that
$d,t\in \omega$ with $d\ge t$,  $r=2^{d-t}$, $a_{0},...,a_{r-1}\in \boldm_{t}$ and
$S=\sum_{i=0}^{r-1} a_{i}2^{i 2^{t}}$. Then for all $i\in r$ we have
$\boldm_{d} \models \sigma(S,i,t)=a_{i}$
\end{lemma}

Proof of Lemma \rref{AA27}. This is an immediate consequence of Lemma
\rref{B0.5} in the $i=j$ special case. \enp{(Lemma \rref{AA27})}  

\begin{lemma} \llabel{KB0} $Q(x,y)$ will denote either
the relation $x=y$ or the relation $x\le y$ among the integers. The following
statement
holds in both cases.
Assume that  $   p(x_{1},\ldots ,x_{k},y,z)  ,q(x_{1},\ldots ,x_{k},y,z) $
are terms of $\calm$ and for each $d\in \omega$, $f_{d}$ is a $k+2$-ary function
on $\boldm_{d}$	defined by: for all $a_{1},\ldots ,a_{k}, u,v\in \boldm_{d}$,\\
\indent if $Q(u,v)$ \ \ then $\boldm_{d}\models f_{d}(a_{1},\ldots ,a_{k},u,v)=p(a
_{1},\ldots ,a
_{k},u,v)$,
and \\
\indent if $\neg Q(u,v)$ then $\boldm_{d}\models f_{d}(a_{1},\ldots ,a_{k},u,v)=q(a
_{1},\ldots ,a
_{k},u,v)$.\\
Then there exists a term $t(x_{1},\ldots ,x_{k},y,z) $ of $\calm$ such that
 for all $d\in \omega$ and for all $a_{1},\ldots ,a_{k}, u,v\in \boldm_{d}$,
$\boldm_{d}\models f_{d}(a_{1},\ldots ,a_{k},u,v)=t(a_{1},\ldots ,a
_{k},u,v)$.
  \end{lemma}

Proof of Lemma \rref{KB0}. Assume the  $Q $ is the relation
$x=y$ and $h,g$ are the terms whose existence are sated in Lemma \rref{B0.55}.
Then $$t(x_{1},\ldots ,x_{k},y,z) =g(y,z)p(x_{1},\ldots ,x_{k},y,z) + (\bfegy -g(y,z))
q(x_{1},\ldots ,x_{k},y,z)$$ If $Q$ is the relation $x\le y$ then the term $h$
is used in a similar way. \enp{(Lemma KB1)}

\begin{lemma} \llabel{MP0} There exist terms $\sigma (x,y), \tau (x,y)$
of the language $\calm$ such that for all $d\in \omega$ if $r,j,k\in \boldm_{d}$,
$k>1$ and
$r=2^{j} $
then the following two conditions are satisfied.
\begin{cond} \llabel{MP0.1} $r^{k+1}\le 2^{2^{d} } $, and
$S=\sum_{i=0}^{k}r^{i} $ implies $S\in \boldm_{d}$ and $\boldm_{d}
\models S=\sigma(j,k)$,
\end{cond}
\begin{cond} \llabel{MP0.2} $(k+1)r^{k}(r-1)
<2^{2^{d} } $, and
$T=\sum_{i=1}^{k}i r^{i-1} $ implies $T\in \boldm_{d}$ and $\boldm_{d}
\models T=\tau(j,k)$.
\end{cond}
\end{lemma}

Proof of Lemma \rref{MP0}.
Condition \rref{MP0.1}). $S=\sum_{i=0}^{k}r^{i}=\frac {r^{k+1} -1 }
{r-1} $, therefore the assumption $r^{k+1}\le 2^{2^{d} } $ implies $S\in \boldm
_{d}$ and $\boldm_{d}\models S=\div(\bfq((k+1)j)- \bfegy, 2^{j}- \bfegy) $.
Condition \rref{MP0.2}). The proof is similar to the previous case, but here we use
that $$
\sum_{i=1}^{k} ir^{i-1}= \frac{1-r^{k+1}}{(1-r)^{2}}-\frac{(k+1)r^{k}}{1-r}
=\frac {(k+1)r^{k}(r-1)-
r^{k+1} -1 }{(r-1)^{2}}$$
\enp{(Lemma \rref{MP0})}

\begin{lemma} \llabel{B0.6}   There exists a term  $   \eta $    of
 $   \calm $    such that for all  $   d,t,m,k\in \omega $    with  $   t\le d $,      $
m\le
2^{d-t} $,       $   k< m $   and for all  $   a\in\boldm_{d}  $    the
following
\xev{212d}
two statements are equivalent:

\begin{cond} \llabel{B0.61}
  $   \boldm_{d} \models a=\eta(t,m,k) $  \end{cond}

\begin{cond} \llabel{B0.62}
 for all  $   i\in 2^{d-t} $,      if  $   i\equiv k $     $   (\mod \ m) $    then
 $   a[i,t]=1 $,      otherwise  $   a[i,t]=0 $.\end{cond}
 \end{lemma}

Proof of Lemma \rref{B0.6}. First we consider the special
case
 $   k=0 $.
Let  $   a_{0}\in \boldm_{d} $    the unique integer so that
condition \rref{B0.62} is
satisfied with  $   a\legyen a_{0} $    and $k=0$. The integer  $   a_{0} $
can
be
expressed as the sum
of a geometric sequence, that is,
 $$   a_{0}=\sum_{j=0}^{\alpha(d,t,m)}2^{jm
2^{t}}
 $$     where  $   \alpha(d,t,m)=\lfloor (2^{d-t}-1)/m \rfloor $.
$\alpha(d,t,m)$ can be written in the form of $\div (2^{d-t}-\bfegy,m) $,
so it is a term of $\calm$. Therefore  Lemma \rref{MP0} implies that
 there exists a term $\xi(x,y)$ of $\calm$ such
that for all $d\in \omega$,  $   \boldm_{d}\models
a_{0}=\xi(m,k)$.

 Let  $   h $    be a term of
 $   \calm $    such that for all  $   d\in \omega $    and for all  $   a,b $    in $
\boldm_{d} $,       $   a<b $    implies  $   \boldm_{d} \models
h(a,b)=\bfegy $
and
 $   a \ge b $    implies  $   \boldm_{d} \models h(a,b)=\bfnull $.
  Lemma \rref{B0.55} implies the existence of such a term.

Suppose now that  $   k\in m $    is arbitrary. We construct a term
 $   \eta_{1} $   which works if  $    k+m\alpha(d,m,t)<2^{d-t} $    and
another
term
 $   \eta_{2} $    which works if  $   k+m\alpha(d,m,t)\ge 2^{d-t}  $    and then
we use Lemma \rref{KB0} to get the term
 $   \eta $.

Assume  $   a $    satisfies condition
\rref{B0.62} with a  $   k\in m $    such that
 $    k+m\alpha(d,m,t)<2^{d} $    and  $   a_{0} $    satisfies condition
\rref{B0.62}
with
 $   k=0 $.     Then  $   a=2^{k2^{t}}a_{0} $, which gives the definition of $\eta
_{1}$.

If   $    k+m\alpha(d,m,t)\ge 2^{d} $    then
 \xev{212e}    $   a=2^{k2^{t}}(a_{0}-2^{\alpha(d,m,t)2^{t}}) $, which defines
$\eta_{2}$.
\enp{(Lemma \rref{B0.6})}

\begin{lemma} \llabel{F1}
For all $d,t  \in \omega$ with $t\le d$, there exists a unique integer  $e\in \boldm_{d}$, such that
 for all  $   i\in 2^{d-t} $,
 $   e[i,t]=1  $. Moreover,  for this  integer $e$, we have $\boldm_{d} \models
 e=\div(-\bfegy, \bfq(\bfq(t))-\bfegy) $. \end{lemma}

Proof of Lemma \rref{F1}. The uniqueness follows for the facts that $e<
2^{^{2^{d} } } $ and the first $2^{d-t} $,  $2^{2^{t} } $-ary digits of $e$
are given. Adding the geometric series representing $e$ as it is done in the proof
of Lemma \rref{MP0} we get  $\boldm_{d} \models
 e=\div(-\bfegy,\bfq(\bfq(t)) -1) $. \enp{(Lemma \rref{B0.6})}

\begin{definition} For all $d,t \in \omega$ the unique element $e \in \boldm_{d}
$ with the
 properties described in Lemma \rref{F1} will be denoted by $e_{d,t}$. The term
$ \div(-\bfegy, \bfq(\bfq(x))-1) $ with the free variable $x$ will be  denoted by
$\bar e(x)$. Therefore  if $t\in \boldm _{d}$, $t\le d$ then   $ \boldm _{d} \models 
\bar
e(t)=e _{d,t}$
\vege\end{definition} \vskip 5pt

In the following definitions we introduce new operations on the elements of 
$\boldm_{d}$. These operations will be defined in the following way. A natural number 
$t\le d$ is given and for each $a\in \boldm_{d}$ we consider the vector whose 
coordinates are the $2^{2^{t}}$-ary digits of $a$. The new operations will be defined 
as vector operations performed on vectors of this type. These operations and the way
they can be uniformly defined in $\boldm$ (e.g., by a propositional formula of $\calm$)
will be  important for the proofs of Circuit Simulation Lemma and the Vector Property.  

\begin{definition} 1. Assume that  $   d,t \in
\omega $,
 $   t\le d $,       $   a\in \boldm_{d} $,      and  $   \delta\in \lbrace
0,1\rbrace
 $.     We
define  $   (a)_{d,t,\delta} $    as the unique integer  $   b\in
\boldm_{d} $    such
that
for all  $   k\in 2^{d-t} $,      if  $   k\equiv \delta $     $   (\mod \ 2)) $
then
 $   b[k,t]=a[k,t] $,      otherwise  $   b[k,t]=0 $.

2. Suppose that  $   d,t \in \omega $    and  $   t\le d $.     We define an
operation  $   a\oplus_{d,t} b $    on  $   \boldm_{d} $.     For all  $   a,b,c
\in
\boldm_{d} $,       $   a\oplus_{d,t} b=c $    iff
for all  \xev{203b}
 $   k\in 2^{d-t} $,       $   a[k,t]+b[k,t]\equiv c[k,t] $     $   (\mod \
2^{2^{t}}) $.

We define a binary operation  $   \odot_{d,t}  $  on $\boldm_{d}$   if  $d\in
\omega$ and $t\in d+1$.
     For all  $a,b\in \boldm_{d}$   $a \odot _{d,t} b$  is defined in the following way.
If $a\notin \boldm _{t}=2^{2^{t}}$ then  $a \odot _{d,t} b=0$. If
$a\in\boldm_{t} $ then for all $c \in \boldm_{d}$,
  $   a\odot_{d,t} b=c $
iff for all
 $   k\in 2^{d-t} $,       $   a \cdot b[k,t]\equiv c[k,t] $,       $   (\mod \
2^{2^{t}}) $,
where the operation {\idez} $   \cdot $   " is  the multiplication between
integers.

3.  Suppose that  $   d,t \in \omega $    and  $   t\le d $.   We define a binary operation  $   \oslash_{d,t}  $  on $\boldm_{d}$.  
     For all  $a,b\in \boldm_{d}$,   $a \oslash _{d,t} b$  is defined in the following way.
If $b=0 $ then  $a \oslash _{d,t} b=0$. If
$b \not= 0  $ then  $   a\oslash_{d,t} b $ is the unique element $c$ of
$ \boldm_{d}$,
 such that for all
 $   k\in 2^{d-t} $,       $   \lfloor a[k,t]/b \rfloor = c[k,t] $.
 \vege\end{definition}

\begin{remark} The operation  $   \oplus_{d,t} $    corresponds to
the modulo  $   2^{2^{t}} $    addition of  $   2^{d-t} $    dimensional
vectors. The
operation  $  a \odot_{d,t} b $, if we restrict $a$ to the set $\boldm _{t}$,
corresponds the multiplication of
 $   2^{d-t} $    dimensional vectors modulo  $   2^{2^{t}} $    by scalars
from
the set
 $   \boldm_{t}=2^{2^{t}}  $.   The operation $a \oslash b$ is the integer division of
each component of $a$ by the scalar $b$. In the case of $b=0$ our definition is compatible with
the interpretation of   $\div$ in $\boldm_{d}$. More precisely we have  the following. 
Suppose $d,t\in \omega$,
$d\le t$,  $a,b \in \boldm_{t}$ and $c=a \oslash_{d,t} b$. Then for all $k\in 
\boldm_{t}\models \div(a[k,t],b)=c[k,t]$. (This holds even for $b=0$.) If $b\in \boldm_{d} \bcks 
\boldm_{t}$, then $a \oslash b =0$ \vege\end{remark}

\begin{lemma} \llabel{M1} Each family of relations
 $   R=\langle
R^{(d)} \mid d\in \omega \rangle $     on
 $   \boldm $,    defined in  the conditions below, is uniformly
propositional on $\boldm$.

\begin{cond} \llabel{M1.0}   $   R^{(d)} $    is the unary relation on  $
\boldm_{d} $
defined
by $R^{(d)}(t)\leftrightarrow d
$    iff $t\le d$,
  \end{cond}

\begin{cond} \llabel{M1.1} for each boolean function  $   f $
with
two variables,  $   R_{f}^{(d)} $    is the ternary relation on  $   \boldm_{d} $
defined
by $$     R_{f}^{(d)}(a,b,c)\leftrightarrow \forall k\in 2^{d},
f(a[k,0],b[k,0])=c[k,0]$$
  \end{cond}

\begin{cond} \llabel{M1.2} for each integer  $   r $,
 $   R_{r}^{(d)} $    is the binary relation on  $   \boldm_{d} $    defined by
   $   R_{r}^{(d)}(a,b) $    iff
 $   \shift_{d,r}(a)=b  $,       \end{cond}

\begin{cond} \llabel{M1.25}
 $   R^{(d)} $    is the binary binary relation on  $   \boldm_{d} $
defined
by
   $   R^{(d)}(a,t) $    iff  $   t\le d $    and for all  $   i\in 2^{d-t} $,
 $   a[i,t]=1  $,       \end{cond}

\begin{cond} \llabel{M1.3} for each integer  $   r $,
 $   R_{r}^{(d)} $    is the ternary relation on  $   \boldm_{d} $    defined
by,  $   R_{r}^{(d)}(a,b,t) $    iff  $   t\le d $    and
 $   \shift_{d,t,r}(a)=b  $,      \end{cond}

\begin{cond} \llabel{M1.4} for each
 $   \delta \in \lbrace 0,1\rbrace  $,
 $   R^{(d)} $    is the ternary relation on  $   \boldm_{d} $    defined
by,  $   R^{(d)}(a,b,t) $    iff   $   t\le d $    and
 $   (a)_{d,t,\delta}=b $,      \end{cond}

\begin{cond} \llabel{M1.5}
   for each  $   \delta\in \lbrace 0,1\rbrace  $,
 $   R^{(d)} $    is the binary relation on  $   \boldm_{d} $    defined by,  $
R^{(d)}(a,t) $    iff  $   t\le d $    and
 $   (2^{2^{d}}-1)_{d,t,\delta}=a $,
\end{cond}

\begin{cond} \llabel{M1.6}
 $   R^{(d)} $    is the quaternary relation on  $   \boldm_{d} $    defined
by
  $   R^{(d)}(a,b,c,t) $    iff    $   t\le d $    and
\xev{190a}
  $   a \oplus_{d,t} b= c $,
 \end{cond}

\begin{cond} \llabel{M1.7}
 $   R^{(d)} $    is the quaternary relation on  $   \boldm_{d} $    defined
by
  $   R^{(d)}(a,b,c,t) $    iff   $   t\le d $,    and
  $   a \odot_{d,t} b=c $,
\end{cond}

\begin{cond} \llabel{M1.10}
$   R^{(d)} $    is the ternary relation on  $   \boldm_{d} $    defined by   $
R^{(d)}(a,b,t) $    iff   $   t\le d $    and  for
all  $   k\in 2^{d-t} $,
$   a[k,t]\equiv -b[k,t] $     $   (\mod \ 2^{2^{t}})  $.
\end{cond}
\end{lemma}

Proof of Lemma \rref{M1}.
\rref{M1.0} For all  $   t\in
\boldm_{d} $,
 $   t\le
d  $    iff  $   \boldm_{d} \models 2^{t}\le \bfn
$. Consequently  $t\le d$ is uniformly propositional on $\boldm$.

In some of the further statements of the
lemma
we use
the assumption  $   t\le d $. Since the conjunction of uniformly propositional
families of relations on $\boldm$
are also uniformly propositional on $\boldm$,
statement  \rref{M1.0}  that we may assume in all of
these cases that
 $   t\le
d $, that is, we prove the equivalence of the given relation and the relation defined by
a
propositional formula with the additional assumption $t\le d$.

\rref{M1.2} and \rref{M1.3} are immediate consequences of
Lemma \rref{B0}.

\rref{M1.1}. This is an immediate
consequence of the following two facts (a) the function
symbols
 $   \cap $    and  $   \caln $    are are interpreted as  \xev{189h}
boolean vector operations  $   \wedge $    and  $   \neg $,      on the
sequences
of
binary bits
 $   a[0,0], a[1,0], \ldots  $    on the elements  $   a $    of the
structure
 $   \boldm_{d} $,
(b) each boolean function  $   f $    can be obtained as a
composition
of the
functions  $   \wedge $    and  $   \neg $.

\rref{M1.25}. The statement follows from Lemma \rref{F1}.

\rref{M1.5}.
We have   $   (2^{2^{d}}-1)[k,0]=1 $    for all  $   k\in2^{d} $, and
$\sum_{i=0}^{2^{t-1} }2^{i}=2^{2^{t} }-1   $.
Therefore
 $$   (2^{2^{d}}-1)_{d,t,0}= (2^{2^{t} }-1 )\sum_{j=0}^{\lfloor (2^{d-t}-1)/2
\rfloor}
2^{2j2^{t} }   $$
Consequently Lemma \rref{MP0}, and
$\boldm_{d}\models 2^{d-t}=\div(\bfn,2^{t})	   $
implies the existence of the required term for $\delta=0$.
To get the term for $\delta=1$ we use the equality
 $ (2^{2^{d}}-1)_{d,t,1}=\shift_{d,t,1} ((2^{2^{d}}-1)_{d,t,0})$ and the already
proven statement \rref{M1.2}.

 \rref{M1.4}. This is a consequence of \rref{M1.5} and the
fact
that
 $   (a)_{d,t,\delta}=a\cap (2^{2^{d}}-1)_{d,t,\delta} $,      where  $   \cap $
is
the
operation defined in  $   \boldm_{d} $,      that is, the vector
operation
 $   \wedge $
performed on  the sequences of binary bits.

\rref{M1.6}. Follows from  $   a\oplus_{d,t}b =( (a)_{d,t,0} +
(b)_{d,t,0}
)_{d,t,0} +
( (a)_{d,t,1} + (b)_{d,t,1}
)_{d,t,1}
     $

\rref{M1.7}. We define $a\odot_{d,t}$ be separately for the cases $a\ge 2^{2^{t}}$ 
and $a<2^{2^{t}}$. We will show that both definition is uniformly proposition, therefore Lemma \rref{KB0} will imply our statement.

If $a\ge 2^{2^{t}}$ than $a\odot_{d,t} b=\bfnull$ which gives a propositional definition.

Assume that 	$a<2^{2^{t}}$. Then    $ a\odot_{d,t}b =( a
(b)_{d,t,0}
)_{d,t,0} +
( a (b)_{d,t,1}
)_{d,t,1}
     $. The already proven statement \rref{M1.4} implies that this is equivalent to
a propositional formula.

\rref{M1.10} The condition
{\idez}for
all  $   k\in 2^{d-t} $,
 $   a[k,t]\equiv -b[k,t] $     $   (\mod \ 2^{2^{t}})  $   " is equivalent
to
 $   a\oplus_{d,t}
b=0 $,      therefore our assertion is a consequence of \xev{195a}statement
\rref{M1.6}.
     \enp{(Lemma \rref{M1})}

\begin{lemma} \llabel{AA32} Assume that $k\in \omega$, $f(x_{0},...,x_{k-1})$
is a boolean function of $k$ variables. Then there exists a term $\tau$ of $\calm$, such 
that for all $d\in \omega$, and for all $a_{0},...,a_{k-1}\in \boldm_{d}$ we have
that for all $i\in 2^{d}$, $$\boldm_{d}\models f(a_{0}[i,0],...,a_{k-1}[i,0])=
\Bigl(\tau(a_{0},...,a_{k-1})\Bigr)[i,0]$$
\end{lemma}

Proof of Lemma \rref{AA32}. The boolean function  $f $
can be expressed using only the boolean operations $\wedge$, $\neg$.
The corresponding expression in $\calm$ using the operations $\cap$ and $\caln$
will be $\tau$. \enp{(Lemma \rref{AA32}})

\begin{lemma} \llabel{KB3}
The family of ternary relations $\langle R^{(d)}
\mid d\in \omega\rangle$,
 is uniformly propositional on  $\boldm$,  where

\begin{cond} \llabel{KB3.1}
 $   R^{(d)} $    is the ternary relation on  $   \boldm_{d} $
defined by  $R^{d)}(a,b,c) $    iff  $a+b=c$ among the integers.
\end{cond}
\end{lemma}

Proof of Lemma \rref{KB3}. For all $a,b,c \in \boldm_{d}$, $a+b=c$ (as integers)
is equivalent to $\boldm_{d} \models a+b=c \wedge a\le c \wedge b\le c$.
\enp{(Lemma \rref{KB3})}

\begin{remark}  So far we have proved about some functions and relations, that we 
defined in terms of the $2^{2^{t}}$-ary digits of integers, that they are propositional.
In particular the operations $\oplus, \odot$ were among these functions. In the 
remaining part of this section we will consider relations that are also defined in terms of 
$2^{2^{t}}$-ary bits of integers but now we will allow in the definitions statement 
which consider inequalities between the corresponding digits of two integers. For 
example such a relation is  ``$R^{(d)}(a,b,t)$ iff  (for all $k\in 2^{d-t}$, $a[k,t]\le 
b[k,t]$)" (see Lemma  \rref{F3}). We will show that this particular family of relations
is uniformly existential, and we will state many similar results which will be useful later 
in proving the Circuit Simulation Lemma.
\vege\end{remark}

\begin{lemma} \llabel{F1.5} The family of binary relations $\langle R^{(d)}
\mid d\in \omega\rangle$ is uniformly propositional on $\boldm$,  where

\begin{cond} \llabel{F1.6} $R^{(d)}$ is the  binary relation defined on $\boldm_{d}$
by $R^{(d)}(a,t)$ iff $t\le d$ and for all $i\in 2^{d-t}$, $a[i,t]$ is either $0$ or 
$2^{2^{t}}-1$.
\end{cond}
\end{lemma}

\begin{remark}
1. In this lemma the  definition of the   relation $R^{(d)}$ contains a universal 
quantification on the set $2^{d-t}$. In spite of that, we have to show that the 
definition is 
equivalent to
a propositional statement. We do this by showing that a vector of length $2^{d-t}$,
namely  a vector consisting of the $2^{2^{t}}$-ary digits of an integer depending on
$a$ is the $0$ vector.  We will use frequently this argument to eliminate of a universal 
quantifier (restricted to $2^{d-
t}$).

 2. The condition  ``for all $i\in 2^{d-t}$, $a[i,t]$ is either $0$ or 
$2^{2^{t}}-1$" is equivalent to the following: the sequence of first $2^{d}$  binary 
bits of $a$ is formed from
blocks of $0$s and $1$s each of length exactly $2^{t}$, or equivalently the value 
$a[i,0]$ depends only on $\lfloor i/ 2^{t}\rfloor	$ for all $i\in 2^{d-t}$. In the proof
we will use the fact that if we consider the same blocks for the binary bits of the 
integer $e_{d,t}$ then the least significant bit in such a block is $1$ and all of the 
other bits are zeros. 
\end{remark}

Proof of Lemma \rref{F1.5}. 
The relation $R^{(d)}(a,t)$ is equivalent to the following:
$t\le d$ and  for all $i\in 2^{d} $, if $i\not\equiv 0$ $(\mod \ 2^{t})$ then
the $i$th binary bit of $a$ is the same as the $i-1$th binary bit of $a$
(which is the $i$th bit of $2a$). Using the observation about $e_{d,t}$ in the previous
remark we get the following.
$R^{(d)}(a,t)$ holds iff $t\le d$ and  for all $i\in 2^{d} $,
  $A(i,t,a)$ holds, where $A(i,t,a)\equiv$ ``if $e_{d,t}
[i,0]\not= 1$ then $a[i,0]= (2a)[i,0]$". By Lemma \rref{F1} and Lemma 
\rref{B0.5} there exists terms 
$\eta,\xi$ of $\calm$ such that $e_{d,t}[i,0]=(\eta(i,t))_{\boldm_{d}}$
and $a[i,0]=(\xi(i,a))_{\boldm_{d}}$. Condition $A$ can be expressed by
boolean vector operations on the binary bits of the integers $e_{d,t}, a$ and $
2a$, in the sense that $A$ holds iff all of the components of the resulting 
vectors are $0$s.  Therefore Lemma \rref{AA32} implies that
the relation $A$ is propositional.
\enp{(Lemma \rref{F1.5})}

\begin{definition} Assume that $t,a,i\in \omega$. We define a function $\bit_{a,t,i}$ 
on $2^{t}$ by
$\bit_{a,t,i}(k)=(a[i,t])[k,0]$ for all $k\in 2^{t}$. 
According to this definition,  $\bit_{a,t,i}(k)$ is the $k$th binary bit of
the $i$th $2^{2^{t}}$-ary digit of the integer $a$.
We define another function $\incr_{a,t,i}(k)$, by $\incr_{a,t,i}(k)=\max_{j=0}^{k}
\bit_{a,t,i}(k) $ for all $k\in 2^{t}$. For fixed $a,t$ and $i$, $\incr_{a,t,i}(k)$ is monotone increasing
in $k$ and taking values in the set $0,1$. 
The function 
$\incr_{a,t,i}$ can be also defined   by recursion, namely,  $\incr_{a,t,i}(0)=\bit_{a,t,i}(0)$ and for all
$k\in 2^{t}-1$, $\incr_{a,t,i}(k+1)=\max \lbrace\bit_{a,t,i}(k+1), \incr_{a,t,i}
(k)\rbrace $.
 We define also a monotone decreasing function $\decr_{a,t,i}$ on 
$2^{t}$ by a similar recursion in the opposite direction
$\decr_{a,t,i}(2^{t}-1)=\bit_{a,t,i}(2^{t}-1)$ and for all
$k\in 2^{t}\bcks \lbrace 0\rbrace $, $\decr_{a,t,i}(k-1)=\max \lbrace \bit_{a,t,i}(k-1), 
\decr_{a,t,i}(k)\rbrace $. Equivalently,  for all $k\in 2^{t}$, $\decr_{a,t,i}(k) 
=\max_{j=k}^{2^{t}-1} \bit_{a,i,t}(j)$. For fixed $a,i,t$,  the function $\decr_{a,t,i}
(k)$ is monotone decreasing in k.
\end{definition}

\begin{lemma} \llabel{F1.7} Each family of relations
 $   R=\langle
R^{(d)} \mid d\in \omega \rangle $     on
 $   \boldm $,    defined in one of the conditions below, is uniformly
propositional on $\boldm$.

\begin{cond} \llabel{F1.8}   $   R^{(d)} $    is the ternary relation on  $
\boldm_{d} $
defined
by $R^{(d)}(a,t,b)$ iff $t\le d$ and for all $i\in 2^{t-d}$, $k\in 2^{t}$ we have 
$\bit_{b,t,i}(k)=\incr_{a,t,i}(k)$.
  \end{cond}

\begin{cond} \llabel{F1.9}   $   R^{(d)} $    is the ternary relation on  $
\boldm_{d} $
defined
by $R^{(d)}(a,t,b)$ iff $t\le d$ and for all $i\in 2^{t-d}$,
 $k\in 2^{t}$ we have 
$\bit_{b,t,i}(k)=\decr_{a,t,i}(k)$.
  \end{cond}
\end{lemma}

\begin{remark} The importance of this lemma is that the result of the recursive process
contained in the definition of the function $\incr_{a,t,i}$ can be verified by a 
propositional statement.  
  \vege\end{remark}

Proof of Lemma \rref{F1.7}. We use similar reasoning in this proof as in the proof
of Lemma \rref{F1.5}. Consider first of condition \rref{F1.8}. For given $d,a,t$
clearly there exists a unique $b\in \boldm_{d}$ such that for all $i\in 2^{d-t}$ and
$k\in 2^{t}$, $\bit_{b,t,i}(k)=\incr_{a,t,i}(k)$, since all the bits of $b$ are determined.
This integer $b$ is also uniquely determined by the following condition:
 ``for all $j\in 2^{d-
t}$ if $e_{d,t}[j,0]=1$, then
$b[j,0]=a[j,0]$, if $e_{d,t}[j,0]=0$ then $b[j,0]=\max \lbrace a[j,0],  b 
[j-1,0] \rbrace= \max \lbrace a[j,0], (2 b 
)[j,0] \rbrace $" using the same argument  as in the case of Lemma \rref{F1.5} 
we get that the family $R^{(d)}$ is propositional. Statement \rref{F1.9} can be proved
in a similar way. 
\enp{(Lemma \rref{F1.5})}

\begin{definition} For all $d,t\in \omega$, $t\le d$, we define two functions
$\Incr^{(d)}_{t}$, $\Decr^{(d)}_{t}$
 on
$\boldm_{d}$ with values in $\boldm_{d}$. For each $a\in \boldm_{d}$,  $\Incr^{(d)}_{t}(a)$ will be the
unique integer $b\in \boldm_{d}$ such that for all $i\in 2^{d-t} $ and $k\in 2^{t}$, we have $\bit_{b,t,i}(k)=\incr_{a,t,i}(k)$.
For each $a\in \boldm_{d}$,  $\Decr^{(d)}_{t}(a)$ will be the
unique integer $b\in \boldm_{d}$ such that for all $i\in 2^{d-t} $ and $k\in 2^{t}$, we have $\bit_{b,t,i}(k)=\decr_{a,t,i}(k)$.
\end{definition}

\begin{lemma} \llabel{AA45} The family of binary functions  $f_{0}^{(d)}$, 
$f_{1}^{(d)}$, $d\in \omega$ defined below are uniformly propositional.

\begin{cond} \llabel{AA45.0} For each $a,t\in \boldm_{d}$, if  $t\le d$ then $f_{0}^{(d)}(a,t)= \Incr^{(d)}_{t}(a)$, otherwise $f_{0}^{(d)}(a,t)=0$.
\end{cond}

\begin{cond} \llabel{AA45.1} For each $a,t\in \boldm_{d}$, if  $t\le d$ then 
$f_{1}^{(d)}(a,t)= \Decr^{(d)}_{t}(a)$, otherwise $f_{1}^{(d)}(a,t)=0$.
\end{cond}
\end{lemma}

Proof of Lemma \rref{AA45}. The lemma is an immediate consequence of Lemma  
\enp{(Lemma \rref{F1.7})}

\begin{lemma} \llabel{F2} Each family of ternary relations $\langle R^{(d)}
\mid d\in \omega\rangle$  defined below is uniformly existential on $\boldm$,  where

\begin{cond} \llabel{F2.1}
 $   R^{(d)} $    is the ternary relation on  $   \boldm_{d} $    defined by  $
R^{(d)}(a,b,t) $    iff  $   t\le d $    and for
all  $   k\in 2^{d-t} $,
 $   b[k,t]=0 $
if   $   a [k,t]=0 $,      and  $   b [k,t]=e_{d,t}[k,t] $    otherwise. \end{cond}

\begin{cond} \llabel{M1.8}
 $   R^{(d)} $    is the ternary relation on  $   \boldm_{d} $    defined by  $
R^{(d)}(a,b,t) $    iff  $   t\le d $    and for
all  $   k\in 2^{d-t} $,
 $   b[k,t]=0 $
if   $   a [k,t]=0 $,      and  $   b [k,t]=1 $    otherwise. \end{cond}
\end{lemma}

Proof of Lemma \rref{F2}. Statement \rref{F2.1}.   Assume that $t\le d$.
The definition of the relation $R^{(d)}$ implies that $R^{(d)}(a,b,t)$ holds
iff
for all $i\in 2^{d-t}$ either each binary bit
of $b[i,t]$ is $1$ or each binary bit of $b[i,t]$ is $0$, (depending on whether $a[i,t]$
 has a nonzero bit or not). 
Therefore  the definitions of the functions $\Incr^{(d)}$ and  
$\Decr^{(d)}$
imply that $R^{(d)}(a,b,t)$ iff $b=\Decr^{(d)}_{t}(\Incr^{(d)}_{t}(a))$. Therefore 
$R^{(d)}(a,b,t)$ iff
there exists a $c\in \boldm_{d}$ such that $b=\Decr^{(d)}_{t}(c)$ and $c=\Incr^{(d)}_{t}
(a)$. By Lemma \rref{AA45} the families of functions $\Incr$ and $\Decr$ are 
uniformly propositional 
therefore the family $R^{(d)}$, $d\in \omega$ is uniformly existential.

Statement \rref{M1.8} is a consequence of statement \rref{F2.1} of the present lemma 
and statement \rref{M1.1}  of Lemma \rref{M1} and Lemma \rref{F1}. Suppose that 
$t\le d$
and let $R_{0}$ be the relation defined in condition \rref{F2.1}, and let $R$ be the 
relation defined in condition \rref{M1.8}, then $R(a,b,t)$ holds iff
there exists a $c\in \boldm_{d}$ such that $R_{0}(a,c,t)$ and  for all $i\in 2^{d}$
we have that $e_{d,t}[i,0]=1$ implies that $b[i,0]=c[i,0]$ and $e_{d,t}[i,0]=0$ 
implies that $b[i,0]=0$. By Lemma \rref{F1} $e_{d,t}$ is the value of a term, and by
statement \rref{M1.1} of Lemma \rref{M1} the fact that a boolean relations holds 
between the $i$th bits of the integers $b$, $c$, and $e_{d,t}$ can be expressed
by a propositional formula. 
 \enp{(Lemma \rref{F2})}

\begin{lemma} \llabel{F3}
The families of  ternary  relations $\langle R_{i}^{(d)}
\mid d\in \omega\rangle$, $i=0,1$
defined below are uniformly existential on $\boldm$, where

\begin{cond} \llabel{M1.9a}
 $   R_{0}^{(d)} $    is the ternary relation on  $   \boldm_{d} $
defined by  $   R_{0}^{(d)}(a,b,t) $    iff  $   t\le d $    and for
all  $   k\in 2^{d-t} $,
 $   a[k,t]\not=b[k,t] $,     \end{cond}

\begin{cond} \llabel{M1.9b}
 $   R_{1}^{(d)} $    is the ternary relation on  $   \boldm_{d} $    defined by  $
R_{1}^{(d)}(a,b,t) $    iff  $   t\le d $    and   for
all  $   k\in 2^{d-t} $,
 $   a[k,t]\le b[k,t] $.
 \end{cond}
\end{lemma}

Proof of Lemma \rref{F3}.
According to statement \rref{M1.0} of Lemma \rref{M1} we may assume that
$t\le d$.

Statement \rref{M1.9a}.   Let  $   w $    be the unique element of  $   \boldm_{d} $    
such
that for all
 $   i\in 2^{d}  $,       $   w [i,0] $    is the exclusive or of    $   a [i,0] $    and  $   b
[i,0] $,
and let  $   w'=\Decr^{(d)}_{t}(Incr ^{(d)}_{t})(a)$.  
Clearly $R_{0}^{(d)}(a,b,t)$ iff $w'=e_{t}$. Therefore Lemma \rref{AA45} and 
Lemma
\rref{F1} with $t=0$  imply that the family $R_{0}^{(d)}$ is uniformly existential.

\rref{M1.9b}.
We claim that
\begin{cond} \llabel{M1.11} $   R_{1}^{(d)}(a,b,t) $     iff
 $   \boldm_{d} \models a\le b \wedge
b=a\oplus_{d,t} (b-a)$.
\end{cond}

According to  statement \rref{M1.6} of Lemma \rref{M1}
this implies our assertion.
We prove now statement  \rref{M1.11}. If  $   R^{(d)}_{1}(a,b,t) $    holds then looking at
 the binary representations of  $   a $    and  $   b $
we get that   $   a\le b $,      and   for all  $   k\in 2^{d-t}  $,  $   b [k,t]=(b [k,t]-
a[k,t])+ a[k,t]  $, where both
terms are nonnegative. Therefore the definition of the operation $\oplus_{d,t}$
implies that  $   \boldm_{d} \models a\le b \wedge
b=a\oplus_{d,t} (b-a)$ holds.
  Assume now that  $   \boldm_{d} \models a\le b \wedge
b=a\oplus_{d,t} (b-a)$  holds and let
$c=b-a$. Since $a\le b$ the integer operation and the operation in $\boldm_{d}$ gives
the same value for $c$.
We perform the integer addition  $   a+c $    in the  $   2^{2^{t} }  $-ary
number system, starting from the least significant
digits,  $   a [0,t], c [0,t] $.     As long as there
is no carryover the condition  $   b=a\oplus_{d,t} c $
 implies that  $   a [k,t]\le  b[k,t] $    as required by $R^{(d)}_{1}(a,b,t)$.
We claim that $   \boldm_{d} \models a\le b \wedge
b=a\oplus_{d,t} (b-a)$ implies that there is no carryover at all.
Assume that the first carryover occurs at the addition
  $   a [k,t]+ c[k,t] $    for some  $   k<2^{d-t}-1  $.
This implies that when we add  $   a [k+1,t]+ c[k+1,t] $
 we have to add the carryover  $   1 $.    On the other hand, because of
 $   b=a\oplus_{d,t} c $,      we have
 $   a [k+1,t]+ c[k+1,t]\equiv b[k+1,t] $     $   (\mod  \ 2^{2^{t} } ) $    so
together
with the
carryover  $   1 $    we do not get  $   b[k+1,t] $
 as the next digit, a contradiction. Assume now
that the carryover occurs at  $   k=2^{d-t}-1  $.      This however, together with
 $   b=a\oplus_{d,t} c $,
 contradicts the assumption
 $   \boldm_{d}\models  a\le b$.
\enp{(Lemma \rref{F3})}

\begin{definition} Assume that  $d,t\in \omega $,       $   d\ge t $.
 The set of all integers  $   a\in \boldm_{d} $    such that for
all  $   j\in 2^{d-t} $,       $   a[j,t]\in \lbrace 0,1\rbrace  $.
We will be called the zero-one set of  $   \boldm $    with
parameters
 $   d,t $    and
will be denoted by  $   \zo(d,t) $.
  \vege\end{definition}

\begin{lemma} \llabel{M6} For each  $   d\in \omega $,      let
 $   R^{(d)}(x,y) $    be the binary relation on  $   \boldm_{d} $
defined
by: for
each  $   a,t\in \boldm_{d} $,       $   R^{(d)}(a,t)$    iff  $   t\le d $    and
 $   a\in\zo(d,t) $.     Then the family  of binary
relations  $   R=\langle R^{(d)} \mid d\in \omega\rangle $
 is uniformly existential on $ \boldm$.      \end{lemma}

Proof of Lemma \rref{M6}. Assume that  $   d,t\in \omega $,      $   d\le t $.
\xev{196a}
Recall that  $   e_{d,t} $    is the unique element of
 $   \boldm_{d} $    such that for all  $   j\in 2^{d-t} $,
 $   e_{d,t}[j,t]=1 $.     For all  $   a\in \boldm_{d} $,      we have  $   a\in
\zo $    iff for all  $   j\in
2^{d-t} $,       $   a[j,t]\le e_{d,t}[j,t] $.
Therefore
 Lemma \rref{F1}  and statement
\rref{M1.9a}
of Lemma \rref{F3}  imply the conclusion of the lemma. \enp({Lemma
\rref{M6}})

\begin{lemma} \llabel{Q2} Assume that $f$ is  boolean function of
two variables. Then the family of relations
 $   R=\langle
R^{(d)} \mid d\in \omega \rangle $,
  is uniformly
existential on $\boldm$, where

\begin{cond} \llabel{Q2.1}  $   R^{(d)} $    is the quaternary relation on  $
\boldm_{d} $
defined
by $     R^{(d)}(a,b,c,t) $  iff $t\le d$, for all $ k\in 2^{d-t}$,
$a[k,t],b[k,t],c[k,t] \in \lbrace 0,1\rbrace$, and
$f(a[k,t],b[k,t])=c[k,t]$.
  \end{cond}
\end{lemma}

Proof of Lemma \rref{Q2}.  Lemma \rref{M6} implies that the relation
$\Phi(a,b,c,t)\equiv$\idez$t\le d$ and
$a[k,t],b[k,t],c[k,t] \in \lbrace 0,1\rbrace$" is uniformly existential on $\boldm$.
	Assume now that for some $d\in \omega$
and  $a,b,c,t\in \boldm_{d}$, and $\Phi(a,b,c,t)$ holds.  Then, using statement
\rref{M1.1} of Lemma \rref{M1} and the fact
that  for all $i\in 2^{d} $,  $e_{d,t}[i,0]=1$ if $i\equiv 0$ $(\mod \ 2^{t} )$ and
 $e_{d,t}[i,0]=0$ otherwise, we may express $f(a[k,t],b[k,t])=c[k,t]$ uniformly on
$\boldm$ by a propositional formula of $\calm$. \enp{(Lemma \rref{Q2})}

\begin{lemma} \llabel{AA30} Assume that $m\in \omega$ 
and $\calb(X_{0},...,X_{m-1})$ is a boolean expression, where $X_{0},...,X_{m-1}$
are boolean variables.
 Then the family of $m+2$-ary relations
 $   R=\langle
R^{(d)} \mid d\in \omega \rangle $     on
 $   \boldm $,     is uniformly
existential on $\boldm$, where

\begin{cond} \llabel{AA30.1}  $   R^{(d)} $    is the $m+2$-ary relation on  $
\boldm_{d} $
defined
by $     R^{(d)}(a_{0},...,a_{m-1},c,t) $  iff
 $t\le d$, for all $ k\in 2^{d-t}$,
$a_{0}[k,t],...,a_{m-1}[k,t],c[k,t] \in \lbrace 0,1\rbrace$, and
$\calb(a_{0}[k,t],...,a_{m-1}[k,t])=c[k,t]$.
  \end{cond}
\end{lemma}

Proof of Lemma \rref{AA30}. The lemma follows from Lemma \rref{Q2} and Lemma 
\rref{M6}. 
Let $\calb=\calb_{0},\calb_{1},...,\calb_{s-1}, \calb_{s},...,\calb_{s+m-1} $ be the 
sequence of all
subformulas of $\calb$, where $\calb_{0},\calb_{1},...,\calb_{s-1} $ are not variables,
 and  $\calb_{s},...,\calb_{s+m-1} $ are variables. Assume that for all $i\in s$, 
$\calb_{i}=f_{i}(\calb_{i_{0}},\calb_{i_{1}})$, where $f_{i}$ is a boolean operation
of two variables.  Suppose  that $t\le d$ and for all $ k\in 2^{d-t}$,
$a_{0}[k,t],...,a_{m-1}[k,t],c[k,t] \in \lbrace 0,1\rbrace$. Then
 $\boldm_{t} \models R^{(d)}(a_{0},...,a_{m-1},c,t) $ iff
exists $u_{0},...,u_{s+m-1}$ such that for all $i=s,...,s+m-1$ $a_{i}=u_{i}$, and 
for all $i\in s$ and $k\in 2^{d-t}$, $u_{i}[k,t]=f_{i}(u_{i_{0}}[k,t],u_{i_{1}}[k,t])$.
Lemma \rref{Q2} and Lemma \rref{M6} imply that this condition can be expressed 
uniformly by an 
existential formula of $\calm$.
 \enp{(Lemma \rref{AA30})}
\vskip 5pt

The following lemma implies the Vector Property (formulated in section \rref{sketch})
for the operations $\max$ and $\min$.

\begin{lemma} \llabel{F4}
The families of quaternary relations $\langle R_{i}^{(d)}
\mid d\in \omega\rangle$,  $i=0,1$
are uniformly existential on $\boldm$, where

\begin{cond} \llabel{F4.1}
 $   R_{0}^{(d)} $    is the quaternary relation on  $   \boldm_{d} $
defined by  $R_{0}^{(d)}(a,b,u,t) $    iff  $   t\le d $    and for
all  $   k\in 2^{d-t} $,
  $ u[k,t]=\min\lbrace a[k,t], b[k,t] \rbrace $,  and

$   R_{1}^{(d)} $    is the quaternary relation on  $   \boldm_{d} $
defined by  $R_{1}^{(d)}(a,b,u,t) $    iff  $   t\le d $    and for
all  $   k\in 2^{d-t} $,
    $u[k,t]=\max\lbrace a[k,t], b[k,t] \rbrace $.
     \end{cond}
\end{lemma}

\begin{remark} For the proof of Lemma \rref{F4} we need two other lemmas.
In these lemmas we show that  we  can define in a uniformly existential way
 $0,1$-valued functions  on the set $2^{d-t}$ which select the values $k\in 2^{d-t}$
where $a[k,t]\not= b[k,t]$ or where $a[k,t]\le b[k,t]$. In the proof of Lemma
\rref{F4} these and similar $0,1$-valued functions, which can be represented by a single element of $\boldm_{d}$, will be existentially quantified.
\vege\end{remark}

\begin{lemma} \llabel{KB1}
The family of quaternary relations $\langle R^{(d)}
\mid d\in \omega\rangle$,
 is uniformly existential on $\boldm$, where

\begin{cond} \llabel{KB1.1}
 $   R^{(d)} $    is the quaternary relation on  $   \boldm_{d} $
defined by  $R^{(d)}(a,b,w,t) $    iff  $   t\le d $    and for
all  $   k\in 2^{d-t} $,
 $   a[k,t]=b[k,t]\rightarrow w[k,t]=0 $, and
$   a[k,t]\not=b[k,t]\rightarrow w[k,t]=1 $
     \end{cond}
\end{lemma}

Proof of Lemma \rref{KB1}. As in the previous lemmas we will assume that $t\le d$.
Let $\beta$ be the unique element of $\boldm_{d}$  such that
for all $   k\in 2^{d-t} $,  $   \beta[k,t]\equiv -b[k,t] $ $(\mod \ 2^{2^{t} } )$,
 and let $c=a \oplus_{d,t}\beta$. Conditions \rref{M1.10} and \rref{M1.6} of
Lemma \rref{M1} imply that $c$ has a uniform propositional definition, that is a 
propositional formula $P(x,y,z,s)$ such that $\boldm_{d} \models \forall x, P(x,a,b,t) 
\leftrightarrow x=c $, where $P$ does not depend on $d,t,a$ or $b$.  
Clearly for all $   k\in 2^{d-t} $,   $   c[k,t]=0 \rightarrow w[k,t]=0 $, and
$   c[k,t]\not=0\rightarrow w[k,t]=1 $ and so $R^{(d)}(a,b,w,t)$ iff  
$$\boldm_{d}\models
t\le d\wedge \exists c, P(c,a,b,t) \wedge   \forall k<2^{t} ,  \Psi(c,w,k,t)$$
where $$ \Psi(c,w,k,t) \equiv
 (c[k,t]=0 \rightarrow w[k,t]=0) \wedge (c[k,t]\not=0\rightarrow w[k,t]=1) $$     
Therefore the statement of the lemma
follows from condition \rref{M1.8} Lemma \rref{F2} with $a\legyen c$ and 
$b\legyen w$. \enp{(Lemma \rref{KB1})}

\begin{lemma} \llabel{KB2}
The family of quaternary relations $\langle R^{(d)}
\mid d\in \omega\rangle$,
 is uniformly existential on $\boldm$, where

\begin{cond} \llabel{KB2.1}
 $   R^{(d)} $    is the quaternary relation on  $   \boldm_{d} $
defined by  $R^{d)}(a,b,w,t) $    iff  $   t\le d $    and for
all  $   k\in 2^{d-t} $,
 $   a[k,t]\le b[k,t]\rightarrow w[k,t]=0 $, and
$   a[k,t]>b[k,t]\rightarrow w[k,t]=1 $.
     \end{cond}
\end{lemma}

Proof of Lemma \rref{KB2}.
As in the previous proofs we  may assume that
$t\le d$.
 First we prove the lemma for the
modified relation $\bar R^{(d)} $ defined by,

\begin{cond} \llabel{KB2.2}
 $\bar R^{(d)} (a,b,w,t)$ iff
$R^{(d)}(a,b,w,t) \wedge (a= (a)_{d,t,0}) \wedge  (b= (b)_{d,t,0})$.
\end{cond}

Let $c$
be the unique element of $\boldm_{d}$ such that $a \oplus_{d,t} c= b$.
As we have seen in the proof of Lemma \rref{KB1} the element $c$ has a uniform
propositional definition. It is a consequence of condition \rref{KB2.2} and the definition
of   $c$ that

\begin{cond} \llabel{KB2.3}  $a+c=b$ (among the integers) iff
for each even $k\in 2^{d-t} $, $a[k,t]\le b[k,t]$.
\end{cond}

Lemma \rref{KB3} implies that \idez$a+c=b$ among the integers" is a uniformly
propositional relation on $\boldm$
therefore
{\idez}for each odd $k\in 2^{d-t} $, $a[k,t]\le b[k,t]$"$\equiv \exists c, a+c=b$ is uniformly existential. This
completes the proof for the family of relations $\bar R^{(d)}$.  The same proof
works also for the relation
$\tilde R^{(d)} (a,b,w,t)$ iff
$R^{d)}(a,b,w,t) \wedge (a= (a)_{d,t,1}) \wedge  (b= (b)_{d,t,1})$.
 We have $a=
(a)_{d,t,0}+ (a)_{d,t,1}$, $b=
(b)_{d,t,0}+ (b)_{d,t,1}$. First we apply the already proven part of the lemma for the
pair $(a)_{d,t,0}, (b)_{d,t,0}$ and get a uniformly existential definition for the
corresponding element  $w$ that we will denote by $\bar w$. Using the pair
$(a)_{d,t,1},
(b)_{d,t,1}$ in a similar way we get an existential definition for $\tilde w$.
We have $R(a,b,w,t)$ iff $\boldm_{d} \models w=\bar w
+\tilde w$ which together with the existential defintions of $\bar w$ and $\tilde w$
gives the existential formula for $R^{(d)}$.
\enp{(Lemma \rref{KB2})}

Proof of Lemma \rref{F4}. 
As in the previous lemmas we may assume that
$t\le d$.
 We consider first the family of relations $R_{0}^{(d)}$. 
 The existential formula
defining the relation
$R_{0}^{(d)}$ will be  equivalent to the following statement. \vskip 5pt 

{\sl
There exists $v,w,w'\in
\boldm_{d}$ such that, \\
(i) for all $k\in 2^{d-t}$, $ a[k,t]\le b[k,t] \rightarrow v[k,t]=0$, and   $ a[k,t]>
b[k,t]
\rightarrow v[k,t]=1$, and \\
(ii) for all $k\in 2^{d-t}$, $ a[k,t]= u[k,t] \rightarrow w[k,t]=0$, and   $ a[k,t]\not
= u[k,t]
\rightarrow w[k,t]=1$, and \\
(iii) for all $k\in 2^{d-t}$, $ b[k,t]= u[k,t] \rightarrow w'[k,t]=0$, and   $ b[k,t]\not
=u[k,t]
\rightarrow w'[k,t]=1$, and \\
(iv) for all $k\in 2^{d-t}$, $ w[k,t]=0 \rightarrow v[k,t]=0$, and   $ w[k,t]=1
\rightarrow (w'[k,t]=0 \wedge v[k,t]=1)$.}
\vskip 5 pt

We claim that this statement is equivalent to $R_{0}(a,b,u,t)$.
First we show that if $u[k,t]=\min \lbrace a[k,t],b[k,t]\rbrace $ for all $k
\in 2^{d-t}$, then there exist $v,w,w'\in \boldm_{d}$ satisfying conditions
(i),(ii),(iii)
and (iv).  We define for each fixed $k\in 2^{d-t}$, the integer $v[k,t]\in \lbrace 
0,1\rbrace $ by condition (i). This gives a $v\in \boldm_{d}$ satisfying condition
(i). In a similar way we define the integer $w\in \boldm_{d}$ by condition (ii) and 
the integer $w'\in \boldm_{d}$ by condition (iii). We have to show that the integers
$v,w,w'$ defined this way satisfy condition (iv). Suppose that a $k\in 2^{d-t}$ is fixed.

If $w[k,t]=0$ then by condition (ii) $a[k,t]=u[k,t]=$ $\min \lbrace a[k,t],b[k,t]\rbrace
\le b[k,t] $ and therefore by condition (i) $v(k,t)=0$.

If $w[k,t]=1$ then by condition (ii) $a[k,t]\not= u[k,t]=\min \lbrace 
a[k,t],b[k,t]\rbrace $. Therefore $u[k,t]=b[k,t]<a[k,t]$. Condition (iii) and 
$u[k,t]=b[k,t]$ implies $w'[k,t]=0$. Condition (i) and $b[k,t]<a[k,t]$ implies
$v[k,t]=1$, which completes the proof of condition (iv) and the fact that there exist 
integers
 $v,w,w'$ satisfying conditions (i),(ii),(iii)
and (iv). 

Assume now that there exist integers $v,w,w'\in \boldm_{d}$ satisfying conditions
 (i),(ii),(iii)
and (iv) and we show that for all $k\in 2^{d-t}$, $u[k,t]= \min \lbrace a[k,t],
b[k,t]\rbrace $. Suppose that a $k\in 2^{d-t}$
is fixed.  Statement (ii)
implies that $w[k,t]$ is either $0$ or $1$. If $w[k,t]=0$ then according to
 (iv) $v(k,t)=0$ and by (ii) $u[k,t]=a[k,t]$.   
$v(k,t)=0$ and (i) implies that $a[k,t]\le b[k,t]$  and therefore $u[k,t]= \min \lbrace a[k,t],
b[k,t]\rbrace $. 

Assume now that $w[k,t]=1$. According to
 (iv) $v(k,t)=1$ and  $w'(k,t)=0$. Therefore by (i) $a[k,t]>b[k,t]$ and by (iii) 
$b[k,t]=u[k,t]$.   
Consequently $u[k,t]= \min \lbrace a[k,t],
b[k,t]\rbrace $.  This completes the proof of the fact that $R_{0}(a,b,u,t)$ holds iff
there exist integers $v,w,w'$ satisfying conditions  (i),(ii),(iii)
and (iv).

All of the four statements (i), (ii), (iii), and (iv) are uniformly existential  on
$\boldm$. This can be proved in each case separately using the following lemmas:
statement (i): Lemma \rref{KB2}, statement (ii): Lemma \rref{KB1},
statement (iii): Lemma \rref{KB1}, statement (iv): Lemma \rref{Q2}.

The statement of the lemma  for the family of relations $R_{1}^{(d)}$  can be proved  
in a similar way, or we may use the fact that 
 $\boldm_{d} \models \forall x,y, \max(x,y)=-\min(-x,-y)$ .
\enp{(Lemma \rref{F4})}

\begin{definition} 1. Let $\bff$ be a $k$-ary function symbol of $\calm$ for some
$k\in \lbrace 0,1,2\rbrace $. For all $d,t\in \omega$ with $d\ge t$, we define
a $k$-ary  function $\Upsilon_{\bff,d,t}$ on the universe $\boldm_{d}$ in the 
following way.
Assume that $d,t\in \omega $ is fixed with $d\ge t$ and $a_{0},...,a_{k-1}\in 
\boldm_{d}$. Then
 $\Upsilon_{\bff,d,t}(a_{0},...,a_{k-1}) $ is the unique element $b\in 
\boldm_{d}$ with the property that for all $i\in 2^{d-t}$, we have $\boldm_{t} 
\models \bff(a_{0}[i,t],...,a_{k-1}[i,t])=b[i,t]$. The function $\Upsilon_{\bff,d,t}$
will be also called the parallel version of the operation $\bff$. In the special cases
$\bff=+,\times$ we will use the notation $\Upsilon_{+,d,t}=\oplus_{d,t}$
 $\Upsilon_{\times,d,t}=\otimes_{d,t}$. For the remaining function symbols $\bff$
we will sometimes write $\bff_{d,t}$ instead of $\Upsilon_{\bff,d,t}$, e.g. we may write $\div_{d,t}$,
 $\min_{d,t}$, $\cap_{d,t}$ etc.
 
2. Let $\bff$ be a $k$-ary function symbol of $\calm$ for some
$k\in \lbrace 0,1,2\rbrace $. 
We define  a family of $k+2$-ary	relations $\bupsilon_{\bff}=\langle 
\bupsilon_{\bff}^{(d)} \mid 
d\in \omega \rangle$  on $\boldm$. For all
$d\in \omega$ and for all $a_{0},...,a_{k-1},b,t \in \boldm_{d} $, $\bupsilon^{(d)}
(a_{0},...,a_{k-1},b,t )
$ iff $t\le d$ and $\Upsilon_{\bff,d,t}(a_{0},...,a_{k-1})=b$.
We will say that the parallel $\bff$ operation is uniformly existential on $\boldm$, if the 
family of relations $\bupsilon_{\bff}$ is uniformly existential on $\boldm$. 
\vege\end{definition}

\begin{lemma} \llabel{AA24} Suppose that $\bff$ is a function symbol of $\calm$
such that
$\bff \notin \lbrace \times, \div, \bfp\rbrace$. Then the parallel $\bff$ operation
is uniformly existential on $\boldm$. 
\end{lemma}

\begin{remark} 1. For some of the function symbol even more is true, in the sense that
the family of relations  $\bupsilon_{f}$ is  uniformly propositional. See e.g., statement 
 \rref{M1.6} of Lemma \rref{M1} about the parallel version of addition, which was 
denoted by $\oplus_{d,t}$.

2. For the three the function symbols $ \times, \div, \bfp  $
we will be able to prove the lemma only in a weaker form, namely
the existential formula defining the relation $\bupsilon_{\bff,d,t}$ will not be
considered in the structure $\boldm_{d}$ but in a larger structure 
$\boldm_{d+c(d-t)}$ for a sufficiently large constant $c\in \omega$.
 This generalized form of the lemma will have a crucial role
in the proof of Theorem \rref{TT3}. \vege\end{remark}

\vskip 5pt
Proof of Lemma \rref{AA24}. We show  separately for each function symbols of 
$\calm$, that the statement of the lemma is true.

Symbol $\bfnull$. $\Upsilon_{\bfnull,d,t}=b$ is equivalent to $\boldm_{d} \models 
b=\bfnull$.

Symbol $\bfegy$. $\Upsilon_{\bfegy,d,t}=b$ is equivalent to $\boldm_{d} \models 
b=e_{d,t}$, therefore Lemma \rref{F1} implies our statement.

Symbol $-\bfegy$. $\Upsilon_{-\bfegy,d,t}=b$ is equivalent to $\boldm_{d} \models 
b=-\bfegy$. (Indeed for all $i\in 2^{t}$, $\boldm_{t}\models b[i,t]=-\bfegy$ implies
that  all of the $2^{t}$ binary bits of $b[i,t]$ is $1$.) 

Symbol $+$. This follows from statement \rref{M1.6} of Lemma \rref{M1}.

Symbol $\bfn.$ $\Upsilon_{\bfn,d,t}=b$ is equivalent to $b[i,t]=2^{t}$ for all $i\in
2^{d-t}$, that is, $b=2^{t}e_{d,t}$. Therefore our statement follows from
Lemma \rref{F1}.

Symbols $\caln$ and $\cap$. The statement is an immediate consequence of the fact 
that these two operations are  boolean vector operations performed on the binary 
forms of the arguments. 

Symbols $\max$ and $\min$. The statement is equivalent to Lemma \rref{F4}.
\enp{(Lemma \rref{AA24})}

\begin{definition} For all $d,t\in \omega$	with $d\ge t$, we define a binary relations 
$<_{d,t}$ on $\boldm_{d}$, by $a<_{d,t} b$ iff for all $i\in 2^{d-t}$, $a[i,t]<b[i,t]$.
\vege\end{definition}

\begin{lemma} \llabel{AA41} The family of ternary relations $Q=\langle Q_{d} \mid d 
\in 
\omega\rangle$ is uniformly existential, where for all $d\in \omega$ and for all
$a,b,t\in \boldm_{d}$, we have $Q_{d}(a,b,t)$ iff  ``$d\ge t $ and $a<_{d,t}b$". 
 \end{lemma}

Proof of Lemma \rref{AA41}. Let $R$ be the quaternary family of relations defined in 
Lemma \rref{KB2}. Then for all $d\in \omega$ and for all $a,b,t\in \boldm_{d}$ we 
have $Q_{d}(a,b,t)$ iff $R_{d}(b,a,e_{d,t},t)$. By Lemma \rref{F1}, $e_{d,t}$ can be 
written as a $0$-ary term, therefore Lemma \rref{KB2} implies our statement.
\enp({Lemma \rref{AA41}}

\begin{lemma} \llabel{AA25}  
For each 
 boolean function $\delta(x,y)$ of two variables the family of quaternary relations
$R=\langle R^{(d)} \mid d\in \omega\rangle$
is uniformly existential where for all $d\in \omega$ with $t\le d$, and for all
$a,b,c,t\in \boldm_{d}$,
$R^{(d)}(a,b,c,t)$ iff $t\le d$ and  for all $i\in 2 ^{d-t}$,
$c[i,t]=\delta(\min (a[i,t],1), \min(b[i,t],1))$.
\end{lemma}

Proof of Lemma \rref{AA25}.  Assuming that $t\le d$, $R^{(d)}(a,b,c,t) $ is equivalent 
to the following.

\begin{cond} \llabel{AA25.1}
There exists $a',b'\in \boldm_{d}$ such that 
$a'=\Decr^{(d)}_{t}(\Incr^{(d)}_{t}(a))$,
$b'=\Decr^{(d)}_{t}(\Incr^{(d)}_{t}(a))$,
and for all $i\in 2^{d}$, either
 ``$e_{d,t}[i,0]=1$ and $c[i,0]=\delta(a'[i,0],b'[i,0])$" or
``$e_{d,t}[i,0]=0$ and $c[i,0]=0$".\end{cond}

 By Lemma \rref{AA45} the definitions of $a'$ and $b'$ are uniformly
propositional, by Lemma \rref{F1}  $e_{d,t}$ is the value of a term,
 therefore  statement \rref{M1.1} of Lemma \rref{M1} imply that
the relation described in condition \rref{AA25.1} is uniformly propositional. 
Consequently the family of relations $R^{(d)}$, $d\in \omega$ is	uniformly 
existential. (We can get another proof by  using Lemma \rref{Q2} and the fact that the 
function $\min_{d,t}$ is uniformly existential, as stated
in Lemma \rref{AA24}.)
\enp{(Lemma \rref{AA25})}

\begin{lemma} \llabel{AA46}The family of quaternary relations $R^{(d)}$, $d\in \omega$ is uniformly existential, where for all $d,a,b,c,t\in \omega$,
$R^{(d)}(a,b,c,t)$ holds iff $t\le d$ and $a\oslash_{d,t} b=c$ 
\end{lemma}

Proof of Lemma \rref{AA46}.  We may assume that $t\le d$. Then 
$a\oslash _{d,t} b=c$ iff either (($b>2^{2^{t}}$ or $b=0$) and $c=0$)  or ``$b\le 2^{2^{t}}$ and
there exists a $w\in \boldm_{d} $, such that $w<_{d,t} b$ and $a=(b\odot_{d,t} c) \oplus_{d,t} w $".
By statement \rref{M1.7} of Lemma \rref{M1} the relation $a=(b\odot_{d,t} c) \oplus_{d,t} w $ is uniformly existential, and by  Lemma \rref{AA41} the relation
$w<_{d,t} b$ is also uniformly existential. This implies that the relation $a \oslash_{d,t} b= c$ is also uniformly existential.
 \enp{(Lemma \rref{AA46})}

\begin{lemma} \llabel{Y3} Assume that  $k,m,l\in \omega$ and  $f ^{(j)}
=\langle f _{d}^{(j)} \mid d\in \omega \rangle$ are families of $k$-ary function on
$\boldm$, for
$j=0,1,\ldots ,m-1$ and  $g=\langle g _{d} \mid d\in \omega \rangle
$ is a family of $m$-ary functions on $\boldm$. Let $h$ be the family
of $k$-ary functions $h=\langle  h_{d} \mid d\in \omega \rangle$ on $\boldm$
defined by $h _{d} (a_{0},\ldots ,a_{k-1})= g_{d} (f_{d}^{(0)}(a_{0},\ldots ,a_{k-1}) ,\ldots ,
f_{d}^{(m-1)}(a_{0},\ldots ,a_{k-1}))$ for all  $d\in \omega$,
$a_{0},\ldots ,a_{k-1}\in \boldm _{d}$.    Suppose further that each of the families
$g, f ^{(0)},\ldots ,f ^{(m-1)}$ are uniformly existential on $\boldm$. Then the
family $h$ is also uniformly existential on $\boldm$.
\end{lemma}

Proof of Lemma \rref{Y3}. We have that for all $d\in \omega$,
and for all $a_{0},...,a_{k-1},b\in \boldm_{d}$,
$$ b=h(a_{0},...,a_{k-1}) \leftrightarrow \Psi(a_{0},...,x_{a-1},b)) $$
where $\Psi(x_{0},...,x_{k-1},y) $ is the formula $$\exists z_{0},...,z_{m-1}, \ 
g(z_{0},...,z_{m-1})=y \wedge \bigwedge_{i=0}^{m-1} z_{i}=f_{i}
(x_{0},...,x_{k-1})$$ 
Writing the existential formulas defining the functions $g,f_{0},...,f_{m-1}$
into the formula $\Psi$ we get the  existential formula of $\calm$ defining the function 
$h$.
\enp {(Lemma \rref{Y3})}

\begin{definition} We will denote by $\call^{(=)}$ the first-order language with
equality which contains the constant symbols $\bfnull$, $\bfegy$ and  does not contain 
any other relation symbols, 
function symbols or constant symbols. For each $m\in \omega$, $\boldn_{d}$ will denote a model of $\call^{(=)}$ with $\universe(\boldn_{m})=m$
and $(\bfnull)_{\boldn_{m}}=0$, $(\bfegy)_{\boldn_{m}}=1$.
\vege\end{definition}

\begin{lemma} \llabel{Y6} Let $k\in \omega$ and let  $P(x_{0},...,x_{k-1})$ be a propositional formula
of $\call^{(=)}$. Then the  family of $k+1$-ary relations $R^{(P)}= \langle R_{d}^{(P)} \mid d\in \omega \rangle$ is uniformly existential, where for all $d\in \omega$, 
$a_{0},...a_{k-1},u\in \boldm_{d}$, $R_{d}^{(P)}(a_{0},...,a_{k-1})$
holds iff $u\le d$ and   for all $i\in 2^{d-u}$, 

$$\boldn_{\bar u}\models  P(a_{0}[i,u],...,a_{k-1}[i,u])  $$
where $\bar u=2^{2^{u}}$
\end{lemma}

Proof of Lemma \rref{Y6}. 
Let $a_{k}=0$, $a_{k+1}=e_{d,u}$, and $\kappa=k+2$.
 For each
$r,s\in \kappa$ let $b_{\kappa r+s}= \min_{d,u}(e_{d,u},a_{r}-a_{s})$. Lemma 
\rref{F1} implies that $e_{u,d}$ is a term of $f$ and by Lemma \rref{AA24} the 
operation $\min_{d,t}$
is uniformly existential. Therefore for each    fixed $r$ and $s$ there exists a
uniform existential definition for the integer $b_{\kappa r+s}$, consequently, there 
exists
an existential formula $\psi(x_{0},...,x_{k-1},y,z_{0},...,z_{\kappa^{2}-1})$ of 
$\calm$
whose choice depends only on $k$, such that
$b_{0},...,b_{\kappa^{2}-1}$ is the unique sequence of length $\kappa^{2}$ from 
the
elements of $\boldm_{d}$ such that $$\boldm_{d}\models 
\psi(a_{0},...,a_{k-1},u,b_{0},...,b_{k^{2}-1})$$

We have $a_{k}[i,u]=0$ and $a_{k+1}[i,u]=1$ for all $i\in 2^{d-u}$.
Therefore for a fixed $i\in 2^{d-u}$ the sequence of $0,1$ values $b_{0}
[i,u],....,b_{\kappa^{2}-1}[i,u]$ determines  the truth values  of  the following 
statements in $\boldn_{\bar u}$:  $a_{j}[i,u]=0$, $a_{j}[i,u]=1$  for all $j\in k$, 
$a_{r}[i,u]= a_{s}[i,u] $ for all $r,s\in k$. Therefore there exists a boolean expression
$\calb(x_{0},....,x_{\kappa^{2}-1})$ such that ``for all $i \in 2^{d-u}$ $\boldn_{\bar u}\models  P(a_{0}[i,u],...,a_{k-1}[i,u])  $"
iff
$$\forall i\in 2^{d-u} \calb(b_{0}[i,u],...,b_{\kappa^{2}-1}[i,u])=0$$
Lemma \rref{AA30} implies that this property
of the sequence $b_{i}$, $i\in \kappa_{2}$
can be expressed by an existential formula whose choice depends only on 
$\kappa$ (and consequently only on $k$). This completes the proof since
we have already seen that the sequence $b_{i}$ is definable by such an existential
formula.
 \enp{(Lemma \rref{Y6})}

\begin{definition}  1. We will denote by $\alpha \circ \beta$ the concatenation of the 
sequences $\alpha$ and $\beta$.

2. Assume that $u,a,i,l \in \omega $, and  $\alpha=\langle
\alpha_{0},...,\alpha_{l-1} \rangle$ is a sequence of natural numbers.  The sequence $ 
a[i+\alpha_{0},u],...,a[i+\alpha_{k-1},u], a[i-\alpha_{0},u],...,a[i-\alpha_{k-1},u]$
will be denoted by $\vec a[i \pm \alpha\wr u]$. 
\vege\end{definition}

\begin{lemma} \llabel{Y9} Let $k,l\in \omega$ and let  $P(x_{0},...,x_{2k l-1})$ be a propositional formula
of $\call^{(=)}$. Then the  family of $k+1$-ary relations $R^{(P)}= \langle R_{d}^{(P)} \mid d\in \omega \rangle$,  is uniformly existential, where for all $d\in \omega$, 
$a_{0},...a_{k-1}, \alpha_{0},...,\alpha_{l-1},u\in \boldm_{d}$, 
$\alpha= \langle \alpha_{0},...,\alpha_{l-1}\rangle$,
$R_{d}^{(P)}(a_{0},...,a_{k-1})$
holds iff  the following conditions are satisfied:

\begin{cond} \llabel{Y9.1} $u\le d$ and for all $j\in l$, $\alpha_{j}< 2^{d-u}$ 
\end{cond}

\begin{cond} \llabel{Y9.2}
and   for all $i\in 2^{d-u}$, 

$$\boldn_{\bar u}\models  P\Bigl(\vec a_{0}[i\pm \alpha \wr u]\circ...\circ a_{k-1}[i \pm \alpha 
\wr u]\Bigr)  $$
where $\bar u=2^{2^{u}}$.   \end{cond}
\end{lemma}

Proof of Lemma \rref{Y9}. 
For each $j\in l$, $r\in k$ we choose an integer $b_{j k +r}\in \boldm_{d}$ such that

\begin{cond} \llabel{Y9.3}
 $\boldm_{d} \models b_{jk+r }= 2^{\alpha_{j} 2^{u}}a_{r}$  \end{cond}

 and with the same $j$ and $r$ another integer that will be denoted by
$b_{jl+r+kl}$ such that

\begin{cond} \llabel{Y9.4}
$\boldm_{d} \models b_{jk+ r+ l k}= \div (a_{r}, 2^{\alpha_{j} 2^{u}})$.
\end{cond}

We apply now Lemma \rref{Y6} with $k\legyen 2kl$, $a_{i}\legyen b_{i}$ for all
$i\in 2kl$. Conditions \rref{Y9.3} and \rref{Y9.4} guarantee that all of the arguments of 
$ P\Bigl(\vec a_{0}[i\pm \alpha \wr u]\circ...\circ a_{k-1}[i \pm \alpha 
\wr u]\Bigr)$ of condition \rref{Y9.2} is of the form $b_{j}[i,u]$ for a suitably
chosen $j\in 2kl$, whose choice does not depend on $i$.

Therefore there exists a propositional formula $P'(x_{0},...,x_{2k l-1})$
of  $\call^{(=)}$ such that for all $i\in 2^{d-u}$, $\boldn_{\bar u} \models P\Bigl(\vec a_{0}[i\pm \alpha \wr u]\circ...\circ a_{k-1}[i \pm \alpha 
\wr u]\Bigr)$ is equivalent to $\boldn_{\bar u} \models P'(b_{0},...,b_{2 k l-1})$, 
(where $P$, $P'$ may differ only in the order of its variables.)
Therefore Lemma \rref{Y6} implies the conclusion of the present lemma.
 \enp{(Lemma \rref{Y9})}

\eject
\subsection{\llabel{largerst} Extending existential formulas to larger structures}

In  section \rref{existential1} we have proved about several families of relations 
$R_{d}$, $d\in \omega$ that there exists a single existential formula of $\phi$ which 
for each $d\in \omega$ defines the relation $R_{d}$ in the structure $\boldm_{d}$. 
Sometimes we will need 
an existential definition for $R_{d}$ not in the structure $\boldm_{d}$ but in another 
larger structure $\boldm_{q}$ with $q\ge d$. The results of this section will show that 
such an existential definition  always exists provided that we can use in it $d$ as a 
parameter. The following lemma considers the special case when the relation $R_{d}$
is defined by a single function symbol of $\calm$.

\begin{lemma} \llabel{J5.1}
 Suppose that
 $   f(x_{0}, \ldots ,x_{j-1}) $    is a function symbol of  $   \calm $.
Then
there exists
a term  $   t(x_{0}, \ldots ,x_{j-1},y) $    of  $   \calm $    such that
for
all
  $   d,q\in \omega $    with  $d\le q$     and for
all  $   a,b_{0}, \ldots ,b_{j-1}\in \boldm_{d} $,      we have
 $   \boldm_{d}
\models
a=f(b_{0}, \ldots ,b_{j-1})  $     iff  $   \boldm_{q} \models
a=t(b_{0}, \ldots ,b_{j-1},d) $.     \end{lemma}

Proof Lemma \rref{J5.1}.
 For the various function symbols  $   f $    of
 $   \calm $    the choice of  $   t $    is the following:

Constant symbols. Assume that  $   f  $    is one of the constant
symbols
 $   \bfnull $    or  $   \bfegy $.
In these \xev{194i} cases  $   t $    is identical to the constant symbol  $   f $.     If
 $   f=-\bfegy $    then  $   t(y)=2^{2^{y}}-\bfegy $.     If  $   f=\bfn $    then
 $   t(y)=2^{y} $.

In the following definitions we will write  $   \mod(x,y) $    for
the
term
 $   x-y\lfloor x/y\rfloor  $

Unary function symbols.  $   f=\caln $,       $   t(x_{0},y)=
\mod(\caln(x_{0}),2^{2^{y}})
  $.      $   f=\bfp $,
 $   t(x_{0},y)=\min(\bfp(x_{0}),2^{2^{y}}-1) $.

Binary function symbols.  $   f=+ $,
 $   t(x_{0},x_{1},y)=\mod(x_{0}+x_{1},2^{2^{y}}) $,
 $   f=\times $,
 $   t(x_{0}x_{1},y)=\mod(x_{0}x_{1},2^{2^{y}}) $,
 $   f=\div $,       $   t(x_{0},x_{1},y)=\div(x_{0},x_{1}) $,
 $   f=\max $,       $   t(x_{0},x_{1},y)=\max(x_{0},x_{1}) $,
 $   f=\min $,       $   t(x_{0},x_{1},y)=\min(x_{0},x_{1}) $,
 $   f=\cap $,
 $   t(x_{0},x_{1},y)=\mod(\cap(x_{0},x_{1}),2^{2^{y}}) $.
\enp{(\rref{J5.1})}.

\vskip 5pt
If we
consider
instead of a function symbol  $   f $    a term  $   \tau $    of  $   \calm $    we
may
replace each function symbol in  $   \tau $    by the term whose
existence is
stated in \rref{J5.1}. This way we get the following:

\begin{corollary} \llabel{J5.2}
 Let
 $   \tau(x_{0}, \ldots ,x_{j}) $  be a term of  $   \calm $.    Then there
exists a term  $   t(x_{0}, \ldots ,x_{j-1},y) $    of  $   \calm $    such
that
for
all sufficiently large  $   d,q\in \omega $,     with $d\le q $,      and
for
all
 $   a,b_{0}, \ldots ,b_{j-1}\in \boldm_{q} $,      we have
 $   \boldm_{d}
\models
a=\tau(b_{0}, \ldots ,b_{j-1})  $     iff  $   \boldm_{q} \models
a=t(b_{0}, \ldots ,b_{j-1},d) $.     \end{corollary}

\begin{lemma} \llabel{AA13} For all $k\in \omega$ and for all propositional
formulas $P(x_{0},...,x_{k-1})$ of $\calm$, there exists a propositional formula
$P'(x_{0},...,x_{k-1},y)$ of $\calm$ with the following property. Assume that
$d,q\in \omega$, $d\le q$. Then the following two conditions are satisfied:

\begin{cond} \llabel{AA13.1}
for all
$a_{0},...,a_{k-1}\in \boldm_{d}$,
$\boldm_{d}\models P(a_{0},...,a_{k-1})$ iff $\boldm_{q}\models 
P'(a_{0},...,a_{k-1},d)$. \end{cond}

\begin{cond} \llabel{AA13.2}
for all
$b_{0},...,b_{k-1}\in \boldm_{q}$,
	$\boldm_{q}\models 
P'(b_{0},...,b_{k-1},d)$ implies $b_{0},...,b_{k-1}\in \boldm_{d}$. \end{cond}
\end{lemma}

Proof of Lemma \rref{AA13}.
The equality is the single relation symbol of the language
 $   \calm $.     Consequently each atomic formula of  $   \calm $    is of
the
form
 $   \tau_{1}=\tau_{2} $,      where  $   \tau_{1},\tau_{2} $    are terms of
 $   \calm $.    We construct a  propositional formula $P''(x_{0},...,x_{k-1}) $
 by substituting in $P$ for each atomic formula 	$\tau_{1}=\tau_{2}$ the atomic 
formula $t_{1}=t_{2}$, where $t_{i}$ is the term whose existence is stated in 
Corollary
\rref{J5.2}  with $\tau\legyen \tau_{i}$, for $i=1,2$.
$P'(x_{0},...,x_{k-1},y) $ will be the formula  $$P''(x_{0},...,x_{k-1}) \wedge 
\bigwedge _{i=0}^{k-1} x_{i} <2^{2^{y}} $$
The statement of the lemma is an immediate consequence of Corollary \rref{J5.2} .
 \enp{(Lemma \rref{AA13})} \vskip 5pt

As we have mentioned already in section \rref{sketch} it is very important in the proof
of the  Collapsing statement that we are able to encode sequences formed from 
the elements of a structure $\boldm_{d}$ by a single element of a larger structure 
$\boldm_{q}$. Here we consider the implication of such an encoding for the number of
existential quantifiers in an existential formula of $\calm$.  
The following lemma states that if 	$\phi(y_{0},\ldots ,y_{m-1})$ is an existential 
first-order formula of $\calm$ containing $k$ existential quantifiers, then there exists 
another existential formula $\psi(y_{0},\ldots ,y_{m-1})$ of $\calm $ containing only a 
single existential quantifier such that 
for all $d\in \omega$, and  for all $a_{0},\ldots ,a_{m-1}\in \boldm_{d}$,  
$\boldm_{d}\models 
\phi(a_{0},\ldots ,a_{m-1})$ is equivalent to $\boldm_{d+p}\models 
\psi(a_{0},\ldots ,a_{m-1})$, where $p$ depends only on $k$ and the number of 
quantifiers in $\phi$. Moreover  the formula $\psi$ can be given in the form of
$\psi (y_{0},\ldots ,y_{m-1})=\psi'(\pi(y_{0},\ldots ,y_{m-1}))$ where $\pi$ is a term 
of 
$\calm$
whose length is linear in $m$.

\begin{lemma} \llabel{J6}  There exists 	a $c_{0}\in \omega$, such that for all  
$m,k\in\omega$ there exist a $p\in \omega$ and a 
term $\pi(z_{0},\ldots ,z_{m-1},w)$ of $\calm$ of length
at most  $c_{0}m$,
 such that for  all  propositional formulas $P(x_{0}\ldots .,x_{k-1},y_{0},\ldots 
,y_{m-1})$ of 
$\calm$, there exists a propositional formula $Q(x,y)$ of 
$\calm$ 
with the property that for all $d\in \omega$, if $q>d+p $, then the following holds:

For all $a_{0},\ldots ,a_{m-1}\in \boldm_{d}$,
$$ \boldm_{d} \models \exists x_{0},\ldots ,x_{k-1} P(x_{0}\ldots .,x_{k-1},a_{0},
\ldots ,a_{m-1}) 
\ \ \leftrightarrow \ \ \boldm_{q} \models \exists x,Q(x,\pi(a_{0},\ldots ,a_{m},d)) $$
\end{lemma}

\begin{remark} In this lemma we replaced several existential quantifiers by a single 
one,
and in the propositional part of the existential formula we replaced several parameters
by a single one. These steps were needed since the indirect assumption in the proof
of Theorem \rref{TT3} is that a formula of the type $\exists x, F(x,y)$ is equivalent to a
propositional formula. In order to apply this indirect assumption we need existential 
formulas with a single quantifier and a single parameter. The upper bounds  on the 
integer $q$ and on the size of the term $\pi$ will be needed when by repeated 
use of the indirect assumption  we will eliminate quantifiers from an arbitrary first-order 
formula
of $\calm$. In each step,  the number of quantifiers in the formula will decrease, but 
the  size of the 
structure where we interprete the formula will grow. The upper bounds are needed to 
keep this
growth within reasonable limits.   
\vege\end{remark} \vskip 5pt

Proof of Lemma \rref{J6}.  We may assume that both $k$ and  $m$ are powers of 
$2$. (Otherwise
we may add new variables to the formula $P$ to make these numbers 
a power of $2$.) Assume that  $m=2^{s}$, $k=2^{r}$. We claim that
 the integer $p=r+s+2$ meets the requirements of the lemma. 
The  term $\pi(y_{0},...,y_{m-1},z)$ is defined by $\pi(y_{0},...,y_{m-1},z)=
z 2^{\div(\bfn,2)}+\sum_{i=0}^{m-1}y_{i}2^{i 2^{z}}$. 
(Recall that $\boldm_{q}\models \bfn= 2^{q}$.)

  If $a_{0},...,a_{k-1}\in
\boldm_{d}$ then we have that  $\boldm_{q}\models 
\pi(a_{0},...,a_{m-1},d)=b=b_{0}+b_{1}$, 
where
$b_{0}=d 2^{2^{q-1}}$  and  $b_{1}=\sum_{i=0}^{m-1}a_{i}2^{i 2^{d}} $.
Since  $a_{i}<2^{2^{d}}$ for $i\in m$ we have   $b_{1} <2^{m2^{d}} < 
2^{2^{d+s}}$. Therefore $q>d+s+2$  implies that $b_{1}<2^{2^{q-1}}$. 
 As a consequence if
$\pi_{0}(y)=\div(y,2^{\div(\bfn,2)}  )$, and
$\pi_{1}(y)=y- 
\pi_{0}(y)2^{\div(\bfn,2)} $,
then, using that  $\boldm_{q}\models 2^{\div(\bfn,2)}=2^{2^{q-1}}$,  we get that

\begin{cond} \llabel{J6.1}\ \ \ \ 
 $\boldm_{q} \models \pi_{0}(\pi(a_{0},...,a_{m-1},d))=d$, and  \vskip 5pt
\centerline{$\boldm_{q} \models \pi_{1}(\pi(a_{0},...,a_{m-1},d))= 
\sum_{i=0}^{m-1} a_{i}2^{i 2^{d}}$}\end{cond}

Motivated by these identities  we define the propositional formula $Q$ in the following
way using the term
$\sigma(x,y,z,w)$ that was defined in Lemma \rref{B0.5}. Here it is used to
extract a single term from the sum $\sum_{i=0}^{m-1} a_{i}2^{i 2^{d}}$ and from
another sum of similar type. Our definition for $Q(x,y)$ is:

$$Q(x,y)\equiv  P'(\kappa_{1} \ldots ,\kappa_{k-1},\lambda_{1},...,\lambda_{k-1},
\pi_{0}(y))$$
where $\kappa_{i}=\sigma(x,i,i,\pi_{0}(y))$ for $i=0,1,...,k-1$, 
$\lambda_{j}=\sigma(\pi_{1}(y),j,j,\pi_{0}(y))$, and 
where $P'$ is the formula defined in Lemma \rref{AA13} if we apply the lemma 
for the present formula $P$ and $k\legyen k+l$.
With this definition we get the truth value of $\boldm_{q} \models \exists x,Q(x,
\pi(a_{0},\ldots ,a_{m},d))$   in the following way. Condition \rref{J6.1} gives the 
values of
$\boldm_{q}\models \pi_{i}(\pi(a_{0},\ldots ,a_{m},d))$, for $i=0,1$. Putting this 
into the 
defining formula of $Q$ and using Lemma \rref{AA13}, we get  that $\boldm_{q} 
\models \exists x,Q(x,
\pi(a_{0},\ldots ,a_{m},d))$ is equivalent to $\boldm_{q} \models \exists x_{0},\ldots 
,x_{k-1}, P'(x_{0}\ldots ,x_{k-1},a_{0},\ldots ,a_{m-1},d) $. Lemma \rref{AA13} and
the related choice of $P'$ implies that the last expression is equivalent to
$ \boldm_{d} \models \exists x_{0},\ldots ,x_{k-1}, P(x_{0}\ldots .,x_{k-1},a_{0},
\ldots ,a_{m-1})$ as claimed in the present lemma.
\enp{(Lemma \rref{J6})}

\eject

\section{\llabel{turing}   Existential definitions and turing machines}

\begin{definition}
 ${\nbl}$ will denote the set of all
 pairs $\langle a_{0},a_{1}\rangle$ with $a_{0},a_{1}\in \omega$
and $a_{0}\ge a_{1}$.
  \vege\end{definition}

\begin{definition}  1. Suppose that $k\in \omega$ and for all $\langle u,t\rangle \in
{\nbl}$,
$R_{u,t}$ is a $k$-ary relation defined on $\boldm_{u}$. We will say that the family
of relations $ R=\langle R_{u,t} \mid \langle u,t\rangle \in {\nbl}
\rangle$ is
polynomially existential in $\boldm$, if there exists an integer  $c\in \omega$ and an existential
first-order formula
$\phi(x_{0},\ldots ,x_{k-1},y,z)$ of the language $\calm$ such that 

\begin{cond} \llabel{AA2}
for all $v,u,t\in
\omega$, if $u\ge t $
and $v\ge c(u-t)+t$, then
for all $a_{0},\ldots ,a_{k-1}\in \boldm_{u}$, $R_{u,t}(a_{0},\ldots ,a_{k-1})$ holds iff
$\boldm_{v} \models \phi(a_{0},\ldots ,a_{k-1},u,t)$. \end{cond}

 In this case we will say that
the formula $\phi $ is a defining formula of the family of relations $R$.
A family of $k$-ary functions $f_{u,t}$, $\langle u,t\rangle \in {\nbl}$ will be called polynomially existential if the family of relations $R_{u,t}$, $\langle u,t\rangle \in {\nbl}$ is polynomially existential, where for each $\langle u,t\rangle \in {\nbl}$, and $a,b_{0},...,b_{k-1}\in \boldm_{u}$,
$R_{u,t}(a,b_{0},...,b_{k-1}) $ iff $\boldm_{u} \models a=f_{u,t}(b_{0},...,b_{k-1})$.

2. Assume that $\bff$ is a $k$-ary function symbol  of $\calm$. We will say that
the function symbol $\bff$ is polynomially existential  if the family of relations
$F_{\bff}=\langle \bff_{d,t}\mid \langle d,t\rangle\in \nbl \rangle$
is polynomially existential. 
\vege\end{definition}

\begin{remark}  The expression ``polynomially existential" is motivated by the
following facts.  We may represent  an element of $\boldm_{u}$
by the sequence of its $2^{2^{t}}$-ary digits, that is, by a sequence of
length $2^{u-t} $ whose elements are from $\boldm_{t}$, provided that $u\ge t$.
In the existential formula defining the relation  $R_{u,t}$ we can existentially quantify
 elements of $\boldm_{v}$ which also can be represented by the sequences of their 
$2^{2^{t}}$-ary  digits. For the smallest integer  $v$ satisfying condition
 \rref{AA2}
 the length of such a sequence  
is  $2^{c(u-t)}$. This number  is a
polynomial of  $2^{u-t}$, that is,   for the definition of the relation $R_{u,t}$ it is 
enough to existentially quantify a sequence whose length is only a polynomial of the 
length of the sequences which represent the elements of $\boldm_{u}$. 
The next lemma shows that in the definition of a polynomially existential family of 
relations we may replace the assumption $v\ge c(u-t)+t$ by $v=c(u-t)+t$, and so 
considering only the smallest choice for the integer $v$ is justified.
 \vege\end{remark}

\begin{lemma} \llabel{AA63}
The definition of a polynomially existential family of relations
remains valid if we replace condition \rref{AA2} by the following condition

  \begin{cond} \llabel{AA63.1}
for all $v,u,t\in
\omega$, if $u\ge t $
and $v= c(u-t)+t$, then
for all $a_{0},\ldots ,a_{k-1}\in \boldm_{u}$, $R_{u,t}(a_{0},\ldots ,a_{k-1})$ holds 
iff
$\boldm_{v} \models \phi(a_{0},\ldots ,a_{k-1},u,t)$.
\end{cond}
\end{lemma}

Proof of Lemma \rref{AA63}. The statement of the lemma is an immediate consequence 
of Lemma \rref{J6}
\enp{(Lemma \rref{AA63})}

\begin{lemma} \llabel{Y4} Assume that  $k,m,l\in \omega$ and  $f ^{(j)}
=\langle f _{d,u}^{(j)} \mid \langle d,u\rangle\in {\nbl} \rangle$ are families of $k$-ary function on
$\boldm$, for
$j=0,1,\ldots ,m-1$ and  $g=\langle g _{d,u} \mid \langle d,u\rangle\in {\nbl}  \rangle
$ is a family of $m$-ary functions on $\boldm$. Let $h$ be the family
of $k$-ary functions $h=\langle  h_{d,u} \mid \langle d,u\rangle\in \nbl
  \rangle$ on $\boldm$
defined by $h _{d,u} (a_{0},\ldots ,a_{k-1})= g_{d,u} (f_{d,u}^{(0)}(a_{0},\ldots 
,a_{k-1}) ,\ldots ,
f_{d,u}^{(m-1)}(a_{0},\ldots ,a_{k-1}))$ for all  $ \langle d,u\rangle\in \nbl
  $,
$a_{0},\ldots ,a_{k-1}\in \boldm _{d}$.    Suppose further that each of the families
$g, f ^{(0)},\ldots ,f ^{(m-1)}$ are polynomially existential on $\boldm$. Then the
family $h$ is also polynomially existential on $\boldm$.
\end{lemma}

Proof of Lemma \rref{Y4}. Assume that $c\in \omega$ is an integer 
and $\phi_{0},...,\phi_{m-1}$, $\gamma$ are existential formulas of $\calm$
such that for all $d,u\in \omega$ and for all $a_{i},a_{i,0},...,a_{i,k-1}\in 
\boldm_{d}$, $i\in m$ and $b,b_{0},...,b_{m-1}\in \boldm_{d}$
we have that  for all $ i\in m$, $\boldm_{u+c(d-u)}\models \phi_{i}
(a_{i},a_{i,0},...,a_{i,k-1},d,u)
$   iff $f^{(i)}_{d,u}(a_{i,0},...,a_{i,k-1})=a_{i}$ and  $\boldm_{u+c(d-u)}\models 
\gamma(b,b_{0},...,b_{m-1},d,u)
$   iff $g_{d,u}(b_{0},...,b_{m-1})=b$. 
Then we have that for all $x_{0},...,x_{k-1},y\in \boldm_{d}$,   $h_{d,u}(x_{0},...,x_{k-1})=y$ 
iff $$\boldm_{u+c(d-u)}\models \exists z_{0},...,z_{m-1}, \Psi_{0}\wedge \Psi_{1}\wedge 
\Psi_{2} $$ where
$\Psi_{0}\equiv\bigwedge_{i\in m} z_{i}<2^{2^{d}}$, $\Psi_{1} \equiv \bigwedge_{i\in 
m} \phi_{i}(z_{i},x_{0},...,x_{k-1},d,u ))$ and
$\Psi_{2}\equiv \gamma(y,z_{0},...,z_{m-1},d,u)$.
Therefore the existential formula  $\exists z_{0},...,z_{m-1}, \Psi_{0}\wedge 
\Psi_{1}\wedge \Psi_{2} $ shows that the family of functions $h$
is polynomially existential.  \enp{(Lemma \rref{Y4})}

\begin{definition} In the following a turing machine will mean a turing machine with a single tape and a single head, whose each cell contain a natural number less than $2^{q}$ for some fixed 
$q\in \omega$. We may also consider the contents of the cells as $0,1$-sequences
$\delta_{0},...,\delta_{q-1}$ of length $q$. If we say that $\calt $ is a turing machine
without specifying the value of $q$ then we assume that $q=2$, that is each cell
contains a $0,1$ bit. 
 The machine has always a finite number of cells, but as the machine works it can always open new cells. Since we will consider only polynomial time
 computation on this machine  the exact way as the input and output is presented is not important.
E.g. if the input consists of several integers we may give their binary bits in even numbered cells, while the bits in the odd numbered cells signal the start of a 
new input number and the end of the input.
We will call  this type of turing machines also unlimited turing machines 
when we want to distinguish them from another class of turing machine, the
 restricted 
turing machines that we will define later. 
(In a restricted turing machine
the number of cells is fixed when the machine 
starts to work, and there are  restrictions on the contents of the cells too.)  
\vege\end{definition}

\begin{definition} Assume that $d,u\in \omega$, $d\ge u$ and $\chi\in \zo(d,u)$,
that is, $ \chi=\sum_{i=0}^{2^{d-u}-1}\delta_{i}2^{i2^{u}}$, where $\delta_{i}
\in \lbrace 0,1\rbrace $ for all $i\in 2^{d-u}$. The the integer $\sum_{i=0}^{2^{d}-1}
\delta_{i}2^{i}$ will be denoted by $\bin(\chi)$, motivated by the fact the we interprete
the $2^{2^{u}}$-ary digits of $a$ as {\it bin}ary	bits.
\vege\end{definition}

\begin{lemma} \llabel{AA73}
Suppose that $c_{0}\in \omega$ and $\calt$
is a turing machine such that for all $n,j\in \omega$ with $j<n$, $h<2^{n}$, the 
machine $\calt$ at an input $\langle n,j,h \rangle $ and at time $n^{c_{0}}$ 
provides the output $\lambda(n,j,h)\in n+1$.
 Then   the  family of relations $R_{d,u}$, $\langle d,u\rangle \in \nbl
$ is polynomially existential, where for all $\langle 
d,u\rangle \in \nbl
$, $R_{d,u}$ is defined in the following way.
Suppose that  $a,b,\chi \in 
\boldm_{d}$. Then $R_{d,u}(a,b,\chi) $  iff the following holds

\begin{cond} \llabel{AA73.a}
 $\chi\in \zo(d,u)$ and
for all $i\in 2^{d-u}$ $$b[i,u]=a[\lambda(2^{d-u},i, \bin(\chi)),u]$$
\end{cond}
\end{lemma}

In Lemma \rref{AA73} 
we are speaking about polynomial time computation on turing machines so
the exact parameters of the machine  the number of tapes,
etc. are irrelevant. From the point of view of our proof however these parameters 
are important so we give a more detailed definition of a turing machine.  Before we start 
the proof of Lemma \rref{AA73} we will prove another  related result Lemma \rref{Y10}.

In the next definition we define a class of turing machines which will be called 
restricted turing machines or shortly r.-turing machines which will differ 
at the following points from the turing machines defined earlier: \smallskip

(i) the number of cells are given when the machine starts to work, new cells
cannot be opened, \smallskip

(ii) each cell contains a $0,1$ sequence  $\langle \delta_{0},...,\delta_{\mu-1}
\rangle $ of length $\mu$, for some fixed $\mu\in \omega$, \smallskip

(iii) the contents of the cells where the head is located,  determine 
in itself that the head is there and
determines also the state of the automaton, \smallskip

(iv) it is possible to tell in the knowledge of the
content of a cell $C$ whether $C$ is at one of the two ends of the tape an if the answer is yes, it is possible to tell whether it is the first or the last cell.

\begin{definition} We define a class of turing machines which will be called 
restricted turing machines or shortly r.-turing machines motivated by the 
fact that the length of the tape and the contents of the cells
cannot be arbitrary,  there are some {\it restrictions} on them.
Such a machine $\calt$ consists of a tape of length
$\ell=\tplength(\calt)$. The cells are denoted by $\cell_{0},...,\cell_{\ell-1}$,
they are given when the machine starts to work, new cells cannot be opened.
Each cell at each time contains a $0,1$-sequence of length $\mu=\Width(\calt)$,
where $\mu\in \omega$, $\mu>3$. $\mu=\Width(\calt)$ will be also called the large width of the machine.
If $\delta_{0},...,\delta_{\mu-1}$ is the content of cell $j$ at time $t$, for some
$j\in \ell$, $t\in \omega$, then for all $i\in \mu$ we will denote $\delta_{i}$ by
$\cont_{t,j,i}$.
A $k\in \omega$, $k<\mu-3$ is fixed. The first $k$ bits that is $\cont_{t,j,0},...,
\cont_{t,j,k-1}$ will be called the work bits of $\cell_{j}$ at time $t$. They will play
the role of the contents of the cells of a turing machine in the traditional sense
the remaining bits, that is, $\cont_{t,j,k},...,
\cont_{t,j,\mu-1}$ will contain information related to requirements (iii) and (iv)
formulated before this definition. The integer  $k$ will be called the small width
of the machine and denoted by $k=\width(\calt)$.
We assume that the movement of the head and the changes of the contents of the 
cells from time $t$ to time $t+1$ is directed by  the finite automaton 
$\aut(\calt)$	with $2^{\nu} $
states where $\nu=\mu-k-3$. The states are identified with the natural numbers 
$0,1,..., 2^{\nu}-1$. The state of the automaton $\aut(\calt) $ at time $t$
will be denoted by $\state_{t}$. At time $0$ the finite automaton
is always in state $0$. 

At each time $t\in \omega$ the head of the machine is at one of the cells. If it is at $\cell_{j}$ then we will write $\head_{t}=j$.
If $\head_{t}=j$ then $\cont_{t,j,k+\nu}=1$, and the sequence $\cont_{t,j,k},...,
\cont_{t,j,k+\nu-1}$ are the binary bits of $\state_{t}(\aut(\calt))$, where  $\state_{t}(\aut(\calt))$ is the state of the automaton $\aut(\calt)$ at time $t$.

Finally  the values $\cont_{t,j,k+\nu+1}$ and  
$\cont_{t,j,k+\nu+2}$ indicate whether $\cell_{j}$ is at an edge of the tape and 
if it is which one. Namely
$\cont_{t,j,k+\nu+1}=1$ iff $j=0$ and
$\cont_{t,j,k+\nu+2}=1$ iff $j=\ell-1$. We will call  $\cont_{t,j,k+\nu+1}$ and  
$\cont_{t,j,k+\nu+2}$ the edge bits.

The change of the contents of the cells from time $t$ to time $t+1$
is done in the following way. Assume  that  at time $t$ the head is at $\cell_{j}$. Then the 
automaton $\aut(\calt)$ gets the  the work bits of $\cell_{j}$ at time $t$,
that is, the sequence, $\cont_{t,j,0},...,\cont_{t,j,k-1}$  as input and depending 
on this and its state at time $t$ it provides an output which determines the following 
three things:
(i) the work bits of $\cell_{j}$ at time $t+1$, (the work bits  of the other cells remain 
unchanged), (ii) the state of $\aut(\calt)$ at time $t+1$, (iii) the movement of the head
from time $t$ to time $t+1$, that is, whether it stays where it is, or it attempts to move
to the neighboring cell on the left or right (if it is at an edge of the tape where the 
desired movement is not possible then it stays where it is). Therefore
(ii) and (iii) together with $\head_{t}$ determine $\head_{t+1}$ and $\state_{t+1}$.
which  uniquely determine
$\cont_{t+1,j,i}$ for all $j\in \ell$, $i=k,k+1,...,\mu-1$.

For a fixed $t\in \omega, j\in \ell$ the sequence $\cont_{t,j,0},...,\cont_{t,j,\mu-1}$,
will be denoted by $\vec\cont_{t,j}$.  This definition does not define
the symbols  $\vec\cont_{t,-1}$, $\vec\cont_{t,\ell}$,
however  we will use these symbols to denote $0,1$-sequences of lengths $\mu$, and 
on each occasion we will tell what are their values. \end{definition}

The advantage of using restricted turing machines is their 
property stated in the following lemma. This lemma will make it easy to define 
the history of a restricted turing machine by an existential formula of $\calm$ in a suitably chosen
structure $\boldm_{v}$.

\begin{lemma} \llabel{Y14} 
Suppose that $\calt$ is a restricted turing machine with 
$\tplength(\calt)=\ell$, $\Width(\calt)=\mu$, $\width(\calt)=k$.
 Then there  exist  boolean
functions	$\calb_{i}$, for all $i\in \mu$, with $3\mu$ variables such that 
for all $t\in \omega$,
$j\in \ell$, $i \in \Width(\calt)$, and for all possible values of the vectors
$\vec\cont_{t+1,-1},
\vec\cont_{t,\ell} \in \lbrace 0,1\rbrace^{\mu} $, we have
$$\cont_{t+1,j,i}=\calb_{i}(\vec\cont_{t,j-1}\circ \vec\cont_{t,j}
\circ \vec\cont_{t,j+1})$$
\end{lemma}

Proof of Lemma \rref{Y14}.  
The values $\vec\cont_{t,j-1},\vec\cont_{t,j},\vec\cont_{t,j+1}$
determine the answer to the following questions.

(i) Is the head located at time $t$ at one of the cells $\cell_{j-1},\cell_{j},\cell_{j+1}$
and if it is which one?

(ii)
If the answer to question (i) is yes then what is  $\state_{t}$
and the what are the contents of the cells $\cell_{j-1}, \cell_{j},\cell_{j+1}$?

(iii) Is  $\cell_{j}$ at the edge of the tape, or  more precisely, which one of the following
equations hold
$j=0$ or $j=\ell-1$?

 The answers to questions (i), (ii), and (iii) uniquely determine what is 
$\cont_{t+1,i,j}$. Moreover using  the answer for question (iii) we can makes sure
that the values $\vec\cont_{t,-1}$ and $\vec\cont_{t,\ell}$ are not used to answer any 
of these question even if they are present among the three sequences  
$\vec\cont_{t,j-1}\circ \vec\cont_{t,j}
\circ \vec\cont_{t,j+1}$. \enp{(Lemma \rref{Y14})}

\begin{definition} Assume that $\calt$ is restricted turing machine,
$\tplength(\calt)=\ell$, $\Width(\calt)=\mu$, $\width(\calt)=k$,
and an $u\in \omega $ is fixed. 
The sequence of integers $a_{0},...,a_{k-1}$, 
will be called the $u$-based input for the machine $\calt$, if
for each $i\in  k$,
$a_{i}=\sum_{j=0}^{\ell-1} \cont_{0,j,i}2^{j2^{u}}$.
 Our definition implies that  the $u$-based
input uniquely determines the complete history of the machine that is all of the values
$\cont_{t,j,i}$ for $t\in \omega$, $j\in \ell$, and $i\in \mu$. 
For all $i\in \mu$, $T\in \omega$ we define an integer $b_{i,T}$ by
$$b_{i,T}=\sum_{t=0}^{T-1} \sum_{j=0}^{\ell-1} \cont_{t,j,i}2^{(t\ell +j) 
2^{u}}$$
 The sequence
$b_{0,T},...,b_{\mu-1,T}$ will be called the $u$-based history of the machine $\calt$
till time $T$. 
\vege\end{definition}

\begin{lemma} \llabel{Y10} Assume that  $c_{0},k,\mu,\in \omega$
and $\cala$ is a finite automaton with $|\cala|=2^{\mu-k-3}$.   Then there exists an
existential formula $\psi$ of $\calm$ such that for all
 $d,u\in \omega$, with $d\ge u$ and for all
$a_{0},...,a_{k-1}\in \boldm_{d}$, $b_{0},...,b_{\mu-1},\in \boldm_{v}$,
where $v=u+(c_{0}+1)(u-d)$ the following two conditions are
equivalent:

\begin{cond} \llabel{Y10.1} $\boldm_{v} \models \psi(a_{0},...,a_{k-1},b_{0},....,
b_{\mu-1},d,u)$
\end{cond}
 
\begin{cond} \llabel{Y10.2} For all $i\in k$, $a_{i}\in \zo(d,u)$, and
if $\calt $ is a restricted turing machine with $\aut(\calt)=\cala$,
$\width(\calt)=k$, $\tplength(\calt)=\ell=2^{d-u}$,  
$\Width(\calt)=\mu$, and
 with the $u$-based input $a_{0},...,a_{k-1}$, then its  $u$-based
history till time $T=\ell^{c_{0}}$
is $b_{0},...,b_{\mu-1}$. 
\end{cond}
\end{lemma}

Proof. 
This lemma is a consequence of Lemma \rref{Y9} and Lemma \rref{Y14}. 
We apply lemma \rref{Y9}
with $k\legyen k+\mu +1$, $l\legyen 4$ $d\legyen v$
$a_{0}\legyen a_{0},...,a_{k-1}\legyen a_{k-1}$, $a_{k}\legyen b_{0},...,a_{k+\mu-1}
\legyen b_{\mu-1}$, $a_{k+\mu}\legyen e_{d,u}$,  $\alpha_{0}\legyen 0$   $\alpha_{1}\legyen \ell-1$,
$\alpha_{2}\legyen \ell$, $\alpha_{3} \legyen \ell +1$.

 In the formulation of the  propositional statement $P$ 
we follow the notation of the present lemma. The propositional formula $P$
will say the following for all $r\in 2^{v-u}$:
 $\Bigl( e_{d,u}[r,u]=1 
\rightarrow \bigwedge_{i\in k} ( b_{r}[i,u]=a_{r}[i,u])\Bigr)$ and  if $e_{d,u}[r,u]=0$,
then 

$$b_{i}[r,u]=\calb_{i}(\vec \beta_{r-\ell-j-1}\circ \vec\beta_{r-\ell-j}
\circ \vec\beta_{r-\ell-j+1})$$
where $\calb_{i}$ is the boolean expression from Lemma \rref{Y14}
and $\vec \beta_{r}=\langle  b_{0}[r,u],...,b_{\mu-1}[r,u]\rangle $.
Lemma \rref{Y14}    implies that if the the propositional formula $P$ holds
for all $r\in 2^{v-d}$ then $b_{0},...,b_{\mu-1} $ is the $u$-based history
of the machine $\calt$. Therefore
Lemma \rref{Y9} implies the existence of the existential formula $\psi$.
\enp{(Lemma \rref{Y10})}

Proof of Lemma \rref{AA73}.  Suppose that  $a,b,\chi \in 
\boldm_{d}$ and  $R_{d,u}(a,b,h) $ holds, that is, 
for all $i\in 2^{d-u}$ $$b[i,u]=a[\lambda(2^{d-u},i, \bin(\chi)),d]$$ 
This means we get  the $2^{2^{u}}$-ary digits of the integer $b$
from the digits of the integer $a$ in the following way. To get the $i$th digit
$b[i,u]$, we have to compute, using  the turing machine $\calt$,
the value of  $\lambda_{i}= \lambda(2^{d-u},i, \bin(\chi))$.
If $\lambda_{i}<2^{d-u}$ then $b[i,u]=a[\lambda_{i},u]$ which is an element of 
$2^{2^{u}}$. This is true also for the $\lambda_{i}=2^{d-u}$ but, since $a\in 
\boldm_{d}$, we have $a[2^{d-u},u]=0$. Therefore each the $2^{2^{u}}$-ary
digit of $b$  is either one of the first $2^{d-u}$ digits of $a$
or it is $0$.

Let $S$ be the set of sequences of length $s$ from the elements of $\boldm_{u}$,
where $s=2^{d-u}$. We define  maps $\eta_{j,\iota}$, $j\in s-1$, $\iota=0,1,2$,
that map $S$ into itself. Suppose that $x=\langle x_{0},...,x_{s-1}\rangle\in S$, then

$\eta_{i,0}(x)=\langle y_{0},...,y_{s-1}\rangle$,
where for all $j \in \lbrace 0,1,...,s-2\rbrace \bcks \lbrace i\rbrace 
$, $y_{j}=x_{j}$, and $y_{i}=0$. That is, we get $\eta_{i,0}
(x)$ from $x$ by  replacing $x_{i}$ with $0$.

 $\eta_{i,1}(x)=\langle y_{0},...,y_{s-1}\rangle$,
where for all $j \in \lbrace 0,1,...,s-2\rbrace \bcks \lbrace i,i+1\rbrace 
$, $y_{j}=x_{j}$, and $y_{i}=x_{i+1}$, $y_{i+1}=x_{i}$. That is, we get $\eta_{i,1}
(x)$ from $x$ by swapping $x_{i}$ and $x_{i+1}$.

$\eta_{i,2}(x)=\langle y_{0},...,y_{s-1}\rangle$,
where for all $j \in \lbrace 0,1,...,s-2\rbrace \bcks \lbrace i\rbrace 
$, $y_{j}=x_{j}$, and $y_{i}=x_{i+1}$. That is, we get $\eta_{i,0}
(x)$ from $x$ by replacing $x_{i}$ with $x_{i+1}$.

Clearly if $d,u,a,\chi,b$ are given as above then there exists a sequence
$J=\langle \eta_{i_{m},\iota_{m}} \mid m=0,1,...,\kappa-1$, where $\kappa<s^{3}$ 
such that 

\begin{cond} \llabel{AA73.2}
if $A=\langle a[0,u],...,a[s-1,u] \rangle$, $B=\langle b[0,u],...,b[s-1,u] \rangle$,
then $$B=\eta_{i_{0},\iota_{0}}(\eta_{i_{1},\iota_{1}}(...\eta_{i_{\kappa},
\iota_{\kappa}}(A)...   ))$$
\end{cond}

Since the function $\lambda $ was
computable by an (unlimited) turing machine $\calt$ in time polynomial in $s$,  the 
sequence $J=\langle \langle i_{m},\iota_{m} \rangle \mid m=0,1,...,\kappa-1\rangle$ is polynomial 
time 
computable as well.
So far we have assumed that $\calt$ is an unlimited turing machine whose each
cell contain a single $0,1$ bit. The same computation can be performed by
an unlimited turing $\calt_{1}$ machine turing whose each cell contains two bits.   
The advantage of using such a machine is that the encoding of the input can be done
in a form which is convenient for definitions by $\calm$ formulas in a structure
$\boldm_{v}$. 

More precisely  
the assumptions of the lemma imply that there exists a constant  $c_{1}\in \omega$  
and an unlimited  
turing machine $\calt_{1}$ such that for each $j\in \omega $, $\cell_{j}$ at time $t$,
contains two bits $\cont_{t,j,0}$ and $\cont_{t,j,1}$
and the machine in time $s^{c_{1}}$ computes a sequence $\langle i_{m},
\iota_{m}\mid m=0,1,...,\kappa-1\rangle$ which satisfies condition \rref{AA73.2}.
We will denote the time by $t_{m}$ when the computation of the pair $\langle i_{m},
\iota_{m} \rangle$ has been completed.
 Moreover we also assume that the input 
$\chi$ is given at time $0$ in the form $\cont_{0,j,0}=\chi[j,2]$,
 for all $j< \lceil \log_{2} \chi\rceil$, and $\cont_{0,j,i}=0$
 for all other values of $j,i\in \omega$, where 
$\cont_{0,j,i}$ is defined. (We may assume that at time $0$ the length of the tape is
determined by the length of the input, and when a new cell is opened its initial
content is always $\langle 0,0 \rangle$.)
 
 We define now
a restricted turing machine  $\calt_{2}$ with $\Width(\calt_{2})=c_{2}$,
$\width(\calt_{2})=k_{0}$ 
 where 
$k_{0},c_{2}\in \omega$, $k_{0}<c_{2}-3$  are  constants that we will fix later and $\tplength(\calt_{2})=\ell=s^{c_{1}}$.
When the machine starts to work $\cont_{0,j,i}$ is the same for $\calt_{1}$ and 
$\calt_{2}$ where both values are defined. If $\cont_{0,j,i}^{(\calt_{1})}$ is not 
defined  and $i<c_{2}-3$ then $\cont_{0,j,i}^{(\calt_{2})}=0$. (For 
$i=c_{2}-3,c_{2}-2,c_{2}-1$
$\cont_{0,j,i}^{(\calt_{2})}$ is determined by the definition of a restricted turing 
machine.) 

  We partition the first 	$k_{0}$ bits of each cell $j$ at time $t$ into subsets 
$X_{t,j},Y_{t,j},Z_{t,j}$. 
We will call the the $X$-bit, $Y$-bit and $Z$-bits. $X_{t,j}$ contain the first
two  bits (corresponding to the bits used by $\calt_{1}$), $Y_{t,j}$ contains the next two bits
and $Z_{t,j}$ contains the remaining work bits.
The computation    done by $\calt_{2}$ will consist of $\kappa$ consecutive time
intervals $I_{0},...,I_{\kappa-1}$. Each interval $I_{m}$
is further divided into four consecutive intervals $J_{m},K_{m},L_{m},M_{m}$.

In the interval $J_{m}$, $m\in \kappa$,
 using only the $X$  bits 
of its cells  $\calt_{2}$ simulates the computation done by $\calt_{1}$
in the time  interval $(t_{m-1},t_{m}]$, (where $t_{-1}=-1$). 
 
After that this simulation is suspended, and during the intervals 
$K_{m},L_{m},M_{m}$ the  $X$ bits of the cells do not change, and consequently  
$\calt_{2}$ will be able to continue the
simulation of $\calt_{1}$ in the time interval $J_{m+1}$.

In the interval $K_{m}$, $\calt_{2}$ does the following, while it leaves the
$X$ bits and $Y$ bits 
  unchanged in each cell.
$\calt_{2}$ takes the head to $\cell_{i_{m}}$, 
where $\langle i_{m},\iota_{m} \rangle$ is the pair computed by $\calt_{1}$ by time
$t_{m}$.  For this no other work bits are used than the $Z$ bits.   Meanwhile 
$\aut(\calt_{2})$ ``remembers" the value of  $\iota_{m}$, that is, it has enough states 
to use different ones depending on the value of $\iota_{m}$. 

During the whole interval $L_{m}$ the head remains at   $\cell_{i_{m}}$. When 
interval $L_{m}$ starts
 say at $\calt_{2}$-time $t_{m}'$, $\calt_{2}$
writes the binary form of $\iota_{m}+1$ into the two bits of   
the set $Y_{t_{m},i_{m}}$.
 That is, $\cont_{t_{m}',i_{m},2}=(\iota_{m}+1)[0,0]$ and 
$\cont_{t_{m}',i_{m},3}=(\iota_{m}+1)[1,0]$.  All of the other 
work bits remain unchanged

At time $t_{m}'+1$  still in interval $L_{m}$, the head  remains at $\cell_{i_{m}}$ and the bits in $Y_{t_{m}'+1,i_{m}}$
are changed into $0$, that is,  the $\cont_{t_{m}'+1,i_{m},
2}=0$ and $\cont_{t_{m}'+1,i_{m},3}=0$.
All of the other work bits remain unchanged.

In the interval $M_{m}$ 
the head goes back to the 
position where it was at the end of interval $K_{m}$. 
During this the $X$ and $Y$ bits  do not change.
(The very last interval of the form $M_{m}$ is exceptional in the sense that is has no end
since according to our definition the machine cannot stop, so the head remains
at the same place and the content of all of the cells remain unchanged.)

This completes the description of the computation done by
$\calt_{2}$. It is easy to see that there  exists a finite automaton $\cala$,
such that if  $\cala=\aut(\calt_{2})$ then $\calt_{2}$ will perform the described computation. $\Width(\calt_{2})=c_{2}$ and $\width(\calt_{2})=k_{0}$ are chosen in a way that is
compatible with  the described computation and the choice of  $\cala$.

Let $t$ be a time according to $\calt_{2}$ and let $j\in \ell$ such that at least on of the bits in $Y_{t,j}$ is not $0$, equivalently
$\cont_{t,j,2}\not=0$ or $\cont_{t,j,3}\not= 0$.  Then we will say that the integer $j$ 
is a critical cell number at time $t$.
We will need later the following immediate consequence of the definition of
$\calt_{2}$: 

\begin{cond} \llabel{AA73.3} the machine $\calt_{2}$ has the property that
for each $t\in \omega$ there exists at most one integer $j\in \ell$ such that
$j$ is a critical cell number at time $t$. If $j$ is a critical cell number at time $t$ then
then there exists a unique $m\in \kappa$ such that the machine $\calt_{2}$
at time $t$, wrote the binary bits of $\iota_{m}+1$ into the $Y_{t,j}$ bits. 
 (In this case $\iota_{m}$ will be called the critical
map-number at time $t$.)   \end{cond}

The total time needed for $\calt_{2}$ to complete the described steps is at most 
$T=s^{c_{3}}$, where $c_{3}\in \omega$ is sufficiently large with respect to $c_{1}$.
We will write $\calt_{2}^{(d,u)}$
instead of $\calt_{2}$ if we want to emphasize its dependence on $d$ and $u$.
Applying Lemma \rref{Y10}
we get the following.

There exists an existential formula $\psi$ of $\calm$ such that for all $d,u\in \omega$
with $d\ge u$ for all $\chi\in \zo_{d,u}$ and for all $b_{0},...,b_{c_{2}-1}\in 
\boldm_{v}$, here $v=u+(c_{3}+c_{1})(d-u)$ the following two conditions are
equivalent

\begin{cond} \llabel{Y15.1}$\chi\in \zo_{d,u}$,  and the $u$-based history
of the machine $\calt_{2}^{(d,u)}$ with $u$-based input $\langle \chi,0,...,0\rangle$
is $b_{0},...,b_{c_{2}-1}$.
\end{cond}

\begin{cond} \llabel{Y15.2} $\boldm_{v}\models \psi(b_{0},...,b_{c_{2-1}},\chi,
d,u)$
\end{cond}

The next step is to give an existential formula $\phi$ of $\calm$ such that 
for all $b_{0},...,b_{c_{2}-1}\in \zo(v,u)$ and for all $a,b, 
\chi \in \boldm_{d}$,
$\boldm_{v} \models \phi(b_{0},...,b_{c_{2}-1},a,b, 
\chi,d,u)  $ iff condition \rref{AA73.a} of the lemma is satisfied.
This together with the equivalence of conditions \rref{Y15.1} and \rref{Y15.2}
clearly implies the conclusion of the Lemma \rref{AA73}, (we also need that
according to Lemma \rref{M6} the conditions $b_{0},...,b_{c_{2}-1}\in \zo(v,u)$,
$\chi\in \zo_{d,u}$ can be described by an existential formula in $\boldm_{v}$.) 

We define an integer $\alpha\in \boldm_{v}$. We for each $j\in \ell,
t\in T$, where $s=2^{d-u}$, $\ell=s^{c_{1}}$, $T=s^{c_{3}}$, $\alpha_{t,j}$ will denote the integer 
$\alpha[t \ell+j,u]$.  We define $\alpha_{t,j}$ by induction on $t$.

For $t=0$,
$\alpha_{0,j}=a[j,u]$. Assume that $\alpha_{t-1,j}$ has been defined for some $t\in \omega\bcks \lbrace 0\rbrace $ and for all $j\in \ell$.
If  $j\ge s$ or there is no critical cell number at time $t$, then $\alpha_{t,j}=\alpha_{t-1,j}$.
If there exists a critical cell number $j_{0}$ at time $t$ and $\iota_{m}$
is the 
 critical map-number at time $t$. Then we define $\alpha_{t,j}$ for all $j<s
$, 
by

$$\langle \alpha_{t,0},...,\alpha_{t,s-1} \rangle =
\eta_{j_{0},\iota_{m}}\Bigl( \langle \alpha_{t-1,0},...,\alpha_{t-1,s-1} \rangle  \Bigr)$$ 
and by $\alpha_{t,j}=0$ for all $j\ge s$. 
This completes the definition of  the integer  $\alpha$.
  The definition implies that  
  we have $\alpha_{T,j}=b[j,u]$ for all $j\in s$. In the  definition we treated
separately the cases $j<s$ and $j\ge 0$. Later we will use the fact that
the integer $w_{v,d,u}=e_{d,u}\sum_{i=0}^{T-1}2^{i \ell 2^{u}}$, has the property that for all $j\in \ell$, $t\in T$, $w[t \ell+j,u]=1$ if $j<s$ and and
$w_{v,d,u}[t \ell+j,u]=0$ otherwise. Moreover by Lemma \rref{MP0} there exists a
term $\sigma$ of $\calm$ whose choice does not depend on anything, such that $\boldm_{v}
\models  w_{v,d,u}=\sigma(v,u,d) $ and as a consequence $w$ is definable by an existential formula in $\calm.$

We show now that the integer $\alpha$ can be defined by an existential formula
in $\boldm_{v}$ (using $b_{0},...,b_{c_{2}-1},a,d,u$ as a parameters). 
We will denote by $\vec\cont_{t,j}$ the sequence $\cont_{t,j,0},...,\cont_{t,j,c_{2}-1}$
with respect to the machine $\calt_{2}$.  As earlier we use the symbols $\vec\cont_{t,-1}$
$\vec\cont_{t,\ell}$ to denote $0,1$ sequences of length $c_{2}$, whose values will
be decided later. In a similar way $\alpha_{t,-1}$, $\alpha_{t,\ell}$ will denote 
integers in $\boldm_{u}$ whose values will be decided later.
The definition of the integers $\alpha_{t,j}$  implies the following:

\begin{cond} \llabel{Y13.1}  
There exists a propositional formula $P$ of the language
$\call^{(=)}$ such that 
for all $t\in s^{c_{3}}$,
$j\in \ell$, $i \in c_{2}=\Width(\calt_{2})$, and for all possible definitions of the 
vectors
$\vec\cont_{t,-1},
\vec\cont_{t,\ell}\in \lbrace 0,1\rbrace ^{c_{2}}$ and the integers $\alpha_{t,-1}, 
\alpha_{t,\ell} \in \boldm_{u}$, we have 
that $\alpha_{t+1,j}$ is the unique integer $A\in \boldn_{\bar u}$ such that
$$\boldn_{\bar u}\models  P(A,  \alpha_{t-1,j}, \alpha_{t,j}, \alpha_{t+1,j},  
\vec\cont_{t,j-1}\circ \vec\cont_{t,j}
\circ \vec\cont_{t,j+1}, w_{v,d,u}[t s^{c_{3}}+j])$$
\end{cond}

Now we use Lemma \rref{Y9} and get that the  $\alpha $ is definable by  an existential
formula in $\boldm_{v}$. 

The definition of $\alpha$ implies that $\boldm_{v} \models b=\div(\alpha, 
2^{\ell(T-1)2^{u}})$
and $a= \mod(\alpha, 2^{s 2^{u}})$, that is, $\boldm_{v} \models a=
\alpha -2^{s 2^{u}} \div (\alpha,2^{s2^{u}}) $, where $\mod(x,y)$ is the least
nonnegative residue of $x$ modulo $y$.
\enp{(Lemma \rref{AA73})}

Lemma \rref{Y10} that we have formulated and proved earlier states
that if 	$\calt$ is  a restricted turing machine  $\calt$ such that $\width(\calt)$ and
$\aut(\calt)$ are constants, then its history can be defined by an existential formula
in $\boldm_{v}$, where $v$ is large enough so that  the $u$-based history can be 
presented as a sequence in $\calm_{v}$. Now we formulate a consequence of the 
lemma which is dealing with not the whole history of $\calt$ but only its
input-output relation.

\begin{lemma} \llabel{Y12} Assume that  $c_{0}, c_{1},k\in \omega$ and 
$\cala$ is a finite automaton.
  Then the 
family of relations $\langle R_{d,u} \mid \langle d,u\rangle \in {\nbl} \rangle$ is polynomially existential,
where for all $d,u\in \omega$, $d\ge u$, and for all $a_{0},...,a_{k-1},b_{0},
...,b_{k-1}\in \boldm_{d} $,  $R_{d,u}( a_{0},...,a_{k-1},b_{0},...,b_{k-1})$
holds iff there exists a restricted turing machine $\calt$ with $\width(\calt)=k$, 
 $\aut(\calt)=\cala$, $\tplength(\calt)=2^{c_{0}(d-
u)}$ such the the following holds:

\begin{cond} \llabel{Y12.1} If the $u$-based input of the machine $\calt$ is the 
sequence $a_{0},...,a_{k-1}$, and $T=2^{c_{1}(d-u)}-1$, then
for all $i\in k$, $b_{i}=\sum_{j=0}^{2^{d-u}-1} \cont_{T,j,i}2^{j 2^{u}}$.
\end{cond}
\end{lemma}

Proof of Lemma \rref{Y12}.  The statement of the lemma
is an immediate consequence of Lemma \rref{Y10}, since we have to say only
by an existential formula  that there exists a history of $\calt$ which is 
compatible with the given $u$-based input and the given
contents of the cells at time $T$. \enp{(Lemma \rref{Y12})}

\eject

\section{\llabel{multiplication} Polynomially existential definition for 
parallel multiplication}

In this section we prove the following

\begin{lemma} \llabel{AA74} 
The function symbol  	$\times $ 
of $\calm$ is  polynomially existential in $\boldm$. Equivalently,
the family of functions $\langle \otimes_{q,u} \mid \langle q,u\rangle \in \nbl \rangle$
is polynomially existential.
\end{lemma}

As a first step of the proof we reformulate the statement of Lemma \rref{AA74}. 

\begin{lemma} \llabel{AA76}The following statement implies Lemma
\rref{AA74}. Let $F=\langle F_{d,u}  \mid \langle d,u\rangle \in {\nbl}
\rangle $
be the the family of ternary functions defined in the following way. For all $d,u\in 
\omega$
with $d\ge u$,  and for all $a,b,q\in \boldm_{d}$,  $\boldm_{d}\models F_{d,u}
(a,b,q)=a\otimes _{q,u}b  $ if  $a,b\in \boldm_{u}$, $q\ge u$ and $d\ge u+2(q-u)+2$, otherwise $F_{d,u}(a,b,q)=0 $.
Then the family of function $F_{d,u}$ is polynomially existential.
\end{lemma}

Proof of Lemma \rref{AA76}.   Let 
$\Gamma$ be the set of all triplets $\langle d,q,u\rangle	 \in \omega^{3}$,
such that $d\ge q\ge u$ and $d\ge u+2(q-u) +2$.
Suppose that the family  functions $F$ is polynomially existential.
This implies that there exists an existential formula $\psi$ of $\calm$ and a $c_{0}
\in \omega$, such that for all $\langle d,q,u\rangle \in \Gamma$,
and for all $a,b,c\in \boldm_{q}$,  and for all  $w\ge u+c_{0}(d-u)$ 
we have 

\begin{cond} \llabel{AA76.1} $\boldm_{w}\models\psi(a,b,c,q,d,u)$
iff $\boldm_{q}\models a \otimes_{q,u} b=c$. 
\end{cond}

 Assume now that we choose 
  $\langle q,u\rangle\in {\nbl}$ arbitrarily with the only restriction
$q>u$,  $a,b,c\in \boldm_{u}$,
and we define $d$ by $d=q+2(q-u)+2$, and so we have
  $\langle d,u,q\rangle \in \Gamma$. Let $c_{1}\in \omega$ be a constant with 
$c_{1}>c_{0}+6$. Then $u+c_{1}(q-u)\ge u+c_{0}(d-u)$  and therefore condition 
\rref{AA76.1} implies that 

\begin{cond} \llabel{AA76.2} $\boldm_{u+c_{1}(q-u)}\models\psi(a,b,c,q,d,u)$
iff $\boldm_{q}\models a \otimes_{q,u} b=c$. 
\end{cond}

This holds for all $\langle q,u\rangle \in \nbl
$, with $q>u$ and for all $a,b,c \in \boldm_{q}$. The $q=u$ case however
is trivial since $\boldm_{q} \models a\otimes_{q,q} b=ab$. Therefore if we define another
existential formula $\psi'$ of $\calm$ with $\psi'(a,b,c,q,u) \equiv (q=u\wedge 
ab=c) \vee \psi(a,b,c,q,q+2(q-u)+2,  u)$, then  we have that for all $\langle q,u\rangle \in {\nbl}$ and for all $a,b,c\in \boldm_{q}$

\begin{cond} \llabel{AA76.3} $\boldm_{q+c_{1}(q-u)}\models\psi'(a,b,c,q,u)$
iff $\boldm_{q}\models a \otimes_{q,u} b=c$. 
\end{cond} 
as required in the definition of a polynomially existential family of functions.
\enp{(Lemma \rref{AA76})}

Proof of Lemma \rref{AA74}.
Let $F=\langle F_{d,u} \mid \langle d,u\rangle \in {\nbl}\rangle$ be the family of functions defined in Lemma \rref{AA76}. We will 
show that $F$ is polynomially existential.
We define four families of functions 
$F_{i}=\langle F_{d,u,i} \mid \langle d,u\rangle \in {\nbl}\rangle$
for $i=0,1,2$ and 
$G=\langle F_{d,u,i} \mid \langle d,u\rangle \in {\nbl}\rangle$.
The functions $F_{d,u,i}$, $i=0,1,2$ will be binary functions and the functions 
$G_{d,u}$ will be  ternary functions.  For each fixed $\langle d,u\rangle \in \nbl
$ we define a ternary function $F'=\langle F'_{d,u} \mid \langle d,u\rangle
\in {\nbl} \rangle$. For all $a,b,q\in \boldm_{d,u}$,
$$F_{d,u}'(a,b,q)=F_{d,u,2}\biggl( G_{d,u}\Bigl(F_{d,u,1}(a,q), F_{d,u,1}(a,q),
q \Bigr),q \biggr)$$

We will  prove the following two statements.

\begin{cond} \llabel{AA76.a}  The families of functions
$F_{0},F_{1},F_{2}, G,$  
are all polynomially existential. \end{cond}

\begin{cond} \llabel{AA76.b}  For all $\langle d,u\rangle \in {\nbl}$ and 
for all  $q\in \boldm_{d}$ and  $a,b\in \boldm_{q}$,   we have
$$F_{d,u}'(a,b,q)=  F_{d,u}(a,b,q)    $$ 
\end{cond}

According to Lemma \rref{Y4}  the composition of polynomially existential families
is polynomially existential. The family $F'$ was defined as a composition,  therefore 
 condition \rref{AA76.a}  implies that the family $F'$
is polynomially existential. The fact that the  family $F'$ is polynomially
existential and condition \rref{AA76.b} together imply that the family $F$ is also
polynomially existential. (The functions $F'_{d,u}$ and $F_{d,u}$ are not necessarily
identical on $\boldm_{d}$ since the equality in condition \rref{AA76.b} is guaranteed only for $a,b\in \boldm_{q}$.)
 Indeed assume that $\psi$ is an existential formula of
$\calm$ and $\boldm_{u + c_{0}(u-d)} \models \psi'(a,b,c,q,d,u)$ is equivalent to
$F'_{d,u}(a,b,q)=c$, for all $a,b,c,q\in \boldm_{d}$. Let $\psi(a,b,c,q,d,u) \equiv
((a\ge 2^{2^{q}}\vee b\ge 2^{2^{q}})\wedge c=0)\vee \psi'(a,b,c,q,d,u)$.
Then $\psi'$ is an existential formula of $\calm$, and  $\boldm_{u + c_{0}(u-d)} 
\models \psi'(a,b,c,q,d,u)$ is equivalent to
$F_{d,u}(a,b,q)=c$. 

Therefore to complete the proof of Lemma \rref{AA74} it is sufficient  to
show that there exists families of functions $F_{0},F_{1},F_{2}, G,$ such that
conditions
 \rref{AA76.a}
and \rref{AA76.b}  are satisfied. (In the latter one $F'$ is defined as a composition $F_{0},F_{1},F_{2}, G,$ as indicated earlier.)

Let 
$\Gamma$ be the set of all triplets $\langle d,q,u\rangle	 \in \omega^{3}$,
such that $d\ge q\ge u$ and $d\ge u+2(q-u) +2$.

We start the definition on the functions  $F_{d,u,i}$, and $G$  by defining their values
in places that are not interesting for us. Suppose that $d,u,q\in \omega$, 
$d\ge u$, $q\in \boldm_{d}$ and $\langle d,q,u\rangle \not\in \Gamma$. Then for
all $x,y\in \boldm_{u}$,
 $F_{d,u,0}(x,q)=F_{d,u,0}(x,q)=F_{d,u,0}(x,q)=G(x,y,q)=0$. Since $\langle d,q,u\rangle \not\in \Gamma$ implies $F_{d,u}(x,y,q)=0$ for all $x,y\in \boldm_{u}$
as well condition \rref{AA76.b} is satisfied if $\langle d,q,u\rangle \not\in \Gamma$.
Therefore, starting from  from this, point we consider only the $\langle d,q,u\rangle \in \Gamma$ case. Since $\langle d,q,u\rangle\in \Gamma$ is a propositional statement in
$\boldm_{d}$, this is sufficient for our purposes.

{\sl The  family of functions $F_{0}$ and $F_{1}$.} Assume that a
$\langle d,q,u\rangle\in
\Gamma$ is fixed,  $s=2^{q-u}$ and $p\in 2^{s2^{u}}=\boldm_{d}$. If 
$p\notin \boldm_{q}$ then $F_{d,u,0}(p,q)=F_{d,u,1}(p,q)=0$. Assume now that  
$p\in \boldm_{q}$.
 We define
 $F_{d,u,i}
(p,q)$ for $i=0,1$.
The integer $p$ can be written  in the form of  $$
p=\sum_{i=0}^{s-1}\pi(i)2^{i 2^{u}}$$ where $\pi (i)\in 2^{2^{u}} $ for all
$i\in s$.
Our definitions are
$$ F_{d,u,0}(p,q)=\sum_{i=0}^{s-1}\pi(i)2^{2i s 2^{u}},
 F_{d,u,1}(p,q)=\sum_{i=0}^{s-1}\pi(i)2^{2i  2^{u}},
$$
that is, in both cases we got the $2^{2^{d}}$-ary digits of $F_{d,u,i}(p,q)$
from the digits of $p$ by moving the digits of $p$ to different places and
putting $0$s in the remaining places. We have that 
$F_{d,u,0}(p,q)[j,u]=\pi(\frac{j}{2s})$ if $2s|j$ and $F_{d,u,0}(p,q)[j,u]=0$
otherwise. We apply Lemma \rref{AA73} with  $n\legyen 2^{d-u}$,  $\chi\legyen 2^{(q-u)2^{u}}  $,   $h=\bin(\chi)= 2^{q-u}$, and 
 the function $\lambda(n,j,h)$  is defined in the following way:
if $j=2 h i$ for some $i\in h$ then $\lambda(n,j,h)=i$ otherwise $\lambda(n,j,h)=n$.

Lemma \rref{AA73} implies that
 the family
$F_{0}$
is polynomially existential. We can show in a similar way that the family $F_{1}$ is also
polynomially existential.  

{\it The family of functions $G$}.  Assume that $\langle d,q,u\rangle \in {\nbl}$
and $x,y\in \boldm_{d}$. Then $G_{d}(x,y)=xy$. Clearly  the family of
 functions $G$ is polynomially existential.

{\sl The definition of the function $F_{d,u,2}$.} Assume that a $\gamma \in \langle
d,q,u\rangle \in
\Gamma$ is fixed and $p\in \boldm_{d}$ and $p[i,d]=\pi(i)$. Then
$F_{d,u,2}(p)=\sum_{i=0}^{s-1}  \pi(2s i+2i)2^{i2^{u}} $.
In the same way as in the case of the family $F_{0}$ we can show using 
Lemma \rref{AA73} that the family $F_{2}$ is polynomially existential.

This completes the definition of the families $F_{i}$ and $G$ and the 
proof of statement \rref{AA76.a}. 
Now we prove statement \rref{AA76.b}.

It is sufficient to show that for all $a,b\in \boldm_{q}$, and for all  $i\in 2^{d-u}$, 
$$F_{d,u}'(a,b,q)[i,u]=  F_{d,u}(a,b,q)[i,u] $$ 
Since the values of $d,u$ and $q$ are fixed now, we will write $F_{0}(x)$ instead of 
$F_{d,u,0}(x,q)$, $F_{2}(x,y)$ instead of $F_{d,u,2}(x,y,q)$, etc.

Let $a=\sum_{j=0}^{s-1}a_{j}2^{j2^{u}}$ and  $b=\sum_{j=0}^{s-1}b_{j}2^{j2^{u}}$, where $s=2^{u-q}$.
 Then,  by the definition of $F$ and the definition of the operation  $\otimes_{q,u}$  we have $F(a,b)[i,u]=(a_{i}b_{i})_{\boldm_{u}} $, that is,  $a_{i}b_{i}$ must be 
computed in $\boldm_{u}$.

If we compute  $F'$,  according to its definition as a composition, we get the 
following.
$F_{0}(a)=\sum_{j=0}^{s-1} a_{j} 4^{s j 2^{u}}$ and 
$F_{1}(b)=\sum_{k=0}^{s-1} b_{k} 4^{k 2^{u}}$.
Therefore if we compute $a b$ in the $4^{2^{u}}$-ary numeral system
then all of the products $a_{j} b_{k}$ will contribute to to a different digits
of the product $a b$, since the sums $sj+k$, $j,k\in s$ are all different.
Moreover $a_{j},b_{k}\in \boldm_{u}$ imply that the product $a_{j} b_{k}$
as the product of integers is less than $4^{2^{u}}$, therefore there is no
carryover, and we have that

\begin{cond} \llabel{AA76.c}
 $F_{0}(a) F_{0}(b)=\sum_{r=0}^{s^{2}-1}\alpha_{r}4^{r 2^{r}}$,
where $\alpha_{r}=a_{j} b_{k}<4^{2^{u}}$, if $r= s j+k$, and $j,k\in s$,
where every operation is performed as among integers. \end{cond}

Moreover the assumptions $a,b\in \boldm_{q}$ and $d\ge u+2(u-q)+2$ imply that
the product $F_{0}(a) F_{0}(b) $ among the integers is the same that  
in $\boldm_{d}$. Since 
$F_{0}(a) F_{0}(b)=G(F_{0}(a), F_{0}(b))$
we have that  $F_{2}(F_{0}(a) F_{0}(b))= F'(a,b)$.
Let $h=F_{0}(a) F_{1}(b) $.    Statement 
\rref{AA76.c} imply that $h[2si+2i,u]= (a_{i} b_{i})_{\boldm_{u}}$, therefore,
according to the definition of $F_{2}$, we get that  $F'(a,b)[i,u]=
(a_{i} b_{i})_{\boldm_{u}}$, where $(a_{i} b_{i})_{\boldm_{u}}$. This  
completes the proof of statement \rref{AA76.b}. \enp{(Lemma \rref{AA74})}

\eject 
\section{\llabel{moperations} The $\calm$ operations are polynomially existential.}

In this section we prove  Theorem \rref{TT7} which is equivalent to  the following lemma.

\begin{lemma} \llabel{E82}  For each   function symbol $\bff$ of $\calm$, 
the parallel $\bff$ operation is polynomially existential in $\boldm$. Equivalently, 
for each function symbol $\bff$
the family of functions $\langle \bff_{d,t} \mid \langle d,t\rangle\in \nbl  \rangle$
is polynomially existential in $\boldm$.
\end{lemma}

We will use the following three lemmas in the proof of  Lemma \rref{E82}.

\begin{lemma} \llabel{Y32} Suppose that $\tau(x_{0},...,x_{k-1})$ is a term
of $\calm$, such that for each function symbol $\bff$ of $\calm$, if $\bff$ occurs in
$\tau$ then $\bff$ is polynomially existential. Then the family of $k$-ary relations
$R=\langle R_{d,t} \mid \langle d,t\rangle \in \nbl \rangle$ is polynomially existential,
where for all $d,t\in \omega$ with $d\ge t$ and for all $a_{0},...,a_{k-1}\in 
\boldm_{d}$, we have $R_{d,t}(a_{0},...,a_{k-1}) $ iff for all $i\in 2^{d-t}$,
$\boldm_{t}\models \tau(a_{0}[i,t],...,a_{k-1}[i,t])=\bfnull$.

Assume now  that for all of the function symbols $\bff$ in $\tau$, $\bff\notin \lbrace \times,\div,
\bfp\rbrace $.  Then the family 
of $k+1$-ary relations $Q=\langle Q_{d} \mid d\in \omega\rangle$  is uniformly 
existential, where for all $d\in \omega$ and for all $a_{0},...,a_{k-1},t\in \boldm_{d}$,
$Q_{d}(a_{0},...,a_{k-1},t)$ iff $t\le d$ and $R_{d,t}(a_{0},...,a_{k-1})$.
\end{lemma}

Proof of Lemma \rref{Y32}. 
To prove the first statement of the lemma we construct an existential formula $\psi$ of $\calm$ such that for all
sufficiently large $c>0$,  for all $d,t\in \omega$ with $d\ge t$ and for all 
$a_{0},...,a_{k-1}\in \boldm_{d}$, we have $R_{d,t}(a_{0},...,a_{k-1}) $ iff
$\boldm_{v} \models \psi(a_{0},...,a_{k-1},d,t)$, where $v=t+c(d-t)$.

Let  $\bff^{(0)},...,\bff^{(l-1)}$ be the function symbols of $\calm$ occurring in 
$\tau$,  and let $\tau_{0},...,\tau_{r-1}$ be the sequence of all subterms of $\tau$
where $\tau_{i}=x_{i}$ for $i\in k$ and $\tau_{r-1}=\tau$.
Assume that
$\tau_{i}=\bff^{(g_{i,0})}(\tau_{g_{i,1}},\tau_{g_{i,2}})$, where $g_{i,0}\in l $ and 
$g_{i,1},g_{i,2}\in r$. (If $\bff^{(g_{i,0})}$ is not binary then one or both arguments of it may be 
missing).
The formula $\psi$ will say the following: there exists $b_{0},...,b_{r-1}\in 
\boldm_{d}$, such that  $$\boldm_{d}\models b_{r-1}=\bfnull \wedge \biggl( 
\bigwedge_{j\in k} b_{j}=a_{j} \biggr) \wedge \bigwedge _{i\in r} 
b_{i}=\bff^{(g_{i,0})}_{d,t}(b_{g_{i,1}},b_{g_{i,2}}) $$ 
Our assumption that all of the function symbols $\bff^{(i)}$, $i\in l$  are polynomially
existential implies that this condition can be expressed by an existential formula in 
$\boldm_{v}$.

For the proof of the second statement of the lemma we note that by
Lemma \rref{AA24}, $\bff\not\in \lbrace \times,\div,\bfp\rbrace $ implies that
the parallel operation $\bff$ is uniformly existential. Using this the  the fact that family 
$Q$ is uniformly
existential can be proved in the same way as the first part of the lemma. 
 \enp{(Lemma \rref{Y32})}
  
\begin{definition} Assume that $P(x_{0},...,x_{k-1},y_{0},...,y_{l-1})$ is a propositional 
formula of $\calm$. We will say that $P$ is 	$k$-sensitive, if the following three
conditions are satisfied:

(i) for each 
occurrence
$ \times(\tau,\sigma)$ of the function symbol $\times $ of $\calm $ in
$P$, where $\tau,\sigma$ are terms of $\calm$, there exists a $j\in l$ such that
$\tau=y_{j}$,

(ii) for each 
occurrence
$ \div(\tau,\sigma)$ of the function symbol $\div $ of $\calm $ in
$P$, where $\tau,\sigma$ are terms of $\calm$, there exists a $j\in l$ such that
$\sigma=y_{j}$,

(iii) the formula $P$ does not contain the function symbol $\bfp$ of $\calm$.
\vege\end{definition}

\begin{lemma} \llabel{Y35} Suppose that $P(x_{0},...,x_{k-1},y_{0},...,y_{l-1})$ is a 
$k$-sensitive propositional formula 
of $\calm$.
 Then the family of $k+l+1$-ary relations
$R=\langle R_{d} \mid  d \in \omega \rangle$ is uniformly existential,
where for all $d\in \omega$, $a_{0},...,a_{k-1}, b_{0},...,
b_{l-1},t\in 
\boldm_{d}$, we have $R_{d}(a_{0},...,a_{k-1},b_{0},...,b_{l-1},t) $ iff $d\ge t$,
$b_{0},...,b_{l-1}\in \boldm_{t}$	and for all $i\in 
2^{d-t}$,
$\boldm_{t}\models P(a_{0}[i,t],...,a_{k-1}[i,t],b_{0},...,b_{l-1})$.
\end{lemma}

Proof of Lemma \rref{Y35}. The proof of the lemma is the essentially the same 
as the proof of Lemma \rref{Y32}. (We have to use now, that according to Lemma 
\rref{M1}, the family of relations $Q_{d}(a,b,c,t)$ is uniformly propositional,
where for all $a,b,c,\in \boldm_{d}$, $Q_{d}(a,b,c,t)$ holds iff $d\le t$
and  $\odot_{d,t}(a,b)=c$.)
 \enp{(Lemma \rref{Y35})}

\begin{lemma} \llabel{Y36} Suppose that $P(x_{0},...,x_{3k-1},y_{0},...,y_{l-1})$ is a 
$3k$-sensitive propositional formula 
of $\calm$.  Then the family of $k+l+1$-ary relations
$R=\langle R_{d} \mid  d \in \omega \rangle$ is uniformly existential,
where for all $d\in \omega$, $a_{0},...,a_{k-1}, b_{0},...,
b_{l-1},t\in 
\boldm_{d}$, we have $R_{d}(a_{0},...,a_{k-1},b_{0},...,b_{l-1},t) $ iff $d\ge t$,
$b_{0},...,b_{l-1}\in \boldm_{t}$	and for all $i\in 
2^{d-t}$,
$$\boldm_{t}\models P(\vec A_{0},...,\vec A_{k-1},b_{0},...,b_{l-1})$$
where $\vec A_{j}$ stands for  the sequence
$a_{j}[i-1,t],a_{j}[i,t],a_{j}[i+1,t]$ for 
$j=0,1,...,k-1$.
\end{lemma}

Proof of Lemma \rref{Y36}.  We use Lemma \rref{Y35} with
$k\legyen 3k$. The statement $\boldm_{t} \models  P(\vec A_{0},...,\vec 
A_{k-1},b_{0},...,b_{l-1})$
is equivalent to $$\boldm _{t}\models
\exists g_{0},...,g_{k-1},h_{0},...,h_{k-1},  \Phi \wedge \bigwedge_{i\in k}
h_{i}=2^{2^{q}}a_{i} \wedge g_{i}=\div(a_{i},2^{2^{q}})$$ 
where $$\Phi\equiv P(\vec B_{0},...,\vec B_{k-1},b_{0},...,b_{l-1})$$
and $B_{j}$ stands for the sequence $g_{j}[i,q],a_{j}[i,q],h_{j}[i,q]$
for $j=0,1,...,k-1$. Therefore applying Lemma \rref{Y35} for the given propositional 
formula $P$ we get the required existential formula.
 \enp{(Lemma \rref{Y36})}

\begin{lemma} \llabel{AA42} Assume that, $k,m\in \omega$ and  $P$ is a propositional 
formula  of $\calm$ with
$3k+m+1$ free variables and  
$\sigma$ does 
not contain any of the function symbols $\times, \div $ or $\bfp$.
 Then $R=\langle R_{d} \mid d\in \omega  
\rangle $ is a uniformly existential family 
of $k+m+1$-ary relations in $\boldm$  , where for each $d\in \omega$, and for each
$a_{0},...,a_{k-1},w_{0},...,w_{m-1},q\in \boldm_{d}$, $R_{d}(a_{0},...a_{k-1},w_{0},...,w_{m-1},q)$ holds
iff  $q\le d$, $w_{0},...,w_{m-1} \in \boldm_{q}$ and  for all $i\in 2^{d-q}$,
 $\boldm_{q} \models P(\vec A_{0},...,\vec A_{k-1},
w_{0},...,w_{m-1},q)$, 
where $\vec A_{j}$ is the sequence $ a_{j}[i-1,q], a_{j}[i,q],a_{j}
[i+1,q]$, for all $j\in 
k$.
\end{lemma}

Proof of Lemma \rref{AA42}.   We may assume that  
$P$ is of the form  $\sigma(x_{0},...,x_{3k+m})=\bfnull$.
 We have $$a_{j}[i-1,q]=(qa_{j})
[i,q] \wedge
a_{j}[i+1,q]=(\lfloor a_{j}/q \rfloor)[i,q] $$ 
Therefore $R_{d}(a_{0},...a_{k-1},w_{0},...,w_{m-1},q)$ 
holds iff for all $i\in 2^{d-q}$, $$\boldm_{q} \models \sigma (\vec B_{0},...,\vec 
B_{k-1}, 
w_{0},...,w_{m-1},q)  =\bfnull$$ where $\vec B_{j}$ is the sequence  $ (q a_{j})
[i,q], a_{j}[i,q], 
(\lfloor a_{j}/q \rfloor)[i,q] $

Since the term $\sigma$ does not contain the function symbols the second part
of Lemma \rref{Y32} implies the conclusion of lemma.
\enp{(Lemma \rref{AA42})}

Proof of Lemma \rref{E82}. 
We know from Lemma \rref{AA24} that the statement of the lemma holds
if $\bff\notin \lbrace \times,\div,\bfp\rbrace  $.  Lemma \rref{AA74} implies the 
statement of the present lemma for $\bff=\times$. Our next goal is to show that the 
function symbol $\div$ is existentially parallel.
The operation $\div(a,b)$ has an existential definition among integers.
Namely  for all $a,b,c\in \omega$,    $\div(a,b)=c$ iff
``($b=0\wedge c=0$) or ($b\not=0$ and
 there exists an  $r\in \omega$, with $a=c b+r$ and 
$r <b$)". This definition is not good if $a,b,c\in \boldm_{d}$ and  we perform the 
arithmetic
operations in $\boldm_{d}$, since there can be many different
$c\in \boldm_{d}$ with $\boldm_{d}\models a-b <c b \le a$ and for all
of them $r\legyen a-c b$ meets the requirement of the definition of $\div(a,b)$.
Therefore we have to add the condition that the  product $ c b$ computed among the 
integers   is the same as the product $c b$ in $\boldm_{d}$.
 The following Lemma says that the parallel
version of this condition is polynomially existential.

\begin{lemma} \llabel{AA7} The family of  ternary relations $R=\langle R_{u,t} \mid \langle u,t\rangle  \in 
{\nbl}\rangle$ is polynomially existential, where $R_{u,t}$ is defined in the 
following way.

Assume that $u,t\in \omega$, $u\ge t$, and $a,b,w \in \boldm_{u}$. Then
$R_{u,t}(a,b,w)$ holds iff for all $i\in 2^{u-t}$,
$w[i,t]=0$ implies that $a[i,t]b[i,t]<2^{2^{t}}$, and 
 $w[i,t]=1$ implies that $a[i,t]b[i,t]\ge 2^{2^{t}}$
\end{lemma}

Proof of Lemma \rref{AA7}. 
For each fixed $t\in \omega $ we define two functions $F_{t}$, $G_{t}$ on $\omega$.
Suppose that $a=\sum_{i=0}^{\infty}\alpha_{i}2^{i2^{t}}\in \omega$, where 
$\alpha_{i}=a[i,t]$. Then
$F_{t}(a)= \sum_{i=0}^{\infty}\alpha_{i}2^{2i2^{t}}=\sum_{i=0}^{\infty}
\alpha_{i}2^{i2^{t+1}}$. This definition implies that every second $2^{2^{t}}$-
ary digits of $F_{t}(a)$ is
$0$ and between the $0$s we have the digits of $a$. That is $F_{t}(a)$ ``stretches" 
out the $2^{2^{t}}$-ary form by a factor of  $2$, and puts $0$s in the odd numbered
places. 
We get the integer $G_{t}(a)$ from $a$ by keeping only its $2^{2^{t}}$-ary digits
at the odd numbered  places and then ``compressing" this sequence by a factor of two.
More precisely if $a\sum_{i=0}^{\infty}\alpha_{i}2^{i2^{t}}$
then $G_{t}(a)= \sum_{j=0}^{\infty} \alpha_{2j+1}2^{j2^{t}}$.

These functions has the following useful property.

\begin{proposition} \llabel{AA7.1}
 Assume $u,t\in \omega$,
$u\ge t$,   $a,b\in \boldm_{u}$, and let $$h_{u,t}(a,b)=G_{t}\Bigl(F_{t}(a) \otimes_{u+1,t+1} 
F_{t}(b)\Bigr) $$
Then for all $i\in 2^{u-t}$, $a[i,t]b[i,t]<2^{2^{t}}$ iff $h[i,t]=0$.
\end{proposition}

Proof of Proposition \rref{AA7.1}. 
When we compute the parallel product $F_{t}(a) \otimes_{u+1,t+1} 
F_{t}(b)$ to get the $i$th component of the result we have to multiply
$a[i,t]$ and $b[i,t]$ modulo $2^{2^{t+1}}$. Suppose that the result 
is $\mu_{i}2^{2^{t}}+\lambda_{i}$, where $\mu_{i},\lambda_{i}\in 2^{2^{t}}$. 
We have  $F_{t}(a) \otimes_{u+1,t+1} 
F_{t}(b)=\sum_{i=0}^{2^{u-t}-1}(\mu_{i}2^{2^{t}}+\lambda_{i}) 2^{i2^{t+1}}$.
Therefore the definition of $G_{t}$ implies that $G_{t}\Bigl(F_{t}(a) \otimes_{u+1,t+1} 
F_{t}(b)\Bigr)= \sum_{i=0}^{2^{u-t}-1}\mu_{i}2^{2^{t}}$. 
We have  $\mu_{i}=(h_{u,t}(a,b))[i,t]$ and so the definition of $\mu_{i}$ implies the 
statement of the proposition. \enp{(Proposition \rref{AA7.1})}

Lemma \rref{AA73} and Lemma \rref{AA74} together  imply that the family of functions $
H= \langle h_{u,t} \mid \langle u,t\rangle \in \nbl\rangle$ is 
polynomially existential, where the function 
$h_{u,t}$ is defined in Proposition \rref{AA7.1}.
The relation $R_{u,t}(a,b,w)$ can be defined by ``there exists a $\rho \in 
\boldm_{u}$ such that $\rho=h_{u,t}(a,b)$ and for all $i\in 2^{u-t}$,
$w=\min_{u,t}(e_{u,t}, \rho) $". By Lemma \rref{AA24} and Proposition \rref{AA7.1}
this shows that family $R_{u,t}$ is polynomially existential.
  \enp{(Lemma \rref{AA7})}

\begin{definition} We define a binary term $\rem(x,y)$ of $\calm$ by
$\rem(x,y)=x+(-\bfegy)y\div(x,y)$.  (Suppose that $d\in \omega$, $a,m,b \in 
\boldm_{d}$, $m\not=0$.
Then $\boldm_{d} \models b=\rem(a,m)$ iff $b $ is the least nonnegative residue
of $a$ modulo $m$.) For all $d,t\in \omega$ with $d\ge t$, $\rem_{d,t}$ will be the
binary function defined on $\boldm_{d}$ in the following way. For all $a,m,b \in 
\boldm_{d}$, $\rem_{d,t}(a,m)=b $  iff  for all $i\in 2^{d-t}$, $\rem(a[i,t],m[i,t])=
\rem(b[i,t])$.
\vege\end{definition}

\begin{lemma} \llabel{AA39} The function symbol $\div$ and the 
family of functions $\langle \rem_{d,t}\mid \langle d,t\rangle \in \nbl \rangle$ are polynomially existential in $\boldm$.
        \end{lemma}

Proof of Lemma \rref{AA39}. 
 Let $R=\langle R_{u,t} \mid \langle
 u,t\rangle \in {\nbl} \rangle$ be the family of ternary relations defined in
Lemma \rref{AA7}. 
For all $d\in \boldm_{d}$, and for all $q,a,b
\in \boldm_{d}$,
$\boldm_{d} \models q =\div_{d,t}(a,b)$ iff $$ R(b,q,e_{d,t}) \ \wedge  \ \exists r \in 
\boldm_{d}, \Bigl(
a=(q  \otimes_{d,t} b) \oplus_{d,t} r \ \wedge \  r<_{d,t} q \Bigr) $$
We have already proved that $\boldm$ is existentially parallel with respect to 
multiplication  and addition, and by 
Lemma 
\rref{AA7} the relation $R$ is polynomially existential, therefore 
this definition of $\div_{d,t}$ implies that the 
function symbol $\div$ is polynomially existential as well.

The  family $\langle \rem_{d,t}\mid \langle d,t\rangle \in \nbl \rangle$
is polynomially existential since the function $\rem_{d,t}$
is defined by a term
which is using on function symbols which are polynomially existential.
\enp{(Lemma \rref{AA39})}.

\begin{lemma} \llabel{Y31} The function symbol $\bfp$ of $\calm$ is polynomially existential. 
\end{lemma}

Proof of Lemma \rref{Y31}.  
We have to construct an existential formula $\psi$ such that if $c\in \omega$ is 
sufficiently large then  for all $d,t\in \omega$
with $d\ge t$, and for all $a,b\in \boldm_{d}$, we have 
$\bfp_{d,t}(a)=b$ iff $\boldm_{v}\models \psi(a,b,d,t)$, where $v=t+c(d-t)$.

Suupose that $K\in \omega$ is a sufficiently large integer and $c$ is sufficiently
large with respect to $K$.

We will construct the formula  $\psi$ in the form of
$$\psi(a,b,d,t)\equiv (t\le K \wedge  \psi_{0}(a,b,d,t)) \vee 
(t>K \wedge  \psi_{1}(a,b,d,t))$$ 
(Here we used $a,b,d,t$ as variables to make the roles of the variables  clear.)

{\it The definition of $\psi _{0}$.} For all integers $i\in [-2^{K},2^{K}] $
we define three elements of $\boldm_{d}$, $a_{i},t_{i}$
and $M_{i}$. If $i\ge 0$ then 
 $$\boldm_{d} \models  a_{i}=2^{i}a \ \wedge \ t_{i}=2^{i}t \ \wedge \ M_{i}=2^{i}e_{d,t}$$   
If $i < 0$ then 
 $$\boldm_{d} \models  a_{i}=\div(a,2^{i}) \ \wedge  \  t_{i}=\div(t,2^{i}) \ \wedge   
\ M_{i}=\div(e_{d,t},2^{i})  $$

This definition implies that for each $i\in [-2^{K},2^{K}]$ there exist terms $\alpha_{i}, \tau_{i},\mu_{i}$ of $\calm$, depending only on $i$, such that for all
choices of $d, K$, $t\in \omega$, $t\le K$, $a\in \boldm_{d}$,  
we have $$\boldm_{d} \models a_{i}=\alpha_{i}(a,d,t) \wedge  t_{i}=\tau_{i}(a,d,t) \wedge  M_{i}=\mu_{i}(a,d,t)  $$

For each $j\in 2^{d}$, the three sequence of integers 
$\langle a_{i}[j,0] \mid j\in [-2^{K},2^{K}]\rangle$,
$\langle t_{i}[j,0] \mid j\in [-2^{K},2^{K}]\rangle$,
$\langle M_{i}[j,0] \mid j\in [-2^{K},2^{K}]\rangle$,
together uniquely determine the integers $a[j,t]$, $t$, and $\rem(j,2^{t})$. 

Consequently there exists a boolean expression $\calb$ with $3 (2^{K+1}+1)$ variables such that for all $j\in 2^{d-t}$ we have $$(\bfp_{d,t}(a))[j,0]=
\calb(\vec a_{i}[j,0], \vec t_{i}[j,0],\vec M_{i}[j,0] ) $$
where $\vec x_{i}[j,0]$, stands for the sequence $\langle x_{i}[j,0] \mid 
i\in [-2^{K},2^{K}]\rangle$ for $x=a,t,M$.
According to Lemma \rref{AA32} this implies that there exists a term 
$\sigma$ of $\calm$ such that  $\boldm_{d}\models \bfp_{d,t}(a)=\sigma(a,t)$.
Based on that, using Corollary \rref{J5.2} of Lemma \rref{J5.1} we can define the formula $\psi_{0}$ with the required properties.

{\it The definition of $\psi_{1}$.} Assume that $d,t\in \omega$, $d\ge t >K$ and $a\in
\boldm_{d}$. 
We define three integers $A_{0},A_{1},A_{2}\in \boldm_{d}$, such that for all
$i\in 2^{d-t}$ exactly on of the following two conditions are satisfied

(i) $A_{0}[i,t]=a[i,t]\ge 2^{t}$ and $A_{1}[i,t]=A_{2}[i,t]=0$.

(ii) $A_{0}[i,t]=0$, $a[i,t]<2^{t}$, $a[i,t]=A_{1}[i,t]2^{q}+A_{2}[i,t]$,
and $A_{2}[i,t]<2^{q}$.

This property uniquely determines the integers $A_{0},A_{1},A_{2}\in \boldm_{d}$,
 moreover Lemma \rref{Y32}
implies that that there exists an existential formula $\phi$ of $\calm$
such that $A_{0},A_{1},A_{2}$ are the unique elements of $\boldm_{d}$
such that $\boldm_{v}\models \phi(a,A_{0},A_{1},A_{2},d,t) $, where $v=t+c(d-t)$.

Let $B_{0}=\odot_{d,t}(2^{2^{t}-1},\min_{d,t}(A_{0},e_{d,t}))$. (That is $B_{0}[i,t]=2^{2^{t-1}}$ if $A_{0}[i,t]\not= 0$, otherwise  $B_{0}[i,t]=0$.)

Then $$\bfp_{d,t}(a)={\max}_{d,t}\biggl(B_{0}, \otimes_{d,t} \Bigl(\bfp_{d,t}[ \odot_{d,t} (2^{q},A_{1})], \bfp_{d,t}[A_{2}] \Bigr)\biggr)$$
We know already that the functions symbols $\max$ and $\times$ are polynomially 
existential, therefore it is sufficient to show that the following two families of relations relations are polynomially existential:

\begin{cond} \llabel{Y33.3}
$\Theta=\langle \Theta_{d,t}  \mid \langle 
d,t\rangle \in \nbl\rangle$, 
where
for each $d,t\in \omega$ with $d\ge t$ and for each $a,b\in \boldm_{d}$, $\Theta_{d,t}(a,b)$ holds iff $t>K$ and  for all $i\in 2^{d-t}$, $a[i,t]< 2^{t}$ and $2^{q} | a[i,t]$
and $b=\bfp_{d,t}(\odot_{d,t}(2^{q},a))$
\end{cond}

\begin{cond} \llabel{Y33.2}
$\Phi=\langle \Phi_{d,t}  \mid \langle 
d,t\rangle \in \nbl\rangle$, 
where
for each $d,t\in \omega$ with $d\ge t$ and for each $a,b\in \boldm_{d}$, 
$\Phi_{d,t}(a,b)$ holds iff $t>K$ and  for all $i\in 2^{d-t}$, $a[i,t] <2^{q}$
and $b=\bfp_{d,t}(a)$
\end{cond}

In other words, the original problem about the existential definability of 
$\bfp_{d,t}(a)$
 has to be solved with some additional assumptions.
The following lemma  will be used to  define the family $\Theta$ in an existential way.

\begin{proposition} \llabel{Y34} The family of relations $\Theta$ defined in condition
\rref{Y33.3} is uniformly  existential.
\end{proposition}

Proof of Lemma \rref{Y34}. For each $d\in \omega$, let $\Theta'_{d}$ be the ternary
relation on $\boldm_{d}$, defined by: for each $t,a,b\in \boldm_{d}$,
$\Theta'_{d}(a,b,t)$ holds iff $t\le d$ and $\Theta_{d,t}(a,b)$.
We show that the family of relations  $\Theta'=\langle \Theta_{d}' \mid d\in 
\omega\rangle$ is uniformly existential. This clearly implies the conclusion of the 
proposition.

By Lemma \rref{Y36} it is sufficient to show that there exists  $3k$-sensitive
 propositional formula $P(x_{0},...,x_{3k-1},y_{0},...,y_{l-1})$ of $\calm$ 
with $k=3$ and $l=1$ 
and with
property.

\begin{cond} \llabel{Y37.1}
For all $d\in \omega$ and for all $t,q,a,b\in \boldm_{d}$, we have $\Theta'_{d}(a,b,t)$ iff there exists
$a_{0},...,a_{3},w_{0}\in \boldm_{d}$ such that $a_{0}=a$, $a_{1}=b$,
$a_{2}=e_{d,t}$,   $d\ge t$, $q=\lfloor \log_{2} t\rfloor$, $w_{0}=2^{q}$  and for all $i\in 
2^{d-q}$, $\boldm_{q}\models  P(\vec A_{0},...,\vec A_{3},w_{0})$,
where $\vec A_{j}$ stands for the sequence 
$a_{j}[i-1,q],a_{j}[i,q],a_{j}[i+1,q]$ for $j=0,1,2,3$.\end{cond}

The condition that there exists $a_{0},...,a_{3},w_{0}\in \boldm_{d}$ such that $a_{0}=a$, $a_{1}=b$,
$a_{2}=e_{d,t}$,   $d\ge t$, $q=\lfloor \log_{2} t\rfloor$, $w_{0}=2^{q}$
can be expressed by an existential formula $\chi$  in $\boldm_{d}$ since $q=\lfloor \log_{2} 
t\rfloor$ is equivalent to $2^{q} \le t < 2^{q+1}$. 
Therefore condition \rref{Y37.1} implies that  $\Theta'(d)(a,b,t)$ is equivalent to
$\boldm_{d}\models
\exists a_{0},...,a_{3},w_{0}, \chi(a,b,t,a_{0},...,a_{3},w_{0})$ and for all
$i\in 2^{d-q}$, $\boldm_{q}\models P(\vec A_{0},...\vec A_{3},w_{0}) $, where $\chi$ is an existential formula of $\calm$.

 Lemma \rref{Y36} implies that
the condition  ``for all $i\in 
2^{d-q} $, $ \boldm_{q} \models  P(\vec A_{0},...,\vec A_{k-1},b_{0},...,b_{l-1})$" can be expressed  by an 
existential formula of $\calm$ in $\boldm_{d}$, therefore we only have to prove
the existence of a $9$-sensitive propositional formula $P$ satisfying condition
\rref{Y37.1}.

In the definition of $P$	the variable $a_{3}$ will be denoted by $h$
and for $a_{0},a_{1},a_{2}$ we will use the symbols $a,b,e_{d,t}$ indicated in condition
\rref{Y37.1}. To make the formulas more understandable we rename 
the variables $x_{i}$ in the following way:
 $x_{3j}=z_{j,-1},x_{3j+1}=z_{j,0},x_{3j+1}=z_{j,1}$. The advantage of this notation is that in condition \rref{Y37.1} the variable  $z_{j,\delta}$
takes the value  $a_{j}[i+\delta]$ for $j=0,1,...,k-1$, $\delta=-1,0,1$.

We define  $P$ by $P\equiv \bigwedge _{r\in 5} \Lambda_{r}$ where the 
propositional formulas   $\Lambda_{r}$ are defined below. First we write 
each formula  $\Lambda_{r}$ with the variables 	$z_{j,\delta}$, this is its 
definition,  
and then as a motivation,  we write the formula $\Lambda_{r}$ in the form that we get
if the variables take the values indicated y condition \rref{Y37.1} and  by the abbreviation
$h=a_{3}$.

  $ \Lambda_{0}\equiv z_{2,0}=1 \rightarrow z_{3,0}=z_{0,0}$\\
\centerline{$e_{d,t}[i,q]=1$ implies $h[i,q]=a[i,q]$} \vskip 5pt

 $\Lambda_{1}\equiv ( z_{2,0} = 0 \wedge z_{3,-1}\not= 0) \rightarrow  z_{3,0}=z_{3,-1}-w_{0}$\\
\centerline{$e_{d,t} = 0 \wedge h[i-1,q]\not= 0$ implies  $h[i,q]=h[i-1,q]-2^{q}$} \vskip 5pt

  $\Lambda_{2}\equiv   (z_{2,0} = 0 \wedge z_{3,-1}= 0) \rightarrow z_{3,0}=0$\\
\centerline{$e_{d,t}[i,q] = 0 \wedge h[i-1,q]= 0$ implies  $h[i,q]=0$} \vskip 5pt

 $\Lambda_{3}\equiv z_{3,0}\not=w_{0} \rightarrow z_{1,0}=0$\\
\centerline{$h[i,q]\not=2^{q}$ implies $b[i,q]=0$} \vskip 5pt

  $\Lambda_{4}\equiv z_{3,0}=w_{0}  \rightarrow z_{1,0}=1$\\
\centerline{$h[i,q]=2^{q}$ implies $b[i,q]=1$} \vskip 5pt

The meaning of these formulas is the following. Assume that an integer
$i_{0}\in 2^{d-q}$ is given with $e_{d,t}[i_{0},q]=1$. It means that $2^{i_{0}2^{q}}=
2^{\nu 2^{t}}$ for some $\nu\in 2^{d-t}$, therefore $i_{0}=\nu 2^{t-q}$.
Because of the assumption $a[\nu,t]<2^{2^{q}}$  have that $a[\nu,t]=a[i_{0},q]$.
We have to show that the described formulas together are equivalent to,
$b[\nu,t]=2^{a[i_{0},q] 2^{q}}$. Since $b[\nu,t]$ is a power of $2^{2^{q}}$, its
$2^{2^{q}}$-ary form contains a single nonzero digit and this digit will be one.

The propositional formula $P$ as we have defined it  determines when 
$b[i,q]=1$ in the following way. The value of $h[i_{0},q]$ is $a[i_{0},q]$,
then as starting from $i=i_{0}$ each time we increase $i$ with $1$ the value
of $h_{i,q}$ will decrease by $2^{q}$. Therefore we will have  $2^{i2^{q}}=
2^{a[\nu,t]2^{q}}$ when $h[i,q]$ becomes $0$ so at that value of $i$ we have
$b[\nu,t]=2^{i 2^{q}}$. The propositional formula $P$ defined above describes
this definition of $h$ and the connections between the values of $a,b,h$ and $e_{d,t}$.
The syntax of the formula shows that it is $3k$-sensitive, so it satisfies
condition  \rref{Y37.1}.
  \enp{(Proposition \rref{Y34})}

\begin{proposition} \llabel{Y38} The family of relations $\Phi$ defined in condition
\rref{Y33.2} is uniformly  existential.
\end{proposition}

Proof of Proposition \rref{Y38}.   The proof is similar to the 
proof of 
Proposition \rref{Y34}.
 For each $d\in \omega$, let $\Phi'_{d}$ be the ternary
relation on $\boldm_{d}$, defined by: for each $t,a,b\in \boldm_{d}$
$\Phi'_{d}(a,b,t)$ holds iff $t\le d$ and $\Phi_{d,t}(a,b)$.
We show that the family of relations  $\Phi'=\langle \Phi_{d}' \mid d\in 
\omega\rangle$ is uniformly existential. This clearly implies the conclusion of the 
proposition.

By Lemma \rref{Y36} it is sufficient to show that there exists  $3k$-sensitive
 propositional formula $P(x_{0},...,x_{3k-1},y_{0},...,y_{l-1})$ of $\calm$ 
with $k=7$ and $l=0$ 
and with
property.

\begin{cond} \llabel{Y38.1}
For all $d\in \omega$ and for all $t,q,a,b\in \boldm_{d}$, we have $\Phi'_{d}(a,b,t)$ iff there exists
$a_{0},...,a_{6}\in \boldm_{d}$ such that $a_{0}=a$, $a_{1}=b$,
$a_{2}=e_{d,t}$,  $a_{3}=e_{d,t}2^{(2^{t-q}-1)2^{q}}$, 
$a_{4}=b2^{(2^{t-q}-1)2^{q}}$  $d\ge t$, $q=\lfloor \log_{2} t\rfloor$,   and for all $i\in 
2^{d-q}$, $\boldm_{q}\models  P(\vec A_{0},...,\vec A_{4})$,
where $\vec A_{j}$ stands for the sequence 
$a_{j}[i-1,q],a_{j}[i,q],a_{j}[i+1,q]$ for $j=0,1,...,6$.\end{cond}

The condition that there exists $a_{0},...,a_{6}\in \boldm_{d}$ such that 
$a_{0}=a$, $a_{1}=b$,
$a_{2}=e_{d,t}$, $a_{3}=e_{d,t}2^{(2^{t-q}-1)2^{q}}$, $a_{4}=
b2^{(2^{t-q}-1)2^{q}}
 $  $d\ge t$, $q=\lfloor \log_{2} t\rfloor$, 
can be expressed by an existential formula $\chi$  in $\boldm_{d}$ since $q=\lfloor \log_{2} 
t\rfloor$ is equivalent to $2^{q} \le t < 2^{q+1}$. 
Therefore condition \rref{Y38.1} implies that  $\Phi'(d)(a,b,t)$ is equivalent to
$\boldm_{d}\models
\exists a_{0},...,a_{6}, \chi(a,b,t,a_{0},...,a_{6})$ and for all
$i\in 2^{d-q}$, $\boldm_{q}\models P(\vec A_{0},...\vec A_{6}) $, where $\chi$ is an existential formula of $\calm$.

 Lemma \rref{Y36} implies that
the condition  ``for all $i\in 
2^{d-q} $, $ \boldm_{q} \models  P(\vec A_{0},...,\vec A_{k-1},b_{0},...,b_{l-1})$" can be expressed  by an 
existential formula of $\calm$ in $\boldm_{d}$, therefore we only have to prove
the existence of a $k$-sensitive propositional formula $P$ satisfying condition
\rref{Y38.1}. (In the present case $k=7$ and $l=0$.)

In the definition of $P$	the integer $a_{5}$ will be denoted by $h$, and the 
integer $a_{6}$ by $g$
and for $a_{0},a_{1},a_{2},a_{3},a_{4}$ we will use the expressions $a,b,e_{d,t}$,
$e_{d,t}2^{(2^{t-q}-1)2^{q}}$, $b 2^{(2^{t-q}-1)2^{q}}$ indicated in condition
\rref{Y38.1}. To make the formulas more understandable we rename 
the variables $x_{i}$ in the following way:
 $x_{3j}=z_{j,-1},x_{3j+1}=z_{j,0},x_{3j+1}=z_{j,1}$. The advantage of this notation is that in condition \rref{Y38.1} the variable  $z_{j,\delta}$
takes the value  $a_{j}[i+\delta]$ for $j=0,1,...,k-1$, $\delta=-1,0,1$.

We define  $P$ by $P\equiv \bigwedge _{r\in 8} \Lambda_{r}$ where the 
propositional formulas   $\Lambda_{r}$ are defined below.
We write the formulas $\Lambda_{r}$, $r\in 8$ in the form when already
$a_{j}[i+\delta,q]$ has been substituted for $z_{j,\delta}$. The formulas 
$\Lambda_{r}$ with the variables $z_{j,\delta}$  (as in Proposition \rref{Y34}), can be derived from this by performing the 
reverse substitutions.

\begin{cond} \llabel{AA48.1} 

  $\Lambda_{0}\equiv e_{d,t}[i,q]=1 \rightarrow (h[i,q]=a \wedge  g[i,q]=1)$
\end{cond}

\begin{cond} \llabel{AA48.2} 
$\Lambda_{1}\equiv (e_{d,t}[i,q]=0 \wedge h[i-1,q]\not=0) \rightarrow 
h[i,q]=h[i-1,q]-1$

$\Lambda_{2}\equiv (e_{d,t}[i,q]=0 \wedge h[i-1,q]=0) \rightarrow 
h[i,q]=0$
\end{cond}

\begin{cond} \llabel{AA48.3}
$\Lambda_{3}\equiv (e_{d,t}[i,q]=0 \wedge h[i,q]< h[i-1,q]=0) \rightarrow g[i,q]=g[i-1,q]+g[i-1,q]$

$\Lambda_{4} \equiv e_{d,t}[i,q]=0 \wedge h[i,q]=h[i-1,q] \rightarrow 
g[i,q]=g[i-1,q]$

\end{cond}

\begin{cond} \llabel{AA48.4} 

$\Lambda_{5} \equiv (e_{d,t}2^{(2^{t-q}-1)2^{q}})[i,q]=1 \rightarrow
 (b 2^{(2^{t-q}-1)2^{q}})[i,q]=g[i,q]$

$\Lambda_{6} \equiv (e_{d,t}2^{(2^{t-q}-1)2^{q}})[i,q]=0 \rightarrow
 (b 2^{(2^{t-q}-1)2^{q}})[i,q]=0$
\end{cond}

The meaning of these formulas is the following. Assume that an integer
$i_{0}\in 2^{d-q}$ is given with $e_{d,t}[i_{0},q]=1$. It means that 
$2^{i_{0}2^{q}}=
2^{\nu 2^{t}}$ for some $\nu\in 2^{d-t}$, therefore $i_{0}=\nu 2^{t-q}$.
Because of the assumption $a[\nu,t]<2^{q}$  have that $a[\nu,t]=a[i_{0},q]$.
We have to show that the described formulas together are equivalent to,
$b[\nu,t]=2^{a[i_{0},q] 2^{q}}$. Since $b[\nu,t]$ is a power of $2^{2^{q}}$, its
$2^{2^{q}}$-ary form contains a single nonzero digit and this digit will be one.

The propositional formula $P$ as we have defined it  determines when 
$b[i,q]=1$ in the following way. 
  We consider the sequence
$h[i,q]$, for  $i=0,..., 2^{t-q}-1$. According to the definition of $\Lambda_{0}$	it starts with 
the integer $a$, and then, according to $\Lambda_{1}$ and $\Lambda_{2}$, each element of the sequence is  
smaller by $1$ 
than the previous one, till it 
reaches the value  $0$ where it remains constant. Since $a<2^{q}= 2^{ \lceil \log_{2} 
t\rceil} \log\le 2t < 2^{t-q}$
the value $0$ will be reached for some $i\in 2^{t-q}$.
The  sequence $g[i,q]$, $i=0,1,...2^{t-q}$ starts with $g[0,q]=1$ according to $
\Lambda_{0}$
and it is increasing by a factor of $2$ at each step where the sequence
$h[i,q]$, $i=0,1,...,2^{t-q}$ is decreasing by $1$, and then remains constant.
 Since the sequence $h[i,q]$,  $i=0,1,...,2^{t-q}$ reaches the value $0$ in $a$ steps, 
the sequence
$g[j,q]$ reaches the value $2^{a}$ in the same $a$ steps and this will be its last value 
as well, that is, if $i_{1}$ is the largest integer  with $ i_{0}+2^{t-q}>  i_{1}>i_{0}$
then
$g[i_{1},q]=2^{a}$. The integer $i_{1}$ is also the unique integer $i$ in the interval
$[i_{0},i_{0}+2^{t-q}-1]$ with the property that $(e_{d,t}2^{(2^{t-q}-1)2^{q}})
[i,q]=1$, so it can be used to identify $i_{1}$ in $\Lambda_{5}$ and $\Lambda_{6}$.
\enp{(Proposition \rref{Y38})}
\enp{(Lemma \rref{Y31})}

\eject
\section{\llabel{rampoly} RAMs and polynomially existential functions,
Proof of Theorem~\rref{TT8}}

Proof of Theorem \rref{TT8}.   Assume that $F=\langle F_{d,u}\mid \langle 
d,u\rangle \in {\nbl}\rangle$  is a family of $k$-ary functions such that
for each $d,u\in \omega $ with $d\ge u$, $F_{d,u}$ is a function defined
on $\boldm_{d}$ with values in $\lbrace 0,1\rbrace $ and the family is polynomial time
computable with respect to $\boldm$.  We have to show that
there exists a $c\in \omega$  and an 
existential  formula $\phi_{0}$
of $\calm$ such that 

 \begin{cond} \llabel{Y20.1}  for all $d,u\in\omega$ with $d\ge 
u$ and for all  
$a_{0},...,a_{k-1}\in \boldm_{d}$, $b\in \lbrace 0,1\rbrace $ we have
	 $F_{d,u}(a_{0},...,a_{k-1})=b$
iff $\boldm_{v} \models \phi_{0}(a_{0},...,a_{k-1},b,d,u)$, where $v=u+c(d-u)$.
\end{cond}

The assumption that the family  $F$ is polynomial time computable with respect to
$\boldm$ implies that there exist a $\gamma_{1}$
and a program $P$ such that  the following holds

\begin{cond} \llabel{Y25.4}
for all sufficiently large $d\in \omega$, for all $u\in \omega$ with $d\ge u$, and for all
$a_{0},...,a_{k-1}\in \boldm_{d}$, machine $N_{m}$ (a RAM with word length $m$), where $m=2^{d}$, 
with  program $P$ and input $k,d,u, a_{0},...,a_{k-1}$, using only the first $2^{\gamma_{1}(d-
u)}$ memory cells in time $2^{\gamma_{1}(d-u)}$ computes  $F_{d,u}(a_{0},...,a_{k-1})$. 
\end{cond}

We will choose the constant  $c\in \omega$  later. 
 We define  now  the existential formula $\phi_{0}$ of $\calm$.
  For this definition we will assume that each
operation of $\calm$ is binary. (The unary operations are  considered binary operations 
which do not depend on their second argument and the constants are considered as 
binary operations which do not depend on any of their arguments.)

Let   $m=2^{d}$, $s=2^{\gamma_{1}(d-u)}$. We will denote by $\calr$  a random access with word length $m$, which has
$s$ memory cells. (That is we get $\calr$ from $N_{m}$ keeping only its
first $s$ memory cells.)  
Our assumption
is that the machine $\calr$ with program $P$ and with input $k,d,u,a_{0},...,a_{k-1}$
in time $s$ computes the value of $F_{d,u}(a_{0},...,a_{k-1})$.
More precisely we assume the following.
 
Suppose that $P$ is the sequence $p_{0},...,p_{c'-1}$, and at time $0$ the content
of $\cell_{i}$ is $\rho_{i}$ for all $i\in s$. Then $\rho_{i}=p_{i}$ for all $i\in c'$,
$\rho_{c'}=k$, $\rho_{c'+1}=d$, $\rho_{c'+2}=u$, $\rho_{c'+3+i}=\alpha_{i}$
for all $i\in k$, and $\rho_{j}=0$ for all $j\in s$ with $j\ge c'+3+k$.
Our assumption is that if the machine starts to work with this initial state then at time
$s-1$  the content of $\cell_{0}$ is $F_{d,u}(a_{0},...,a_{k-1})$.

 Let $\rho=\sum_{i=0}^{s-1}\rho_{i}2^{i2^{u}}$. With this notation
at time $0$ the content of $\cell_{i}$ is $\rho[i,u]$ for all $i\in s$. 
We will say that the integer $\rho$ is the unified input of the machine $\calr$. (The motivation for this expression is that $\rho$ determines  all of the integers 
in the input sequence and the program $P$ as well.)	 It is important that 

\begin{cond} \llabel{Y22.1}
there exists a term
 $\xi_{0}$
of $\calm$ depending only on $P$  and $k$ such that
$\boldm_{v} \models \rho=\xi_{0}(a_{0},...,a_{k-1},d,u) $.
\end{cond}
This implies that in the formula $\phi_{0}$ to be defined we can use $\rho$ as 
an argument.

Now we will use the following trivial fact. If $M$ is a random access machine which
works with $\nu$ memory cells till time $\nu$ then the output of $M$ can be computed
by   a circuit $C$ of size $\nu^{c_{3}}$ where $c_{3}$ is a constant, and the circuit
$C$  is given
independently of the contents of the memory cells at time $0$. The gates of the
circuit are performing the $\calm$ operations. For each memory cell $x$ of $\calm$ there 
is an input node of $C$ where the input is the content of cell $x$ at time $0$.
Moreover the circuit $C$ can be constructed by a turing machine with the input $\nu$ 
in time $\nu^{c_{4}}$, where $c_{4}\in \omega$ is a constant.

We apply this for the present situation with $M\legyen \calr$, $\nu\legyen s$ and 
get the following, where $\bfk$ denotes the number of $\calm$ operations.

\begin{proposition} \llabel{Y26}
There exists a sequence
of triplets $T=\langle \langle \alpha_{i},\kappa_{i},\lambda_{i}\rangle \mid i\in 
s^{c_{3}}\rangle$  with the following properties:

\begin{cond} \llabel{Y11.3} for all $i\in s$, $\alpha_{i}=\kappa_{i}=\lambda_{i}
=0$, and for all $i\in s^{c_{3}}\bcks s$,
 $\alpha_{i}\in \bfk$, $\kappa_{i}\in i$, $\lambda_{i}\in i$,
\end{cond}

\begin{cond} \llabel{Y11.4}
 there exists  a turing machine $\calt'$ such that if the machine  $\calt'$ gets $s$  
 as input,  then it computes in time
$s^{c_{4}}$, the sequence $\langle\alpha_{i},\kappa_{i},\lambda_{i}\rangle$,
\end{cond}

\begin{cond} \llabel{Y11.5} if the sequence  $\bfb=\langle \beta_{i} \mid i\in s^{c_{3}} 
\rangle$ satisfies conditions (i) and (ii), then it also satisfies condition (iii), where

(i) for all $i\in s$, $\beta_{i}=\rho[i,u]$, that is, $\beta_{i}$ is the content of 
$\cell_{i}$, at time $0$
 in the machine $\calr$ with  unified
input $\rho$,

(ii) for all $i\in  s^{c_{3}}\bcks s$,
 $\boldm_{m} \models \beta_{i}=\bff_{\alpha_{i}}(\beta_{\kappa_{i}},\beta_{\lambda_{i}})$,

(iii)
$\beta_{s^{c_{3}}-1}$ is the output of $\calr$ at time
$s^{c}$ at unified input $\rho$. \end{cond}
\end{proposition}

In other words the circuit $C$ has a node $x_{i}$ for each $i\in s^{c_{3}}$.
The nodes $x_{i}, i\in s$ are the input nodes, the node $x_{s^{c_{3}}-1}$
is the output node, and for each $i\in s^{c_{3}}\bcks s$,  at node $x_{i}$
there is a gate performing the operation $\bff_{\alpha_{i}}$ on the arguments
which are the outputs of gates (or input nodes)  at nodes $x_{\kappa_{i}}$ and 
$x_{\lambda_{i}}$. 

Assume now that a turing machine $\calt'$ is fixed that determines a sequence
$T$ with the properties described above. The  unified input $\rho$
and the program $P$ are also fixed.
The formula $\phi_{0}$ in $\boldm_{v}$ will be equivalent to the following: ``there exists a sequence
$B=\langle B_{i} \mid i\in s^{c_{3}}\rangle $ which satisfies conditions (i) and (ii)
of Proposition \rref{Y26}
with $\beta_{i}\legyen B_{i}$	 and  for this sequence 
$B$ we have $B_{s^{c_{3}}-1}=b$".  For such a formula $\phi_{0}$ we clearly have 
 $F_{d,u}(a_{0},...,a_{k-1})=b$
iff $\boldm_{v} \models \phi_{0}(a_{0},...,a_{k-1},b,d,u)$.

As a first step we reformulate the properties of (i) and (ii) of Proposition
\rref{Y26}. The goal of this reformulation is to make these properties more easily 
expressible by existential formulas in $\boldm_{v}$. 
  We will use the following notation: $s_{0}=s^{c_{3}}$,
$h=\lceil \log_{2} \bfk \rceil$.

\begin{proposition} \llabel{Y27} Assume that $B=\langle B_{i} \mid i\in s_{0} \rangle$ is a sequence
with $B_{i}\in \boldm_{m}$. Then conditions (i) and (ii) of Proposition \rref{Y26}
are satisfied with $\bfb\legyen B$ iff there exist $2+\bfk+h$ sequences $K$, $L$, $S^{(j)}$,
$j\in \bfk$, $Q^{(r)}$, $r\in h$, all them of length $s_{0}$, satisfying the following
conditions

\begin{cond} \llabel{Y11.6} for all $i\in s$, $B_{i}$ is the content of $\cell_{i}$ 
at time $0$
 in the machine $\calr$ with program $P$ and unified input
input $\rho$,
\end{cond}

\begin{cond} \llabel{Y11.7} for all $i\in s_{0}\bcks s$, 
$K_{i}=B_{\kappa_{i}}$, $L_{i}=B_{\lambda_{i}}$, and  for all $i\in s$, $K_{i}=L_{i}=0$.
\end{cond}

\begin{cond} \llabel{Y11.8} for all $i\in s_{0}$, $r\in h$,
 $Q^{(r)}_{i}= \alpha_{i}[r,2] $ (and consequently $Q^{(r)}(i)\in \lbrace 0,1\rbrace 
$).
\end{cond}

\begin{cond} \llabel{Y11.9} for all $i\in s_{0}$, $j\in \bfk$,
 $\boldm_{m} \models S^{(j)}_{i}= \bff_{j}(K_{i},L_{i})  $.
\end{cond}

\begin{cond} \llabel{Y11.10} for all $i\in s_{0}$, 
$B_{i}=S_{i}^{(j)}$, where $j$ is the unique integer with  $j\in \bfk$, and
$\forall r\in h, j[r,s]=Q_{i}^{(r)}$. 
 \end{cond}
\end{proposition}

Proof of Proposition \rref{Y27}
Since we fixed the turing machine $\calt'$ the sequence $\langle  \langle \alpha_{i}, 
\kappa_{i},
\lambda_{i} \rangle \mid i\in s^{c_{3}}\rangle $ is given.
Assume first that $\bfb\legyen B$  satisfies conditions (i) and (ii) of property
\rref{Y11.5}. Then 
the sequences $B=\langle \beta_{i} \mid i\in s^{c_{3}}\rangle$,
$K=\langle \beta_{\kappa_{i}} \mid i\in s^{c_{3}}\rangle$,
$L=\langle \beta_{\lambda_{i}} \mid i\in s^{c_{3}}\rangle$, obviously satisfy
conditions \rref{Y11.6} and \rref{Y11.7}. Condition \rref{Y11.8} and \rref{Y11.9} define 
the sequences
$Q^{(r)} $ and $ S^{(j)}$, $r\in h$, $j\in \bfk$, and condition
condition \rref{Y11.10} is the same as condition (ii) of property \rref{Y11.5}.

In the other direction assume that the sequences $B,K,L$, etc, satisfy conditions
 \rref{Y11.6},
 \rref{Y11.7}, \rref{Y11.8},  \rref{Y11.9}, and we show that $\bfb\legyen B$ satisfies 
conditions (i),(ii) of property \rref{Y11.5}. Condition \rref{Y11.6} imply condition (i). 
By condition \rref{Y11.10} and \rref{Y11.8} we have   $B_{i}=S_{i}^{(\alpha_{i})}$,
and therefore by condition \rref{Y11.9}, $B_{i}=\bff_{\alpha_{i}}(K_{i},L_{i})$
which implies condition (ii).
\enp{(Proposition \rref{Y27})}

Let $c\in \omega$ be constant sufficiently large with respect to $
\gamma_{1},k,\length(P),\bfk, c_{3},c_{4}$, and let $v=u+c(d-u)$.
We show that for each  
of the  conditions  \rref{Y11.6},...,\rref{Y11.10}
 there exists an existential formula $\psi$ such that the condition holds iff $\boldm_{v}\models 
\psi(\bar B,\bar K,\bar L, \vec {\bar S},
\vec {\bar Q},\rho,d,u)$, where for each sequence $A=\langle a_{i} \mid i\in s^{c_{3}}\rangle$, $\bar A$ is the 
integer $\sum_{i=0}^{s^{c_{3}-1}}a_{i}2^{i 2^{u}}$, and $\vec {\bar S}=\langle \bar 
S^{(0)},...,S^{(\bfk-1)}\rangle$, $\vec {\bar Q}=\langle \bar 
Q^{(0)},...,Q^{(h-1)}\rangle$.

Condition \rref{Y11.6} is equivalent to 
$\boldm_{v} \models  \overline{\mod}(\bar B,2^{s^{c_{3}}}2^{u})=\rho(a_{0},...,a_{k-1},d,u)$,
where $\overline{\mod}(x,y)$ is a term of $\calm$ such that 
for all $w\in \omega$ and for all $x,y,z\in w$
$\boldm_{w}\models z=\overline{\mod}(x,y)$  iff the least nonnegative residue of $x$ modulo $y$
is $z$, that is, $z=\mod(x,y)$. 

To show that condition \rref{Y11.8} is equivalent to an existential formula in 
$\boldm_{v}$ we use Lemma \rref{Y12}.   We have that
$\alpha_{i}\le \bfk$, $\bfk$ is a constant and the turing machine $\calt'$ computes the
bits of $\alpha_{i}$ in time $2^{c_{4}(d-u)}$ where $c_{4}\ll c$  (and $v=u+c
(d-u)$). Lemma \rref{Y12} is about
 restricted turing machines. $\calt'$ is not restricted but
can make a restricted machine from it with the definition  $\tplength(\calt')=2^{c_{4}
(d-u)}$. The integer $\bar Q$ is defined already in a way that Lemma \rref{Y12} is 
applicable with $b_{0}\legyen \bar Q$.  
Therefore the existential formula whose existence is stated in Lemma 
\rref{Y12} 
 will describe condition \rref{Y11.8}.

The fact that condition \rref{Y11.7} can be expressed by an existential formula 
is a consequence of Lemma \rref{AA73}. We get the sequences $K_{i}$, $L_{i}$
form the sequence $B_{i}$ by moving its elements to (possibly several)  new places
and inserting $0$s. The destinations of the elements and the places of zeros are 
computed by a turing machine as required by Lemma \rref{AA73}.  

The fact that condition \rref{Y11.9}  can be expressed by an existential formula
in $\boldm_{v}$ is an immediate consequence of Lemma \rref{E82}.

We show now that condition
  \rref{Y11.10} can be described by an existential formula
whose existence is guaranteed by Lemma \rref{Y9}. We have to show that the
condition can be expressed by a propositional formula $\calp$ of $\call^{(=)}$. 
For each $j\in \bfk$, let  $\calb_{j}(x_{0},...,x_{h-1})$ be a boolean expression so
 that if $a_{i}=j[i,2]$ that $\calb_{j}(a_{0},...,a_{h-1})=1$, and otherwise
$\calb_{j}(a_{0},...,a_{h-1})=1$. We define the propositional formula $\calp$ of 
$\call^{(=)}$, by $$\calp(X,Y_{0},...,Y_{\bfk-1},Z_{0},...,Z_{h-1})\equiv
\bigwedge_{j\in \bfk} \calb_{j}(Z_{0},...,Z_{h-1})\rightarrow X=Y_{j}
$$  

Clearly for each $i\in s^{c_{3}}$, 
$\boldn_{\bar u} \models \calp(
 B_{i},S_{i}^{(0)},...,S_{i}^{(\bfk-1)},Q^{(0)},...,Q^{(h-1)}) $, where $\bar 
u=2^{2^{u}}$, iff condition \rref{Y11.10} holds for this particular integer $i$.
Therefore by Lemma \rref{Y9} there exists an existential formula which is 
true in $\boldm_{v}$ iff condition \rref{Y11.10} is satisfied.

We define $\phi_{0}$ as the conjunction of all of the existential formulas that 
expresses the various conditions and the formula $b\in \lbrace 0,1\rbrace \wedge  B_{s^{c_{3}}-1}=b$,
which can be written in $\boldm_{v}$ as $b=\min(\bfegy,  b)\wedge  \div(\boldb, 2^{s^{c_{3}-1}2^{u}})=b$. The formula $\phi_{0}$ defined this way clearly
meets all of our requirements.
\enp{(Theorem \rref{TT8})}

\eject
\section{\llabel{circuits} Circuits}

In this section we will evaluate algebraic circuits by first-order existential formulas.
We consider circuits whose gates are computing
the functions of $\calm$ in a structure $\boldm_{t}$, that is, the
circuit evaluates a term $\mu$ of $\calm$ in the structure $\boldm_{t}$.

Both the structure of the circuit computing the value of the term $\mu$
and the sequence of its inputs are encoded by elements of $\boldm_{v}$, where $v>
t$. We show
in  Lemma \rref{E33} that there exists an existential formula $\phi$, which do not 
 depend on anything, and   which decides whether an element of 
$\boldm_{t}$ is the output of the circuit, provided that $v>t+c \log(|C_{\mu}|)$,
where $C_{\mu}$ is a circuit computing the value of the term $\mu$, and $c$
is a sufficiently large constant. This result
will be used in section \rref{csuklas} where we formulate and prove the ``collapsing 
statement" mentioned in the introduction. In fact the application of Lemma \rref{E33}
will be the key step in that proof.

To give a rigorous formulation of the mentioned result we have to tell how the circuits
computing the values of terms of $\calm$ are encoded in $\boldm_{v}$. We also have 
to describe the method of encoding  the sequence of inputs for such a circuit. This
latter encoding is simpler. If the circuit has $k$ inputs $a_{0},...,a_{k-1}\in 
\boldm_{t}$, then they will be represented by the unique $2^{2^{t}}$-ary natural 
number
whose $2^{2^{t}}$-ary digits are $a_{0},...,a_{k-1}$. This natural number will be 
denoted by $\enc_{k,t}(a_{0},....,a_{k-1})$, that is, $\enc_{k,t}(a_{0},....,a_{k-1})=
\sum_{i=0}^{k-1} a_{i}2^{i2^{t}} $.

This method of encoding a sequence by a single integer will be used in encoding a
circuit. We consider a circuit as a directed acyclic graph with labelings on its vertices
and edges which defines the gates and the flow of information in the circuit. The 
following definition describes the details of the encoding of such a circuit. After that we
will describe some basic properties of the defined encoding and then formulate Lemma
\rref{E33} and sketch of its proof.

\begin{definition} 
We will always assume that all of the function symbols of $\calm$ are $\bff_{0},...,
\bff_{s-1}$. We include a new unary function symbol $\id$  among the function symbols 
of $\calm$ whose  interpretation is always the identity function, that is, for
all $d\in \omega$, $a\in \boldm_{d}$, we have $\boldm_{d}\models \id(a)=a$.
(This will correspond to a gate whose output is the same as its input which will be
useful in circuit constructions.) We define the notion of a
$\xcalm$-circuit.
 (Essentially this will be a finite algebraic circuit
 whose each gate $a$ is associated with one of the
 function symbols 
$\xbff_{i}$, say $\xbff_{i_{a}}$. If an interpretation $\boldm_{t}$ of $\calm$ is fixed, then 
the  gate $a$
performs the  operation $\xbff_{i_{a}}$) in the structure $\boldm_{t}$.

Suppose that $m\in \omega$.  We will say that $C$ is a   $\xcalm$-circuit of size
$m$ if $C$ is 
 a  vertex-labeled and edge-labeled directed acyclic graph with multiple edges  on
the set of vertices $m=\lbrace 0,1,\ldots .,m-1\rbrace $, 	satisfying the
following conditions:

{\sl
(i) Each node has either $0,1$ or $2$ incoming edges, the nodes with $0$ incoming
edges will be called the input nodes.  (The two incoming edges may have a common 
tail). The node $m-1$ will be called the output node.

(ii) If a node has two incoming edges then exactly one of them is labeled by $0$ and
the other is labeled by $1$. If a node has a single incoming edge then this edge is labeled by $0$.

(iii)
Each input node is labeled by the integer $s$, and all of the other nodes are labeled 
by
an element of
the set $\lbrace 0,1,\ldots ,s-1\rbrace $, where $s$ is the number of function symbols
in the language $\xcalm$. If the label of the node $a$ is $i$, where $i\in s$, then
the arity of the function symbol $\bff_{i}$ is identical to the number of incoming
edges at node $a$.

(iv) The input nodes form an initial segment of the ordered set $\lbrace 0,1,\ldots ,
m-1
\rbrace $.} 

We define an evaluation $\chi=\chi_{C}$	of the $\xcalm$-circuit $C$ in a structure $\boldm_{t}$ in the following 
way. A
function $g$ defined on the set of  input nodes with values in  $\boldm_{t}$ will 
be
called an
input. If an input $g$ is given we assign an element  $\chi^{(g)}(a)$ of 
$\boldm_{t}$ to each
node $a$ of the circuit in the following way. If $a$ is an input node then 
$\chi^{(g)}(a)=g(a)$.
Assume
now that $a$ is not an input node 
it is labeled by the integer $j\in s$
and there are  two incoming edges at $a$, say,
 $e_{0}$ labeled by $0$ and $e_{1}$ labeled by $1$.
This implies that the arity of $\bff_{j}$ is $2$.
 If $e_{i}$ starts from the node
$b_{i}$, for $i=0,1$,  then we define $\chi^{(g)} (a)$ by
$\boldm_{t} \models \chi^{(g)}(a)=\xbff_{j}(\chi^{(g)}(b_{0}), \chi^{(g)}(b_{1}))$.
If there is exactly one incoming edge with tail $b $ then the  the arity of
$\bff_{j}$ is $1$ and $\boldm_{t} \models \chi^{(g)}(a)=\xbff_{j}(\chi^{(g)}(b)))$. Finally, if the are no incoming edges at all then $\bff_{j}$ is a constant symbol and $\boldm_{t} \models \chi^{(g)}(a)=\xbff_{j}^{(\tau)}$.

 Our 
assumptions imply that, for a
given input $g$, this
defines a unique
function $\chi^{(g)}$ on the set of nodes of the circuit $C$. The value of the 
function
$\chi^{(g)}$ at
the single output node $m-1$ is called the output of the circuit at input $g$. The
function $\chi^{(g)}$ will be called the evaluation function of the circuit at input $g$.

For later use we define the depth of an element  $a$ of the circuit $C$ as the largest 
natural number $i$ such that there exists  a path of length $i$ starting at an input node 
and ending in $a$. Therefore the depth of each input node is $0$, and the depths of all 
other nodes are positive integers.  The set of all nodes with depth at most $i$ will be 
denoted by $\Start_{i}(C)$. The restriction of the function
$\chi^{(g)}$ to the set $\Start_{i}(C)$ will be denoted by $\chi^{(g,i)}$.
 If we want to make explicit the dependence of $\chi^{(g,i)}$ on the circuit $C$, we 
will write $\chi_{C}^{(g,i)}$.

Assume that $C$ is a $\xcalm$-circuit of size $m$.
The circuit $C$ is uniquely determined by the following three sequences
each of length $m$: \vskip 5 pt

(i) \  the sequence $\langle \alpha_{0,0},\alpha_{0,1},\ldots , \alpha_{0,m-1} 
\rangle$,
where $\alpha_{0,i}\in m$ is the
tail of the incoming edge labeled with $0$,	whose head is node $i$, provided that 
such an
incoming edge exists,  and $\alpha_{0,i}=i$ otherwise,
 \vskip 5pt

(ii) \ the sequence $\langle \alpha_{1,0},\alpha_{1,1},\ldots , \alpha_{1,m-1} \rangle$, 
where
$\alpha_{1,i} \in m$ is the
tail of the incoming edge labeled with $1$, whose head is node $i$, provided that 
such an incoming edge
exists, and $\alpha_{1,i}=i$ otherwise,
\vskip 5 pt

(iii) the sequence $\langle \alpha_{2,0},\alpha_{2,1},\ldots , \alpha_{2,m-1} \rangle $, 
where
$\alpha_{2,i}\in s+1$ is the
label of node $i$. \vskip 5 pt

Our next goal is to encode an $\calm$-circuit $C$ of size $m$ with an integer.
The encoding will depend also on a parameter $d\in \omega$. 
So the circuit will be represented by a pair of integers.
Suppose that a
 $d\in \omega$ is fixed with $m<2^{2^{d}}$.
We also assume that
 $s+1< m$ and therefore $\alpha_{i,j}<
m<2^{2^{d}}$ for all $i\in 3, j\in m$. For each 	$i=0,1,2$, we define
an integer   $\balpha_{i}^{(d)}=\sum_{j=0}^{m-1}\alpha_{i,j}2^{j2^{d}}$.
Since $s+1< \alpha_{i,j}<m$ for all $i\in 3, j\in m$, we have that
    for $i=0,1,2$, the integer $\balpha_{i}^{(d)}$ uniquely determines the sequence 
$\langle \alpha_{i,0},\ldots ,
\alpha_{i,m-1}\rangle$. In fact $\alpha_{i,j}=\balpha_{i}^{(d)}[j,d]$ for all $i\in 3$,
$j\in m$.
We will write $\balpha_{i}^{(d,C)}$ instead of $\balpha_{i}^{(d)}$  if we want to 
make 
explicit the dependence of $\balpha_{i}^{(d)}$ on $C$.

For a given $d\in \omega $ with $m<2^{2^{d}}$, we  encode the circuit $C$
by a single integer  $\circode_{d}(C)$  defined by $\circode_{d}(C)= 
2^{2^{d}}(\sum_{i=0}^{2}\balpha_{i}^{(d)}2^{im2^{d}})+m
$.  The inequalities $\balpha_{i}^{(d)} < 2^{m2^{d}}  $, $i\in 3$  and 
$m<2^{2^{d}}$  imply 
that 
for a fixed language $\xcalm$ and a fixed $d\in \omega $, $\circode_{d}(C)$ uniquely
determines the circuit $C$. Indeed, $m$ is the least nonnegative residue of $\circode
_{d}(C)$ modulo $2^{2^{d}}$. The $2^{2^{m d}}$-ary digits of 
$\circode_{s}(C)-m$ give the integers $\balpha_{i}^{(d)}$ for $i=0,1,2$
and these determine the sequence $\alpha_{i,j}$, $i=0,1,2$, $j\in m$.

The number of nodes in an $\calm$-circuit $C$ will be denoted by $|C|$. \vege\end{definition}

\begin{definition} 
We define two functions $\Circ_{0}$ and $\Circ_{1}$ on the set
of all $\calm$-circuits. If  $C$ is an $\calm$-circuit
and   $d$ is the smallest natural number such that
 $|C|<2^{2^{d}}$, then
$|C|=d$ and $\Circ_{1}(C)=\circode_{d}(C)$.
\vege\end{definition}

The following lemma says that from the integer $\circode_{d}(C)$ we can get 
back all of the elements of the circuit $C$ by computing the values of a term
$\tau$ in a suitably chosen structure $\boldm_{u}$.

\begin{lemma} \llabel{A70}  There exist terms $\sigma(x,y), \tau(x,y,z,w), 
\kappa_{i}(x,y)$, $i=0,1,2$
of $\calm$ such that if $\xcalm$ has $s$ function
symbols, $d,u\in \omega $, $d\le u$,  $s<m<2^{2^{d}}$ and $C$ is a $\xcalm$-circuit 
with
$m$  nodes then  $\circode_{d}(C)\in \boldm_{u} $  implies that 
$\boldm_{u}\models m=\sigma(\circode_{d}(C),d)$     and  also implies that for all 
$i\in 3$, $j\in m$,
$$\boldm_{u}\models  \alpha_{i,j}=\tau(\circode_{d}(C),d,i,j) \wedge \balpha_{i}^{(d,C)}= \kappa_{i}(\circode_{d}(C),d)
$$ where the integers
$\alpha_{i,j}$  are defined  in the definition of $\circode_{d}(C)$. Moreover
$$ |C|=  \circode_{d}(C)- 2^{2^{d}}\lfloor 
\circode(C)/2^{2^{d}}\rfloor $$

\end{lemma}

Proof of Lemma \rref{A70}.
Clearly $m$ is the least nonnegative residue of $\circode_{d}(C)$ modulo 
$2^{2^{d}}$ and so $m= \circode_{d}(C)- 2^{2^{d}}\lfloor 
\circode(C)/2^{2^{d}}\rfloor$. We also have $\balpha_{i}^{(d)}=(\lfloor \circode_{d}
(C)/2^{2^{d}}\rfloor )[i,d]$ for $i\in 3$. Therefore using Lemma \rref{B0.5} and the 
fact that 
$\alpha_{i,j}=\balpha_{i}^{(d)}[j,d]$ for all $i\in 3$,
$j\in m$ we   get the term $\tau$.
\enp{(Lemma \rref{A70})}

\begin{definition}  
If the number of input nodes of  an $\calm$-circuit $C$ 
is $k $   then we will say that  $C$ is a $k$-ary circuit.
 Assume that $g$ is an input of the $k$-ary $\xcalm$-circuit $C$ evaluated according to 
the interpretation $\boldm_{t}$ of $\xcalm$. 
By the 
definition
of an input  this means that $g$ is a function with values in $\boldm_{t}$,
and  defined on the set of input nodes, that is, on the set $k$.  In this case to express 
the fact that the output of the circuit  $C$ at input  $g$ is $a$ we will write
$\boldm_{t}\models a=C(g)$ or $\boldm_{t}\models a=C(g(0),\ldots ,g(k-1))$.
 
Suppose that we evaluate the $\calm$-circuit $C$ in a structure $\boldm_{t}$.
We will encode an
 input sequence  $\langle g(0),\ldots ,g(k-1)
\rangle$ by the integer  $\enc_{k,t}(g(0),\ldots ,g(k-1))= \sum_{i=0}^{k-1} 
2^{i2^{t}}g(i)$.
This number will be called an encoded input of the $\xcalm$-circuit $C$ with respect to 
the 
interpretation $\tau$ of $\xcalm$. Clearly, for a given $\xcalm$-circuit $C$, and a given 
interpretation $\boldm_{t}$ of $\calm$, the encoded 
input uniquely determines the
corresponding input sequence $g$. (This is a consequence of the fact that the length 
of the input $k$ is uniquely  
determined by the circuit $C$, while the integer $t$ is uniquely determined by the 
structure $\boldm_{t}$.) 
 \vege\end{definition}

\begin{lemma} \llabel{W50} For all sufficiently large  $m\in 
\omega$ if $t\in \omega$, if $t\in \omega$, $m<2^{2^{t}}$, and $C$ is 
a $\calm$-circuit of size $m$, then $\circode_{t}(C)< m+2^{(3m +1)2^{t}} $.
\end{lemma}

Proof of Lemma \rref{W50}.  This is an immediate consequence of the definition
of $\circode_{t}(C)$.
 \enp{(\rref{W50})}

\begin{definition} 
Suppose now  that $k\in \omega$, $k\ge1$ and   $\tau(x_{0},\ldots ,x_{k-1})$ is a term of $\calm$, where we allow that 
some of the variables $x_{i}$ does not occur in $\tau$ (but all of the variables of 
$\tau$ is from the set $\lbrace x_{0},...,x_{k-1}\rbrace $).
 In this definition we will 
consider all of variables $ x_{0},...,x_{k-1} $  as  subterms of $\tau(x_{0},\ldots 
,x_{k-1})$,
even those variables which do not actually occur in $\tau$. (For example, if 
$\tau(x_{0},x_{1})$
is the term
 $x_{0}$ then $x_{1}$ is a subterm of $\tau(x_{0},x_{1})$.) 
We construct an $\calm$-circuit of $C_{\tau}$ based on $\tau$. Let
$\sigma=\langle \sigma_{0},\ldots ,\sigma_{m-1}\rangle$ be
a sequence which consists of all of the pairwise distinct 
subterms of $\tau(x_{0},\ldots ,x_{k-1})$. A subterm with several occurrences in $\tau$ 
is represented only once in the sequence $\sigma$.  We also assume that  
$\sigma_{m-1}$ is the term $\tau$, and  for all $i\in k$,
$\sigma_{i}$ is the term $x_{i}$. The set of nodes of the 
circuit
$C_{\tau}$ will be $m$, the label of each node $i\in m$ will be  $j$ if the outmost
$\calm$-operation of the term $\sigma_{i}$ is $\bff_{j}$.  If such an operation does 
not 
exists, that 
is, 
$\tau$ is a variable then the label is $s$. The edges of $C$ are defined in the following 
way. If $i$ is labeled by $j$ then we distinguish three 
cases according to the arity of the function symbol $\bff_{j}$.  If $ \bff_{j}$  is a 
binary function symbol and
$\tau_{i}=\bff_{j}(\tau_{i'},\tau_{i''})$, then an edge labeled with $0$ 
points from
$i'$ to $i$, and an edge labeled by $1$ points from $i''$ to $i$.
If $\bff_{j}$ is  a unary function symbol  and $\tau_{i}=\bff_{j}(\tau_{i'})$, 
then
an edge points from $i'$ to $i$, and it is labeled
 by $1$. 
 Finally if $\bff_{j}$ is a
constant symbol and $\tau_{i}=\bff_{j}$ then are no edges ending at 
 $i$.

The circuit $C_{\tau}$ will be called the 
circuit associated with the term 
$\tau$.	 The size of the circuit $C_{\tau}$, that is, $m$ will be called the circuit size
of $\tau$ and will be denoted by $\csize(\tau)$. 

In this definition the order of the subterms in the sequence 
$\langle \sigma_{0},...,\sigma_{m-1}\rangle$ was arbitrary apart from the
choices of  $\sigma_{0},...,\sigma_{k-1}$ and $\sigma_{m-1}$. Therefore
the circuit $C_{\tau}$ depends on an arbitrary choice in the definition. This choice 
however is important only for the order of the nodes.

The fact the we considered each $x_{i}$, $i\in k$ as a subterm of 
$\tau(x_{0},...,x_{k-1})$ implies that
that the circuit $C_{\tau}$ has exactly $k$ input nodes even if in the evaluation
of the circuit $C_{\tau}$ the input provided at some of the nodes is not used.

We   define the followng functions  on the set
of all $\calm$-terms: $\Circ_{0}(\tau)=\Circ_{0}(C_{\tau})$, $\Circ_{1}(\tau)=
\Circ_{1}(C_{\tau})$ and for all $d\in \omega$, $\circode_{d}(\tau)=\circode_{d}
(C_{\tau})$.

Suppose that $C$ is an $\calm$-circuit with $k$ input nodes,
and $\mu(x_{0},...,x_{k-1})$ is a term of $\calm$. We say that the circuit $C$ computes the term $\calm$ iff for all $d\in \omega$, and for all $a_{0},...,a_{k-1}
\in \boldm_{d}$ we have $\boldm_{d}\models \mu(a_{0},...,a_{k-1})= C(a_{0},
...,a_{k-1})$.  
\vege\end{definition}

In the next lemma logarithm means logarithm of base two.

\begin{lemma} \llabel{EW51} Assume that $C$ is an
 $\calm$-circuit.  Then  $$\log  \log  |C| -1 \le \Circ_{0}(C)\le  
\log \log  |C|$$
 and $$\Circ_{1}(C)\le 
|C|^{8|C|} $$
\end{lemma}

Proof of Lemma \rref{EW51}.  By its definition $\Circ_{0}(C)$ is the smallest natural 
number $d$ with $|C|<2^{2^{d}}$. This implies the bounds on
$\Circ_{0}(C)$. 
According to  Lemma \rref{W50},
if $\Circ_{0}(C)=t$ then $\Circ_{1}(C) =
\circode_{t}(C)\le |C|+2^{(3|C|+1)2^{t}} $. Using the already 
proven upper bound on $t=\Circ_{0}(C)$ we get the claimed 
inequality.
\enp({Lemma \rref{EW51}}).

\begin{lemma} \llabel{E33} 
There exists an existential formula $\phi(x_{0},...,x_{4})$ of $\calm$  with the 
following 
property. For  all sufficiently large $c\in \omega$, for all
$\calm$-circuits $C$, and  for all $t,v\in \omega$,
if the number of inputs of $C$ is $k$ and
$v> t+c \log |C|$ then for all $a_{0},\ldots ,a_{k},b\in \boldm_{t}$,  we 
have
that $\Circ_{0}(C)\in \boldm_{v} $,  $\Circ_{1}(C)\in \boldm_{v} $,
$\enc_{k,t}(a_{0},\ldots ,a_{k-1})=
\sum_{i=0}^{k-1}a_{i}2^{i2^{t}} \in \boldm_{v}$, and
$$\boldm_{t} \models C(a_{0},\ldots ,a_{k-1})=b \ \ \leftrightarrow \ \
\boldm_{v} \models \phi(\enc_{k,t}(a_{0},\ldots ,a_{k-1}),b ,t,\Circ_{0}(C),
\Circ_{1}(C)  ) $$
\end{lemma}

Proof of Lemma \rref{E33}. 
Let $F= \langle F_{d,t} \mid d,t\in {\nbl} \rangle $ be the family of 
quaternary functions defined on $\boldm_{d}$ in the following way. Assume that
$d,t\in \omega$, $d\ge t$, and 
 $w,b, C_{1},C_{2}\in \boldm_{d}$. Then $F_{d,t}(w, b, C_{0},C_{1})\in \lbrace 0,1\rbrace $, and $F_{d,t}(w, b, C_{0},C_{1})=1$ iff the following three 
conditions are satisfied:

\begin{cond} \llabel{Y24.1} there exists an 
 $\calm $-circuit $C$ 
such that
 $d\ge \max(C_{0},t)$, 
$\Circ_{0}(C)=
C_{0}$, $\Circ_{1}(C)=C_{1}$, $2^{d-t}>|C|$,
 \end{cond} 

\begin{cond} \llabel{Y24.15} if condition \rref{Y24.1} holds and $k$ is the number of input nodes of $C$,
then there exists a sequence $a_{0},...,a_{k-1}\in \boldm_{t}$ such that
$w=\enc_{k,t}
(a_{0},...,a_{k-1})=w$, 
\end{cond}

\begin{cond} \llabel{Y24.2} if both conditions  \rref{Y24.1},  \rref{Y24.15} hold
then  
$\boldm_{t} \models C(a_{0},...,a_{k-1})=b$.
\end{cond}

We claim that the family of functions $F$ is polynomial time computable 
with respect to $\boldm$. Let $\gamma_{1}\in \omega$ be a sufficiently large constant.
We will show  that there exists a program $P$
such that the following statement, needed for polynomial time computability
with respect to $\boldm$,  is true (note that the variable $k$ of that
definition has the value $4$ now):

\begin{cond} \llabel{Y28.2}
for all sufficiently large $d\in \omega $, for all $t\in \omega$ with $d\ge t$, and for all
$w,b,C_{0},C_{1}\in \boldm_{d}$, the following holds.   
 The machine $N_{m}$ (a RAM with word length $m$), where $m=2^{d}$, 
with  program $P$ and input $4,d,t, w,b,C_{0},C_{1}$, using only the first 
$2^{\gamma_{1}(d-
t)}$ memory cells in time $2^{\gamma_{1}(d-t)}$ computes  $F_{d,t}( 
w,b,C_{0},C_{1})$. 
\end{cond}

 First assume that there exists a circuit
$C$ satisfying conditions \rref{Y24.1} and \rref{Y24.15}. Then the assumption $2^{d-
t}>|C|$ implies
that  machine $N_{m}$ can determine the underlying directed graph of $C$, and the 
labellings
of its nodes and edges in  time polynomial in $2^{d-t}$.  The assumption  
$m=2^{d}$ implies that each $\calm$-operation  in $\boldm_{t}$	can 
be performed in 
constant time on $N_{m}$ and since the number of nodes of $C$ is at most $2^{d-
t}$, $P$ can determine the integers $a_{i}\in \boldm_{t}$, $i\in k$ and can evaluate 
the
circuit $C$ time polynomial in $2^{d-t}$, and comparing the output to $b$ can 
determine the value  of $F_{d,t}(w,b,C_{0},C_{1})$ in polynomial time.

The same  computation can be performed also by $P$ on an arbitrary input.
If the construction of the graph of $C$ does not terminate in time, or the result
 contradicts
to conditions \rref{Y24.1},  or \rref{Y24.15} then the  value of
$F_{d,t}$ is $0$, otherwise the machine gets the value of $F_{d,t}$ as described 
above. This completes the proof of the fact that the family $F$ is polynomial time 
computable with respect to $\boldm$.

Theorem \rref{TT8} implies that the family $F$ is polynomially existential, that is, 
there exists a  there exists a $c_{0}\in \omega$  and an 
existential  formula $\phi$
of $\calm$ such that 

 \begin{cond} \llabel{Y28.3}  for all $d,t\in\omega$ with $d\ge 
t$ and for all  $w, b, C_{0},C_{1}\in \boldm_{d}$
 and for all
$v\ge t+c_{0}(d-t)$ we have
	 $F_{d,t}(w, b, C_{0},C_{1})=1$
iff $\boldm_{v} \models \phi(w, b, C_{0},C_{1},d,t)$.
\end{cond}

We define an existential formula $\psi(x_{0},x_{1},y_{0},y_{1},w)$
of $\calm$ by $$\psi(x_{0},x_{1},y_{0},y_{1},w) \equiv 
\phi(x_{0},x_{1},y_{0},y_{1},w+c_{2}\bfp(y_{0}),w)$$
where $c_{2}$ is a sufficiently large constant. (The meaning of the expression 
$w+c_{2}\bfp(y_{0})$ is that we want to define $d$, in order to choose a member
 $F_{d,t}$ of the family $F$, by
 $d=t+c_{2}2^{C_{0}}$.)

Assume now that  $t\in \omega$, $C$ is an $\calm$-circuit, with
$k$ inputs, $a_{0},...,a_{k-1},b\in \boldm_{t}$, $\enc_{k,t}(a_{0},...,a_{k-1})=w$,
and let $c_{1}$ be sufficiently large with respect to $c_{0}$. We claim that for all
$v>t+c_{1}\log |C|$,
$\boldm_{t} \models C(a_{0},...,a_{k-1})=b$ iff
$\boldm_{v}\models \psi(w,b,C_{0},C_{1},t)$.

Let $d=t+c_{2} 2^{C_{0}}$. First
we show that
$w,b,C_{0},C_{1}\in
\boldm_{d}$.  By the definition of the function $\enc_{k,t}$ we have 
  $w\le 2^{ |C| 2^{t}}$, therefore it is sufficient to show that $|C|2^{t} <2^{d}$, 
or equivalently
$\log |C|+ t<d$.  By Lemma \rref{EW51} $\frac{1}{2}\log |C|\le 2^{C_{0}}$, and so 
the definition of $d$ implies the claimed inequality.

Since $b\in \boldm_{t}$ and $t\le d$ we have  $b\in \boldm_{t}$. $C_{0}\le d$ and so
$C_{0}\in \boldm_{d}$. Finally by Lemma \rref{EW51}
$C_{1}<|C|^{8|C|}$, therefore it is sufficient to show that $\log\log(|C|^{8|C|}) <d$.
We have $\log\log(|C|^{8|C|})=3+\log |C|+\log\log |C|$. According Lemma 
\rref{EW51} $\log |C| \le 2^{ C_{0}+1}$ so if $c_{2}$ is a sufficiently large constant then 
$C_{1}\in 
\boldm_{d}$. 

Since  $w,b,C_{0},C_{1}\in \boldm_{d}$,
 the definition of the function $F_{d,t}$ implies that $\boldm_{t} \models 
C(a_{0},...,a_{k-1})=b$ iff $F_{d,t}(w, b, C_{0},C_{1})=1$

 Condition \rref{Y28.3} implies that if $v\ge t+c_{0}(d-t)$
then 
$\boldm_{t} \models C(a_{0},...,a_{k-1})=b$ iff 
$\boldm_{v}\models \phi(w,b,C_{0},C_{1},d,t)$
which is equivalent to 
$\boldm_{v}\models \psi(w,b,C_{0},C_{1},t)$.

This is true if $v\ge t+c_{0}(d-t)$. Assume now that we know
only that $v>t+c \log |C|$, as required in the present lemma, where $c\in \omega$
is sufficiently large with respect to $c_{0}$ and $c_{2}$.
We have that $ d-t=c_{2}2^{C_{0}}$, and so by Lemma \rref{EW51} 
$d-t\le c_{2} \log |C|$  and therefore $t+c_{0}(d-t)\le t+ c \log |C|$ and consequently
for all $v\ge t+c \log |C|$,
$\boldm_{t} \models 
C(a_{0},...,a_{k-1})=b$ iff $\boldm_{v}\models \psi(w,b,C_{0},C_{1},t)$.
  \enp{(Lemma \rref{E33})}

 %%%%%%%%eeeeeeeeeeeeeeeeeeeeeeeeeeeeeeeeeeeeeeeeeee

\eject
\section{\llabel{csuklas}  Collapsing and Predictivity}

{\it Notation.} In this section $\log$ will always mean logarithm with base  $2$
unless we explicitly state it otherwise.

\subsection{\llabel{truthvalue}  Expressing the truth value with terms}

This section contains the ``collapsing" argument. We show that if Theorem \rref{TT3}
is not true then the hierarchy of first-order formulas of $\calm$, interpreted in the structures $\boldm_{d}$, collapses in a quantitative sense. For each $d\in \omega$ 
we define a class of first-order formulas $\Theta_{d}$ by giving some bounds on the 
number of their quantifiers, which may depend on $d$. We also define a function 
$\bfg$ on $\Theta_{d}$ with values in $\boldm_{d}$, and a term $\tau$
of $\calm$   with an upper bound on its size, also depending on $d$, such that if $q$
is about $d+\log d$, then for each formula $\phi\in \Theta_{d}$ and for each
$xa\in \boldm_{d}$ we have $$\boldm_{d} \models \phi(a) \leftrightarrow  \boldm_{q}\models \tau(a,\bfg(\phi))=\bfnull $$

That is, our indirect assumption implies that   we are able to express the truth value of a not too large first-order formula in $\boldm_{d}$ as the value of a term
in $\boldm_{q}$, where $q$ is not very much larger then $d$. This will lead,
in the following sections, to the final
diagonalization argument after we also prove the ``simulation statement ", namely, that  each not too large term in $\boldm_{q}$ can be 
evaluated by a  first-order formula $\psi$ in $\boldm_{d}$, moreover 
these formulas  can be chosen from the class $\Theta_{d}$.

The situation will be slightly more complicated than the picture given in the preceding,
paragraphs, since the class $\Theta_{d}$ will depend on other parameters as well, which will make it easier to
choose the first-order formula mentioned above in a way that it meets all of our requirements.

We give now a rigorous formulation of the collapsing statement as outlined above and then we will sketch its proof.

\begin{definition} 
1. Suppose that $\phi(x_{0},\ldots ,x_{k-1})= Q_{0}x_{0},\ldots ,Q_{k-1}x_{k-1}, 
P(x_{0},\ldots ,x_{k-1})$ is a first-order prefix formula of $\calm$, where 
$P(x_{0},\ldots ,x_{k-1})$ is a propositional formula and $Q_{i}$, $i\in k$ are quantifiers. 
In this section if we say that $\phi$ is a prefix formula of $\calm$ we will always
assume, unless we explicitly state it otherwise, that $P(x_{0},\ldots ,x_{k-1})$ is of the 
form $t(x_{0},\ldots ,x_{k-1})=\bfnull$,
where $t$ is a term. (It is easy to see that there exists a $c>0$, such that 
 for each $k\in \omega$, and for each propositional formula
$P(x_{0},\ldots ,x_{k-1})$, there exists a term $t(x_{0},\ldots ,x_{k-1})$ such that
$\length(t)\le c \length (P)$, and  for all $d\in \omega$, and for all 
$a_{0},\ldots a,_{k-1}\in \boldm_{d}$, $\boldm_{d} \models P(x_{0},\ldots ,x_{k-1})
\leftrightarrow t(x_{0},\ldots ,x_{k-1})=\bfnull$.) 
Suppose that $\phi(x_{0},\ldots ,x_{k-1})\equiv Q_{0}x_{0},\ldots ,Q_{k-1}x_{k-1}, 
t(x_{0},\ldots ,x_{k-1})=\bfnull$.  Then $\term(\phi)$ will denote the term 
$t(x_{0},\ldots ,x_{k-1})$.

2. If $\phi(x_{0},\ldots ,x_{k-1})\equiv Q_{0}x_{0},\ldots ,Q_{k-1}x_{k-1}, 
t(x_{0},\ldots ,x_{k-1})=\bfnull$ is a first-order formula of $\calm$, 
then the $\calm$-circuit 
$C_{t}$ associated with the term $t$
will be also denoted by $C_{\phi}$.
\vege\end{definition}

\begin{definition} 
Assume that $\call$ is a first-order language  
and $\call'$ is the second-order extension of $\call$. The set of all second-order
formulas $\Psi$ of $\call'$ which satisfies the following conditions will be denoted by
$\SForm (\call)$:

(i)  $\Psi$ does not contain second-order quantifiers.

(ii) the only second-order variables that may be contained in
 $\Psi$ are variables for $k$-ary relations for some $k\in \omega$.  
(Such a variable represents a $k$-ary relation between the
elements of the universe.)  

Usually we will write such a formula $\Psi$ in the form of $\Psi(x_{0},\ldots ,x_{k-1},
Y_{0},\ldots ,Y_{l-1})$ where  $x_{0},\ldots ,x_{k-1}$ are all of the first-order variables 
contained in $\Psi$,  and $Y_{0},\ldots ,Y_{l-1}$ are all of the  second-order variables 
contained in $\Psi$. According to our
definition the variables $x_{0},\ldots ,x_{k-1}$ represent elements of the universe and
the variables  $Y_{0},\ldots ,Y_{l-1}$ represent $k_{0},\ldots ,k_{l-1}$-ary relations on the 
universe.
\vege\end{definition}

\semmi{YYYY   Motivacio a quantifier elimination assumption hoz}

We formulate below a statement that we will call the $\cald$-quantifier elimination assumption, where $\cald$ can be a  real-valued function defined on $\omega$.
In the case $\cald(x)=\epsilon (\log x)^{\frac{1}{2}}$  the $\cald$-quantifier
elimination assumption follows from the assumption that Theorem \rref{TT3} is not
true.

\begin{definition} 1. Assume that $\cald$ is a 
function. The conjunction of the following two conditions will be
called the $\cald$-quantifier elimination
assumption for $\boldm$ or shortly $\cald$-elimination assumption:

\begin{cond} \llabel{EH51.1}  $\cald$ is a monotone increasing 
function
defined on an interval $[r,\infty)$ of the real numbers with positive real values
 for a suitably chosen $r\ge 0$,  \end{cond}

\begin{cond} \llabel{EH51.2} for all propositional formulas $P(x,y)$ of $\calm$ and  
for
all sufficiently large
 $d\in \omega$, there exists a term $\tau$ of $\calm$, such that $
\depth(\tau)\le \cald(d)$, and for all $a\in \boldm_{d}$,
$\boldm_{d}\models \exists x, P(x,a)$
iff $\boldm_{d}\models \tau (a)=\bfnull$.
\vege\end{cond}
\end{definition}

\begin{definition} Suppose that $\phi$ is a prenex first-order formula of $\calm$.
 The
total number of quantifiers, both existential and universal in $\phi $ will be denoted by 
$\quant(\phi)$.\vege\end{definition}

\vskip 5pt
{\it Notation.} In a first-order formula if a sequence of quantifiers of the same type 
occurs for example $\exists z_{0},\ldots ,\exists z_{k-1}$ then, sometimes, we will 
abbreviate 
it by writing $\exists \vec z$, where $\vec z$ is the sequence of variables 
$z_{0},\ldots ,z_{k-1}$.

\begin{definition} Assume that $\phi$ is a first-order prenex formula of $\calm$,
and $\langle j_{m},...,j_{1}\rangle$
is a sequence of positive integers. 
We will say that the quantifier pattern of $\phi$ is $\langle j_{m},...,j_{1}\rangle$ if 
the following conditions are satisfied.

\begin{cond} \llabel{AA21.1}
$\phi \equiv Q_{m}\vx_{m},\ldots ,Q_{1}\vx_{1},P(\vx_{1},\ldots ,\vx_{m})$, 
 $\vx_{i}$ is a sequence $x_{i,0},\ldots ,x_{i,j_{i}-1}$  of variables of $\calm$,
and $Q_{i}$ is a quantifier binding the variables  in $\vx_{i}$.  
\end{cond}

\begin{cond} \llabel{AA21.2}  There exists a $\delta\in \lbrace 0,1\rbrace $
such that,  for each  for $i=1,...,m$,  
 $Q_{i}$ is universal iff $i\equiv \delta $  $(\mod \ 2)$. 
\end{cond}

We will refer to the expression $Q_{i} \vx_{i}$ in the formula $\phi$ as a block or as a block of
quantifiers. 
We define the notion of quantifier pattern in the same way for prenex formulas
in $\SForm(\calm)$ as well. Since these formulas contain only first-order
 quantification  the definition remains unchanged.
\vege\end{definition}

\begin{remark}  We write in the quantifier pattern $\langle 
j_{m},\ldots ,j_{1}\rangle$ the elements of the sequence $j_{1},...,j_{m}$ in reverse 
order since we will have an inductive proof about formulas with a given quantifier 
pattern which starts with eliminating the innermost  block of quantifiers. 
  This will simplify the notation in the inductive proof. 
\end{remark}

\begin{definition} Assume that $M,j_{1},\ldots ,j_{m}$ are positive integers.
$\Form (M,j_{m},\ldots ,j_{1})$ denotes the set of all prenex first-order formulas  
$\phi$ of 
$\calm
$ such that $\csize(\phi)\le M$ and the quantifier pattern of $\phi$ is $\langle 
j_{m},\ldots ,j_{1}\rangle$. 
\vege\end{definition}

The main result of section \rref{truthvalue} is the following  Lemma  \rref{EC12} which 
is the ``collapsing" statement. The remaining part of this section contains
the proof of  Lemma  \rref{EC12}.

\semmi{YYYY magyarazat, mit jelent Lemma \rref{EC12}}

\begin{lemma} \llabel{EC12}  For all  $c\in \omega\bcks \lbrace 0\rbrace $,  if  $\epsilon>0$ is sufficiently small with respect $c$  then  the following holds. 
Assume that

\begin{cond} \llabel{EC12.0} the
 $\cald$-quantifier elimination assumption holds for $\boldm$, where
 $\cald(x)=\epsilon (\log x)^{\rcp{2}}$,   \end{cond}

\begin{cond} \llabel{EC12.1}
 $d \in \omega$ is sufficiently large with respect to $\epsilon$,
\end{cond}

\begin{cond} \llabel{EC12.2}  $\delta=\lfloor \cald (d+\log d)\rfloor= \lfloor \epsilon (\log (d+\log d))^{\rcp{2}}\rfloor  $,
$m\in \omega $, $m \le c \delta $,   and $\iota_{m},...,\iota_{1}$ are positive integers 
with
$\iota_{m}+ \ldots +\iota_{1} \le c^{\delta}$, $\iota_{m}\le c\delta $.
 \end{cond}

Then there exists a function  $\bfg$ which assigns to each prenex
formula $$\phi \in \Form 
(c^{\delta},\iota_{m},\ldots ,\iota_{1})$$ a natural number 
$\bfg(\phi) <2^{2^{d-1}}$, and there exists a
term $\tau(x,y)$ of $\calm$ 
 such that the following conditions are satisfied:

\begin{cond} \llabel{EC13.3} $\csize(\tau) \le  3\cdot 2^{d+\log d}
$ 
 \end{cond}

\begin{cond} \llabel{EC13.4}  for each prenex  formula $\phi \in \Form( 
c^{\delta},\iota_{m},\ldots ,\iota_{1})$ and 
 for  each $a\in 
\boldm_{d}$,  if $q=d+m \lfloor \frac{\log d}{m}\rfloor$ then $$\boldm_{d}
 \models  \phi (a) \ \leftrightarrow \
\boldm_{q}\models \tau( a, \bfg(\phi)
 )=\bfnull$$
 \end{cond}
\end{lemma}

\begin{remark} The function $\bfg$ plays the role of G\"odel numbering in our
proof. Apart from the upper bound given in the Lemma we do not need
\vege\end{remark}

{\it Sketch of the proof of Lemma \rref{EC12}}. The  $\cald$-quantifier elimination 
 says that  for each    first-order formula  $ \exists x, 
P(x,y)$, of $\calm$, where $P$ is propositional, and for each  $d\in \omega$, there exists a $\tau$ with 
$\depth(\tau)<d$ such that for all $a\in \boldm_{d}$, $\boldm_{d} \models \exists x, 
P(x,y)$ iff $\boldm_{d} \models \tau(a)=\bfnull$.
In in Lemma \rref{EC12} instead of the formula $\exists x, P(x,y)=0$ which does not 
depend on $d$ we have an arbitrary first-order formula $\phi$ whose size may grow 
with $d$. We reach a similar conclusion, namely $\boldm_{d}
 \models  \phi (a) \ \leftrightarrow \
\boldm_{q}\models \tau( a, \bfg(\phi)
 )=\bfnull$. It will help in finding such a term $\tau $ that, (a) the structure 
$\boldm_{q}$
is may be somewhat larger than $\boldm_{d}$, 
and (b) $\tau$ may contain a parameter $\bfg(\phi)$ which encodes the formula
$\phi$ by an integer in $\boldm_{d}$.

The structure of the proof is the following. First we try to eliminate the innermost block
of quantifiers in  $\phi \equiv Q_{m}\vx_{m},\ldots ,Q_{1}\vx_{1},P(\vx_{1},\ldots ,
\vx_{m})$,  namely the block
 $Q_{1} \vx_{1}$. 
We may assume without the loss of generality that $Q_{1}$ is existential (otherwise we 
work with formula $\neg \phi$).
 We want to accomplish the quantifier elimination by using the $\cald$-elimination 
assumption.
We consider the formula without the other quantifiers namely the formula 
$\psi\equiv  \exists \vx_{1},  P(\vx_{1},\vx_{2},\ldots ,\vx_{m})$. Here 
the  $ \vx_{2},\ldots ,\vx_{m}$ are free variables their role is the same that the role
of the variable $y$ in the formula $\exists x, P(x,y)$. If we can show that $\psi$
is equivalent to a propositional formula   $\tau(\vx_{2},\ldots ,\vx_{m})=0 $, in the 
sense that they are equivalent for all choices of the values of the variables
$\vx_{2},\ldots ,\vx_{m}$, then we may replace the original formula $\phi $
by the simpler formula    $Q_{m}\vx_{m},\ldots ,Q_{2}\vx_{2},\tau(\vx_{2},\ldots ,
\vx_{m})=\bfnull$ and continue the elimination with the next block of quantifiers.

As a first step we consider only the elimination of the first block of quantifiers $Q_{1}\vx_{1}$. There are three problems that prevents us from using directly
the $\cald$-uantifier elimination assumption.

(i) In the  $\cald$-elimination assumption there is only one parameter the variable $y$   in the formula
$P(x,y)$, while we now we have all of the variables $\vx_{2},\ldots ,\vx_{m}  $

(ii) In the $\cald$-elimination assumption there is only one existential quantifier,
the  quantifier  $\exists x$, while now we have the whole block $\exists \vx_{1}$, where the number of variables may even depend on $d$.

(iii) In the $\cald$-elimination assumption  the propositional formula
$P(x,y)$ does not depend on $d$ while now $P(\vx_{1},\ldots ,
\vx_{m})$ may depend on $d$

What may help in overcoming the problems caused by this changes is that the 
assumptions of the lemma
imply upper bounds on the number of parameters, the number of existential qauntifiers,
and the size of the formula $P$. These upper bounds are in condition 
\rref{EC12.2} and in the assumption $\phi\in \Form(c^{\delta}, \iota_{m},...,
\iota_{1})$.

We will be able to reduce all of the numbers mentioned in problems (i),(ii), and (iii) to 
one (the value needed in the $\cald$-elimination assumption) by considering the formula $\phi $
not in the structure $\boldm_{d}$ but in a larger structure $\boldm_{v}$. In such a larger structure $\boldm_{v}$ we may encode a sequence of elements of $\boldm_{d}$
by a single integer of $\boldm_{v}$. (The same way as it was done in \cite{Ajt0}.)
This encoding will solve problem (i) and problem (ii). For the solution
of problem (iii) we use Lemma \rref{E33} about the evaluation of circuits with 
existential formulas. The propositional formula $P(\vx_{1},\ldots ,
\vx_{m})$ can be written in the form of $ \xi(\vx_{1},\ldots ,
\vx_{m})=\bfnull $, where $\xi$ is a term of $\calm$. We may consider the algebraic circuit
corresponding to $\xi$. As Lemma \rref{E33} states, this circuit defined over $\boldm_{v}$, can be evaluated by a first-order formula in a structure $\boldm_{v'}$, where $v'>v$,
 provided that its input
is encoded by a single integer, and the circuit itself is also encoded by two integers.
Lemma \rref{E33} also gives an upper bound $v'$.

This way we will be able to substitute the propositional formula  $P$ in problem (iii) by
an existential formula of constant size. (See Lemma \rref{ES1} later in this section.) The new existential quantifiers can be merged
by the already existing existential quantifiers mentioned in problem (i) and all of them can  be reduced to
a single quantifier  by going to a larger structure. 

This quantifier elimination that we described for the first block of quantifiers, can be recursively repeated
and gradually eliminate all of the quantifiers while we have to evaluate the formulas
in larger and larger structures. Later we will sketch further details of the proof as
we are getting to the definitions and lemmas which describe the specific parts of the
proof. {\it End of Sketch}

\begin{definition} Let $\tau$ be a term of $\calm$. We will say that $\tau $
is a $0,1$-term if for all $d\in \omega$ and for all $a\in a$ we have
$\boldm_{d} \models \tau(a)=\bfnull \vee \tau(a)=\bfegy$
\vege\end{definition}

In the following definition, starting with a formula $\phi  $ $\equiv$ $ Q_{m}\vx_{m},\ldots 
,Q_{1}\vx_{1},$ $P(\vx_{1},\ldots ,
\vx_{m},x)$ that we have at the beginning of the inductive proof of Lemma
\rref{EC12},
we describe  the sequence of formulas that we derive from $\phi$ as we eliminate its
blocks of quantifiers one-by-one.

\begin{definition} 
Assume that $m\in \omega \bcks \lbrace 0,1\rbrace $, $j_{m},\ldots ,j_{1}$ are
positive integers,  $\vx_{i}$ is the sequence of variables  $x_{i,0},\ldots ,x_{i,j_{i}}$ 
of
$\calm$, and $\phi$ is a first-order prenex formula of $\calm$,
$\phi \equiv Q_{m}\vx_{m},\ldots ,Q_{1}\vx_{1},P(\vx_{1},\ldots ,
\vx_{m},x)$, with quantifier pattern  $\langle j_{m},\ldots ,j_{1}\rangle$, where $P$ is 
a 
propositional formula  of $\calm$, moreover
$Q_{i}$ is  the universal quantifier for all even $i\in \lbrace 1,\ldots ,m \rbrace$, 
and $Q_{i}$ is the existential quantifier for all odd $i \in \lbrace 1,\ldots , m \rbrace$. 

We define a sequence of formulas $\phi_{i}$ of $\calm$ for $i=0,1,\ldots ,m$, by 
recursion 
on $i$. The free variables of the formula $\phi_{i}$ will be $  
\vx_{i+1},\ldots ,\vx_{m},x$.
For $i=0$, $\phi_{0}(\vx_{1},\ldots ,
\vx_{m},x)\equiv \neg P(\vx_{1},\ldots ,
\vx_{m},x)$. Assume that $\phi_{i-1}(\vx_{i},\ldots ,\vx_{m},x) $   has been 
already
defined for some $i=1,\ldots ,m$.
Then $\phi_{i}$ is defined by $\phi_{i} (\vx_{i+1},\ldots ,\vx_{m},x)\equiv \exists 
\vx_{i},
 \neg \phi_{i-1}(\vx_{i},\vx_{i+1},\ldots ,\vx_{m},x )$. The formula $\phi_{i}$ 
defined 
this way will be called the $i$th segment of the formula $\phi$.

Clearly if $ m$ is odd  then $\phi_{m}\equiv \phi$, and if $m$ is even, then
$\phi_{m}\equiv \neg \phi$. (We get this by replacing the quantifiers $\forall \vx_{i},
(\ldots )$ in
the definition of $\phi$ by $\neg \exists \vx_{i}, \neg(\ldots )$ for all even $i\in 
[1,m]$.)
 \vege\end{definition}

\begin{definition} Suppose that $\cald$ is a function so that the $\cald$-quantifier
elimination assumption holds for $\boldm$.  Then $\cals=\cals_{\cald}$ will denote the 
function $2^{\cald}$.
\vege\end{definition}

\begin{remark} The functions $\cals_{\cald}$ will be useful for us since for every term
$\tau$ of $\calm$ if $\depth(\tau) \le \cald(d)$ for some $d\in \omega$, then
$\csize(\tau)\le 2 \cals_{\cald}(d)$. This is a consequence of the fact that the arities
of the function symbols of $\calm$ are at most two. \semmi{eredetileg hibasan 
$\csize(\tau)\le  \cals_{\cald}(d)$ volt. Lehet, hogy kesobb valhol ez igy van 
felhasznalva}
\vege\end{remark}

The following Lemma \rref{ES1} solves  the problems (i), (ii), and (iii) mentioned in the 
sketch of the proof of Lemma \rref{EC12}. (The remaining problem, reducing the 
number of existential quantifiers from a constant to one, will be solved  Lemma \rref{ES3}.)
Lemma \rref{ES1} will be used in the inductive proof of Lemma \rref{EC12}. Applying 
Lemma  \rref{ES1} we will be able to
eliminate a block of quantifiers in the inductive step.

\begin{lemma} \llabel{ES1} For all sufficiently large $c\in \omega$ the following 
holds.
Assume that 

\begin{cond} \llabel{ES0.1}
 $\cald$ is a function and the $\cald$-quantifier elimination assumption
holds, 
\end{cond}

	\begin{cond} \llabel{ES0.2}
 $\psi$ is an existential formula of $\calm$ of the form
$$\psi\equiv \exists x_{0},\ldots ,x_{k-1}, 
 \xi(x_{0},\ldots ,x_{k-1},y_{0},\ldots ,y_{l-1})=\bfnull$$
where $\xi $ is a $0,1$-term of  $\calm$,
\end{cond}

\begin{cond} \llabel{ES0.3}
 $r,v\in \omega$
and $v\ge r+ c \lceil\log (\csize(\xi)) \rceil $.
\end{cond}

Then there exists a  $0,1$-term $\eta 
(x_{0},\ldots ,x_{l-1},w_{0},w_{1},w_{2}) $ of 
$\calm$
such that the following conditions are satisfied:

\begin{cond} \llabel{ES1.1} $\csize(\eta)\le 2^{\cald(v)}+c(k+l)$
\end{cond}

\begin{cond} \llabel{ES1.2}
for all $a_{0},\ldots ,a_{l-1} \in \boldm_{r}$,   $$\boldm_{r} \models \psi 
(a_{0},\ldots ,a_{l-1})  \ \ \leftrightarrow \ \
\boldm_{v} \models \eta (a_{0},\ldots ,a_{l-1},\Circ_{0}(\xi),\Circ_{1}(\xi),r 
)=
\bfnull$$  
\end{cond}
\end{lemma}

We will prove the lemma in three steps. First, in Lemma \rref{ES2} instead of the 
propositional statement $\eta=\bfnull$ we will have an existential statement  of constant 
size, but with possibly more than one existential quantifiers. In lemma \rref{ES3} we 
reduce the number of existential quantifiers to one. Then, using the $\cald$-quantifier 
elimination assumption, we complete the proof of Lemma \rref{ES1}.

Proof of Lemma \rref{ES1}. Assume that $\cald,$ are fixed satisfying condition
\rref{ES0.1} of the lemma. Sometimes we will write $\cals(x)$ instead of 
$2^{\cald(x)}$. 
As a first step we prove the following Lemma \rref{ES2} (without the assumption of $\cald$ quantifier 
elimination). In this lemma is a similar statement to Lemma \rref{ES1} but now we 
express the truth value of $\boldm_{r}\models \psi$, not by a term $\eta$  in $\boldm_{v}$
but by  constant size first-order  existential formula $\phi$ in $\boldm_{v}$.  So we are saying less because because $\phi$ has quantifiers, but at the same time also saying 
more since $\phi$ is of constant size.

\begin{lemma} \llabel{ES2} There exists an existential first-order formula
$\phi(x_{0},\ldots ,x_{5})$ of $\calm$ such that for all  sufficiently large $c_{1}>0$ 
and 
for all integers $k,l$, for all
formulas 
$\psi$ and terms $\xi$  of $\calm$ satisfying condition \rref{ES0.2} of Lemma,
\rref{ES1},  and 
for all
$r,v'\in \omega$ with 
 $v'\ge r+ c_{1} \lceil\log (\csize(\xi)) \rceil $,
we have

\begin{cond} \llabel{ES2.1}
for all $a_{0},\ldots ,a_{l-1} \in \boldm_{r}$,   $$\boldm_{r} \models \psi 
(a_{0},\ldots ,a_{l-1})  \ \ \leftrightarrow \ \
\boldm_{v'} \models \phi (A,r,\Circ_{0}(\xi),
\Circ_{1}(\xi),k,l 
)$$ 
 where $A=\enc_{l,r}(a_{0},\ldots ,a_{l-1}) $. 
\end{cond}
\end{lemma}

\begin{remark} The important point in this lemma is that
the formula $\phi$ does not depend on anything. Therefore we  replaced the formula 
$\psi$ of arbitrary size 	with a fixed formula
 $\phi$ of constant size, while $k,l,\xi,r$ can be arbitrarily large.   
\vege\end{remark} \vskip 5pt

Proof of Lemma \rref{ES2}.
First we describe the formula $\phi$ as a mathematical statement, and then we show
using Lemma \rref{E33} that this statement can be expressed by an existential formula 
$\phi$ of $\calm$, as required by the lemma. 

The formula $\phi$
will say the following:

\begin{cond} \llabel{ES2.2}
 there exists an  element $u\in \boldm_{v'}$, with
$u<2^{k2^{r}}$ such that if $u_{i}=u[i,r]$, for $i=0,...,k-1$,  
 then $\xi(u_{0},...,u_{k-1},a_{0},...,a_{l-1})=0$ \end{cond}

If we describe the statement in \rref{ES2.2} as it is by a first-order formula of $\calm$, 
then the 
size of the formula will depend on $k$ and $l$ so it is not suitable for our purposes.
Lemma \rref{E33} however provides an existential first-order formula of constant size 
which 
decides whether a term $\mu$ in the structure $\boldm_{t}$ takes a given value $b$, 
at a given evaluation of the variables of the term $\mu$. The evaluation of the 
variables
is given by a single integer, and the term $\mu$ is given by the two integers
$\Circ_{0}(\mu)$ and $\Circ_{1}(\mu)$.
 Lemma \rref{E33} is applicable for the present case with $\mu \legyen \xi$, $k\legyen 
k+l$, 
$t\legyen r$,
$v\legyen v' $, $a_{i}\legyen a_{i}$, for $i=0,1,...,l-1$ and $a_{l+j}\legyen u_{j}$
for $j=0,1,...,k-1$, $b\legyen 0$. Let $\phi'(y_{0},...,y_{4})$ be the existential formula whose existence is guaranteed by lemma \rref{E33} with this choices of the parameters.
The definition of the function  $\enc$ implies that  $\enc_{k+l,r}(a_{0},...,a_{l-1},u_{0},...,u_{l-1})=\enc_{l,r}(a_{0},...,a_{l-1})
+u2^{l2^{r}}=A+u2^{l2^{r}}$.  Therefore the formula  
$\phi(x_{0},...,x_{5}) \equiv \exists u,
\phi'(x_{0}u 2^{x_{5} 2^{x_{1}}},x_{1},x_{3},x_{4})$ meets our requirements,
since with $x_{0}\legyen A$, $x_{1}\legyen r$ $x_{2} \legyen \Circ_{0}(\xi)$,
$x_{3}\legyen  \Circ_{0}(\xi), x_{4}\legyen k$, $x_{5}\legyen l$ we get that

$\boldm_{v'}\models \phi (\enc_{l,r}(A,r,\Circ_{0}(\xi),\Circ_{1}(\xi),k,l 
) $ iff there ``exists an $u=\sum_{i=0}^{k}u_{i}2^{i2^{r}}  <2^{k2^{r}}\in \boldm_{v'}$
with $\boldm_{v'} \models \phi'(A+u 2^{l r}, \bfnull,r,\Circ_{0}(\xi),\Circ_{1}(\xi) ) $".
This last statement by Lemma \rref{E33} is equivalent to  	condition
\rref{ES2.2}. Therefore the formula $\phi$ our requirements.
 \enp{(Lemma \rref{ES2})}

The existential formula $\phi$ in Lemma \rref{ES2} may have more than 
one existential quantifier. Suppose that $\phi \equiv \exists y_{0},...,y_{s-1},
P(y_{0},...,y_{s-1},x_{0},...,x_{5})$, where $P$ is a propositional formula
 of $\calm$.	To reduce the number of existential quantifiers in the formula
$\phi$	to one, and also to replace the six  parameters $A, r,\Circ_{0}(\xi),
\Circ_{1}(\xi),k,l        $
of $\phi$ by a single parameter, we use Lemma \rref{J6} with the propositional 
formula  $P$ occurring  in $\phi$.
We get the following stronger version of Lemma \rref{ES2}.

\begin{lemma} \llabel{ES3} There exists  a term $\pi(x_{0},\ldots ,x_{5})$ of 
$\calm$ and there exists an existential first-order formula
$\phi(x)$ of $\calm$, containing a single existential quantifier,  such that for all  
sufficiently large $c_{1}>0$, and for all integers $k,l$, for all
formulas 
$\psi$ and terms $\xi$  satisfying condition \rref{ES0.2} of  Lemma  \rref{ES1},
 and for all
$r,v\in \omega$ with 
 $v\ge r+ c_{1} \lceil\log (\csize(\xi)) \rceil $,
we have that

\begin{cond} \llabel{ES3.1}
for all $a_{0},\ldots ,a_{l-1} \in \boldm_{r}$,   $$\boldm_{r} \models \psi 
(a_{0},\ldots ,a_{l-1})  \ \ \leftrightarrow \ \
\boldm_{v'} \models \phi (\pi(A,r,\Circ_{0}(\mu),
\Circ_{1}(\mu),k,l) 
)$$ 
 where $A=\enc_{l,r}(a_{0},\ldots ,a_{l-1}) $. 
\end{cond}
\end{lemma}

Proof of Lemma \rref{ES3}. With the choice $v=v'+c_{2}$ where $v'$ is the integer 
whose existence is stated in Lemma \rref{ES2}
the statement of the present lemma is an immediate consequence of Lemma 
\rref{J6}  and Lemma
\rref{ES2}.  \enp{(Lemma \rref{ES3})}

To complete the proof of Lemma \rref{ES1} we use the $\cald$-quantifier elimination
 assumption with the existential formula $\phi$ whose existence is stated in Lemma
\rref{ES3}.
We get
that there exists a term $\eta'$ of $\calm$ with $\csize(\eta')\le 2^{\cald(v)}$
such that

\begin{cond} \llabel{ES3.2}
for all $a_{0},\ldots ,a_{l-1} \in \boldm_{r}$,   $$\boldm_{r} \models \psi 
(a_{0},\ldots ,a_{l-1})  \ \ \leftrightarrow \ \
\boldm_{v} \models \eta' (\pi(A,r,\Circ_{0}(\mu),
\Circ_{1}(\mu),k,l) 
)=0 $$ 
 where $A=\enc_{l,r}(a_{0},\ldots ,a_{l-1}) $. 
\end{cond}

The definition of $A$ implies that there exists a term $\sigma$ of $\calm$ with
length at most $c_{2}l$, where $c_{2}$ is a constant, such that $
\boldm_{v} \models A=\sigma(a_{0},\ldots ,a_{l-1})$. There exist also terms   
$\sigma'$, $\sigma''$ (without any free variables)
of $\calm$ of lengths at most $c_{2}(k+l)$ such that  $
\boldm_{v} \models k=\sigma' \wedge l=\sigma''$. Therefore the term
$\eta= \eta'(\pi(\sigma(x_{0},\ldots ,x_{l-1}),w_{0},w_{1},w_{2},\sigma',\sigma''))$
meets our
requirements. \enp{(Lemma \rref{ES1})}

\begin{lemma} \llabel{AA22} Assume that $k,m\in \omega$, $m\ge k$,
$\langle j_{k},...,j_{1} \rangle $, $ \langle \iota_{m},...,\iota_{1} \rangle$ are 
sequences of positive integers, $j_{k-i}\le \iota_{m-i}$ 
for all $i=0,...,k-1$, and  $\phi $ is 
a prenex formula
of $\calm$, with quantifier pattern $\langle j_{m},...,j_{1} \rangle$. 
Then there exists a prenex
formula $\psi$ of $\calm$ with quantifier pattern 
 $ \langle \iota_{m},...,\iota_{1} \rangle$
such that the propositional parts of $\phi$ and $\psi$ are identical,
and $\vdash \phi \leftrightarrow \psi$.
\end{lemma}

Proof of Lemma \rref{AA22}.
We may add new quantified variables to $\phi $ which do not occur in the propositional
part of $\phi$. By ``padding" $\phi$ with such new variables and quantifiers we
may change its quantifier pattern into $ \langle \iota_{m},...,\iota_{1} \rangle$ in a
way that the obtained prenex formula remains logically equivalent to $\phi$.
\enp{(Lemma \rref{AA22})}

With Lemma \rref{ES1} and \rref{ES3} we have everything that we need to carry out 
the inductive step in the proof of Lemma \rref{EC12}. The following Lemma \rref{ES12}
says exactly what we have to prove at an inductive step, in terms of the quantitative
bounds on the various parameters. It also defines integers denoted by $\gamma_{i,j}$
in Lemma \rref{ES12} that will be used to define the ``G\"odel numbers" $\bfg(\phi)$.
 The role of the 
sequence $\rho_{0}<...<\rho_{m}$ to be defined in Lemma  \rref{ES12}  will be that 
at the $i$th step in the inductive proof 
we will show that $\boldm_{d}\models \phi_{i}(...)$  is equivalent to  $ 
\boldm_{\rho_{i}}\models \tau_{i}(...)=\bfnull$, where $\phi_{i}$ is the $i$the 
segment of the 
formula $\phi$ as defined earlier.
  After the proof of Lemma \rref{ES12} we will return to the proof of Lemma
\rref{EC12}.

\begin{definition} The expression ``$\beta$ is sufficiently large with respect to
$\alpha$"
will be written as  $\alpha \ll \beta$. \vege\end{definition}

\begin{lemma} \llabel{ES12} For all  $c,\alpha_{0}\in \omega\bcks \lbrace 0\rbrace $,   and for $\epsilon>0$, if $c \ll \alpha_{0} \ll \rcp{\epsilon}$ then  the following holds.
Assume that

\begin{cond} \llabel{ES12.0}
 the $\cald$ quantifier elimination assumption holds, where $\cald(x)=\epsilon 
(\log x)^{\rcp{2}}$,
 \end{cond}

\begin{cond} \llabel{ES12.1}
 $d \in \omega$ is sufficiently large with respect to $\epsilon$,
\end{cond}

\begin{cond} \llabel{ES12.2}  $\delta=\lfloor \cald (d+\log d)\rfloor$,
$m\in \omega $, $m \le c \delta$,  and $\iota_{m},...,\iota_{1}$ are positive integers 
with
$\iota_{m}+ \ldots \iota_{1} \le c^{\delta} $, $\iota_{m}\le c\delta $,
 \end{cond}

\begin{cond} \llabel{ES12.3}
       $\rho_{0},\ldots ,\rho_{m}$ is a sequence of natural numbers 
defined by $\rho_{i}=d+i D$, for $i=0,1,...,m$, where $D= \lfloor \frac {\log d}{m} 
\rfloor$,
\end{cond}

\begin{cond} \llabel{ES12.4}
 $\phi \equiv Q_{m}\vx_{m},\ldots ,Q_{1}\vx_{1}, \mu(\vx_{1},\ldots ,
\vx_{m},x)=\bfnull$  is a prenex
first-order formula
of $\calm$, with quantifier pattern $\langle \iota_{m},\ldots ,\iota_{1}\rangle$ 
of $\calm$, 
 where   $\mu$ is a $0,1$-term of $\calm$ with $\csize(\mu)\le  c^{\delta}$, 
$\vx_{i}$ is the 
sequence of
variables $x_{i,0},\ldots ,x_{i,\iota_{i}-1}$, and $Q_{1}$ is an existential quantifier. 
\end{cond}

Then there exist $3(m+1)$	natural numbers   $ 
\gamma_{i,j}
$, $i\in m+1$, $j\in 3$,
and there exists a sequence of   terms $\langle \tau_{0},\ldots ,
\tau_{m} \rangle$ of $\calm$ such 
that
for each $i\in m+1$ the following 
conditions are satisfied:

\begin{cond} \llabel{ES13.1}  
 $\gamma_{i,0}=\Circ_{0}(\tau_{i})$, 
$\gamma_{i,1}=\Circ_{1}(\tau_{i})$,  $\gamma_{i,2}=\rho_{i}$
and $\max \lbrace \gamma_{i,0},\gamma_{i,1},\gamma_{i,2} \rbrace < 
2^{d}$,
\end{cond} 

\begin{cond} \llabel{ES13.2} $\tau_{i}$ has arity $1+3i
+\sum_{j=i+1}^{m}\iota_{j}$,
 \end{cond}

\begin{cond} \llabel{ES13.3}
$\csize(\tau_{m})\le \cals(\rho_{m})$, and
 if $i>0$ then 
$\csize(\tau_{i}) \le  (\cals(\rho_{m}))^{\alpha_{0}} 
$, 
 \end{cond}

\begin{cond} \llabel{ES13.4}
 for  all $ 
\va_{i+1} \in (\boldm_{d})^{\iota_{i+1}} ,\ldots ,\va_{m}\in 
(\boldm_{d})^{\iota_{m}},a\in 
\boldm_{d}$, 
$$\boldm_{d}
 \models  \phi_{0} (  \va_{i+1},\ldots , \va_{m},a) \ \leftrightarrow \
\boldm_{\rho_{0}}\models \tau_{0} ( \va_{i+1},\ldots ,\va_{m},a )=\bfnull$$
and
 if $i>0$ then  $$\boldm_{d}
 \models  \phi_{i} (  \va_{i+1},\ldots , \va_{m},a) \ \leftrightarrow \
\boldm_{\rho_{i}}\models \tau_{i} ( \va_{i+1},\ldots ,
\va_{m},a,\vgamma_{0},\ldots ,
\vgamma_{i}
 )=\bfnull$$
 where the 
formula $\phi_{i}$ is the  $i${\sl th} segment of the formula $\phi$, and
$\vgamma_{r}$ is the
 sequence $\gamma_{r,0},\gamma_{r,1},\gamma_{r,2}$ for all $r\in m$. 
\end{cond}
\end{lemma}

Proof of Lemma \rref{ES12}.  
Assume that $c\ll \rcp{\epsilon}$ and $d,\cals,m,\nu,\iota_{0},\ldots ,\iota_{m},\phi$ 
are given and they satisfy
conditions \rref{ES12.0},\ldots ,\rref{ES12.4} of the lemma.
We construct the sequences $\tau_{i}$, $\vgamma_{i}$, $i=0,1,\ldots ,m$
by recursion on $i$ and at the same time we prove their required properties by 
induction on $i$.  

$i=0$.  We define the term $\tau_{0}$  by $\tau_{0}=\mu $.
 The sequence $\vec\gamma_{0}=\langle \gamma_{0,0},\gamma_{0,1},
\gamma_{0,2} \rangle$
is  defined by condition
 \rref{ES13.1} of the lemma. We check all of the 
conditions that must be satisfied.

Condition \rref{ES13.1}. The first three equalities follows from the definition 
 of $\vgamma_{0}$. 

The upper bound on the integer 
$\gamma_{0,2}$
holds, since $\gamma_{0,2}=\rho_{0}\le \rho_{m}\le d+\log d <2^{d}$. 
According to  Lemma \rref{EW51} we have $\gamma_{0,0}=\Circ_{0}(\mu)\le 
\log\log (c^{\delta})<2^{d}$
and   $\gamma_{0,1}=\Circ_{0,1}(\mu)\le (c^{\delta})^{8c^{\delta}}=2^{8\delta 
c^{\delta} \log c}$.
Since $\delta = \lfloor \cald+\log d\rfloor \le 2\epsilon (\log d)^{\rcp{2}}$ we have 
that $\Circ_{1}(\mu) < 2^{d}$.   

Condition \rref{ES13.2}. The arity of $\mu$ is $1+ \sum_{j=1}^{m}\iota_{i}$.

Condition \rref{ES13.3}.  For $i=0$ this does not state anything. 

Condition \rref{ES13.4}. Since $\rho_{0}=d$ and $\tau_{0}=\mu$ the two
statements whose equivalence is claimed are identical.

$i>0$. Assume that $\tau_{0},\ldots ,\tau_{i-1}$, $\vgamma_{0},\ldots ,
\vgamma_{i-1}$
has been already defined and they meet the requirements of the lemma with $i\legyen 
i-1$.
 For the definition of $\tau_{i}$ we use Lemma \rref{ES1} with $k\legyen \iota_{i}$,
$l\legyen 1+3i+\sum_{j=i+1}^{m}\iota_{j}$, $\xi\legyen \bfegy - \tau_{i-1}$,
$\psi \legyen \exists \vx_{i},\bfegy-\tau_{i-1} (\vx_{i},\vx_{i+1},\ldots ,\vx_{m},x,
\vy_{0},\ldots ,
\vy_{i-1})=\bfnull$,
where $\vx_{j}$ is the sequence of variables $x_{j,0},\ldots ,x_{j,\iota_{j}-1}$, for 
$j=i-1,\ldots ,m $ and $\vy_{j}$ is the sequence of variables $y_{j,0},y_{j,1},y_{j,2} 
$,
$r\legyen \rho_{i-1}$, $v\legyen \rho_{i}$.  We assume that $\rcp{\epsilon}$ is 
sufficiently 
large with respect to the constant $c$ of Lemma \rref{ES1}.
We have to check that the assumption of Lemma 
\rref{ES1} are satisfied by this choice of its parameters. Conditions \rref{ES0.1} and
\rref{ES0.2} are immediate consequences of the definitions and the assumptions of 
Lemma \rref{ES12}.

Condition \rref{ES0.3} of Lemma \rref{ES1}.  Here we separately consider the $i=1$
and the $i>1$ case. Assume first that $i=1$. Then $\xi=\bfegy-\tau_{0}=\bfegy
-\mu$. We have $\csize(\mu)\le c^{\delta}$, 
$r=\rho_{0}$, $v=\rho_{1}$,  and by the definition of the sequence $\rho_{j}$,
$\rho_{1}\ge \rho_{0}+ \frac{\log d}{m}$, where $m\le c \delta $, $\delta=\lfloor 
\cald (d+\log d)\rfloor \le \epsilon (\log (d+\log 
d))^{1\over 2}$.
(We denote the constant $c$ of Lemma \rref{ES1} by $c'$.)
Therefore $r+c'\log (\size(\xi) )=\rho_{0}+c'\log (c^{\delta})\le \rho_{0}+c'\delta 
\log c $. Since $\epsilon>0$ is 
sufficiently small with respect to both $c$ and $c'$, $m\le c \delta$,  we have 
$c'\epsilon \delta \log c<\frac{\log d}{m}$ and therefore $r+c' \log (\size(\xi) )  
\le \rho_{1}=v$
as required.

In the $i>1$ case (of condition \rref{ES0.3} of Lemma \rref{ES1}) the upper bound on 
$\csize(\xi)=\csize(\tau_{i-1})$ follows from
conditions	 \rref{ES13.3}, namely $\csize(\tau_{i-1})\le (\cals(\rho 
_{m}))^{\alpha_{0}}
$. Therefore $\log(\csize(\tau_{i-1})) \le \alpha_{0}\epsilon (\log (d+\log 
d))^{1\over 2}
$. This differs from the same upper 
bound in the $i=1$ case only by a constant factor which is
sufficiently small with respect to $1/\epsilon$, so we may complete the proof in the 
same way as in the $i=1$ 
case. 
This completes the proof of the fact that the assumptions of Lemma \rref{ES1}
hold, and we continue the definitions in the inductive proof of Lemma \rref{ES12} in
the $i>0$ case.

We define $\tau_{i}$ by $\tau_{i}=\eta$, where $\eta$ is the term whose existence is 
guaranteed  by Lemma \rref{ES1}. $\vgamma_{i}$ is defined by \rref{ES13.1}.
We show now that the sequences $\tau_{0},\ldots ,\tau_{i}$, 
$\vgamma_{0},\ldots ,\vgamma_{i}$ satisfy conditions \rref{ES13.1}, \rref{ES13.2},
 \rref{ES13.3}, \rref{ES13.4}
of Lemma \rref{ES12}.
 
Condition \rref{ES13.1}. The first three equalities are  the definitions of 
$\gamma_{i,j}$,
$j\in 3$.
We get the upper bounds on $\gamma_{i,j}$, $j=0,1,2$ in the same way as in the 
$i=0 $ case.

Condition \rref{ES13.2}. The arity of $\eta$ in Lemma \rref{ES1}  is the number of 
free 
variables of $\psi$ plus $3$. The number of free variables of the formula $\exists 
\vx_{i},\tau_{i-1} (\vx_{i},\vx_{i+1},\ldots ,\vx_{m},x,\vy_{0},\ldots ,
\vy_{i-1})$ is $1+3(i-1)+\sum_{j=i+1}^{m}\iota_{j}$. Increasing it by three we get 
the value claimed in  condition \rref{ES13.2}.

Condition \rref{ES13.3}.  
In this proof we will use the following trivial inequality containing the function 
$\cald(x)=\epsilon (\log x)^{1\over 2}$: $$\cald (d+\log d)\le
 2\cald  \Bigl (d+m \Bigl\lfloor 
\frac{\log d}{m}\Bigr\rfloor \Bigr)=2 \cald(\rho_{m})$$

According to condition \rref{ES1.1} of Lemma \rref{ES1} 
$\csize(\tau_{i})=\csize(\eta)\le \cals(v)+c'(k+l)= \cals(\rho_{i}) 
+c'(\iota_{i} +  1+3i+\sum_{j=i+1}^{m}\iota_{j} )\le   \cals(\rho_{i})+ 
c' c^{\delta} +1+3 c\delta  +c^{\delta}$ $\le  c''\cals(\rho_{m}) c^{\delta}$, where
$c''$ is a suitably chosen constant.  (We used here the inequality $\cals(\rho_{i})\le 
\cals(\rho_{m})$.) The definition of $\delta$ implies that $c^{\delta}\le 
2^{\cald(d+ \log d) \log c}\le $ 
$ 2^{2\cald(\rho_{m})\log c} =$ $(\cals(\rho_{m}))^{2\log c}  $.
Since $c \ll \alpha_{0}$ this implies $\csize(\tau_{i})\le \cals(\rho_{m}) 
^{\alpha_{0}}$

In the $i=m$ case we use that fact that the values $k$ and $l$ from 
Lemma \rref{ES1} are smaller than in the general case. Namely $k=\iota_{m}$ and 
$l=1+3m$. Therefore by \rref{ES12.2}  $\csize (\tau_{m})\le \cals(\rho_{m})+ 
c'(c\delta+1+3 c\delta)\le$
$\cals(\rho_{m})+4 c c' \log(\cals(\rho_{m})) \le $ $2 \cals(\rho_{m})$.

Condition \rref{ES13.4}. According to the inductive assumption  for  all $ 
\va_{i} \in (\boldm_{d})^{\iota_{i+1}} ,\ldots ,\va_{m}\in 
(\boldm_{d})^{\iota_{m}},a\in 
\boldm_{d}$, 
$\boldm_{d}\models \phi_{i-1}(\va_{i},\ldots ,\va_{m},a)$
is equivalent to $\tau_{i} ( \va_{i},\ldots ,\va_{m},a,\vgamma_{0},\ldots ,
\vgamma_{i-1}
 )=\bfnull $. This fact, the definition of $\phi_{i}$, and Lemma \rref{ES1} imply that 
for 
all 
$ 
\va_{i+1} \in (\boldm_{d})^{\iota_{i+1}} ,\ldots ,\va_{m}\in 
(\boldm_{d})^{\iota_{m}},a\in 
\boldm_{d}$, the following statements are equivalent

$\boldm_{d}\models  \phi_{i}(\va_{i+1},\ldots ,\va_{m},a)$

$\boldm_{d}\models  \exists \vx_{i}, \neg\phi_{i-1}(\vx_{i},\va_{i+1}\ldots ,
\va_{m},a)$

$\boldm_{\rho_{i-1}} \models\exists \vx_{i}, \  \bfegy -\tau_{i-1} ( \vx_{i},
\va_{i+1},\ldots ,\va_{m},a,\vgamma_{0},\ldots ,
\vgamma_{i-1}
 )=\bfnull $.

$\boldm_{\rho_{i-1}} \models\exists \vx_{i} , \xi ( \vx_{i},\va_{i+1},\ldots ,
\va_{m},a,
\vgamma_{0},\ldots ,
\vgamma_{i-1}
 )=\bfnull $

$\boldm_{\rho_{i}} \models\eta (\va_{i+1},\ldots ,\va_{m},a,\vgamma_{0},\ldots ,
\vgamma_{i-1},\gamma_{i}
 )=\bfnull $

$\boldm_{\rho_{i}} \models\tau_{i} (\va_{i+1},\ldots ,\va_{m},a,\vgamma_{0},\ldots 
,
\vgamma_{i-1}, \gamma_{i}
 )=\bfnull $ \\
The equivalence of the first and last statements of this sequence is claimed in condition 
\rref{ES13.4}
of the present lemma. \enp{(Lemma \rref{ES12})} \vskip 5pt

Proof of Lemma  \rref{EC12}. Assume that $\cals,d,m,\delta, \iota_{1},\ldots ,\iota_{m}$, $\rho_{0},\ldots ,
\rho_{m}$
are fixed with the properties described in the assumptions of Lemma \rref{ES12}.
We define first the function $\bfg$.

Assume that a first-order
 formula $\phi\in \Form(\cals(\rho_{m}),\iota_{1},\ldots ,\iota_{m})$ is  
is given. We apply now Lemma \rref{ES12} for $\phi$
with the given values of the  parameters. Let $\gamma_{i,j}$, $i\in m,j\in 3$
be the natural numbers and let $\tau_{0},\ldots ,\tau_{m}$ be the terms
whose existence is guaranteed by Lemma \rref{ES12}.
We define now $\bfg(\phi)$ by $$\bfg(\phi)=d 
2^{\rho_{m}}+2^{2\rho_{m}}\sum_{i=0}^{m} \sum_{j=0}^{2}
\gamma_{i,j}2^{(3i+j)d} $$
By condition \rref{ES12.2},  $m\le c\delta \le \log d $ and according to condition 
\rref{ES13.1}
$\max\lbrace \gamma_{i,j} | i  \in m+1, j\in 3\rbrace <2^{d}$, so we have that 
 $\bfg(\phi)<2^{2^{d-1}}$ as stated in the lemma. We also claim that

\begin{cond} \llabel{EC14}
$\bfg(\phi)$ uniquely determines all of the integers $\gamma_{i,j}$.\end{cond}
 
This is true since $d2^{\rho_{m}}$ is the residue of $\bfg(\phi)$ divided by
$2^{2\rho_{m}}$. This uniquely determines both $d$ and  $\sum_{i=0}^{m} 
\sum_{j=0}^{2}
\gamma_{i,j}2^{(3i+j)d} $.  According to \rref{ES13.1} $\gamma_{i,j}<2^{d}$,
therefore this sum uniquely determines all of the integers $\gamma_{i,j}$, $ i  \in 
m+1, j\in 3$.

This process as we got the integers $\gamma_{i,j}$ for $\bfg(\phi) $
can be implemented by a term of $\calm$, which is evaluated in $\boldm_{\rho_{m}}$. 
Indeed we have
$\boldm_{\rho_{m}}\models \div(\bfg(\phi), \bfn)=b$, where $b=\sum_{i=0}^{m} 
\sum_{j=0}^{2}
\gamma_{i,j}2^{(3i+j)d} $ and $\boldm_{\rho_{m}}\models d=\div (\bfg(\phi)-b,
\bfn)$. Finally from $b$ and $d$ we can compute each $\gamma_{i,j}$ using Lemma
\rref{Q0}.
This implies that there exist a term $\chi(x,y,z,w)$, 
of $\calm$  (which does not depend on anything so its length 
is a constant  $c_{1}$), such that
 for each possible choice of $\phi$
with the described properties we have  that for all $i\in m$, $j\in 3$,  
$\boldm_{\rho_{m}}\models  \gamma_{i,j}=\chi(\bfg(\phi),m,i,j)$,
for $i\in m, j\in 3$. 

 We want to define the term $\tau$ such that  for all $a,b\in \boldm_{\rho_{m}}$, 
$$\boldm_{\rho_{m}} \models \tau(a,b)=\tau_{m}\Bigl(a,\vchi(b,m_{0},0)
,\vchi(b,m_{0},1),\ldots , \vchi(b,m_{0},m-1) \Bigr)$$ where $\vchi(b,m_{0},i)$ is the 
sequence $\chi(b,
m_{0},i,0),\chi(b,
m_{0},i,1),\chi(b,
m_{0},i,2)
$ for $i=0,1,\ldots ,m-1$.  We can achieve this by a term $\tau$ whose
 circuit-size is at most $\csize(\tau_{m}) + c' m$, where $c'\in \omega$ is a constant. 
We prove the existence of such a term $\tau$ by
 constructing first an $\calm$-circuit  $C$, which computes the same function that is 
expected
from $\tau$.
The  $\calm$-circuit at the input $a,b$
will compute first the numbers $0,...,m-1$ using 
$m $ nodes.
For each fixed $i\in m$, $j\in 3$	there will be at most $c_{1}$ nodes in the circuit
$C$ to evaluate 
$\chi(a,b,i,j)$ and finally $C$ contains $\csize(\tau_{m})$ nodes to evaluate
 $\tau_{m}$ at the 
 input   $a,\vchi(b,m_{0},0)
,\vchi(b,m_{0},1),\ldots , \vchi(b,m_{0},m-1) $.  The term $\tau$ whose existence is 
stated in the lemma will be a
term of $\calm$ which computes the same value at each input as the circuit $C$
constructed above.

We show now that the function $\bfg$, and the term $ \tau$ satisfies conditions
\rref{EC13.3} and \rref{EC13.4} of the 
 lemma.

Condition \rref{EC13.3}. The definition of the term $\tau$ implies that
$\csize(\tau)\le  \csize(\tau_{m})+ c' m$ for some constant $c'$. Therefore
$m\le c\delta$ and the upper bound on $\csize (\tau_{m})$ given in \rref{ES13.3}
implies that	$\csize(\tau)\le 3\cals(d+\log d)$.

Condition \rref{EC13.4}. The definition of the formula $\tau$ and the terms $
\chi_{i,j}$ implies that
$\boldm_{\rho_{m}}\models \tau(a,\bfg(\phi))=\tau_{m}(a, \vgamma_{0},\ldots ,
\vgamma_{m})$ and therefore condition \rref{ES13.4} with $i=m$ implies our
statement.
\enp{(Lemma \rref{EC12})}

\eject
\subsection{\llabel{predictivity} The predictivity of $\boldm$}

\begin{definition} 1. The set of functions symbols of $\calm$ (including the constant
symbols)  will be denoted by $\fsymb(\calm)$

2. Let $\calj$ be a function. We will say
that $\boldm$ is $\calj$-predictive if the following conditions are satisfied.

\begin{cond} \llabel{EH50.1} The function $\calj$ is a monotone increasing function 
defined on $\omega$  and with
values in $\omega$. 
\end{cond}

\begin{cond} \llabel{EH50.2} 
For all  sufficiently large $d \in \omega$, $\calj(d)\in \boldm_{d}$ and $\calj(d)>d $. 
\end{cond}

\begin{cond} \llabel{EH50.3}
There exists a function  defined on $\fsymb(\calm)$ assigning to each function
symbol
$f(x_{0},\ldots ,x_{k-1})$ of $\calm$,
a formula $\Phi_{f}(x,y,z,Y_{0},\ldots ,Y_{k-1})\in \SForm(\calm)$,
where $x,y,z$ are  free first-order variables and $Y_{0},\ldots ,Y_{k-1}$ are
free variables for binary relations, such that  the following holds.
For all  $d,r\in \omega$ with $d+ r\le \calj({d})$
\xev{092f}
there exists a map $\eta_{d,r}$ of
$\universe(\boldm_{d+r})$ into the set of binary relations on
$\universe(\boldm_{d})$ with the following properties:

(i) For each  $a,u,v\in \boldm_{d}$,  we have $(\eta_{d,r}(a))(u,v)$
iff ``$u=0$ and $v=a$".

(ii) Suppose that $f(x_{0},\ldots ,x_{k-1})$ is a $k$-ary function symbol
of
$\calm$, for some $k=0,1,2$ (including the constant symbols for $k=0$)
and $a_{0},\ldots ,a_{k-1} \in \boldm_{d+r} $.  Then for all $
u,v \in \boldm_{d} $, \xev{092g}
$(\eta_{d,r}(f^{(d+r)}(a_{0},\ldots ,a_{k-1}))(u,v)$ iff $\boldm_{d} \models
\Phi_{f}(u, v, r,\eta_{d,r}(a_{0}),\ldots ,\eta_{d,r}(a_{k-1})) $,
where $f^{(d+r)} = (f)_{\boldm_{d+r}}$.
\vege\end{cond}
\end{definition}

\begin{lemma} \llabel{EF50} Assume that $c>0$ is a real, and  $\calj(x)= \lfloor 
x+c\log x \rfloor $. Then $\boldm$ is $\calj$ predictive. 
\end{lemma}

Proof.  In \cite{Ajt3}  a weaker result of similar nature  is proved which implies 
that there exists a function $g(x)$ with $\lim_{x\rightarrow \infty }g(x)=\infty$,
such that if $\calj_{0}=x +g(x)$  then $\boldm$  is $\calj_{0}$-predictive. 
Some of the partial results of the proof given there were stronger 
than 
 what was needed for the theorem  formulated in \cite{Ajt3}. We get Lemma 
\rref{EF50} by 
using the full strength of these partial results in particular about the first-order 
definability  of the bits of the results of  multiplication and
division between large numbers.

Here we  give only  
the outline of the proof together with  with those details that has to be changed for the 
present
purposes.

We define the function $\calj$ by
$\calj(x)= \lfloor 
x+c\log 
x \rfloor $.
 Assume that $d\in \omega$ is
sufficiently large $\chi\in \omega$ and $d+\chi\le \calj(d)$. First we define the map 
$\eta_{d,\chi}$ whose existence
is required by the definition of predictivity. To make our notation more
concise we will write $\eta_{d,\chi}^{(a)}$ instead of $\eta_{d,\chi}(a)$.

Assume that $a\in \boldm_{d+\chi}$, $2^{d}=n$, $\nu=2^{\chi}$. Let
$a_{i}=\coeff_{i}(a,2^{n})$ for $i=0,1,\ldots ,\nu-1$. We define
$\eta_{d,\chi}$ by:
``for all $u,v\in \boldm_{d}$, $\eta_{d,\chi}^{(a)}(u,v)$ iff $u\in \nu$
and $v=a_{u}$".
This definition implies that if $a\in \boldm_{d}$ then
\xev{119e}
for all $u,v\in \boldm_{d}$, $\eta_{d,\chi}^{(a)}(u,v)$ iff $u=0$ and
$v=a$, that is, our definition satisfies condition \rref{EH50.3}/(i)
\xev{118a}
from the definition of predictivity.

We define now the  formula $\Phi_{f}(x,y,z,Y_{0},\ldots ,Y_{k-1})$
for each function symbol $f$ of $\calm$. (According to the definition of
 $\calj$-predictivity the formula $\Phi_{f}$ cannot depend on the choices of $d$ or 
$\chi$.)

 If $f=\bfc$ is a constant symbol
of $\calm$
then $\Phi_{\bfc} \equiv x=\bfnull \wedge y=\bfc$. By the definition of
$\eta_{d,\chi}$, the formula  $\Phi_{\bfc}$ satisfies condition
\rref{EH50.3}/(ii) from the definition of predictivity, for all constant
symbols $\bfc$ of $\calm$.

We will not  use  the relation
$\eta_{d,\chi}^{(a)}$ directly
in the definition of $\Phi_{f}$,
for the remaining function symbols $f$ of $\calm$,
 but we first define another
 binary relation $\xi_{d,\chi}^{(a)}$ on
\xev{119f}
$\boldm_{d}$ and use this relation.

\begin{definition} 1. For each positive integer $k$ and  $u=\langle
u_{0},\ldots ,u_{k-1} \rangle \in (\boldm_{d})^{k}$, $u\wr_{n}$ will denote
the integer $u_{k-1}n^{k-1}+u_{k-2}n^{k-2}+\ldots +u_{1}n +u_{0} $.

2. Assume that $R$ is a $k$-ary relation
on the set $n=\lbrace 0,1,\ldots ,n-1\rbrace $, where $n=2^{d} $.
$\integer_{k}(R)$ will denote the integer
$\sum\lbrace 2^{u\wr_{n}} \mid  u\in \boldm_{d}^{k} \wedge R(u)\rbrace
$. Clearly $R\rightarrow \integer_{k}(R)$  is a one-to-one map from the
set of all $k$-ary relation on $n$ to the set of all natural
numbers less then $2^{n^{k}}$. If $a\in [0, 2^{n^{k}}-1]$ is a natural
number then the unique $k$-ary relation $R$ on $n$ with
$\integer_{k}(R)=a$ will be denoted by $\integer_{k}^{-1}(R)$.
\vege\end{definition}

\begin{definition} 1. Suppose that $R$ is a $k$-ary relation on
$\boldm_{d}$. We will say that the relation $R$ is $n$-restricted
if for all $u=\langle u_{0},\ldots ,u_{k-1}\rangle \in \boldm_{d}^{k}$,
$ R(u_{0},\ldots ,u_{k-1})$ implies that for all
$i=0,1,\ldots ,k-1$ with
$u_{i}\in n$.

 2. Assume that $d,\chi$ are positive
integers and  $a<2^{n^{2}}$. Then
$\xi_{d}^{(a)}$
is the unique binary relation on $\boldm_{d}$ which satisfies the
following two conditions: (a) The relation $
\xi_{d}^{(a)}$
  is $n$-restricted,
and (b) $\integer_{2}( \xi_{d}^{(a)} )=a$. \vege\end{definition}

\begin{lemma} \llabel{EB6} There exists a first-order formula
$\phi(x,y,z)$ of $\calm $  such that for all
$d\in \omega$ and  for all $a,b\in 2^{2^{d}}$  and $i\in 2^{d}$ we
have that   $b=\coeff_{i}(a,2)$ iff $\boldm_{d}\models \phi(a,b,i) $.
\end{lemma}

Proof. The statement of the lemma follows from  Lemma \rref{B0.5},
 \enp{(Lemma \rref{EB6})}

The following Lemma states that the relations $\xi_{d}^{(a)}$
and
$\eta_{d,\chi}^{(a)}$ can be defined from each other in a first-order way.
It is important that for the definition of the  value
$\xi_{d}^{(a)}(u,v)$ for a fixed pair $u,v$ we may need the
values $\eta_{d,\chi}^{(a)}(x,y)$ for all $x,y\in \boldm_{d}$ and vice
versa.

\begin{lemma} \llabel{EB7} There exist  formulas
$\Psi_{i}(x,y,z,Z)\in \SForm(\calm)$, $i=0,1$, where $x,y,z$ are
\xev{119g}
first-order variables and $Z$ is a variable for a binary relation such
that for all for all sufficiently large $d\in \omega$, for all $\chi \in 2^{d}$
and for all $a\in
\boldm_{d+\chi}$ the following holds: $\boldm_{d} \models \forall u,v,
[\xi_{d}^{(a)}(u,v) \leftrightarrow \Psi_{0}(u,v,\chi,\eta_{d,\chi}^{(a)})]
$ and $\boldm_{d} \models \forall u,v,
[\eta_{d,\chi}^{(a)}(u,v) \leftrightarrow \Psi_{1}(u,v,\chi,\xi_{d}^{(a)})]
$ \end{lemma}

Proof.  Assume $a\in 2^{2^{d+\chi}}$ and
$a=\sum_{i=0}^{(\nu-1)}a_{i}(2^{2^{d}})^{\nu}$.
The formula $\Psi_{1}$ have to express the statement
 $u\le \nu \wedge v\le n \wedge  \coeff_{un+v}(a,2)=1$.  $
\coeff_{un+v}(a,2)=1
 $ is
equivalent to $\coeff_{v}(a_{u},2)=1$. Using the relation
$\eta_{d,\chi}^{(a)}$ we can define $a_{u}$ in a first-order way in
$\boldm_{d}$, namely $x=a_{u}$ iff $\boldm_{d} \models
\eta_{d,\chi}^{(a)}(u,x)$. If $a_{u}$ is given then, by Lemma
\rref{EB6}, $\coeff_{v}(a_{u},2)$
has a first-order definition in $\boldm_{d}$.
 This
\xev{119h}
completes the definition of $\Psi_{0}$. In the first-order formula
$\Psi_{1}$  we have to define $a_{u}$ form its binary coefficients which
can be done by using again  Lemma \rref{EB6}.
 \enp({Lemma
\rref{EB7}})

Lemma \rref{EB7}  implies that it is sufficient
to prove that condition \rref{EH50.3} of the definition of predictivity
holds in the following modified form. For the sake of notational
simplicity we consider here all of the function symbols of $\boldm$
as binary function symbols. In the case of the constant symbols 
$\bfnull,\bfegy, -\bfegy$, and $\bfn$ the interpretation of these symbols is a binary
function which does not depend on its variables.
For the unary functions symbols $\caln$ and $\bfp$ their interpretation
is a binary function which depends only on its first variable. 

\begin{cond} \llabel{EB7.1}
Suppose that $f$ is one of the function symbols  
$\bfnull,\bfegy, -\bfegy$,  $\bfn, \cap,\caln$,\\
$+,\times,\bfp,\div$, $\max, \min, \cap,\caln $ of $\calm$. Then there
exists
a  formula $\Phi_{f}'(x,y,z, Y_{0},Y_{1})\in \SForm(\calm)$, where $x,y,z$ are
first-order variables and $Y_{0},Y_{1}$ are variables for binary
relations  such that for all $c\in \omega$, for all sufficiently large
$d\in \omega$,  and for all
$a,b \in \boldm_{d+\chi} $, and for all $ u,v \in \boldm_{d} $, \xev{170a}
$\xi_{d}^{(f^{(d+\chi)}(a,b))}(u,v)$ is true iff $\boldm_{d}
\models \Phi_{f}'(u, v, \chi,\xi_{d}^{(a)},\xi_{d}^{(b)}) $,
where $f^{(d+\chi)} = (f)_{\boldm_{d+\chi}}$.
  \end{cond}

In other words given the binary bits of $a,b\in 2^{2^{d+\chi}}$, each by
a binary relation on $\universe(\boldm_{d})$,
we have to define in $\boldm_{d}$ in a first-order way the binary bits of
$0,1,2^{2^{d+\chi}}-1 $,  $d+\chi , 2^{a},  \caln(a),$ $ a+b,ab,
a\div b=\lfloor a/b \rfloor, $, $\min(a,b)$, $\max(a,b)$, $a\cap
b$,  where the operations are defined in the structure
$\boldm_{d+\chi}$.  The task is trivial for $0$ and $1$. In the case of 
$2^{2^{d+\chi}}-1$ all of the $2^{d+\chi}$ bits are $1$s. We get the bits of
$d+\chi$ by computing $d+\chi$ with an addition in $\boldm_{d}$, where $d$ can
be defined by a first-order formula using the constant symbol $\bfn$.
 Since $a\cap b$  and $\caln (a)$ are defined by bitwise
operations $\Phi_{f}$ obviously can be easily defined for these two
operations. Therefore we have to prove that condition \rref{EB7.1} holds only for the 
remaining function symbols.

Using the function $\integer_{k}^{-1}$  we can represent natural
numbers from the interval $[0,2^{n^{k}}-1]$ by $k$-ary relations on $n$.
Our next goal is to represent sequences of natural numbers by relation
\xev{170b} on $n$, (where we have a bound both on the length of the
sequence and the sizes of its elements).

\begin{definition}
  1.
The set of all sequences of length $i$, whose
elements are from the set $A$ will be denoted by, $\seq(i,A)$. For
example   the set of all sequences of length $n^{l}$ whose elements are
integers in the interval $[0,2^{n^{k}}-1]$ is $\seq(n^{l}, 2^{n^{k}})$.

2. Assume that
$a=\langle a_{0},\ldots ,a_{j-1}\rangle \in
\seq(n^{l},2^{n^{k}})$.  We will represent this
sequence by a $k+l$-ary relation $
R^{(a)} $ on $n$  defined in the following way.  For
all $i\le j-1$, and for all $u_{0},\ldots ,u_{k-1}, v_{0},\ldots ,v_{l-1}\in n$,
$R^{(a)}(
u_{0},\ldots ,u_{k-1},
v_{0},\ldots ,v_{l-1}
 )$ iff
$(\integer_{k}^{(-1)}(a_{t}))(u_{0},\ldots ,u_{k-1})$, \xev{170c}
where $t=\sum_{i=0}^{l-1} v_{i}n^{i}$. Since in this representation the
length of the sequence cannot be arbitrarily chosen it must be
$n^{l}$, for some $l\in \omega$, we will call this representation a
representation of the sequence without its length.

3. The definition above provides representation only for sequences
with exactly $n^{l}$ elements for some natural number $l$. A sequence
$a=\langle a_{0},\ldots ,a_{j-1}\rangle $  where $j<n^{l}$, $a_{i}\in
[0,2^{n^{k}}-1]$ will be represented in the following way.
 We attach the number $j$ as the first element
to the sequence $a$  and  attach a sequence
of $0$s to its end, so that the total length of the
sequence $a'=\langle j,a_{0},\ldots ,a_{j-1},0,\ldots ,0\rangle $ obtained this
way is $n^{l}$.
The representation of the sequence $a$ together with its length will be
the same as the representation of the sequence $a'$ without its length,
as defined earlier.
 In the following the
representation of a sequence will always mean a representation
of the sequence together with its length unless we explicitly state
otherwise.

4. Assume that $d$ is a positive integer and $n=2^{d}$. We will say that
the set $X$ is $\boldm_{d}$-representable                   \xev{170d}
if there exists natural numbers $k,l$ such that either $X=
\lbrace 0,1,\ldots ,2^{n^{k}}-1 \rbrace
 $ or $X=\seq_{n}(n^{l},2^{n^{k}})$.  If $X$ is
an $\boldm_{d}$ representable set  and
$X=
\lbrace 0,1,\ldots ,2^{n^{k}}-1 \rbrace
 $
then we define its weight by $\weight(X)=k$, if $X=\seq_{n}
(n^{l},2^{n^{k}})
 $ then
we define its weight by $\weight(X)=k+l$.   If $a\in X$, where $X$ is an
$\boldm_{d}$ representable set, then $\relation_{a,n}$ will denote the
$k$-ary or $k+l$-ary relation on $n$ representing the element $a$.
 \vege\end{definition}

  We will consider now families of functions $f^{(d)}$, $d\in \omega$ so
that for each  $d\in \omega $,  $f^{(d)}\in\func(X^{(d)},Y^{(d)}) $
where
both $X^{(d)}$ and $Y^{(d)}$ are
$\boldm_{d}$-representable sets with weight less then $w$ for a
constant $w$.
We are interested in the case when such a family of functions can be
defined by
a first-order formula in $\boldm_{d}$ without using any parameters. The
world ``strongly" that we will use in the definition below refers to
mentioned the lack of parameters.

\begin{definition} 1. Assume that $w_{i}\in \omega$ for $i=0,1$ and  for
all $d\in \omega $, $A_{i}^{(d)}$
are $\boldm_{d}$ representable sets of weight  $w_{i}$ for $i=0,1$,  and
$f^{(d)}\in
\func(A_{0}^{(d)},A_{1}^{(d)})$.
 We
will say that the family of functions $f^{(d)}$ is a strongly first-order
definable family function or a s.f.d.-family in $\boldm$
if there exists a formula $\Gamma(x_{0},\ldots ,x_{w_{1}-1},
 Z)\in \SForm$, where $x_{i}$, $i=0,1,\ldots ,w_{1}-1$ are individual variables
and $Z$ is a variable for $k_{0}$-ary relations such that for all
sufficiently large $d\in \omega$ and   \xev{170e} for all $a\in
A_{0}^{(d)}$, and $b\in A_{1}^{(d)}$
with $f(a)=b$,  we have that for all $u_{0},\ldots ,u_{w_{1}-1}\in n$, $
\relation_{b,n}(u_{0},\ldots ,u_{w_{1-1}}) $ iff
$\boldm_{d}
\models \Gamma
(u_{0},\ldots ,u_{w_{1}-1},\relation_{a,n})$.
\vege\end{definition}

We prove now that condition
\rref{EB7.1}  is satisfied by
each function symbol of $\calm$. As we mentioned already this statement
trivially holds for some of the function symbols.  For the remaining
ones we show now that the corresponding families of functions are
 \xev{119i}
are strongly first-order definable
in $\boldm$.

For $f=\min$ and $f=\max$ the statement is trivial since  $a\le  b$ iff
$\integer_{2}^{-1}(a) \le \integer_{2}^{-1}(b) $ according to the
lexicographic ordering which clearly can be defined in $\boldm_{d}$ in a
first-order way.

The function symbol $f=``+"$. If two integers are given in binary form
each with $m$
bits then the bits of their sum can be defined by a simple well-known
constant depth circuit whose size is linear in $m$. This circuit is
defined in a uniform way \xev{169e}  which makes it possible to
translate it into
a first-order formula interpreted in $\boldm_{d}$.   For later use we
also consider now the case where we have to
add  a sequence of integers. This question has been also studied for
circuits, and it is known that if we have at most $(\log m)^{c_{0}}$
integers with  $m^{c_{1}} $ binary bits then their sum can be computed
by an unlimited fan-in boolean circuit with size $m^{c_{2}}$ and depth
$c_{3}$,
where $c_{2},c_{3}$ depend only on $c_{0}$ and $c_{1}$, see \cite{AHU}.
The construction
of the circuit is uniform, in this case too, and can be translated into
a first-order formulas, that we need for our present purposes, over a
structure containing the arithmetic
operations. 

\begin{definition}  If $b$ is a finite sequence of integers then
$\bolds b$ will denote the sum of its elements.
                   \vege\end{definition}

The following Lemma is proved in \cite{Ajt3}

\begin{lemma} \llabel{EAR1} Assume that $c_{0},c_{1}\in \omega$. Then
there exists a strongly first-order definable family of functions
$f^{(d)}$, $d\in \omega$, such that for all sufficiently large $d$ if
$n=2^{d}$, $j\le n^{c_{0}}$
and  $a$ is  sequence of length $j$, from elements of the set
$2^{n^{k}}$, that is, $a\in\seq(j,2^{n^{k}})$, then $\bolds a
=f^{(d)}(a)$.
 \end{lemma}

We prove condition \rref{EB7.1} for $f=\times$ in a more general form
then needed, namely we will consider products with more than two
factors. This will be useful in the proof of \rref{EB7.1}.

\begin{definition}  \xev{171b} Assume that $a=\langle
a_{0},a_{1},\ldots ,a_{j-1}\rangle
$  is a sequence of integers. Then $\boldp a$ will denote the number
$\prod_{i=0}^{j-1} a_{i}$. \vege\end{definition}

\begin{definition} Assume that $\alpha(x),\beta(x)$ are functions
defined on $\omega$  with real values. We will say that
the pair
$\langle \alpha(x),\beta(x) \rangle$ is acceptable
if there exists a strongly first-order definable family of functions
$f^{(d)}$, $d\in \omega$, such that for all sufficiently large integers
$d\in \omega$, for all nonnegative integers
 $j\le \alpha(d)$, and for all
 $
a\in\seq(j,2^{\beta(d)})$, we have $\boldp a =f^{(d)}(a)$.
 \vege\end{definition}

The following two lemmas are proved in \cite{Ajt3}. The second lemma
is a special case of the first one.

\begin{lemma} \llabel{EAR2} For each fixed $c >0,\epsilon>0$
 the pair
$\alpha(x)=x^{c}$, $\beta(x)=2^{x+x^{1-\epsilon}}$  is acceptable.
 \end{lemma}

\begin{lemma} \llabel{EAR6.4} For all $\epsilon >0$
 there exists a
family of functions $f^{(d)}$, $d\in \omega$, such that, for all
sufficiently large $d\in \omega$ if $a=\langle a_{0},a_{1}\rangle \in
\seq(2, 2^{2^{d+d^{1-\epsilon}}}) $, \xev{171l}  then
$a_{0}a_{1}=f^{(d)}(a)$.  \end{lemma}

Using Lemma \rref{EAR6.4} we can show that condition \rref{EB7.1} is
satisfied by $f=\times$.
If $d$ is sufficiently large and   $d+\chi\le \calj(d)\le d+ c \log\log d  $     then 
$d+d^{1\over
2}>d+\chi$ and therefor Lemma \rref{EAR6.4} implies that,  multiplication in 
$\boldm_{d+\chi}$ can be defined in
$\boldm_{d}$ in the sense of \rref{EB7.1}. This completes the proof of
\rref{EB7.1} for $f=\times.$

Now we prove condition \rref{EB7.1} for $f=\div$.  
We follow the technique used by Beame, Cook, and Hoover (see \cite{BCH})
for performing integer division by small depth circuits. Namely, we reduce integer
division to multiplication and addition by approximating the function $\frac{1}{1-x}$
with an initial segment of its Taylor series.

Assume that  $d$ is sufficiently large, $d+\chi \le \calj(d) \le 
d+c\log\log d$, $a,b\in \boldm_{d+\chi}$, and we want to define $\lfloor
a/b\rfloor $ in $\boldm_{d}$ in a first-order way. First we describe a
way, using general mathematical language, to compute $\lfloor a/b\rfloor
$ and then we show that this can be translated into the formula 	$\Phi'_{f}$
required in \rref{EB7.1}.  We will use the notation $2^{d}=n$ and $2^{\chi}=\nu$.

(i) First we note that it is sufficient to find integers $t,l$  such
that ${1\over b} -t2^{l}<2^{-\nu n-1}$. The reason for this is that in the
possession of the integers $t,l$ we can compute  $\alpha =a t2^{l}$ and
\xev{176b}
$|\alpha -\lfloor a/b\rfloor |<a 2^{-\nu n -1}< 2^{\nu n}2^{-\nu n -1}\le {1\over 
2}$ so we get $\lfloor a/b\rfloor
$ by rounding.

(ii) Let $k$ be an integer so that $1>2^{-k}b>1/2$. If there exists no
integer with this property then the problem is trivial, since we can get
the binary bits of $\lfloor a/b\rfloor $ form the bits of $a$  simply by
shift and the erasure of a block of consecutive bits.  Let $u=2^{-k}b$.
Since $1>u>{1\over 2}$, we have $1<{1\over u} < 2$. We may write
${1\over u}$
in the form of $ v2^{-(n+2)}+R$, where $v\in [0,2^{n+2}]$ is an integer
and
$0\le R< 2^{-n-1}$. ($v$ will be determined by the first $n+1$ bits of
${1\over u}$, and $R$ is what remains from ${1\over u}$ after erasing
these bits.) Let $z=v2^{-(n+2)}$.  The definition of $v$ implies that
$0\le z \le 2$.

(iii) We have  $zb=1+Rz=1+r$, where $|r|<2^{-n+1}$.  We consider the
series
${1\over zb}={1\over 1-(1-zb)}={1\over 1-(-r)}=1-r+r^{2}-r^{3}+\ldots $.
Let $w$ be the sum of the
first    \xev{176c}
  $4\nu$ terms of this geometric series. Clearly $w={1\over
zb}+R_{1}$, where  $|R_{1}|<2^{-3 \nu n}$.
Consequently
${1\over b}=z{1\over zb}=z(w-R_{1})=zw+R_{2}$, where $|R_{2}|<2^{-2 \nu n}$.

Now we show that all of the quantities in this computation can be defined in
a first-order way in $\boldm_{d}$.

Stage (i). The definition of $t$ and $l$ will be described later.
However if we have $t$ and $l $ Lemma \rref{EAR6.4} implies
that we may define  the product  $a t2^{l}$ in a first-order way in
$\boldm_{d}$. The rounding  also can be done in a first-order way.

Stage (ii). The integer $v$ has only $n+2$ bits. In $\boldm_{d}$ we can
quantify $n$ bits with a single existential quantifier, therefore $v$
with the given property is first-order definable in $\boldm_{d}$.

Stage (iii).
Lemma \rref{EAR6.4}
implies that the  product $zb$  can be defined in $\boldm_{d}$. Using
Lemma \rref{EAR1} we get that  $r$  can be defined as well. Each
\xev{176d}
needed terms of the geometric series can be  defined in $\boldm_{d}$, we
define the $i$th term as a product with $i$ factors. 
Since $\nu=2^{\chi}\le 2^{c\log\log n}\le (\log n)^{c}$,
Lemma \rref{EAR2}
implies that the bits of such a product can be defined in $\boldm_{d}$ and by
Lemma \rref{EAR1} the bits of  the sum of the first $4 \nu$  terms can be defined as
well. Therefore we defined $w$ and by Lemma \rref{EAR6.4}  we
can define $zw$ as well. This completes the proof of the fact that
condition \rref{EB7.1}  is satisfied by $f=\div$, and also the proof of
$\calj$-predictivity of $\boldm$.  \enp{(Lemma \rref{EF50})}

\eject

\section{\llabel{conclusion} The Conclusion of the Proof of Theorem \rref{TT3}}

\begin{definition}  1. Assume that $\alpha,k$ are positive integers.
 The  geometric sequence $\langle \alpha, \alpha^{2},\ldots ,\alpha^{k} \rangle$
will be denoted by $\gseq(k,\alpha)$.

2. Assume that $M,j_{m},...,j_{1}$ are positive integers. The set of all prenex 
first-order formulas
 $\phi $	 of $\calm$  satisfying the following two conditions  will be denoted by $\boldl(M,j_{m},...,j_{1})$. 

\begin{cond} \llabel{AA23}   if the quantifier pattern of $\phi$ is $\langle \iota_{k},...,
\iota_{1}\rangle$ then $k\le m$ and $\iota_{k-i}\le j_{m-i}$ for all i=0,...,k-1.
\end{cond}

\begin{cond} \llabel{AA23.1}   if  $\phi\equiv Q_{r}x_{r},...,Q_{1}x_{1}P(x_{r},
...,x_{1})$, where $Q_{r},...,Q_{1}$ are quantifiers and $P$ is a propositional formula
of $\calm$ then $\length(P(x_{1},...,x_{r}))\le M$
\end{cond}

 The set of all prenex formulas 
$\phi \in \SForm(\calm)$ satisfying these two conditions will be denoted by
$\bar\boldl(M,j_{m},...,j_{1})$
\vege\end{definition}

\begin{remark} The definitions of sets $\Form(M,j_{m},...,j_{1})$ and
 $\boldl(M,j_{m},...,j_{1})$ are similar but they are not the same. The set 
$\Form(M,j_{m},...,j_{1})$ contains 
prenex formulas  $\phi$ whose quantifier pattern is exactly $\langle 
j_{m},...,j_{1}\rangle$,
while in the case of  $\boldl(M,j_{m},...,j_{1})$, the sequence $\langle 
j_{m},...,j_{1}\rangle$ is only an upper bound, in some sense, on the quantifier pattern 
of $\phi$. Apart from that, in the case of $\Form$, $M$ is an upper bound on the circuit 
size  of the propositional part of $\phi$, and in the case of $\boldl$ it is an upper 
bound on the length of the propositional part.  
\end{remark}

\begin{lemma} \llabel{EAA15} There exists a $c>0$ such that if  $\Phi_{0},
\Phi_{1} $ are prenex first-order formulas of $\calm$, 
$m\in \omega $, $M\ge 1$, $\beta \ge 2$,
$\Phi_{0},\Phi_{1} \in \boldl(M,\gseq(r, \beta))$ and $\phi$ is one of the formulas 
formulas $\Phi_{0} \wedge \Phi_{1}$, $\Phi_{0} \vee \Phi_{1}$,  $\neg \Phi_{0}$ 
then there exists
a prenex first-order formula  $\psi \in  \boldl(2M+c, \gseq(r+4, \beta ))$ such that 
$ \vdash \phi \leftrightarrow \psi $.
\end{lemma}

Proof of Lemma \rref{EAA15}. We consider only the $\phi \equiv \Phi_{0}\wedge 
\Phi_{1}$ case,
the other logical connectives can be handled in a similar way.	Assume that for
$i=0,1$, 
 $$\Phi_{i}\equiv Q_{m_{i},i}\vx_{m_{i},i},...,Q_{1}\vx_{1,i},
P(\vx_{m_{i},i},...,\vx_{1,i} )  $$ where $\vx_{j,i}$, $j=m_{i},...,1$ is  sequence of
 variables, and $Q_{k,i}$ are quantifiers for $k=m_{i},...,1$. The length of the 
sequence of variables $\vx_{j,i}$ will be denoted by $l_{i,j}$.

Our assumptions 
imply
that  $l_{i,j}\le\beta^{m_{i}-j+1}$  for $i=0,1$, $j\in m_{i},...,
1$. 
First we choose a $c'\in \lbrace 1,2\rbrace $ such that 
for all integers $j\in \omega$, if $Q_{j,0}$ and $Q_{j+c',1}$ are defined, 
then they are  
quantifiers of the 
same type. 
When forming the prenex form of $\Phi_{0} \wedge 
\Phi_{1}$, we will combine the quantifiers $Q_{j,0}$ and $Q_{j+c',1}$  and the 
variables bound by them into a single block
for all $j\in \omega$, provided that both blocks are defined.
 If one of these blocks is not defined then we use the other block 
alone. The assumption $\beta \ge 2 $ implies that 
if the prenex form $\Phi$  of $\Phi_{0}\wedge \Phi_{1}$ constructed this way has a 
quantifier pattern $j_{m},...,j_{1}$, then $\length(\Phi)\le 2M+c$ and 
$j_{m-r}\le \beta^{r}$ for $r=0,...,m-1$. 
\enp{(Lemma \rref{EAA15})}

\begin{lemma} \llabel{EW81} 
For all $\alpha,\beta\in \omega$ there exists a $\gamma\in \omega$  such that
the following holds. Assume that

\begin{cond} \llabel{EW81.1} 
 $\Phi(x_{0},\ldots ,x_{k-1},Y_{0},\ldots ,Y_{l-1})\in \SForm(\calm) $, with 
$\length(\Phi)\le 
\alpha$, where $x_{0},\ldots ,x_{k-1}$, are first-order variables, and $Y_{0},\ldots 
,Y_{l-1} 
$ are second-order
variables, for $k$-ary relations, \end{cond}

\begin{cond} \llabel{EW81.2}
$m,r\in \omega$, $m>0$, and $\Psi_{0}(x_{0},\ldots ,x_{k-1}),\ldots ,
\Psi_{l-1}(x_{0},\ldots ,x_{k-1})\in \boldl(m, 
\gseq(r,\beta))$,\end{cond}

Then there exists a first-order prenex formula $\Theta(x_{0},...,x_{k-1})\in  
\boldl(\gamma m, 
\gseq(r+\gamma,\beta))$.
of $\calm$ such that

\begin{cond} \llabel{EW81.3}
$\Theta(x_{0},\ldots ,x_{k-1})$ is logically equivalent to 
the formula that we get from $\Phi$ by substituting  $\Psi_{i}$
for $Y_{i}$ for all $i\in l$, that is, $$\vdash \Theta \leftrightarrow \Phi(x_{0},\ldots 
,x_{k-1},\Psi_{0}(x_{0},\ldots ,x_{k-1}),
\ldots ,
\Psi_{l-1}(x_{0},\ldots ,x_{k-1}))$$ \end{cond}

\end{lemma}

Proof of Lemma \rref{EW81}.  Assume that 
$\Phi(x_{0},...,x_{k-1},Y_{0},...,Y_{l-1})\equiv Q_{0} y_{0},...,Q_{t-1}y_{t}, 
P(y_{0},...,y_{t-1},x_{0},...,x_{k-1},Y_{0},...,Y_{l-1}) $, where $Q_{0},...,Q_{t-1}$ 
are quantifiers,
$P$ is a propositional formula, and $t\le \alpha$. It is sufficient to show  
that 

\begin{cond} \llabel{EW81.4}
 there exists a prenex formula $\Theta'$ with 	$$\vdash \Theta' \leftrightarrow  
P(y_{0},...,y_{t-1},x_{0},\ldots 
,x_{k-1},\Psi_{0},
\ldots ,
\Psi_{l-1}) $$ such that $\Theta' \in \boldl(\gamma' m, 
\gseq(r+\gamma',\beta))$ for a suitably chosen $\gamma'\in \omega$ which depends 
only on
$\alpha$ and $\beta$. \end{cond}

We prove condition \rref{EW81.4} by induction on the depth  $d$ of the formula $P$.
We will denote by $\gamma_{d}'$ the integer $\gamma'$ which satisfies 
condition \rref{EW81.4} if the depth of $P$ is $d$. (Since $d\le \alpha$, the integer  
$\gamma_{d}$  remains below a bound depending only on $\alpha$ and $\beta$.) 
Assume that  our statement is true for formulas of depth at most $d-1$,  and for 
example,
$P(\vy, \vx, \vY)\equiv 
P_{0}(\vy, \vx, \vY) \wedge
P_{1}(\vy, \vx, \vY)$, 
 and $\Theta_{i}'$, is a  prenex form of  
$ P_{i}(\vy,\vx,\Psi_{0},
\ldots ,
\Psi_{l-1}) $ for $i=0,1$, where  $\Theta_{i} \in \boldl(\gamma_{d-1}' m, 
\gseq(r+\gamma_{d-1},\beta))$.
Lemma \rref{EAA15} implies that $ P_{i}(\vy,\vx,\Psi_{0},
\ldots ,
\Psi_{l-1}) $ has a prenex form $\Theta$ with $\Theta\in \boldl
(2\gamma_{d-1} m+c, 
\gseq(r+\gamma_{d-1}+4,\beta)) \subseteq \Theta\in \boldl(\gamma_{d} m, 
\gseq(r+\gamma_{d},\beta))  $, where $\gamma_{d}=2c\gamma_{d-1}+4$.
The recursive definition of $\gamma$ starting with $\gamma_{0}=1$ implies
that $\gamma_{d}\le 2^{c_{1}d}$ for a suitable chosen constant  $c_{1}\in \omega$.
Since $d$ remains below a bound depending only on $\alpha$,
condition \rref{EW81.4} is satisfied by
$\gamma'=\gamma_{d}$.  
\enp{(Lemma \rref{EW81})}

\begin{lemma} \llabel{EW20} Assume that $\calj$ 
is a 
function, and $\boldm$
is  $\calj$-predictive. For all sufficiently large 
$c_{3},c_{4}\in \omega$,  if
$d\in \omega$ is sufficiently large, $r,k\in \omega$, $d\le r \le \calj(d)$ 	and 
$\tau(x_{0},\ldots ,x_{k-1})$ is a 
term 
of $\calm$, and  $\delta=\depth(\tau)$,  then
there exists a first-order  formula $$\lambda(x_{0},\ldots ,x_{k-1},y,z) \in 
\boldl\Bigl(c_{4}^{\delta},\gseq(c_{3}\delta,c_{4})\Bigr)$$ 
with the 
property that for all 
$a_{0},\ldots ,a_{k-1},b\in
 \boldm_{d}$,  $$\boldm_{r}\models b=\tau(a_{0},\ldots ,a_{k-1})  \ \ \ 
\leftrightarrow
\ \ \ \boldm_{d}\models \lambda(a_{0},\ldots ,a_{k-1},b,r)$$
\end{lemma}

To make an inductive proof possible we prove the lemma in a slightly
stronger form stated in the following lemma. That lemma says that the terms
of $\calm$ has the property which was formulated only for the operations
of $\calm$ in the definition of predictivity. Moreover we also state an upper bound on 
the  quantifier patterns of the formulas involved in this property.

\begin{lemma} \llabel{EW21} Assume that $\calj$ 
is a 
function, and $\boldm$    is  $\calj$-predictive. Then for all sufficiently large 
 $c_{3},c_{4}\in \omega$ the following holds.
 Suppose that $d,r,k\in \omega$ with $d\le r \le \calj(d)$,
$\eta_{d,r}$ is the function whose existence is stated in the definition of 
$\calj$-predictivity, and 
$\tau(x_{0},\ldots ,x_{k-1})$ is a 
term 
of $\calm$ with $\depth(\tau)=\delta$. Then
there exists a  formula  $\Psi_{\tau}(x,y,z,Z_{0},\ldots ,Z_{k-1})\in \SForm(\calm)$,
where $x,y,z$ are free first-order variables and $Z_{0},\ldots ,Z_{k-1}$ are free 
variables
for binary relations, such that
$$\Psi_{\tau}(x,y,z,Z_{0},\ldots ,Z_{k-1})  \in 
\bar\boldl\Bigl(c_{4}^{\delta},\gseq(c_{3}\delta,c_{4})\Bigr) $$
and  the following condition is satisfied:

\begin{cond} \llabel{EW21.1} for all 
$a_{0},\ldots ,a_{k-1},
u,v\in  \boldm_{d}$,  the following two 
statements are equivalent:\\
(i) \ \ $(\eta_{d,r}
(b ) )
(u,v) $, where $b$ is the unique element of $\boldm_{r}$ 
with
$\boldm_{r}\models b=\tau(a_{0},\ldots 
,a_{k-1} )   $,  \vskip 5 pt \noindent
(ii) \ \ $\boldm_{d} \models 
\Psi_{\tau}(u,v,r, \eta_{d,r}(a_{0}),\ldots ,\eta_{d,r}(a_{k-1}))$. \end{cond}
\end{lemma}

Proof of Lemma \rref{EW21}.
We prove the lemma by induction on $\depth(\tau)$. If $\depth(\tau)=0$, then $\tau$
is either a constant symbol  $\bfc$ or a variable $x_{i}$ for some $i\in k$. In the 
former case the formula  $\Psi_{\tau}$ is identical to the formula
$\Phi_{\bfc}$ whose existence is guaranteed in the definition of $\calj$-predictivity.
If $\tau=x_{i}$ then $\Psi_{\tau} (x,y,z,Z_{0},\ldots ,Z_{k-1}) \equiv Z_{i}(x,y)$.

 Assume now that $i>0$ and the Lemma
is true if the depth of $\tau $ is at most $i-1$.  We may assume that all of the function 
symbols of $\calm $ are binary (e.g., a unary function symbol can be replaced by a 
binary which does not depend on its second variable).  Suppose that the term $\tau$
is of the form $\bff(\tau_{0}(x_{0},\ldots ,x_{k}),\tau_{1}(x_{0},\ldots ,x_{k}))$,
 where $\bff$ is a binary function symbol of $\calm$. Then
$\Psi_{\tau} (x,y,z,Z_{0},\ldots ,Z_{k-1})$ is defined in the following way. We will use 
the 
notation $\Phi_{\bff}$ from the definition of $\calj$-predictivity, if $\bff$ is a function 
symbol 
of $\calm$. For all $i=0,1$,	 the relation symbol $Y_{i}$ may occur in the formula 
$\Phi_{\bff}
(x,y,z,Y_{0},Y_{1})$ several times. Assume that the $j$th	 occurrence of the variable  
$Y_{i}$ 
is contained in a subformula of  the form $Y_{i}(\sigma_{j,0},
\sigma_{j,1})$, where $\sigma_{j,0},\sigma_{j,1}$ are terms of $\calm$. 
 We replace each subformula $Y_{i}(\sigma_{j,0},
\sigma_{j,1})$ of $\Phi_{\bff}
(x,y,z,Y_{0},Y_{1})$ by the formula $\Psi_{\tau_{i}}(\sigma_{j,0},
\sigma_{j,1},z,Z_{0},\ldots ,Z_{k-1}) $. The formula obtained this way will be 
$\Psi_{\tau} (x,y,z,Z_{0},\ldots ,Z_{k-1})$. The definition of the formula $\Phi_{\bff}$
and the inductive assumption together imply that the formula $\Psi_{\tau}$ satisfy
condition \rref{EW21.1}.  The property $\Psi_{\tau}(x,y,z,Z_{0},\ldots ,Z_{k-1})  \in 
\bar \boldl \Bigl(c_{4}^{\delta},\gseq(c_{3}\delta,c_{4})\Bigr) $
 follows from the 
inductive assumption and Lemma \rref{EW81}.
\enp{(Lemma \rref{EW21})}

Proof of Lemma \rref{EW20}. The  lemma is a consequence of Lemma 
\rref{EW21}. In the conclusion \rref{EW21.1} of Lemma \rref{EW21} we have the 
formula
$\Psi_{\tau}(u,v,r, \eta_{d,r}(a_{0}),\ldots ,\eta_{d,r}(a_{k-1}))$. Since $a_{i}\in 
\boldm_{d}$ for $i\in k$, the definition of $\calj$-predictivity implies that 
 $\eta_{d,r}(a_{i})(x,y)\equiv x=\bfnull\wedge y=a_{i}$.  Therefore we have a 
first-order formula $\psi_{\tau}$ of $\calm $ such that for all $a_{0}),\ldots 
,a_{k-1}\in 
\boldm_{d}$, 

\begin{cond} \llabel{EW21.2}
for all	$u,v\in  \boldm_{d}$, $$\biggl(\eta_{d,r}  \Bigl(\tau(a_{0},\ldots ,a_{k-1} ) 
\Bigr)
\biggr)(u,v) \ \
\leftrightarrow 
\ \ \boldm_{d} \models 
\psi_{\tau}(u,v,r, a_{0},\ldots ,a_{k-1})$$ \end{cond}

Therefore $\lambda(x_{0},\ldots ,x_{k-1},y,z)\equiv \forall u,v,  \psi_{\tau}
(u,v,z,x_{0},\ldots ,x_{k-1}) \leftrightarrow (u=0 \wedge v=y) $ meets the 
requirements of
Lemma \rref{EW20}.  \enp{(Lemma \rref{EW20})}

\begin{lemma} \llabel{AA20} For all sufficiently small $\epsilon>0$ if
$\cald$ is the function defined by 
$\cald(x)=\epsilon (\log x)^{\rcp{2}}$ for all $x>0$, then the $\cald$ quantifier 
elimination assumption does not hold for $\boldm$.
\end{lemma}

Proof of Lemma \rref{AA20}.  Let $\calj$ be the function $\lfloor x+ \log x \rfloor$.
According to Lemma \rref{EF50}, $\boldm$ is $\calj$-predictive. Therefore
Lemma \rref{EW20} is applicable for the function $\calj$. Let  $c,\alpha_{0} 
\in \omega$, such that $1\ll c \ll \alpha_{0} \ll \rcp{\epsilon}$. We may suppose that 
statement of Lemma \rref{EW20} holds with
a choice of $c_{3}$ and $c_{4}$ such that  $c_{3},c_{4}\ll c$. We may also 
assume that Lemma \rref{EC12}  holds with the present choice of  $c,\epsilon$ and 
$\alpha_{0}$.

Assume that contrary to the statement of the present lemma the $\cald$ quantifier
elimination assumption holds for $\boldm$. Then condition \rref{EC12.0} of Lemma
\rref{EC12} is satisfied.  We choose a $d\in \omega$ such that $\rcp{\epsilon} \ll d$,
that is, condition \rref{EC12.1} of Lemma
\rref{EC12} is also satisfied by the present choices of the parameters.
Let  $\delta=  \lfloor \cald(d+\log d)\rfloor $,  $m= \lfloor c\delta \rfloor
$,    $\iota_{m-i}=c^{i}$ if $i<  \lfloor \delta/2\rfloor$ and $\iota_{m-i}=
c^{ \lfloor \delta/2\rfloor }$ otherwise. 
Clearly these choices satisfy condition  \rref{EC12.2} of Lemma \rref{EC12}.

Since all of the assumptions of Lemma \rref{EC12} are valid for the present choices 
of the parameters,  its conclusion also holds. Let $\bfg$ be the function and
let $\tau(x,y)$ be the term whose existence is stated in Lemma \rref{EC12}.

We have that if $\phi \in \Form(c^{\delta},\iota_{m},...,\iota_{1})$ and $q=\rho 
_{m}$ then

\begin{cond} \llabel{EW22.9}
for all $a\in \boldm_{d}$,  $$\boldm_{d} \models \phi(a) \ \ \leftrightarrow \ \
\boldm_{q} \models \tau(a, \bfg(\phi))= \bfnull $$ 
\end{cond}

We apply now  Lemma \rref{EW20} with $d$, $c_{3}=c_{4} \ll c$,
$r\legyen q$, 	$k\legyen 2$, 
$\tau(x_{0},\ldots ,x_{k-1})\legyen \tau(x_{0},x_{1})$. Let $\lambda(x_{0},x_{1},y,z)$ be the 
first-order formula whose existence is guaranteed by Lemma \rref{EW20}. The 
conclusion of Lemma \rref{EW20} and  condition \rref{EW22.9}  imply that

\begin{cond} \llabel{EW23.1}
for all $a\in \boldm_{d}$,  $$\boldm_{d} \models \phi(a) \ \ \leftrightarrow 
\boldm_{d} \models \lambda(a, \bfg(\phi),\bfnull,q) $$ 
\end{cond}

Let $\sigma_{0}(x)$, $\sigma_{1}(x)$  be terms of $\calm$ such that for all
$h\in \boldm_{d} $ we have $$\boldm_{d} \models \max(\sigma_{0}(h), 
\sigma_{1}(h))< 2^{2^{d-1}} \wedge \sigma_{0}(h)+\sigma_{1}
(h)2^{2^{d-1}}=h$$
For example the terms  $\sigma_{0}(x)=\div(x,2^{2^{d-1}})$, $\sigma_{1}(x)=
x-\sigma_{0}(x)$ meet this requirement.

Let $\mu(x,y)\equiv \lambda(x, \sigma_{0}(y), \bfnull, \sigma_{1}(y))$. 
Recall that $\phi$ was an arbitrary element of the set 
$H=\Form(c^{\delta 
},
\iota_{m},...,\iota_{1})$. For each  and  $\phi \in
H$ let $\boldg(\phi)=\bfg(\phi)+q2^{2^{d-1}} $.  
According to the  definition of the function $\bfg$ in Lemma \rref{EC12} 
$\bfg(\phi)<2^{2^{d-1}}$. Since $q\le d+ \log d$ we have 
 $\boldg(\phi)\in \boldm_{d}$.
 Condition \rref{EW23.1}
and the definition of $\boldg$ imply that

\begin{cond} \llabel{EW23.2}
for all $\phi \in H$ and for all $a\in \boldm_{d}$,  $$\boldm_{d} \models \phi
(a) \ \ 
\leftrightarrow \ \
\boldm_{d} \models \mu(a, \boldg(\phi)) $$ 
\end{cond}

Let $\psi(x)$ be the formula of $\calm$ defined   by $\psi(x)\equiv \neg \mu(x,x)$.
We claim that the formula $\psi$
is in the set $H$.
 Indeed  according to Lemma \rref{EW20}
$\lambda(x_{0},x_{1},y,z) \in 
\boldl\Bigl(c_{4}^{\delta},\gseq(c_{3}\delta,c_{4})\Bigr)$
and therefore $\psi(x) \equiv \neg\mu(x,x) \in \boldl\Bigl(c_{4}^{\delta} +c_{5},
\gseq(c_{3}\delta,c_{4})\Bigr)  $ \, where $c_{5}\in \omega$ is an absolute constant.
 The upper bound
$c_{3}=c_{4} \ll c$,  and the definition of the integers $\iota_{m},...,\iota_{m}$,
implies that $c_{4}^{k}<\iota_{m-k+1}$ for $k=1,...,c_{3}\delta$. Therefore
Lemma \rref{AA22} implies that $\boldl\Bigl(c_{4}^{\delta} +c_{5},
\gseq(c_{3}\delta,c_{4})\Bigr) \subseteq 
\Form(c^{\delta 
},
\iota_{m},...,\iota_{1})
$. (Here we also used that $\csize(\kappa) \le \length(\kappa)$ if $\kappa$ is a term of
$\calm$.)

The fact $\psi\in H$ and condition \rref{EW23.2} leads to a contradiction using
G\"odel's diagonalization argument. Namely, we have by the definition of $\psi$
that $\boldm_{d} \models \psi(\boldg(\psi))\leftrightarrow \neg \mu(\boldg(\psi),
\boldg(\psi))$.
On the other hand condition \rref{EW23.2} with $\phi\legyen \psi$, and
$a\legyen \boldg(\psi)$ yields $\boldm_{d}\models \psi(\boldg(\psi)) \leftrightarrow
\mu(\boldg(\psi),\boldg(\psi)) $, that is we have $\boldm_{d}\models 
\mu(\boldg(\psi),\boldg(\psi)) \leftrightarrow
\neg \mu(\boldg(\psi),\boldg(\psi))  $ a contradiction.
\enp{(Lemma \rref{AA20})}

\vskip5 pt
Proof of Theorem \rref{TT3}. 
Assume that the statement of the theorem is not true. This implies that for all
 $\epsilon>0$, and for all  terms $F(x,y)$ of $\calm$ there exists a sequence of terms 
$G=\langle G_{d}(y) \mid y\in \omega
\rangle $ such that  $G$ decides whether there exists a solution for $F$ and the depth
of $G_{d}$ is smaller than $\epsilon (\log d)^{\rcp{2}}$ for all sufficiently large
$d\in \omega$. Since for each propositional formula $P(x,y)$ of $\calm$ there exists a
term $F(x,y)$ of calm such that for all $d\in \omega $, $\boldm_{d} \models \forall
x,y, P(x,y) \leftrightarrow F(x,y)=0$ we get that for all $\epsilon >0$ if $\cald(x)= 
\epsilon 
(\log d)^{\rcp{2}}$ then the $\cald$ quantifier elimination assumption holds for 
$\boldm$. This however contradicts to Lemma \rref{AA20}. 
\enp{(Theorem \rref{TT3})}

\eject

\section{\llabel{RAM} Random Access Machines}

A detailed description of the random access machines $N_{n}$ is given in \cite{Ajt1}.

Proofs of Theorems \rref{TT1} and \rref{TT2}.
Theorems \rref{TT1} and \rref{TT2}
are simple consequences of Theorem \rref{TT3}. We describe here  the proof 
Theorem \rref{TT1} and indicate  the only place where it has to be changed to get a 
proof  of  Theorem \rref{TT2}. In this description 
if $d\in \omega$ the symbol $n$ always  will denote the integer
$2^{d}$ even if we do not say it explicitly.
We assume that Theorem \rref{TT1} is not true and show that Theorem \rref{TT3}
cannot be true either.  

We will consider programs $R$ running on $N_{n}$ which get only $k$ integers
in $2^{n}$ as input, where $k=1$ or $k=2$. For the sake of simplicity we assume that 
these  integers are already given as the contents of memory cells $\bar c$ and 
(possibly)
$\bar c +1$  at time $0$ when the machine start working, where $\bar c$ is a 
constant.

Assume that $F(x,y)$ is an arbitrary term of $\calm$ and $\epsilon>0$. Using the 
assumption
that Theorem \rref{TT1} is not true, we construct a sequence of terms $G=\langle 
G_{d} \mid 
d\in \omega\rangle$, such that $G$ decides whether there exists a solution for $F$, 
and 
for all 
sufficiently large $d\in \omega $, $\depth(G_{d}) \le \epsilon (\log d)^{\frac{1}{2}}$.

The definitions of the $\calm $ operations in the structures $\boldm_{d}$ imply that 
there exists a $c>0$ and $c$-size binary test $P$, with time requirement $c$ on each 
machine 
$N_{n}$, such that for all 
$d\in \omega$, and $a,b\in 2^{n}$, $P_{n}(b,a)=0$ 
if $\boldm_{d} \models F(a,b)=0$, and $P_{n}(b,a)=1$ otherwise.
(The program $P$ computes the value of $F(a,b)$ and checks whether it is 
$0$.)
If Theorem \rref{TT1} is not true then there exists a $c'\in \omega$, and a $c'$-size 
unary test 
$Q$ such that for all sufficiently large $d\in \omega$, the time requirement of $Q$ on 
$N_{n}$ is 
at most
$\epsilon' (\log d)^{\frac{1}{2}}(\log d)^{-1}$, and for all sufficiently large 
$d\in \omega$,
$Q_{n}(a)=0$ iff $\exists x \in 2^{n}, P_{n}(x,a)=0$, where $\epsilon'>0$ is a
sufficiently small constant with respect to $\epsilon$.
We construct, for each sufficiently large $d\in \omega$,  an 
$\calm$-circuit 
$C_{d}$ such that at the input $a$, $C_{d}$
gives the same output as the program $Q$ on $N_{n}$, 
and the depth of $C_{d}$ is at most $c_{1}\epsilon (\log d)^{\frac{1}{2}}$, where 
$c_{1}$ is a constant which does not depend on $\epsilon$ or $\epsilon'$.
The existence of such an $\calm$ circuit $C_{d}$  implies a term $G_{d}$ with the 
same 
depth which meets our requirements. (Each $\calm $ circuit can be transformed into a 
functionally equivalent $\calm$ term without an increase in the depth, of course the 
size of the term may be much larger than the size of the circuit).
For the construction of $C_{d}$ first we replace $Q$ by another program $Q'$ which 
has the 
same input-output behavior as $Q$ and satisfies the following condition. There exists a 
$c_{2}\in\omega$ such that for all $n\in N$:\\
\indent (a) the size of $Q'$ is less than $c_{2}$, \\
\indent (b)
if $Q'$, while   running on
$N_{n}$, executes an instruction $I$ which involves the memory cell $i$ for some $i\ge c_{2}$, 
then instruction $I$ is either a write instruction or a read instruction. (A memory cell is 
involved an an instruction if either its 
content influences what happens when the  instruction is executed or its content may  
change 
when the
instruction is executed.) \\
\indent (c) the time requirements of $Q'$ on $N_{n}$ is larger that the time 
requirement of $Q$ 
on 
$N_{n} $ at most by a factor of $c_{2}$. 

It is easy to see, using only the definition of a RAM, that such a program $Q'$ exists.
(In the case of Theorem \rref{TT2}  the program  $Q$ is of length $l$, where
the integer $l$	may depend on $n$. In this case we  substitute first $Q$ by a program 
$Q_{0}$ of constant length which gets an input of length $l$ which is written in the 
memory at time $0$.)
We claim now that if $n=2^{d}$ and the time requirement of $Q'$ on $N_{n}$ is 
$t_{n}$ then 
we can
simulate $Q'$ running on $N_{n}$, by an $\calm$ circuit of $C_{d} $ of depth at most
$O(t_{n} \log t_{n})$, whose gates perform operations in $\boldm_{d}$.

The set of nodes of the $\calm$-circuit $C_{d}$  will be denoted 
by 
$\calq$.  For a given input $a$ of the circuit $C_{d}$, we may evaluate the circuit, and 
this
evaluation assigns a value $\chi(a,u)$ for each node $u$ of $\calq$, which is the value 
computed 
by the gate at node $u$ if the input is $a$.
For each  $s\in t_{n}$ and $i\in c_{2}$ the set $\calq$ will have an element 
$u_{s,i}$, and for each $\delta\in \lbrace 0,1,2\rbrace $ and $s\in t_{n}$ the set 
$\calq$
will have an element $v_{s,\delta}$.
 (The set $\calq $ will have other elements as well.) We 
will 
define the circuit in a way that if at input $a$ and at time $s\le t_{n}$, while $Q'$ is 
running on 
$N_{n}$, the content of cell
$i$ for some $i<c_{2}$ is $w$
then $\chi(a,u_{s,i})=w$.
 The nodes $v_{s,\delta}$ for $i>c_{2}$, $s\in t_{n}$ will be used in 
the following
way. If at time $s$ while $Q'$	is running on $N_{n}$, the machine  performs a write 
write instruction, and it writes the integer 
$x$ in cell
$j$ then  $\chi(a,v_{s,0})=x$, $\chi(a,v_{s,1})=j$, and $\chi(a,v_{s,2})=1$. If at time   
$s$ the machine does not perform a write instruction  then  
$\chi(a,v_{s,0}))= 
\chi(a,v_{s,1})=\chi(a,v_{s,2})=0$.

First we note that the existence of a circuit $C_{d}$ with these properties and the
required bound on its depth implies the theorem. Indeed since at the nodes
$u_{s,i}$ we have the contents of the first $c_{2}$ memory cells at each time,
we have the output of the program $Q'$ as well. 

We claim that for all $s\in t_{n}$ there exists an $\calm$ circuit $D_{s}$ of depth at 
most 
$O(t_{n} \log t_{n})$ such that given $\chi(a,u_{s,i})$, $\chi(a,v_{r,\delta})$, $i\in 
c_{2},r\in 
[0,s]$, $\delta \in \lbrace 0,1,2\rbrace $
as input the circuit
gives as output the values  $\chi(a,u_{s+1,i})$, $\chi(a,v_{s+1,\delta})$, $i\in c_{2}
$, $\delta \in \lbrace 0,1,2\rbrace $. Clearly the existence of such circuits $D_{s}$ 
imply the 
existence of the circuit $C_{d}$ with the required properties.

Assume that at time $s$ instruction $I$ is executed. We distinguish two cases 
according to whether $I$ is a read instruction or not. \\ 
\indent (i) if $I$ is not a read instruction
then it is easy to see that condition (b) implies that for each fixed
$i\in c_{2}$, $\delta\in \lbrace 0,1,2\rbrace $ the earlier specified values of
$\chi(a,u_{s+1,i})$ and $\chi(a,v_{s+1,\delta})$ can be computed
by a constant depth $\calm $ circuit  $B_{s,i,\delta}$  from  the input 
$\chi(a,u_{s,i})$.
$i\in c_{2}$, $\delta\in \lbrace 0,1,2\rbrace $.
\\
\indent (ii) if $I$ is a read instruction which reads the content of cell
$j$ then we need a circuit which determines which is the largest integer $r\le s$ 
such that $\chi(a,v_{r,1})=j$, $\chi(a,v_{s,2})=1$, and for this integer $r$,
$\chi(a,v_{r,1})$ will be the current content of cell $j$. (If there is no such $r$ then
it will be the content of cell $j$ at time $0$). Since the total number of nodes needed 
for this
is at most $O(t_{n})$, this can be done by a circuit of depth at most $O(\log t)$.
The circuits in case (i)
and case (ii)
can be combined into a single circuit which first checks whether $I$ is a read
instruction. According to condition (b)
this can be done in constant depth.

Since the number of possible values for $s$ is at most $t_{n}$ this construction
gives the required circuit with depth $O(t_{n} \log t_{n})$.
\enp{(Theorem \rref{TT1} and Theorem \rref{TT2})}

\eject
\section{\llabel{NP} $NP$-completeness, proof of Theorem \rref{TT5}}

Proof of Theorem \rref{TT5}. The theorem is a consequence
of Lemma \rref{Y12}. 
The following problem is  clearly $NP$-complete with a suitable choice of the finite 
automaton 
$\cala$.  The size of the problem is $m$.
 Let $u=\lfloor \log_{2}( m/3)\rfloor$, $d=u$.  Let $\calt$ be a restricted turing machine with $\aut(\calt)=\cala$, $\width(\calt)=1$, $\tplength(\calt)=2^{u}$.

Suppose that an
$b\in 2^{2^{d}-1}$ is given, decide whether there exists an $a\in 2^{2^{d}-1}$ such 
that the $a$ is a possible $u$-based input for the turing machine $\calt$  and  if restricted turing machine $\calt=\langle A,2^{t}\rangle$  starts to work
with $u$-based input $a$, then 
$b= \sum_{j=0}^{2^{d-u}-1}\cont_{T,j,0}2^{j2^{u}}$, where 
$T=2^{d-u}$. (That is,  the machine with 
the input described by $a$ reaches a state described by  $b$ at time $T$.)

Lemma \rref{Y12} implies that there exists an existential formula $\psi$ of $\calm$
such that for all $u\in \omega$, and for all $b\in 2^{2^{d}-1}$ the problem described 
above has a solution
iff $\boldm_{c' u}\models \psi(b,u)$, where  $c'\in \omega$ is
a sufficiently large constant.
The condition $\boldm_{c' u}\models \psi(b,u)$ is equivalent to 
$\boldm_{c' u} \models \exists x,  2^{x}=\bfn \wedge \psi(b, \div(x,c'))  $.
Therefore there exists an existential formula $\psi'$ of $\calm$
such that for all $u\in \omega$, and for all $b\in 2^{2^{d}-1}$ the problem described 
above has a solutioniff $\boldm_{c' u}\models \psi'(b)$.

 The formula $\psi'$ may have more than one
existential quantifiers. However Lemma \rref{J6} implies that
there exists  an existential formula $\phi$ of $\calm$ with a
single existential  quantifier  such that if $c=2c'$, then  $\boldm_{c' u} \psi'(b)$ is equivalent
to $\boldm_{c u} \models \phi(b)$.  Therefore for a suitable chosen term $\tau$ of
$\calm $ this can be written in the form of $\boldm_{c u} \models \exists x, \tau(x,b)$.
Therefore we have reduced  our the NP-complete problem about
turing machines to an instance  of the problem that we called ``the solution
of the equation $\tau(x,b)=0$ in $x$". Moreover, since we are looking for a solution
in $\boldm_{c u}$ the size of the problem is $ 2^{c u}\le 2 ^{c \log_{2} m}=m^{c}$.
\enp{(Theorem \rref{TT5})}

\eject
%%%%%%%%%%%%vvvvvvvvvvvvvvvvv


\begin{thebibliography}{99}


\bibitem{AHU} A.V.~Aho, J.E.~Hopcroft, J.D.~Ullman, {\em The Design and
Analysis of Computer Algorithms}, Addison-Wesley, 1974.

\bibitem{Ajt0} M.~Ajtai. {\em Detreminism versus Nondeterminism with Arithmetic 
Tests and Computation},
Proceedings of the 44th ACM Symposium on Theory of Computing, STOC 2012,
New York, NY, USA,  June 2012, pages 249-268  ACM,
2012.

\bibitem{Ajt} M.~Ajtai, {\em Determinism versus Nondeterminism for
Linear Time RAMs with Memory Restrictions}, Journal of Computer and
Systems Science, 65(1): 2-37, (2002)


\bibitem{Ajt2}
M.~Ajtai,
 {\em Oblivious {$RAM$}s without cryptographic assumptions},
 Electronic Colloquium on Computational Complexity (ECCC),
17:28, 2010.

\bibitem{Ajt1} M.~Ajtai. {\em Oblivious {$RAM$}s without
cryptographic assumptions},
Proceedings of the 42nd ACM Symposium on Theory of Computing, STOC 2010,
Cambridge, Massachusetts, USA, 5-8 June 2010, pages 181--190.  ACM,
2010.


\bibitem{Ajt3} M.~Ajtai. {\em Determinism versus Nondeterminism with Arithmetic
Tests and Computation},
Proceedings of the 44nd ACM Symposium on Theory of Computing, STOC 2012,
New York, NY, USA, May 2012, pages 249-268.  ACM, 1012.

\bibitem{AG} M.~Ajtai, Y.~Gurevich, {\em Monotone versus positive},
 Journal of the ACM (JACM), Vol. 34, Issue 4, Oct. 1987, pp. 1004-1015.

\bibitem{Artin}
E.~Artin. {\em Galois Theory}, Dover Publications, 1998.
(Reprinting of second revised edition of 1944, The University of Notre
Dame Press).

\bibitem{BCH}
P.~Beame, S.~A.~Cook, and H.~J.~Hoover,
\newblock Log Depth Circuits for Division and Related Problems,
\newblock SIAM Journal on Computing, 15(4):994-1003, November 1986.



\bibitem{BJS1} P.~Beame, T.~S.~Jayram, M.~Sacks, {\em Time-space
tradeoffs
for branching programs}, Journal of Computer and Systems Science,
63(4):542-572, December 2001.

\bibitem{BJS2} P.~Beame,  M.~Sacks, Xiadong Sun, E.~Vee, {\em Time-space
trade-off lower bounds for randomized computation of decision problems},
Journal of ACM, 50(2):154-195,
2003.


\bibitem{DPR}  M.~Davis, H.~Putnam and J.~Robinson, {\em The
decision problem
for exponential diophantine equations}, Ann. of Math.  (2) 74 (1961),
425-436. 


\bibitem{HPV} J.E.~Hopcroft, W.~Paul, L.~Valiant. {\it On Time versus Space.}
Journal of ACM, Vol. 24 Issue 2, April 1977, pp. 332-337.

%%%%%%%%%%%%%%%%%%%%%

\bibitem{Magid}  A.~Magid, {\em Differential Galois theory},
Notices of the
American Mathematical Society 46 (9): 1999.


\bibitem{Fortnow}  L.~Fortnow, {\em Time-space tradoffs for
satisfiability},
Journal of Computer and System Sciences, 60:337-353, 2000.


\bibitem{Matijasevic}
Ju.~V.~Matijasevic, {\em Enumerable sets are diophantine}, Dokl.  Akad.
Nauk SSSR 191 (1970), 279-282.  English transi.: Soviet Math.  Doklady
11 (1970), 354-358.


\bibitem{PPST} W.~Paul, N.~Pippenger, E.~Szemer\'edi, and W.~Trotter.
{\em On determinism versus nondeterminism and related problems},
In Proceedings of the 24th IEEE Symposium on Foundations of Computer
Science, pages 429-438.  IEEE, New York, 1983.


\bibitem{Takeuti} G.~Takeuti, \newblock{\em Proof Theory},
North-Holland, Studies in
Logic and the Foundations of Mathematics, Vol. 81, Second edition, 1987.

\bibitem{Yao} A.~Yao \newblock{\em Separating
the polynomial time hierarchy by oracles}, Proc.
26th Annu.  IEEE Symp.  Found.  Comp.  Sci. 1-10
(1985).



\end{thebibliography}
\end{document}

\begin{definition}  1. Assume that  $   a,b\in\omega $,       $   b\ge 2 $.     The natural
number  $   a $    can be written in a unique way in the form of
 $   \sum_{i=0}^{\infty}\alpha_{i}b^{i} $,      where  $   \alpha_{i}\in
\lbrace
0,1, \ldots ,b-1\rbrace  $    for  $   i\in \omega $.     The integer
 $   \alpha_{i} $    will
be denoted by  $   \coeff_{i}(a,b) $.     In other words
 $   \coeff_{i}(a,b) $    is
the  $   i $th ``digit"   
of  $   a $    in the numeral system with base  $   b $,  where the $i$th digit is defined
for all $i\in \omega$.

2. The set of functions symbols of $\calm$ (including the constant
symbols)  will be denoted by $\fsymb(\calm)$

3. Let $\calj$ be a function. We will say
that $\boldm$ is $\calj$-predictive if the following conditions are satisfied.

\begin{cond} \llabel{*EH50.1} The function $\calj$ is a monotone increasing function 
defined on $\omega$  and with
values in $\omega$. 
\end{cond}

\begin{cond} \llabel{*EH50.2} 
For all  sufficiently large $d \in \omega$, $\calj(d)\in \boldm_{d}$ and $\calj(d)>d $. 
\end{cond}

\begin{cond} \llabel{*EH50.3}
There exists a function  defined on $\fsymb(\calm)$ assigning to each function
symbol
$f(x_{0},\ldots ,x_{k-1})$ of $\calm$,
a  first-order formula $\Phi_{f}(x,y,z,Y_{0},\ldots ,Y_{k-1})$ of $\calm$,
where $x,y,z$ are  free first-order variables and $Y_{0},\ldots ,Y_{k-1}$ are
free variables for binary relations, such that  the following holds.
For all  $d,r\in \omega$ with $d+ r\le \calj({d})$
\xev{092f}
there exists a map $a \rightarrow\eta_{d,r}^{(a)}$ of
$\universe(\boldm_{d+r})$ into the set of binary relations on
$\universe(\boldm_{d})$ with the following properties:

(i) For each  $a,u,v\in \boldm_{d}$,  we have $\eta_{d,r}^{(a)}(u,v)$
iff ``$u=0$ and $\coeff_{v}(a,2)=1$".

(ii) Suppose that $f(x_{0},\ldots ,x_{k-1})$ is a $k$-ary function symbol
of
$\calm$,   for some $k=0,1,2$ (including the constant symbols for $k=0$),
$\bar f = (f)_{\boldm_{d+r}}$, and 
$a_{0},\ldots ,a_{k-1} \in \boldm_{d+r} $.  Then for all $
u,v \in \boldm_{d} $, \xev{092g}
$\eta_{d,r}^{(\bar f(a_{0},\ldots ,a_{k-1}))}(u,v)$ iff  \\ \centerline{$\boldm_{d} \models
\Phi_{f}(u, v, r,\eta_{d,r}^{(a_{0})},\ldots ,\eta_{d,r}^{(a_{k-1})}) $.}
\vege\end{cond}
\end{definition}

The proof of the simulation statement is based on the following lemma. 

\begin{lemma} \llabel{*EF50}
{\sl Assume that $c>0$ is a real, and  $\calj(x)= \lfloor 
x+c\log x \rfloor $. Then $\boldm$ is $\calj$ predictive.} \end{lemma}

In \cite{Ajt0} a weaker result of similar nature is proved which implies 
that there exists a function $g(x)$ with $\lim_{x\rightarrow \infty }g(x)=\infty$,
such that if $\calj_{0}=x +g(x)$ then $\boldm$ is $\calj_{0}$-predictive. 
Some of the partial results of the proof given there were stronger 
than 
what was needed for the theorem formulated in \cite{Ajt0}. We get Lemma 
\ref{*EF50} by 
using the full strength of these partial results in particular about the first-order 
definability of the bits of the results of multiplication and
division between large numbers.
Here we give only 
the outline of the proof together with those details that has to be changed for the 
present
purposes.
We define the function $\calj$ by
$\calj(x)= \lfloor 
x+c\log 
x \rfloor $.
Assume that $d\in \omega$ is
sufficiently large, $\chi\in \omega$, and $d+\chi\le \calj(d)$. First we define the map 
$\eta_{d,\chi}$ whose existence
is required by the definition of predictivity. 
Assume that $a\in \boldm_{d+\chi}$, $2^{d}=n$, $\nu=2^{\chi}$. Let
$a_{i}=\coeff_{i}(a,2^{n})$ for $i=0,1,\ldots ,\nu-1$. We define
$\eta_{d,\chi}$ by:
``for all $u,v\in \boldm_{d}$, $\eta_{d,\chi}^{(a)}(u,v)$ iff $u\in \nu$
and  $\coeff_{v}(a_{u},2)=1$. Equivalently $\eta_{d,\chi}^{(a)}(u,v)$ holds iff $u\in \nu$,
 $v\in n$ and  $\coeff_{nu+v}(a,2)=1$".

This definition implies that the relation $\eta_{d,\chi}^{(a)}$  satisfies condition \ref{*EH50.3}/(i)
from the definition of predictivity.
We define now the formula $\Phi_{f}(x,y,z,Y_{0},\ldots ,Y_{k-1})$
for each function symbol $f$ of $\calm$. (According to the definition of
$\calj$-predictivity the formula $\Phi_{f}$ cannot depend on the choices of $d$ or 
$\chi$.)
First we consider the case when $f=\bfc$ is a constant symbol. It is easy to see that the choices  
$\Phi_{\bfnull}\equiv \downarrow$, $\Phi_{\bfegy}\equiv x=0 \wedge y=0$, $\Phi_{-\bfegy}\equiv
x<2^{z},y<\bfn$
 meet our requirements. The formula $\Phi_{\bfn}$ has to say that $x=0$ and $2^{y}$ is  
$(\bfn)_{\boldm_{d+z}}$, but  the symbol $\bfn$ used  in this formula is interpreted 
$(\bfn)_{\boldm_{d}}$. Therefore
 $\Phi_{\bfn} \equiv x=0 \wedge  \exists w, 2^{w}=\bfn \wedge w+z=y$.

Since $a\cap b$ and $\caln (a)$ are defined by bitwise
operations on the binary forms of $a$ and $b$ , the first-order formulas  $\Phi_{\cap}$ and 
$\Phi_{\caln}$ can be
defined in the following way: $\Phi_{\cap}(x,y,z,Y_{0},Y_{1})\equiv Y_{0}(x,y) \wedge 
Y_{1}(x,y)  $, $\Phi_{\neg}(x,y,z,Y_{0})\equiv x<z \wedge y<n \wedge \neg Y_{0}(x,y)$. For the operation 
$\bfp(a)$  (with the meaning $\bfp(a)=\min \lbrace 2^{a}, 2^{n}-1\rbrace $) $\Phi_{\bfp}$ also
can be easily defined, so what remains are the binary operations, $\min,\max,+,\times $, and $\div$.

For binary operations condition (ii) of the definition of predictivity can be written as follows:

\begin{cond} \llabel{*EB7.1}
Suppose that $f$ is one of the function symbols,
$+,\times,\bfp,\div$, $\max, \min $ of $\calm$. Then there
exists
a formula $\Phi_{f}(x,y,z, Y_{0},Y_{1})\in \SForm(\calm)$, where $x,y,z$ are
first-order variables and $Y_{0},Y_{1}$ are variables for binary
relations such that  for all sufficiently large
$d\in \omega$, and for all $\chi\in [0, c\log d]$,
$a,b \in \boldm_{d+\chi} $, and for all $ u,v \in \boldm_{d} $, \xev{170a}
$\eta_{d,\chi}^{(f^{(d+\chi)}(a,b))}(u,v)$ holds iff $\boldm_{d}
\models \Phi_{f}'(u, v, \chi,\eta_{d,\chi}^{(a)},\eta_{d,\chi}^{(b)}) $,
where $f^{(d+\chi)} = (f)_{\boldm_{d+\chi}}$.
\end{cond}

The definition of $\eta_{d,\chi}$ implies that we get each relation $\eta_{d,\chi}^{(a)}$ by
extending a binary relation on $n$ to binary relation on $2^{n}=\universe(\boldm_{d})$, in the  
natural way. Therefore the elements of $\boldm_{d+\chi}$ are encoded by binary relations on
$n$. In the proof Lemma  \ref{*EF50} we will encode sequences from the elements of 
$\boldm_{d+\chi}$ by $k$-ary relations on $n$ where $k$ may be greater than $2$. The following 
definitions are needed for this encoding. 

\begin{definition} 1. For each positive integer $k$ and $u=\langle
u_{0},\ldots ,u_{k-1} \rangle \in (\boldm_{d})^{k}$, $u\wr_{n}$ will denote
the integer $u_{k-1}n^{k-1}+u_{k-2}n^{k-2}+\ldots +u_{1}n +u_{0} $.
2. Assume that $R$ is a $k$-ary relation
on the set $n=\lbrace 0,1,\ldots ,n-1\rbrace $, where $n=2^{d} $.
$\integer_{k}(R)$ will denote the integer
$\sum\lbrace 2^{u\wr_{n}} \mid u\in \boldm_{d}^{k} \wedge R(u)\rbrace
$. Clearly $R\rightarrow \integer_{k}(R)$ is a one-to-one map from the
set of all $k$-ary relation on $n$ to the set of all natural
numbers less then $2^{n^{k}}$. If $a\in [0, 2^{n^{k}}-1]$ is a natural
number then the unique $k$-ary relation $R$ on $n$ with
$\integer_{k}(R)=a$ will be denoted by $\integer_{k}^{-1}(a)$.
\vege\end{definition}

\begin{definition} 1. Suppose that $R$ is a $k$-ary relation on
$\boldm_{d}$. We will say that the relation $R$ is $n$-restricted
if for all $u=\langle u_{0},\ldots ,u_{k-1}\rangle \in \boldm_{d}^{k}$,
$ R(u_{0},\ldots ,u_{k-1})$ implies that for all
$i=0,1,\ldots ,k-1$ with
$u_{i}\in n$.
\vege\end{definition}

Using the function $\integer_{k}^{-1}$ we can represent natural
numbers from the interval $[0,2^{n^{k}}-1]$ by $k$-ary relations on $n$.
Our next goal is to represent sequences of natural numbers by relation
\xev{170b} on $n$, (where we have a bound both on the length of the
sequence and the sizes of its elements).
\begin{definition}
1. The set of all sequences of length $i$, whose
elements are from the set $A$ will be denoted by, $\seq(i,A)$. For
example the set of all sequences of length $n^{l}$ whose elements are
integers in the interval $[0,2^{n^{k}}-1]$ is $\seq(n^{l}, 2^{n^{k}})$.

2. Assume that
$a=\langle a_{0},\ldots ,a_{j-1}\rangle \in
\seq(n^{l},2^{n^{k}})$. We will represent this
sequence by a $k+l$-ary relation $
R^{(a)} $ on $n$ defined in the following way. For
all $i\le j-1$, and for all $u_{0},\ldots ,u_{k-1}, v_{0},\ldots ,v_{l-1}\in n$,
$R^{(a)}(
u_{0},\ldots ,u_{k-1},
v_{0},\ldots ,v_{l-1}
)$ iff
$(\integer_{k}^{(-1)}(a_{t}))(u_{0},\ldots ,u_{k-1})$, \xev{170c}
where $t=\sum_{i=0}^{l-1} v_{i}n^{i}$. Since in this representation the
length of the sequence cannot be arbitrarily chosen it must be
$n^{l}$, for some $l\in \omega$, we will call this representation a
representation of the sequence without its length.

3. The definition above provides representation only for sequences
with exactly $n^{l}$ elements for some natural number $l$. A sequence
$a=\langle a_{0},\ldots ,a_{j-1}\rangle $ where $j<n^{l}$, $a_{i}\in
[0,2^{n^{k}}-1]$ will be represented in the following way.
We attach the number $j$ as the first element
to the sequence $a$ and attach a sequence
of $0$s to its end, so that the total length of the
sequence $a'=\langle j,a_{0},\ldots ,a_{j-1},0,\ldots ,0\rangle $ obtained this
way is $n^{l}$.
The representation of the sequence $a$ together with its length will be
the same as the representation of the sequence $a'$ without its length,
as defined earlier.
In the following the
representation of a sequence will always mean a representation
of the sequence together with its length unless we explicitly state
otherwise.

4. Assume that $d$ is a positive integer and $n=2^{d}$. We will say that
the set $X$ is $\boldm_{d}$-representable \xev{170d}
if there exists natural numbers $k,l$ such that either $X=
\lbrace 0,1,\ldots ,2^{n^{k}}-1 \rbrace
$ or $X=\seq_{n}(n^{l},2^{n^{k}})$. If $X$ is
an $\boldm_{d}$ representable set and
$X=
\lbrace 0,1,\ldots ,2^{n^{k}}-1 \rbrace
$
then we define its weight by $\weight(X)=k$, if $X=\seq_{n}
(n^{l},2^{n^{k}})
$ then
we define its weight by $\weight(X)=k+l$. If $a\in X$, where $X$ is an
$\boldm_{d}$ representable set, then $\relation_{a,n}$ will denote the
$k$-ary or $k+l$-ary relation on $n$ representing the element $a$.
\vege\end{definition}

We will consider now families of functions $f^{(d)}$, $d\in \omega$ so
that for each $d\in \omega $, $f^{(d)}\in\func(X^{(d)},Y^{(d)}) $
where
both $X^{(d)}$ and $Y^{(d)}$ are
$\boldm_{d}$-representable sets with weight less than $w$ for a
constant $w$.
We are interested in the case when such a family of functions can be
defined by
a first-order formula in $\boldm_{d}$ without using any parameters. The
world ``strongly" that we will use in the definition below refers to the
mentioned  lack of parameters.
\begin{definition} 1. Assume that $w_{i}\in \omega$ for $i=0,1$ and for
all $d\in \omega $, $A_{i}^{(d)}$
are $\boldm_{d}$ representable sets of weight $w_{i}$ for $i=0,1$, and
$f^{(d)}\in
\func(A_{0}^{(d)},A_{1}^{(d)})$.
We
will say that the family of functions $f^{(d)}$ is a strongly first-order
definable family function or a s.f.d.-family in $\boldm$
if there exists a first-order formula $\Gamma(x_{0},\ldots ,x_{w_{1}-1},
Z)$, where $x_{i}$, $i=0,1,\ldots ,w_{1}-1$ are individual variables
and $Z$ is a variable for $k_{0}$-ary relations such that for all
sufficiently large $d\in \omega$ and \xev{170e} for all $a\in
A_{0}^{(d)}$, and $b\in A_{1}^{(d)}$
with $f(a)=b$, we have that for all $u_{0},\ldots ,u_{w_{1}-1}\in n$, $
\relation_{b,n}(u_{0},\ldots ,u_{w_{1-1}}) $ iff
$\boldm_{d}
\models \Gamma
(u_{0},\ldots ,u_{w_{1}-1},\relation_{a,n})$.
\vege\end{definition}

We prove now that condition
\ref{*EB7.1} is satisfied by
each function symbol  $\min, \max, +,\times, \div$.  We show now that the corresponding families of 
functions are
are strongly first-order definable
in $\boldm$.
For $f=\min$ and $f=\max$ the statement is trivial since $a\le b$ iff
$\integer_{2}^{-1}(a) \le \integer_{2}^{-1}(b) $ according to the
lexicographic ordering which clearly can be defined in $\boldm_{d}$ in a
first-order way.

The function symbol $f=``+"$. If two integers are given in binary form
each with $m$
bits then the bits of their sum can be defined by a simple well-known
constant depth circuit whose size is linear in $m$. This circuit is
defined in a uniform way \xev{169e} which makes it possible to
translate it into
a first-order formula interpreted in $\boldm_{d}$. For later use we
also consider now the case where we have to
add a sequence of integers. This question has been also studied for
circuits, and it is known that if we have at most $(\log m)^{c_{0}}$
integers with $m^{c_{1}} $ binary bits then their sum can be computed
by an unlimited fan-in boolean circuit with size $m^{c_{2}}$ and depth
$c_{3}$,
where $c_{2},c_{3}$ depend only on $c_{0}$ and $c_{1}$, see \cite{AHU}.
(We may use these results with $m=2^{n}$, since we are quantifying on a set of size $2^{n}$.)
The construction
of the circuit is uniform, in this case too, and can be translated into
a first-order formulas, that we need for our present purposes, over a
structure containing the arithmetic
operations. (We may use these results with $m=2^{n}$, since we are quantifying on a set of size 
$2^{n}$.)
 
\begin{definition} If $b$ is a finite sequence of integers then
$\bolds b$ will denote the sum of its elements.
\vege\end{definition}

The following Lemma is proved in \cite{Ajt0}

\begin{lemma} \llabel{*EAR1} Assume that $c_{0},c_{1}\in \omega$. Then
there exists a strongly first-order definable family of functions
$f^{(d)}$, $d\in \omega$, such that for all sufficiently large $d$ if
$n=2^{d}$, $j\le n^{c_{0}}$
and $a$ is sequence of length $j$, from elements of the set
$2^{n^{k}}$, that is, $a\in\seq(j,2^{n^{k}})$, then $\bolds a
=f^{(d)}(a)$.
\end{lemma}

We prove condition \ref{*EB7.1} for $f=\times$ in a more general form
then needed, namely we will consider products with more than two
factors. This will be useful in the proof of \ref{*EB7.1} for $f=\div$.

\begin{definition} \xev{171b} Assume that $a=\langle
a_{0},a_{1},\ldots ,a_{j-1}\rangle
$ is a sequence of integers. Then $\boldp a$ will denote the number
$\prod_{i=0}^{j-1} a_{i}$. \vege\end{definition}
\begin{definition} Assume that $\alpha(x),\beta(x)$ are functions
defined on $\omega$ with real values. We will say that
the pair
$\langle \alpha(x),\beta(x) \rangle$ is acceptable
if there exists a strongly first-order definable family of functions
$f^{(d)}$, $d\in \omega$, such that for all sufficiently large integers
$d\in \omega$, for all nonnegative integers
$j\le \alpha(d)$, and for all
$
a\in\seq(j,2^{\beta(d)})$, we have $\boldp a =f^{(d)}(a)$.
\vege\end{definition}

The following two lemmas are proved in \cite{Ajt0}. The second lemma
is a special case of the first one.

\begin{lemma} \llabel{*EAR2} For each fixed $c >0,\epsilon>0$
the pair
$\alpha(x)=x^{c}$, $\beta(x)=2^{x+x^{1-\epsilon}}$ is acceptable.
\end{lemma}

\begin{lemma} \llabel{*EAR6.4} For all $\epsilon >0$
there exists a
family of functions $f^{(d)}$, $d\in \omega$, such that, for all
sufficiently large $d\in \omega$ if $a=\langle a_{0},a_{1}\rangle \in
\seq(2, 2^{2^{d+d^{1-\epsilon}}}) $, \xev{171l} then
$a_{0}a_{1}=f^{(d)}(a)$. \end{lemma}

Using Lemma \ref{*EAR6.4} we can show that condition \ref{*EB7.1} is
satisfied by $f=\times$.
If $d$ is sufficiently large and $d+\chi\le \calj(d)\le d+ c \log d $ then 
$d+d^{1\over
2}>d+\chi$ and therefore Lemma \ref{*EAR6.4} implies that, multiplication in 
$\boldm_{d+\chi}$ can be defined in
$\boldm_{d}$ in the sense of \ref{*EB7.1}. This completes the proof of
\ref{*EB7.1} for $f=\times$.

Now we prove condition \ref{*EB7.1} for $f=\div$. 
Assume that $d$ is sufficiently large, $d+\chi \le \calj(d) \le 
d+c\log d$, $a,b\in \boldm_{d+\chi}$, and we want to define $\lfloor
a/b\rfloor $ in $\boldm_{d}$ in a first-order way.  We follow the same 
steps that have been used in \cite{Ajt0}, and earlier  in \cite{BCH} for different 
purposes.  

First we describe,  using general mathematical language, a way to compute $\lfloor a/b\rfloor
$ and then we show that this can be translated into the formula $\Phi_{f}$
required in \ref{*EB7.1}. We will use the notation $2^{d}=n$ and $2^{\chi}=\nu$.

(i) First we note that it is sufficient to find integers $t,l$ such
that ${1\over b} -t2^{l}<2^{-\nu n-1}$. The reason for this is that in the
possession of the integers $t,l$ we can compute $\alpha =a t2^{l}$ and
\xev{176b}
$|\alpha -\lfloor a/b\rfloor |<a 2^{-\nu n -1}< 2^{\nu n}2^{-\nu n -1}\le {1\over 
2}$ so we get $\lfloor a/b\rfloor
$ by rounding.

(ii) Let $k$ be an integer so that $1>2^{-k}b>1/2$. If there exists no
integer with this property then the problem is trivial, since we can get
the binary bits of $\lfloor a/b\rfloor $ form the bits of $a$ simply by
shift and the erasure of a block of consecutive bits. Let $u=2^{-k}b$.
Since $1>u>{1\over 2}$, we have $1<{1\over u} < 2$. We may write
${1\over u}$
in the form of $ v2^{-(n+2)}+R$, where $v\in [0,2^{n+2}]$ is an integer
and
$0\le R< 2^{-n-1}$. ($v$ will be determined by the first $n+1$ bits of
${1\over u}$, and $R$ is what remains from ${1\over u}$ after erasing
these bits.) Let $z=v2^{-(n+2)}$. The definition of $v$ implies that
$0\le z \le 2$.

(iii) We have $zb=1+Rz=1+r$, where $|r|<2^{-n+1}$. We consider the
series
${1\over zb}={1\over 1-(1-zb)}={1\over 1-(-r)}=1-r+r^{2}-r^{3}+\ldots $.
Let $w$ be the sum of the
first \xev{176c}
$4\nu$ terms of this geometric series. Clearly $w={1\over
zb}+R_{1}$, where $|R_{1}|<2^{-3 \nu n}$.
Consequently
${1\over b}=z{1\over zb}=z(w-R_{1})=zw+R_{2}$, where $|R_{2}|<2^{-2 \nu n}$.
Now we show that all of the quantities in this computation can be defined in
a first-order way in $\boldm_{d}$.

Stage (i). The definition of $t$ and $l$ will be described later.
However if we have $t$ and $l $ Lemma \ref{*EAR6.4} implies
that we may define the product $a t2^{l}$ in a first-order way in
$\boldm_{d}$. The rounding also can be done in a first-order way.

Stage (ii). The integer $v$ has only $n+2$ bits. In $\boldm_{d}$ we can
quantify $n$ bits with a single existential quantifier, therefore $v$
with the given property is first-order definable in $\boldm_{d}$.

Stage (iii).
Lemma \ref{*EAR6.4}
implies that the product $zb$ can be defined in $\boldm_{d}$, and so the number
$r$ can be defined as well.
 Each
needed terms of the geometric series can be defined in $\boldm_{d}$, we
define the $i$th term as a product with $i$ factors. 
Since $\nu=2^{\chi}\le 2^{c\log\log n}\le (\log n)^{c}$,
Lemma \ref{*EAR2}
implies that the bits of such a product can be defined in $\boldm_{d}$ and by
Lemma \ref{*EAR1} the bits of the sum of the first $4 \nu$ terms can be defined as
well. Therefore we defined $w$ and by Lemma \ref{*EAR6.4} we
can also define $zw$. This completes the proof of the fact that
condition \ref{*EB7.1} is satisfied by $f=\div$, and also the proof of
$\calj$-predictivity of $\boldm$. \enp{(Lemma \ref{*EF50})}

%%%%%%%%%%%%%%
%%%%%%%%%%%%%
%%%%%%%%%%%%
%%Simulation Statement continued

The  Lemma \ref{*EF50}
proves the ``Simulation" statement in the special case when the term $\tau$
is of depth $1$, that is, it is a single function symbol of $\calm$, e.g., $x+y$, $x\times y$,
$\lfloor x/y\rfloor$ etc. In this case the Predictivity Lemma implies that the formula $\lambda_{\tau}
(x,y)$
can be chosen independently of $d$, and so it is a first-order formula of constant size.

In the general case when the size of $\tau$
may depend on $d$
we construct the formula $\lambda_{\tau}(x,y)$ by recursion on $w$, where $w$ is the depth of the 
term $\tau$. By the 
definition of the set $\calt_{n}$ we have an upper bound on the size of $\tau$ and this provides an 
upper bound on its depth as well.  
It is very 
important that during this recursive construction we need to maintain an upper bound on the quantifier 
pattern on the formula $\lambda_{\tau}(x,y)$ in order to ensure that at the $\lambda_{\tau}\in 
\calh_{n}$.
Assume  for example that $\tau(x,y)=f(\tau_{0}(x,y),\tau_{1}(x,y))$ where $f$
is a binary operation of $\calm$ and $\tau_{0},\tau_{1}$ are terms of $\calm$ of depth less than $w$. 
By the inductive 
assumption we know that there are first-order formulas $\lambda_{f}$, $\lambda_{\tau_{0}}$,
$\lambda_{\tau_{1}}$ with the required properties. Using these formulas we may construct a 
formula
$\lambda_{\tau}$  meeting our requirements.  It is easy to see that this recursive construction implies
that if the depth of $\tau$ is at most $\delta$, then  $\lambda_{\tau}$ can be written in a form 
such that its quantifier pattern is   $\langle j_{0},...,j_{m}\rangle$, where $m\le c_{0}\delta$ and
$j_{i}\le c_{1}^{i}$ for suitably chosen constants $c_{0},c_{1}>1$.  This upper bound on the 
elements in the quantifier pattern sequence motivates the definitions that we have provided for
$\calh_{n}$ and $\calh_{n}'$. The upper bound also implies that $\lambda_{\tau}\in \calh_{n}'$ as 
required by the simulation statement.   
\eject

\begin{definition}  1. Assume that  $   a,b\in\omega $,       $   b\ge 2 $.     The natural
number  $   a $    can be written in a unique way in the form of
 $   \sum_{i=0}^{\infty}\alpha_{i}b^{i} $,      where  $   \alpha_{i}\in
\lbrace
0,1, \ldots ,b-1\rbrace  $    for  $   i\in \omega $.     The integer
 $   \alpha_{i} $    will
be denoted by  $   \coeff_{i}(a,b) $.     In other words
 $   \coeff_{i}(a,b) $    is
the  $   i $th ``digit"   
of  $   a $    in the numeral system with base  $   b $,  where the $i$th digit is defined
for all $i\in \omega$.

2. The set of functions symbols of $\calm$ (including the constant
symbols)  will be denoted by $\fsymb(\calm)$

3. Let $\calj$ be a function. We will say
that $\boldm$ is $\calj$-predictive if the following conditions are satisfied.

\begin{cond} \llabel{*EH50.1} The function $\calj$ is a monotone increasing function 
defined on $\omega$  and with
values in $\omega$. 
\end{cond}

\begin{cond} \llabel{*EH50.2} 
For all  sufficiently large $d \in \omega$, $\calj(d)\in \boldm_{d}$ and $\calj(d)>d $. 
\end{cond}

\begin{cond} \llabel{*EH50.3}
There exists a function  defined on $\fsymb(\calm)$ assigning to each function
symbol
$f(x_{0},\ldots ,x_{k-1})$ of $\calm$,
a  first-order formula $\Phi_{f}(x,y,z,Y_{0},\ldots ,Y_{k-1})$ of $\calm$,
where $x,y,z$ are  free first-order variables and $Y_{0},\ldots ,Y_{k-1}$ are
free variables for binary relations, such that  the following holds.
For all  $d,r\in \omega$ with $d+ r\le \calj({d})$
\xev{092f}
there exists a map $a \rightarrow\eta_{d,r}^{(a)}$ of
$\universe(\boldm_{d+r})$ into the set of binary relations on
$\universe(\boldm_{d})$ with the following properties:

(i) For each  $a,u,v\in \boldm_{d}$,  we have $\eta_{d,r}^{(a)}(u,v)$
iff ``$u=0$ and $\coeff_{v}(a,2)=1$".

(ii) Suppose that $f(x_{0},\ldots ,x_{k-1})$ is a $k$-ary function symbol
of
$\calm$,   for some $k=0,1,2$ (including the constant symbols for $k=0$),
$\bar f = (f)_{\boldm_{d+r}}$, and 
$a_{0},\ldots ,a_{k-1} \in \boldm_{d+r} $.  Then for all $
u,v \in \boldm_{d} $, \xev{092g}
$\eta_{d,r}^{(\bar f(a_{0},\ldots ,a_{k-1}))}(u,v)$ iff  \\ \centerline{$\boldm_{d} \models
\Phi_{f}(u, v, r,\eta_{d,r}^{(a_{0})},\ldots ,\eta_{d,r}^{(a_{k-1})}) $.}
\vege\end{cond}
\end{definition}

The proof of the simulation statement is based on the following lemma. 

\begin{lemma} \llabel{*EF50}
{\sl Assume that $c>0$ is a real, and  $\calj(x)= \lfloor 
x+c\log x \rfloor $. Then $\boldm$ is $\calj$ predictive.} \end{lemma}

In \cite{Ajt0} a weaker result of similar nature is proved which implies 
that there exists a function $g(x)$ with $\lim_{x\rightarrow \infty }g(x)=\infty$,
such that if $\calj_{0}=x +g(x)$ then $\boldm$ is $\calj_{0}$-predictive. 
Some of the partial results of the proof given there were stronger 
than 
what was needed for the theorem formulated in \cite{Ajt0}. We get Lemma 
\ref{*EF50} by 
using the full strength of these partial results in particular about the first-order 
definability of the bits of the results of multiplication and
division between large numbers.
Here we give only 
the outline of the proof together with those details that has to be changed for the 
present
purposes.
We define the function $\calj$ by
$\calj(x)= \lfloor 
x+c\log 
x \rfloor $.
Assume that $d\in \omega$ is
sufficiently large, $\chi\in \omega$, and $d+\chi\le \calj(d)$. First we define the map 
$\eta_{d,\chi}$ whose existence
is required by the definition of predictivity. 
Assume that $a\in \boldm_{d+\chi}$, $2^{d}=n$, $\nu=2^{\chi}$. Let
$a_{i}=\coeff_{i}(a,2^{n})$ for $i=0,1,\ldots ,\nu-1$. We define
$\eta_{d,\chi}$ by:
``for all $u,v\in \boldm_{d}$, $\eta_{d,\chi}^{(a)}(u,v)$ iff $u\in \nu$
and  $\coeff_{v}(a_{u},2)=1$. Equivalently $\eta_{d,\chi}^{(a)}(u,v)$ holds iff $u\in \nu$,
 $v\in n$ and  $\coeff_{nu+v}(a,2)=1$".

This definition implies that the relation $\eta_{d,\chi}^{(a)}$  satisfies condition \ref{*EH50.3}/(i)
from the definition of predictivity.
We define now the formula $\Phi_{f}(x,y,z,Y_{0},\ldots ,Y_{k-1})$
for each function symbol $f$ of $\calm$. (According to the definition of
$\calj$-predictivity the formula $\Phi_{f}$ cannot depend on the choices of $d$ or 
$\chi$.)
First we consider the case when $f=\bfc$ is a constant symbol. It is easy to see that the choices  
$\Phi_{\bfnull}\equiv \downarrow$, $\Phi_{\bfegy}\equiv x=0 \wedge y=0$, $\Phi_{-\bfegy}\equiv
x<2^{z},y<\bfn$
 meet our requirements. The formula $\Phi_{\bfn}$ has to say that $x=0$ and $2^{y}$ is  
$(\bfn)_{\boldm_{d+z}}$, but  the symbol $\bfn$ used  in this formula is interpreted 
$(\bfn)_{\boldm_{d}}$. Therefore
 $\Phi_{\bfn} \equiv x=0 \wedge  \exists w, 2^{w}=\bfn \wedge w+z=y$.

Since $a\cap b$ and $\caln (a)$ are defined by bitwise
operations on the binary forms of $a$ and $b$ , the first-order formulas  $\Phi_{\cap}$ and 
$\Phi_{\caln}$ can be
defined in the following way: $\Phi_{\cap}(x,y,z,Y_{0},Y_{1})\equiv Y_{0}(x,y) \wedge 
Y_{1}(x,y)  $, $\Phi_{\neg}(x,y,z,Y_{0})\equiv x<z \wedge y<n \wedge \neg Y_{0}(x,y)$. For the operation 
$\bfp(a)$  (with the meaning $\bfp(a)=\min \lbrace 2^{a}, 2^{n}-1\rbrace $) $\Phi_{\bfp}$ also
can be easily defined, so what remains are the binary operations, $\min,\max,+,\times $, and $\div$.

For binary operations condition (ii) of the definition of predictivity can be written as follows:

\begin{cond} \llabel{*EB7.1}
Suppose that $f$ is one of the function symbols,
$+,\times,\bfp,\div$, $\max, \min $ of $\calm$. Then there
exists
a formula $\Phi_{f}(x,y,z, Y_{0},Y_{1})\in \SForm(\calm)$, where $x,y,z$ are
first-order variables and $Y_{0},Y_{1}$ are variables for binary
relations such that  for all sufficiently large
$d\in \omega$, and for all $\chi\in [0, c\log d]$,
$a,b \in \boldm_{d+\chi} $, and for all $ u,v \in \boldm_{d} $, \xev{170a}
$\eta_{d,\chi}^{(f^{(d+\chi)}(a,b))}(u,v)$ holds iff $\boldm_{d}
\models \Phi_{f}'(u, v, \chi,\eta_{d,\chi}^{(a)},\eta_{d,\chi}^{(b)}) $,
where $f^{(d+\chi)} = (f)_{\boldm_{d+\chi}}$.
\end{cond}

The definition of $\eta_{d,\chi}$ implies that we get each relation $\eta_{d,\chi}^{(a)}$ by
extending a binary relation on $n$ to binary relation on $2^{n}=\universe(\boldm_{d})$, in the  
natural way. Therefore the elements of $\boldm_{d+\chi}$ are encoded by binary relations on
$n$. In the proof Lemma  \ref{*EF50} we will encode sequences from the elements of 
$\boldm_{d+\chi}$ by $k$-ary relations on $n$ where $k$ may be greater than $2$. The following 
definitions are needed for this encoding. 

\begin{definition} 1. For each positive integer $k$ and $u=\langle
u_{0},\ldots ,u_{k-1} \rangle \in (\boldm_{d})^{k}$, $u\wr_{n}$ will denote
the integer $u_{k-1}n^{k-1}+u_{k-2}n^{k-2}+\ldots +u_{1}n +u_{0} $.
2. Assume that $R$ is a $k$-ary relation
on the set $n=\lbrace 0,1,\ldots ,n-1\rbrace $, where $n=2^{d} $.
$\integer_{k}(R)$ will denote the integer
$\sum\lbrace 2^{u\wr_{n}} \mid u\in \boldm_{d}^{k} \wedge R(u)\rbrace
$. Clearly $R\rightarrow \integer_{k}(R)$ is a one-to-one map from the
set of all $k$-ary relation on $n$ to the set of all natural
numbers less then $2^{n^{k}}$. If $a\in [0, 2^{n^{k}}-1]$ is a natural
number then the unique $k$-ary relation $R$ on $n$ with
$\integer_{k}(R)=a$ will be denoted by $\integer_{k}^{-1}(a)$.
\vege\end{definition}

\begin{definition} 1. Suppose that $R$ is a $k$-ary relation on
$\boldm_{d}$. We will say that the relation $R$ is $n$-restricted
if for all $u=\langle u_{0},\ldots ,u_{k-1}\rangle \in \boldm_{d}^{k}$,
$ R(u_{0},\ldots ,u_{k-1})$ implies that for all
$i=0,1,\ldots ,k-1$ with
$u_{i}\in n$.
\vege\end{definition}

Using the function $\integer_{k}^{-1}$ we can represent natural
numbers from the interval $[0,2^{n^{k}}-1]$ by $k$-ary relations on $n$.
Our next goal is to represent sequences of natural numbers by relation
\xev{170b} on $n$, (where we have a bound both on the length of the
sequence and the sizes of its elements).
\begin{definition}
1. The set of all sequences of length $i$, whose
elements are from the set $A$ will be denoted by, $\seq(i,A)$. For
example the set of all sequences of length $n^{l}$ whose elements are
integers in the interval $[0,2^{n^{k}}-1]$ is $\seq(n^{l}, 2^{n^{k}})$.

2. Assume that
$a=\langle a_{0},\ldots ,a_{j-1}\rangle \in
\seq(n^{l},2^{n^{k}})$. We will represent this
sequence by a $k+l$-ary relation $
R^{(a)} $ on $n$ defined in the following way. For
all $i\le j-1$, and for all $u_{0},\ldots ,u_{k-1}, v_{0},\ldots ,v_{l-1}\in n$,
$R^{(a)}(
u_{0},\ldots ,u_{k-1},
v_{0},\ldots ,v_{l-1}
)$ iff
$(\integer_{k}^{(-1)}(a_{t}))(u_{0},\ldots ,u_{k-1})$, \xev{170c}
where $t=\sum_{i=0}^{l-1} v_{i}n^{i}$. Since in this representation the
length of the sequence cannot be arbitrarily chosen it must be
$n^{l}$, for some $l\in \omega$, we will call this representation a
representation of the sequence without its length.

3. The definition above provides representation only for sequences
with exactly $n^{l}$ elements for some natural number $l$. A sequence
$a=\langle a_{0},\ldots ,a_{j-1}\rangle $ where $j<n^{l}$, $a_{i}\in
[0,2^{n^{k}}-1]$ will be represented in the following way.
We attach the number $j$ as the first element
to the sequence $a$ and attach a sequence
of $0$s to its end, so that the total length of the
sequence $a'=\langle j,a_{0},\ldots ,a_{j-1},0,\ldots ,0\rangle $ obtained this
way is $n^{l}$.
The representation of the sequence $a$ together with its length will be
the same as the representation of the sequence $a'$ without its length,
as defined earlier.
In the following the
representation of a sequence will always mean a representation
of the sequence together with its length unless we explicitly state
otherwise.

4. Assume that $d$ is a positive integer and $n=2^{d}$. We will say that
the set $X$ is $\boldm_{d}$-representable \xev{170d}
if there exists natural numbers $k,l$ such that either $X=
\lbrace 0,1,\ldots ,2^{n^{k}}-1 \rbrace
$ or $X=\seq_{n}(n^{l},2^{n^{k}})$. If $X$ is
an $\boldm_{d}$ representable set and
$X=
\lbrace 0,1,\ldots ,2^{n^{k}}-1 \rbrace
$
then we define its weight by $\weight(X)=k$, if $X=\seq_{n}
(n^{l},2^{n^{k}})
$ then
we define its weight by $\weight(X)=k+l$. If $a\in X$, where $X$ is an
$\boldm_{d}$ representable set, then $\relation_{a,n}$ will denote the
$k$-ary or $k+l$-ary relation on $n$ representing the element $a$.
\vege\end{definition}

We will consider now families of functions $f^{(d)}$, $d\in \omega$ so
that for each $d\in \omega $, $f^{(d)}\in\func(X^{(d)},Y^{(d)}) $
where
both $X^{(d)}$ and $Y^{(d)}$ are
$\boldm_{d}$-representable sets with weight less than $w$ for a
constant $w$.
We are interested in the case when such a family of functions can be
defined by
a first-order formula in $\boldm_{d}$ without using any parameters. The
world ``strongly" that we will use in the definition below refers to the
mentioned  lack of parameters.
\begin{definition} 1. Assume that $w_{i}\in \omega$ for $i=0,1$ and for
all $d\in \omega $, $A_{i}^{(d)}$
are $\boldm_{d}$ representable sets of weight $w_{i}$ for $i=0,1$, and
$f^{(d)}\in
\func(A_{0}^{(d)},A_{1}^{(d)})$.
We
will say that the family of functions $f^{(d)}$ is a strongly first-order
definable family function or a s.f.d.-family in $\boldm$
if there exists a first-order formula $\Gamma(x_{0},\ldots ,x_{w_{1}-1},
Z)$, where $x_{i}$, $i=0,1,\ldots ,w_{1}-1$ are individual variables
and $Z$ is a variable for $k_{0}$-ary relations such that for all
sufficiently large $d\in \omega$ and \xev{170e} for all $a\in
A_{0}^{(d)}$, and $b\in A_{1}^{(d)}$
with $f(a)=b$, we have that for all $u_{0},\ldots ,u_{w_{1}-1}\in n$, $
\relation_{b,n}(u_{0},\ldots ,u_{w_{1-1}}) $ iff
$\boldm_{d}
\models \Gamma
(u_{0},\ldots ,u_{w_{1}-1},\relation_{a,n})$.
\vege\end{definition}

We prove now that condition
\ref{*EB7.1} is satisfied by
each function symbol  $\min, \max, +,\times, \div$.  We show now that the corresponding families of 
functions are
are strongly first-order definable
in $\boldm$.
For $f=\min$ and $f=\max$ the statement is trivial since $a\le b$ iff
$\integer_{2}^{-1}(a) \le \integer_{2}^{-1}(b) $ according to the
lexicographic ordering which clearly can be defined in $\boldm_{d}$ in a
first-order way.

The function symbol $f=``+"$. If two integers are given in binary form
each with $m$
bits then the bits of their sum can be defined by a simple well-known
constant depth circuit whose size is linear in $m$. This circuit is
defined in a uniform way \xev{169e} which makes it possible to
translate it into
a first-order formula interpreted in $\boldm_{d}$. For later use we
also consider now the case where we have to
add a sequence of integers. This question has been also studied for
circuits, and it is known that if we have at most $(\log m)^{c_{0}}$
integers with $m^{c_{1}} $ binary bits then their sum can be computed
by an unlimited fan-in boolean circuit with size $m^{c_{2}}$ and depth
$c_{3}$,
where $c_{2},c_{3}$ depend only on $c_{0}$ and $c_{1}$, see \cite{AHU}.
(We may use these results with $m=2^{n}$, since we are quantifying on a set of size $2^{n}$.)
The construction
of the circuit is uniform, in this case too, and can be translated into
a first-order formulas, that we need for our present purposes, over a
structure containing the arithmetic
operations. (We may use these results with $m=2^{n}$, since we are quantifying on a set of size 
$2^{n}$.)
 
\begin{definition} If $b$ is a finite sequence of integers then
$\bolds b$ will denote the sum of its elements.
\vege\end{definition}

The following Lemma is proved in \cite{Ajt0}

\begin{lemma} \llabel{*EAR1} Assume that $c_{0},c_{1}\in \omega$. Then
there exists a strongly first-order definable family of functions
$f^{(d)}$, $d\in \omega$, such that for all sufficiently large $d$ if
$n=2^{d}$, $j\le n^{c_{0}}$
and $a$ is sequence of length $j$, from elements of the set
$2^{n^{k}}$, that is, $a\in\seq(j,2^{n^{k}})$, then $\bolds a
=f^{(d)}(a)$.
\end{lemma}

We prove condition \ref{*EB7.1} for $f=\times$ in a more general form
then needed, namely we will consider products with more than two
factors. This will be useful in the proof of \ref{*EB7.1} for $f=\div$.

\begin{definition} \xev{171b} Assume that $a=\langle
a_{0},a_{1},\ldots ,a_{j-1}\rangle
$ is a sequence of integers. Then $\boldp a$ will denote the number
$\prod_{i=0}^{j-1} a_{i}$. \vege\end{definition}
\begin{definition} Assume that $\alpha(x),\beta(x)$ are functions
defined on $\omega$ with real values. We will say that
the pair
$\langle \alpha(x),\beta(x) \rangle$ is acceptable
if there exists a strongly first-order definable family of functions
$f^{(d)}$, $d\in \omega$, such that for all sufficiently large integers
$d\in \omega$, for all nonnegative integers
$j\le \alpha(d)$, and for all
$
a\in\seq(j,2^{\beta(d)})$, we have $\boldp a =f^{(d)}(a)$.
\vege\end{definition}

The following two lemmas are proved in \cite{Ajt0}. The second lemma
is a special case of the first one.

\begin{lemma} \llabel{*EAR2} For each fixed $c >0,\epsilon>0$
the pair
$\alpha(x)=x^{c}$, $\beta(x)=2^{x+x^{1-\epsilon}}$ is acceptable.
\end{lemma}

\begin{lemma} \llabel{*EAR6.4} For all $\epsilon >0$
there exists a
family of functions $f^{(d)}$, $d\in \omega$, such that, for all
sufficiently large $d\in \omega$ if $a=\langle a_{0},a_{1}\rangle \in
\seq(2, 2^{2^{d+d^{1-\epsilon}}}) $, \xev{171l} then
$a_{0}a_{1}=f^{(d)}(a)$. \end{lemma}

Using Lemma \ref{*EAR6.4} we can show that condition \ref{*EB7.1} is
satisfied by $f=\times$.
If $d$ is sufficiently large and $d+\chi\le \calj(d)\le d+ c \log d $ then 
$d+d^{1\over
2}>d+\chi$ and therefore Lemma \ref{*EAR6.4} implies that, multiplication in 
$\boldm_{d+\chi}$ can be defined in
$\boldm_{d}$ in the sense of \ref{*EB7.1}. This completes the proof of
\ref{*EB7.1} for $f=\times$.

Now we prove condition \ref{*EB7.1} for $f=\div$. 
Assume that $d$ is sufficiently large, $d+\chi \le \calj(d) \le 
d+c\log d$, $a,b\in \boldm_{d+\chi}$, and we want to define $\lfloor
a/b\rfloor $ in $\boldm_{d}$ in a first-order way.  We follow the same 
steps that have been used in \cite{Ajt0}, and earlier  in \cite{BCH} for different 
purposes.  

First we describe,  using general mathematical language, a way to compute $\lfloor a/b\rfloor
$ and then we show that this can be translated into the formula $\Phi_{f}$
required in \ref{*EB7.1}. We will use the notation $2^{d}=n$ and $2^{\chi}=\nu$.

(i) First we note that it is sufficient to find integers $t,l$ such
that ${1\over b} -t2^{l}<2^{-\nu n-1}$. The reason for this is that in the
possession of the integers $t,l$ we can compute $\alpha =a t2^{l}$ and
\xev{176b}
$|\alpha -\lfloor a/b\rfloor |<a 2^{-\nu n -1}< 2^{\nu n}2^{-\nu n -1}\le {1\over 
2}$ so we get $\lfloor a/b\rfloor
$ by rounding.

(ii) Let $k$ be an integer so that $1>2^{-k}b>1/2$. If there exists no
integer with this property then the problem is trivial, since we can get
the binary bits of $\lfloor a/b\rfloor $ form the bits of $a$ simply by
shift and the erasure of a block of consecutive bits. Let $u=2^{-k}b$.
Since $1>u>{1\over 2}$, we have $1<{1\over u} < 2$. We may write
${1\over u}$
in the form of $ v2^{-(n+2)}+R$, where $v\in [0,2^{n+2}]$ is an integer
and
$0\le R< 2^{-n-1}$. ($v$ will be determined by the first $n+1$ bits of
${1\over u}$, and $R$ is what remains from ${1\over u}$ after erasing
these bits.) Let $z=v2^{-(n+2)}$. The definition of $v$ implies that
$0\le z \le 2$.

(iii) We have $zb=1+Rz=1+r$, where $|r|<2^{-n+1}$. We consider the
series
${1\over zb}={1\over 1-(1-zb)}={1\over 1-(-r)}=1-r+r^{2}-r^{3}+\ldots $.
Let $w$ be the sum of the
first \xev{176c}
$4\nu$ terms of this geometric series. Clearly $w={1\over
zb}+R_{1}$, where $|R_{1}|<2^{-3 \nu n}$.
Consequently
${1\over b}=z{1\over zb}=z(w-R_{1})=zw+R_{2}$, where $|R_{2}|<2^{-2 \nu n}$.
Now we show that all of the quantities in this computation can be defined in
a first-order way in $\boldm_{d}$.

Stage (i). The definition of $t$ and $l$ will be described later.
However if we have $t$ and $l $ Lemma \ref{*EAR6.4} implies
that we may define the product $a t2^{l}$ in a first-order way in
$\boldm_{d}$. The rounding also can be done in a first-order way.

Stage (ii). The integer $v$ has only $n+2$ bits. In $\boldm_{d}$ we can
quantify $n$ bits with a single existential quantifier, therefore $v$
with the given property is first-order definable in $\boldm_{d}$.

Stage (iii).
Lemma \ref{*EAR6.4}
implies that the product $zb$ can be defined in $\boldm_{d}$, and so the number
$r$ can be defined as well.
 Each
needed terms of the geometric series can be defined in $\boldm_{d}$, we
define the $i$th term as a product with $i$ factors. 
Since $\nu=2^{\chi}\le 2^{c\log\log n}\le (\log n)^{c}$,
Lemma \ref{*EAR2}
implies that the bits of such a product can be defined in $\boldm_{d}$ and by
Lemma \ref{*EAR1} the bits of the sum of the first $4 \nu$ terms can be defined as
well. Therefore we defined $w$ and by Lemma \ref{*EAR6.4} we
can also define $zw$. This completes the proof of the fact that
condition \ref{*EB7.1} is satisfied by $f=\div$, and also the proof of
$\calj$-predictivity of $\boldm$. \enp{(Lemma \ref{*EF50})}

%%%%%%%%%%%%%%
%%%%%%%%%%%%%
%%%%%%%%%%%%
%%Simulation Statement continued

The  Lemma \ref{*EF50}
proves the ``Simulation" statement in the special case when the term $\tau$
is of depth $1$, that is, it is a single function symbol of $\calm$, e.g., $x+y$, $x\times y$,
$\lfloor x/y\rfloor$ etc. In this case the Predictivity Lemma implies that the formula $\lambda_{\tau}
(x,y)$
can be chosen independently of $d$, and so it is a first-order formula of constant size.

In the general case when the size of $\tau$
may depend on $d$
we construct the formula $\lambda_{\tau}(x,y)$ by recursion on $w$, where $w$ is the depth of the 
term $\tau$. By the 
definition of the set $\calt_{n}$ we have an upper bound on the size of $\tau$ and this provides an 
upper bound on its depth as well.  
It is very 
important that during this recursive construction we need to maintain an upper bound on the quantifier 
pattern on the formula $\lambda_{\tau}(x,y)$ in order to ensure that at the $\lambda_{\tau}\in 
\calh_{n}$.
Assume  for example that $\tau(x,y)=f(\tau_{0}(x,y),\tau_{1}(x,y))$ where $f$
is a binary operation of $\calm$ and $\tau_{0},\tau_{1}$ are terms of $\calm$ of depth less than $w$. 
By the inductive 
assumption we know that there are first-order formulas $\lambda_{f}$, $\lambda_{\tau_{0}}$,
$\lambda_{\tau_{1}}$ with the required properties. Using these formulas we may construct a 
formula
$\lambda_{\tau}$  meeting our requirements.  It is easy to see that this recursive construction implies
that if the depth of $\tau$ is at most $\delta$, then  $\lambda_{\tau}$ can be written in a form 
such that its quantifier pattern is   $\langle j_{0},...,j_{m}\rangle$, where $m\le c_{0}\delta$ and
$j_{i}\le c_{1}^{i}$ for suitably chosen constants $c_{0},c_{1}>1$.  This upper bound on the 
elements in the quantifier pattern sequence motivates the definitions that we have provided for
$\calh_{n}$ and $\calh_{n}'$. The upper bound also implies that $\lambda_{\tau}\in \calh_{n}'$ as 
required by the simulation statement.   
\eject

%%%%%\cikk\fsize\fsi
